\documentclass[11pt]{article}

\usepackage[top=1in,bottom=1in,left=1in,right=1in]{geometry}
\usepackage{amsmath,amssymb,amsthm}
\usepackage{graphicx}
\usepackage{accsupp}
\newcommand{\FigureOneAlt}{Diagram comparing the naive indicator rule and the proposed smoothed decision function as functions of the estimated treatment contrast. The naive rule jumps from zero to one when the contrast crosses zero. The smoothed rule increases linearly from zero to one over an interval of width h sub n, extending from minus t sub zero times h sub n to (one minus t sub zero) times h sub n, and is constant outside this interval. The parameter t sub zero determines the asymmetry of the interval around zero.}
\newcommand{\FigureTwoAlt}{Two-panel diagram showing the proposed adaptive smoothing decision function for two estimated propensity scores. The horizontal axis is the estimated treatment contrast and the vertical axis is the decision weight. In both panels, the naive rule jumps from zero to one at zero. The smoothed rule increases linearly from zero to one over an interval of width h sub n, from minus the estimated propensity times h sub n to (one minus the estimated propensity) times h sub n. Its value at zero equals the estimated propensity. Panel a shows a lower propensity, whereas panel b shows a higher propensity, which shifts the smoothing interval to the left.}
\newcommand{\FigureThreeAlt}{A dashed vertical line separates regularity from nonregularity. Efficient regular estimators form a red region within both the regular-estimator region and the robust asymptotically linear unbiased-estimator region. Only the latter region extends into the nonregularity side.}

\graphicspath{{./}}
\usepackage{array,multirow,makecell}
\usepackage{float}
\usepackage{algorithm}
\usepackage{algpseudocode}
\usepackage[authoryear,round]{natbib}
\usepackage{setspace}
\usepackage{xurl}
\usepackage[colorlinks,allcolors=blue]{hyperref}

\allowdisplaybreaks
\newtheoremstyle{referenceplain}
  {\topsep}{\topsep}{\itshape}{}{\bfseries}{}{0.5em}{}
\newtheoremstyle{referencedefinition}
  {\topsep}{\topsep}{\normalfont}{}{\bfseries}{}{0.5em}{}
\newtheoremstyle{referenceremark}
  {\topsep}{\topsep}{\normalfont}{}{\itshape}{}{0.5em}{}

\theoremstyle{referenceplain}
\newtheorem{theorem}{Theorem}

\newtheorem{Corollary}{Corollary}
\newtheorem{Proposition}{Proposition}
\newtheorem{Assumption}{Assumption}

\newtheorem{proposition}{Proposition}[section]
\newtheorem{lemma}{Lemma}[section]
\newtheorem{corollary}{Corollary}[section]
\newtheorem{assumption}{Assumption}
\theoremstyle{referencedefinition}
\newtheorem{definition}{Definition}
\newtheorem{example}{Example}

\theoremstyle{referenceremark}

\newcommand{\tauh}{\widehat{\tau}}
\newcommand{\cI}{\mathcal{I}}
\newcommand{\bX}{\mathbf{X}}

\title{Optimal Inference of the Mean Outcome under Optimal Treatment Regime}
\author{
Shuoxun Xu\thanks{Department of Mathematics, The Hong Kong University of Science and Technology, Clear Water Bay, Kowloon, Hong Kong, China; Division of Biostatistics, School of Public Health, University of California, Berkeley, California 94720-7360, USA.}
\and
Xinzhou Guo\thanks{Department of Mathematics, The Hong Kong University of Science and Technology, Clear Water Bay, Kowloon, Hong Kong, China. Corresponding author: \href{mailto:xinzhoug@ust.hk}{xinzhoug@ust.hk}.}
}
\date{September 15, 2026}

\begin{document}

\maketitle

\begin{abstract}
When an optimal treatment regime (OTR) is considered, we need to evaluate the OTR in a valid and efficient way. The classical inference applied to the mean outcome under OTR, assuming the OTR is the same as the estimated OTR, might be biased when the regularity assumption that OTR is unique is violated. Although several methods have been proposed to allow nonregularity in such inference, its optimality is unclear due to challenges in deriving semiparametric efficiency bounds under potential nonregularity. In this paper, we address the bias issue via adaptive smoothing over the estimated OTR and develop a valid inference procedure on the mean outcome under OTR regardless of whether regularity is satisfied. We establish the optimality of the proposed method by deriving a lower bound of the asymptotic variance for the robust asymptotically linear unbiased estimator to the mean outcome under OTR and showing that our proposed estimator achieves the variance lower bound. The considered estimator class is general and the derived variance lower bound paves a novel way to establish efficiency optimality theories for OTR in a more general scenario allowing nonregularity. The merit of the proposed method is demonstrated by re-analyzing the ACTG 175 trial.

\end{abstract}

\doublespacing

\section{Introduction}
The optimal treatment regime (OTR) refers to a personalized treatment assignment strategy designed to maximize the average clinical benefit across a population \citep{li2023optimal}. In the era of precision medicine, the study of OTR plays an increasingly important role as targeted therapies often exhibit treatment effect heterogeneity and the ``one-size-fits-all" strategy might be suboptimal; see the discussion in \citet{malone2014good}. One condition where OTR has been intensively studied is the human immunodeficiency virus (HIV) infection \citep{farahmandpour2023personalized}. For example, \citet{fan2017concordance} found that ZDV+ddl is more favorable than ZDV+zal for HIV-infected patients aged over 37.5 and vice versa, and age is a crucial factor in the clinical practice of prescribing drugs for HIV-infected patients \citep{cardoso2013aging}. OTR has also been widely studied in  other conditions, such as breast cancer \citep{sparano2018adjuvant}, type II diabetes \citep{gaede2008effect} and hypertension \citep{sprint2015randomized}.  A global market of 614.22 Billion USD is projected for precision medicine in 2024 \citep{precedence2024}.

Given the practical importance of OTR, various methodologies have been developed to estimate the OTR over the past decades, which typically fall into two categories, direct and indirect approaches \citep{chakraborty2014inference}. The direct approach first specifies a class of treatment regimes (TR) and estimates the corresponding value function, and then the OTR is selected as the TR maximizing the value function \citep{orellana2010dynamic1, orellana2010dynamic2}. The indirect approach first models  the stage-specific conditional mean outcomes or contrasts, and then the OTR is found
via maximization of these estimated conditional means or contrasts \citep{moodie2012q, zhao2012estimating}.  By 2026, there have been 1,540 publications about ``optimal treatment regime" in google scholar \citep{OTRGoogleScholar2026}.

Besides the estimation of OTR, how to conduct valid and efficient statistical inference on the mean outcome under OTR is a critical question in drug development and regulatory decision-making \citep{chakraborty2014dynamic}.   This is because the mean outcome under OTR measures how good an OTR is and helps decide whether the OTR should be further pursued and implemented \citep{wang2012evaluation}.  For example, \citet{lavori2004dynamic}  evaluated and compared the mean outcomes of two treatment regimes for schizophrenic prodrome, and the better one is later suggested in clinical practices. To infer the mean outcome under OTR, \citet{zhang2012robust} and \citet{zhang2013robust} proposed plug-in methods, and \citet{van2014targeted} and \citet{luedtke2016statistical} established semiparametric optimality. However, the validity and the optimality of existing methods typically rely on the regularity assumption that the OTR is unique.

In practice, the regularity assumption might be violated much more frequently than we expect. In particular, nonregularity arises when the treatments have no individualized clinical differences in a non-negligible subgroup. For example,  \citet{sax2009abacavir} found that among a subgroup of patients with screening HIV-1 RNA levels equal to or above 100,000 copies per milliliter, the compared two treatments, Abacavir-Lamivudine and Tenofovir-Emtricitabine, have no clinical differences w.r.t. the change from the baseline CD4 cell count at week 48. Such nonregularity can break the validity and optimality of the classical approach for inference on the mean outcome under OTR. Specifically, under nonregularity, the classical approach, which simply plugs in the estimated OTR by assuming it is the same as the OTR, often suffers from bias issues because the estimated OTR is inconsistent when the OTR is not uniquely defined as discussed in  \citet{shi2020breaking}. Moreover, nonregularity hinders the establishment of efficiency optimality of the estimators because the classical semiparametric efficiency bound is established for regular estimators which cannot be appropriately defined or naturally extended due to the lack of path-wise differentiability under nonregularity \citep{van2011targeted,luedtke2016statistical}.

Some attempts have been made to address the bias issue and improve efficiency in inference on the mean outcome under OTR when nonregularity is allowed.  \citet{wager2018estimation} and \citet{wang2018learning} considered simple sample split methods, which clearly suffer from efficiency loss. \citet{chakraborty2010inference,chakraborty2014inference} proposed to use $m$-out-of-$n$ bootstrap to correctly approximate the distribution of the plug-in estimator, but the length of the proposed confidence interval (CI) shrinks at the rate of $m^{-1/2}$. \citet{luedtke2016statistical} proposed a one-step online estimator to achieve $\sqrt{n}$-consistency but due to the sequential way of constructing the estimator, this approach is sensitive to how the data are processed and the length of the proposed CI is longer than that of the oracle estimator which assumes we know the true OTR. Based on  the idea of subbagging and refitted cross-validation, a current state-of-the-art approach was proposed by \citet{shi2020breaking} which can achieve better efficiency than that of the oracle estimator via intensive computation. Although existing works keep improving the efficiency of inference on the mean outcome under OTR, the optimality theory is still lacking under potential nonregularity. It remains unknown where the pursuit of efficiency improvement ends and how efficiency-optimal inference on the mean outcome under OTR can be achieved when nonregularity is allowed.

To bridge the gap, we propose an adaptive smoothing approach to infer the mean outcome under OTR. We show that with appropriate smoothing to the estimated decision function of OTR, we can address the bias issue induced by potential nonregularity without intensive computation and deliver valid and efficient inference on the mean outcome under OTR regardless of whether it is regular or not. To establish the efficiency optimality of the proposed method, we define a class of robust asymptotically linear unbiased estimators, derive its lower asymptotic variance bound and show that the proposed method achieves the bound under both homoscedasticity and heteroscedasticity. The considered class is free of path-wise differentiability conditions and includes the efficient regular, the current state-of-the-art and many other valid estimators under both regularity and nonregularity, and the derived variance lower bound provides the foundation for efficiency optimality theory for OTR in a more general scenario allowing nonregularity where classical semiparametric theories fail.

Smoothing approaches have been considered in several related scenarios. \citet{levis2025covariate} employs a global smoothing strategy via the log-sum-exp approximation for inference on covariate-adjusted Balke-Pearl bounds, but directly applying their global smoothing to inference on the mean outcome of OTR would introduce approximation error everywhere and might not fully resolve bias. \citet{chen2023inference} proposes a softmax-based smoothing to deliver valid inference on the optimal mean outcome when nonregularity is allowed, but relies on parametric structural models and does not establish efficiency optimality. \citet{whitehouse2025inference} extends \citet{chen2023inference} to a model-free setting and to more general maximum-type nonregular values than the optimal mean outcome, yet efficiency optimality remains unresolved. \citet{wu2021resampling} applies kernel smoothing to infer the index parameter indexing the optimal regime within a parametric class; however, their smoothing mainly targets the decision rule rather than the value inference and is restricted to the regular case. Moreover, a common limitation across these approaches is that they typically use a single tuning parameter controlling smoothness and assign symmetric weights at the decision boundary, which might ensure validity but might not help them or their simple variants achieve efficiency optimality. By contrast, our proposed method introduces an additional degree of freedom that adapts the boundary weight to optimize efficiency without impacting validity, while a separate smoothing parameter guarantees inference validity. This two-parameter smoothing scheme substantially distinguishes our approach from existing smoothing methods and is precisely what enables us to establish and achieve efficiency optimality under nonregularity, a long-standing open problem in this area noted by \citet{van2011targeted} and \citet{whitehouse2025inference}.

In summary, the contributions of the paper lie in both methodology and theory. Methodologically, we provide a valid and efficient way to assess the OTR when nonregularity is allowed, outperforming existing methods in terms of efficiency. Theoretically, we establish the efficiency optimality of the proposed method within a broad class of estimators allowing nonregularity, paving a novel way to establish optimality theories in nonregular inference where path-wise differentiability is lacking and classical semiparametric theories are not applicable. 

The rest of the paper is organized as follows. In Section 2, we state the problem setting, illustrate the bias under nonregularity, propose the adaptive smoothing method and investigate the theoretical validity of the proposed method. In Section 3, we discuss the challenges of establishing efficiency optimality under nonregularity, introduce the class of robust asymptotically linear unbiased estimators and show that the proposed method is optimal within the class. In Section 4, we compare the proposed method with the competing methods via simulation studies. In Section 5, we re-analyze the ACTG 175 trial with the proposed method and show how much efficiency can be gained with the proposed method in practice.  A discussion is provided in Section 6.

\section{Methodology}
In this section, we start with mathematical formulations of the problem we consider and the bias issue under nonregularity. Then, we introduce the proposed method based on an adaptive smoothing on the estimated decision function, and investigate its theoretical validity and practical implementation in the single time point case of optimal treatment regimes. Extensions to multiple time point cases; i.e., dynamic treatment regimes, can be found in  Section B of the supplementary material.

\subsection{Problem Setting}
Consider an observational study of $n$ patients with observations $\mathcal{O}_n=\{(\mathbf{X}_i,A_i,Y_i),i=1,\dots,n\}$, $n$ i.i.d. copies of $\mathbf{O}=(\mathbf{X}, A, Y)$, where $\mathbf{X} \in \mathcal{X}$ denotes a patient's baseline covariates, $A \in\{0,1\}$ denotes the treatment a patient receives and $Y$ denotes a patient's clinical outcome. Let $Y(0)$ and $Y(1)$ be the potential outcomes under treatment 0 and 1 respectively and $\pi(a,\mathbf{x}):=P(A=a|\mathbf{X}=\mathbf{x})$ be the propensity score. Throughout the paper, we assume $E[Y^2]< \infty$ and make the following classical identifiability assumptions in observational studies \citep{ding2023first}.
\begin{Assumption}[Stable Unit Treatment Value Assumption]\label{A1}
    $Y=\left(1-A\right) Y(0)+A Y(1)$.
\end{Assumption}
\begin{Assumption}[No Unmeasured Confounders]\label{A2}
    $(Y(0), Y(1)) \perp A \mid \mathbf{X}$.
\end{Assumption}
\begin{Assumption}[Overlap]\label{A3}
    $\exists$ $c\in(0,1/2)$ s.t. $c< \pi(a,\mathbf{x})<1-c$.
\end{Assumption}
A treatment regime $d(\cdot)$ is a deterministic function that maps $\mathcal{X}$ to $\{0,1\}$. Given a treatment regime $d$, the potential outcome of a subject is $Y(d)=Y(0)\left\{1-d\left(\mathbf{X}\right)\right\}+Y(1) d\left(\mathbf{X}\right)$ and the mean outcome of a population is $V(d)=E\left\{Y(d)\right\}$. An optimal treatment regime (OTR) $d^{o p t}$ is defined as a regime maximizing the mean outcome; i.e, $d^{ {opt }} \in \arg \max_{d:\mathcal{X}\to\{0,1\}}  V(d) $. In this paper, we aim to construct valid and efficiency-optimal confidence intervals (CIs) for the mean outcome under OTR; i.e., 
\begin{align}
    V_0=V(d^{ {opt }})=\max _d V(d),
\end{align}
as statistical inference on the mean outcome under OTR is crucial for validating the efficacy of OTR, guiding clinical decisions and informing healthcare policies \citep{lavori2004dynamic}.

Note that the OTR, $d^{o p t}$, may not be unique. Denote the conditional mean functions of outcome by $\mu(a, \mathbf{x}) \equiv E\left(Y \mid A=a, \mathbf{X}=\mathbf{x}\right)$ and the contrast function by $\tau(\mathbf{x}) \equiv \mu(1, \mathbf{x})-\mu(0, \mathbf{x})$. When $P(\tau(\mathbf{X})=0)>0$; i.e., the treatments have no differences in a non-negligible subgroup, the set of OTR, $ \mathcal{D}^{ {opt }}=\{d^{o p t}: V\left(d^{o p t}\right)=\max _d V(d)\}$, is not a singleton because for the subjects in $\{\mathbf{X}:\tau(\mathbf{X})=0\}$, taking either treatment or control  would not change the mean outcome. This commonly encountered scenario is referred to as nonregularity, which classical inference often excludes to avoid bias issues \citep{shi2020breaking} and we aim to accommodate here.

In this paper, we focus on the deterministic regime but, indeed, non-deterministic regimes \citep{robins2004optimal} can be naturally accommodated. Specifically, allowing non-deterministic regimes will not change the quantity of interest, $V_0$, and similar bias issues arise when classical inference is applied to the non-deterministic regime under nonregularity as indicated in the development and proof of Theorem \ref{thm1}. In addition, we consider a problem setting of the single time point case; i.e., optimal treatment regime, here for the sake of simplicity, but the proposed method can be naturally extended to the multiple time point case; i.e., dynamic treatment regime, which can be found in Section B of the supplementary material.

To facilitate the presentation of this paper, we let $\widehat{F}$ denote an estimator based on the full dataset $\mathcal{O}_n$ and  $\widehat{F}_{\mathcal{S}}$ denote the same estimator which is nevertheless based on a subset $\mathcal{S}\subset [n]:=\{1,\dots,n\}$. Let $|\mathcal{S}|$ denote the size of the set $\mathcal{S}$ and $\mathcal{S}^c$ denote the complement of $\mathcal{S}$ in $[n]$. Let {$z_{\eta}$} be the upper {$\eta$}-quantile of standard normal distribution. For any two sequences $\left\{a_n\right\},\left\{b_n\right\}$, $a_n \asymp b_n$ implies that there exist some universal constants $c, C>0$ such that $c b_n \leq a_n \leq C b_n$, $a_n \asymp_p b_n$ implies $P(c b_n \leq a_n \leq C b_n)\to 1$ as $n\to\infty$, $a_n\prec b_n$ implies $a_n/b_n\to 0$ as $n\to\infty$ and $a_n\prec_p b_n$ implies $a_n/b_n\to_p 0$ as $n\to\infty$.

\subsection{Bias under Nonregularity}
Note that regardless of uniqueness, one of the OTRs is to assign treatment to the patients with positive contrast outcomes; i.e., $d^{o p t}_0(\mathbf{X}):=\mathbb{I}\{\tau(\mathbf{X})>0\}\in \mathcal{D}^{ {opt }}$, where $\mathbb{I}(\cdot)$ stands for the indicator function. Thus, a doubly-robust type identification of the mean outcome under OTR, $V_0$, is
\begin{align}\label{identification}
    V_0=V\left(d^{o p t}_0\right)=E\psi_i\left(d^{opt}_0, \pi, \mu\right)
\end{align}
where the influence function $\psi_i$ is defined as
\begin{align}\label{AIPWE-IF}
    \begin{aligned}
\psi_i\left(d^*, \pi^*, \mu^*\right) &=\frac{A_i d^*\left(\mathbf{X}_i\right)+\left(1-A_i\right)\left\{1-d^*\left(\mathbf{X}_i\right)\right\}}{\pi^*\left(A_i, \mathbf{X}_i\right)}\left\{Y_i-\mu^*\left(A_i, \mathbf{X}_i\right)\right\} \\
&+d^*\left(\mathbf{X}_i\right) \mu^*\left(1, \mathbf{X}_i\right)+\left\{1-d^*\left(\mathbf{X}_i\right)\right\} \mu^*\left(0, \mathbf{X}_i\right),
\end{aligned}
\end{align}
and $\mu^*$, $\pi^*$ and $d^*$ are any given working models.

Following the identification of the mean outcome under OTR, $V_0$, in Eq. \eqref{identification},  \citet{zhang2012robust} and \citet{zhang2013robust} consider plug-in estimators with double robustness for $V_0$. Let $\widehat{\pi}$, $\widehat{\mu}$ and $\widehat{\tau}$ be the estimators of $\pi$, $\mu$ and $\tau$ respectively and $\widehat{d}(\mathbf{X}):=\mathbb{I}\{\widehat{\tau}(\mathbf{X})>0\}$ be the estimated decision function of OTR. A cross-fitting version of the plug-in estimator is  
\begin{align}\label{plug-in estimator}
    \widehat{V}_0=\frac{1}{n}\sum_{i\in\mathcal{I}} \psi_i\left(\widehat{d}_{\mathcal{I}^c}, \widehat{\pi}_{\mathcal{I}^c}, \widehat{\mu}_{\mathcal{I}^c}\right)+\frac{1}{n}\sum_{i\in\mathcal{I}^c} \psi_i\left(\widehat{d}_{\mathcal{I}}, \widehat{\pi}_{\mathcal{I}}, \widehat{\mu}_{\mathcal{I}}\right),\quad \mathcal{I}\subset[n] \text{ and } |\mathcal{I}|=\lfloor n/2\rfloor
\end{align}
where the cross-fitting aims to remove the self correlation bias induced by the estimation of $d^{o p t}_0$, $\pi$ and $\mu$ \citep{chernozhukov2018double}. Under regularity that $P(\tau(\mathbf{X})=0)=0$ and some mild assumptions including Assumption \ref{A7}, $\sqrt{n}(\widehat{V}_{0}-V_0)$ is asymptotically normal with mean 0 and variance achieving the semiparametric efficiency bound, following similar derivations in \citet{luedtke2016statistical} and \citet{van2014targeted}. Thus, the plug-in estimator $\widehat{V}_0$ leads to valid and optimal inference on the mean outcome under OTR under regularity. Here, Assumption \ref{A7} is a classical doubly robust rate requirement of the estimated propensity score and the estimated outcome function required for inference in existing causal literature \citep{chernozhukov2018double}. The details of the derivation under regularity is provided in Section A.1 of the supplementary material.

\begin{Assumption}[Doubly Robust Convergence Rate of $\widehat{\pi}$ and $\widehat{\mu}$]\label{A7}
    For any subset $\mathcal{I}\subset[n]$, $a\in\{0,1\}$ and asymptotically with probability 1, we have 
\begin{align}
&E\left[\left|\widehat{\pi}_{\mathcal{I}}\left(a, \mathbf{X}\right)-\pi\left(a, \mathbf{X}\right)\right|^2\right]\cdot E\left[\left|\widehat{\mu}_{\mathcal{I}}\left(a, \mathbf{X}\right)-\mu\left(a, \mathbf{X}\right)\right|^2\right]=o(|\mathcal{I}|^{-1}),\label{DR-rate}\\
&\max\left\{E\left[\left|\widehat{\pi}_{\mathcal{I}}\left(a, \mathbf{X}\right)-\pi\left(a, \mathbf{X}\right)\right|^2\right], E\left[\left|\widehat{\mu}_{\mathcal{I}}\left(a, \mathbf{X}\right)-\mu\left(a, \mathbf{X}\right)\right|^2\right]\right\}\to 0 \text{ as $|\mathcal{I}|\to\infty$}.\label{super-consist}
\end{align}
\end{Assumption}
However, under nonregularity where $P(\tau(\mathbf{X})=0)>0$; i.e., there exists a non-negligible subgroup that the treatments have no differences, the plug-in estimator $\widehat{V}_0$ fails to deliver valid and optimal inference on the mean outcome under OTR. This is because the estimated OTR, $\widehat{d}_{\mathcal{I}}$, in Eq. \eqref{plug-in estimator} is inconsistent when the OTR is not uniquely defined. Specifically, under nonregularity, even when the estimated contrast function $\widehat{\tau}_{\mathcal{I}}$ converges to the contrast function $\tau$, after being discretized to $\{0,1\}$, the estimated decision function of OTR, $\widehat{d}_{\mathcal{I}}(\mathbf{X}):=\mathbb{I}\{\widehat{\tau}_{\mathcal{I}}(\mathbf{X})>0\}$, might not converge to the OTR, $d^{o p t}_0:=\mathbb{I}\{\tau(\mathbf{X})>0\}$ due to the nonsmoothness of the indicator function $\mathbb{I}\{\cdot > 0\}$ as illustrated in Example 1 at the end of Section A.1 in the supplementary materials. 

To address the inconsistency issue, a natural idea is to approximate the value function of the OTR $d^{o p t}_0$, an indicator function, by the value functions of a sequence of (pseudo) smoothing decision functions. To enable such approximations, we consider Assumption \ref{A4} which characterizes the complexity of the neighbourhood of the contrast function $\tau(\mathbf{X})$ around 0 and allows nonregularity.
\begin{Assumption}[Assumption (A5) in \citet{shi2020breaking}]\label{A4}
Assume there exist some positive constants $\bar{c}, \alpha$ and $\delta$ such that $P\left(0<\left|\tau\left(\mathbf{X}\right)\right|<t\right) \leq \bar{c} t^\alpha,~\forall 0<t \leq \delta$.
\end{Assumption}
Assumption~\ref{A4} is a classical margin condition in nonregular inference 
\citep{audibert2007fast,luedtke2016statistical,shi2020breaking}, 
where a smaller $\alpha$ indicates a looser bound on the density 
of $\tau(\mathbf{X})$ near zero. In particular, $\alpha<1$ 
accommodates distributions with bounded or even polynomially 
unbounded density, including but not limited to uniform, normal and beta distributions. 
The specific value of $\alpha$ is not needed in our method.

\subsection{Warm-up: Smoothing Decision Function}
\begin{figure}
    \centering
    \BeginAccSupp{method=escape,Alt={\FigureOneAlt},ActualText={\FigureOneAlt}}%
    \includegraphics[width=5cm]{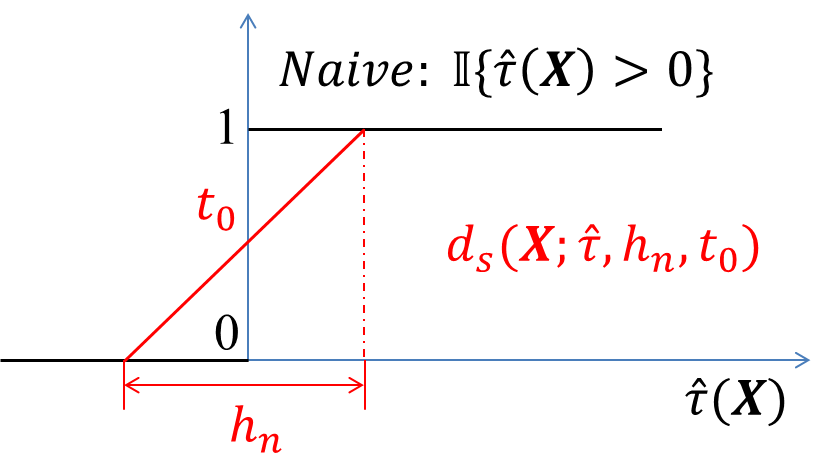}%
    \EndAccSupp{}%
    \caption{Illustration of the proposed smoothing decision function $d_{s}$.}
    \label{as_idea1}
\end{figure}

Let $\mathcal{I}_j$ be a given bipartition of $n$ patients and $\mathcal{I}_{j,k}$ be a given bipartition of $\mathcal{I}_j$ for $j,k=1,2$. We consider the smoothing estimator 
\begin{align}\label{smoothing-estimator}
      \widehat{V}_{s}:=\frac{1}{4}\sum_{j=1,2}\sum_{k=1,2}\frac{1}{|\mathcal{I}_{j,k}|}\sum_{i\in\mathcal{I}_{j,k}}\psi_i(d_{s}(\mathbf{X}_i;\widehat{\tau}_{\mathcal{I}_{3-j}},h_{n,\mathcal{I}_{3-j}},t_0),\widehat{\pi}_{\mathcal{I}_{j,k}^c},\widehat{\mu}_{\mathcal{I}_{j,k}^c})
\end{align}
for $V_0$ based on a (pseudo) smoothing decision function $d_{s}$ used for inference only
\begin{align}
\label{smoothing-function}
\begin{aligned}
    d_{s}(\mathbf{X};\widehat{\tau}_{\mathcal{I}},h_n,t_0):=\mathbb{I}\{\widehat{\tau}_{\mathcal{I}}(\mathbf{X})\in(-t_0 h_n, (1-t_0)h_n)\}\cdot \left(\frac{\widehat{\tau}_{\mathcal{I}}(\mathbf{X})}{h_n}+t_0\right)\\
    +\mathbb{I}\{\widehat{\tau}_{\mathcal{I}}(\mathbf{X})\ge (1-t_0)h_n\}
\end{aligned}
\end{align}
where $h_n$ controls the smoothness of $d_{s}$ and ensures the consistency and $t_0\in (0,1)$ determines the asymmetry of $d_{s}$ as illustrated in Fig. \ref{as_idea1}. The local nature of $d_s$ characterized by $h_n$ ensures 
    zero approximation error for observations with $|\widehat{\tau}(\mathbf{X})|$ bounded below by $h_n$, 
    a property that distinguishes the proposed local smoothing from the global smoothing \citep{levis2025covariate} and is more compatible with the margin condition in Assumption \ref{A4}. The asymmetric extent of $d_s$ characterized by $t_0$ helps distinguish the proposed method from the other smoothing \citep{wu2021resampling,chen2023inference,whitehouse2025inference}, which typically focuses on a single smoothing parameter. As detailed in Sections 2.4 and 3, the choice of $t_0$ does not impact validity but plays a critical role in optimizing efficiency, which existing methods fail to achieve due to the lack of such an additional degree of freedom in smoothing. 
    
    Note that the smoothing decision function $d_s$ is only used for inference on the mean outcome of OTR. To estimate OTR, one should directly apply existing OTR estimation approaches instead of using $d_s$ as $d_s$ is not an appropriately defined decision rule with mapping space $\{0,1\}$. In addition, besides the linear smoothing in Fig. \ref{as_idea1}, other shapes of $d_s$, such as a quadratic one, are possible but given the smoothing scale $h_n$ and the asymmetric extent $t_0$, $d_s$ of any (smoothing) shape is asymptotically equivalent as implied by Lemma A.1 of the supplementary material. Therefore, we consider a simple form of linear smoothing here to help simplify the implementation, which nevertheless would not sacrifice validity or efficiency of the proposed method.  


To facilitate our theoretical analysis, we allow the smoothing tuning parameter $h_{n,\mathcal{I}_{j}}$ to be different across different subsets $\mathcal{I}_j$ in the smoothing estimator $\widehat{V}_s$. It is clear that not any choice of the smoothing tuning parameter $h_{n,\mathcal{I}_{j}}$ leads to valid inference on $V_0$. When $h_{n,\mathcal{I}_{j}}$ is too small, the decision function $d_{s}$ is undersmoothed and is inconsistent; e.g. $h_{n,\mathcal{I}_{j}}=0$ corresponds to the plug-in decision function $d_s=\widehat{d}$. When $h_{n,\mathcal{I}_{j}}$ is too large, the decision function $d_{s}$ is oversmoothed and the difference between $d_{s}$ and $d_0^{opt}$ is too large to guarantee the estimated value function $\widehat{V}_{s}$ converges to $V_0$; e.g., $h_{n,\mathcal{I}_{j}}=+\infty$ corresponds to a flat and noninformative decision function $d_{s}=t_0$.  
To guarantee the validity of the smoothing estimator $\widehat{V}_s$, we derive an appropriate range of the smoothing tuning parameter $h_{n,\mathcal{I}_{j}}$ as stated in Assumptions \ref{A5} and \ref{A6}. Assumption \ref{A5} requires that $h_{n,\mathcal{I}_{j}}$ cannot be too small in order to make sure the proposed smoothing decision function $d_{s}$ is consistent, and Assumption \ref{A6} requires that $h_{n,\mathcal{I}_j}$ cannot be too large in order to make sure the difference between the value function of $d_{s}$ and that of the OTR, $d_0^{opt}$, is negligible. The expectation $E[\cdot]$ in the rate conditions is taken over the distribution of $\mathbf{X}$ only, conditional on the training data in $\mathcal{I}_j$. 

Assumptions \ref{A5} and \ref{A6} are mild assumptions and can be satisfied simultaneously with appropriate choices of the smoothing tuning parameter $h_{n,\mathcal{I}_{j}}$ and certain mild nonparametric rates of the estimated contrast function $\widehat{\tau}_{\mathcal{I}_j}(\mathbf{X})$; i.e., 
\begin{align}\label{sufficient-A6&7}
    E[|\widehat{\tau}_{\mathcal{I}_j}(\mathbf{X})-\tau(\mathbf{X})|^2]\prec_p n^{-\frac{2+\frac{2}{3}\alpha}{2(1+\alpha)}}, \text{ for any $j$,}
\end{align}
where a smaller $\alpha$ in Assumption \ref{A4} imposes a tighter rate requirement for $\widehat{\tau}_{\mathcal{I}_j}$. Eq.~\eqref{sufficient-A6&7} is weaker than the rate requirement for the estimated contrast function $\widehat{\tau}$ in \citet{shi2020breaking}. This is mainly because our proofs adopt Markov's inequality at the third moment rather than the second adopted there. Subject to a moment-comparability condition assumed throughout the paper
\begin{align}
\label{moment-comparability}
E\bigl[|\widehat{\tau}_{\mathcal{I}}(\mathbf{X})-\tau(\mathbf{X})|^{q}\bigr]
=O_p\left(\left(E\bigl[|\widehat{\tau}_{\mathcal{I}}(\mathbf{X})-\tau(\mathbf{X})|^{2}\bigr]\right)^{q/2}\right)
\end{align}
for any fold $\mathcal{I}$ with $|\mathcal{I}|\asymp n$ and $q=3$, the $L^3$ Markov inequality is sharper and thus leads to the weaker rate requirement in Eq.~\eqref{sufficient-A6&7}. Eq.~\eqref{moment-comparability} with $q=3$ is mild and can be satisfied by common parametric and undersmoothed nonparametric estimators as detailed in Section A.5 of the supplementary material. More generally, the same argument can be adapted to any $q\in[2,+\infty)$, with a larger $q$ yielding a weaker rate requirement at the cost of a stronger moment condition. At $q=2$, Eq.~\eqref{moment-comparability} is automatic, and the resulting rate requirement coincides with that of \citet{shi2020breaking}. In practice, Eq. \eqref{sufficient-A6&7} can be satisfied in broad 
scenarios. For example,  
when $\tau$ is estimated by B-spline \citep{shen1998local} 
or kernel ridge regression \citep{zhang2013divide} appropriately, $\alpha<1$ is allowed in Assumption \ref{A4} and thus the density of $\tau(\mathbf{X})$ can be unbounded. 
More examples where $\alpha$ satisfies both Assumption \ref{A4} and Eq. \eqref{sufficient-A6&7} (or equivalently, Assumptions \ref{A5} and \ref{A6}) can be found at Table 1 in \citet{shi2020breaking}.
We defer the choice of $h_{n,\mathcal{I}_j}$ to Section~2.5. Under Assumptions \ref{A1} to \ref{A6} and the moment-comparability condition in Eq.~\eqref{moment-comparability} with $q=3$, we have Theorem \ref{thm1}.


\begin{Assumption}[Lower Bound of $h_{n,\mathcal{I}_j}$]\label{A5}
    $n^{1/4}\left(E[|\widehat{\tau}_{\mathcal{I}_j}(\mathbf{X})-\tau(\mathbf{X})|^2]\right)^{3/4}\prec_p h_{n,\mathcal{I}_j}$  for any $j$.
\end{Assumption}
\begin{Assumption}[Upper Bound of $h_{n,\mathcal{I}_j}$]\label{A6}
    $h_{n,\mathcal{I}_j}\prec n^{-\frac{1}{2(1+\alpha)}}$, where the constant $\alpha$ comes from Assumption \ref{A4}, for any $j$.
\end{Assumption}
\begin{theorem}[Validity of $\widehat{V}_{s}$]\label{thm1}
    Under Assumptions \ref{A1} to \ref{A6} and Eq.~\eqref{moment-comparability} with $q=3$, $\sqrt{n}(\widehat{V}_{s}-V_0)/\widehat{\sigma}_{s}\to_d N(0,1)$, where \begin{align*}
     \widehat{\sigma}_{s}:=\left(\frac{1}{n-1}\sum_{j=1,2}\sum_{k=1,2}\sum_{i\in\mathcal{I}_{j,k}}\left(\psi_i(d_{s}(\mathbf{X}_i;\widehat{\tau}_{\mathcal{I}_{3-j}},h_{n,\mathcal{I}_{3-j}},t_0),\widehat{\pi}_{\mathcal{I}_{j,k}^c},\widehat{\mu}_{\mathcal{I}_{j,k}^c})-\widehat{V}_{s}\right)^2\right)^{1/2}
 \end{align*} 
 and $ \widehat{\sigma}_{s}^2$ converges in probability to
 \begin{align*}
     &\underbrace{\operatorname{Var}\left\{\mu\left(d_0^{o p t}, \mathbf{X}\right)\right\} + E\left(\frac{\mathbb{I}\left\{\tau\left(\mathbf{X}\right)>0\right\}}{\pi\left(1, \mathbf{X}\right)}\operatorname{Var}(Y|\mathbf{X},A=1)+\frac{\mathbb{I}\left\{\tau\left(\mathbf{X}\right)<0\right\}}{\pi\left(0, \mathbf{X}\right)}\operatorname{Var}(Y|\mathbf{X},A=0)\right)}_{\text{Induced by the regular part}}\\
    &+ \underbrace{E\left[\left(\frac{t_0^2}{\pi\left(1, \mathbf{X}\right)}\operatorname{Var}(Y|\mathbf{X},A=1)+\frac{(1-t_0)^2}{\pi\left(0, \mathbf{X}\right)}\operatorname{Var}(Y|\mathbf{X},A=0)\right) \mathbb{I}\left\{\tau\left(\mathbf{X}\right)=0\right\}\right]}_{\text{Induced by the nonregular part and related with $t_0$}}.
 \end{align*}
\end{theorem}
Theorem \ref{thm1} establishes the asymptotic normality of our proposed smoothing estimator $\widehat{V}_s$ and justifies the use of the the two-sided level-{$\eta$} confidence intervals:
 \begin{align}
     \left[\widehat{V}_{s}-{z_{\eta/2}}\cdot\frac{\widehat{\sigma}_{s}}{\sqrt{n}},\widehat{V}_{s}+{z_{\eta/2}}\cdot\frac{\widehat{\sigma}_{s}}{\sqrt{n}}\right]
 \end{align}
 in inferring $V_0$ regardless of whether the regularity assumption is satisfied or not. From Theorem \ref{thm1}, we see that the asymptotic variance of $\widehat{V}_s$ consists of two parts, the regular part which is unavoidable and the nonregular part which depends on $t_0$. This implies that the smoothing tuning parameter $h_{n,\mathcal{I}_{j}}$ will not affect the asymptotic efficiency, while the asymmetric tuning parameter $t_0$ will. Therefore, a natural question to ask is whether we can minimize the variance by adjusting $t_0$ adaptively, and if yes, how. The details of the adaptive smoothing estimator are provided in the coming subsection.

\subsection{Proposed Method: Adaptive Smoothing}
Built on the smoothing estimator, we introduce the proposed method, adaptive smoothing, for inference on $V_0$, the mean outcome under OTR. The key element of the proposed method is an adaptive adjustment of the asymmetric tuning parameter $t_0$ to minimize the variance, or equivalently, optimize the efficiency in the smoothing estimator $\widehat{V}_s$ in Eq. \eqref{smoothing-estimator}. For simplicity, we focus on the case of homoscedasticity that conditional on the baseline covariate $\mathbf{X}$, the outcome under treatment and that under control have the same variance with probability tending to 1 as stated in Assumption \ref{A8}, while similar optimal method and theory can be established in the general case allowing heteroscedasticity and the details are included in Section A.12 and Section D of the supplementary material.

\begin{figure}
\centering
\BeginAccSupp{method=escape,Alt={\FigureTwoAlt},ActualText={\FigureTwoAlt}}%
\begin{minipage}[t]{0.45\textwidth}
\centering
\textbf{(a)}\par\smallskip
\includegraphics[width=5cm]{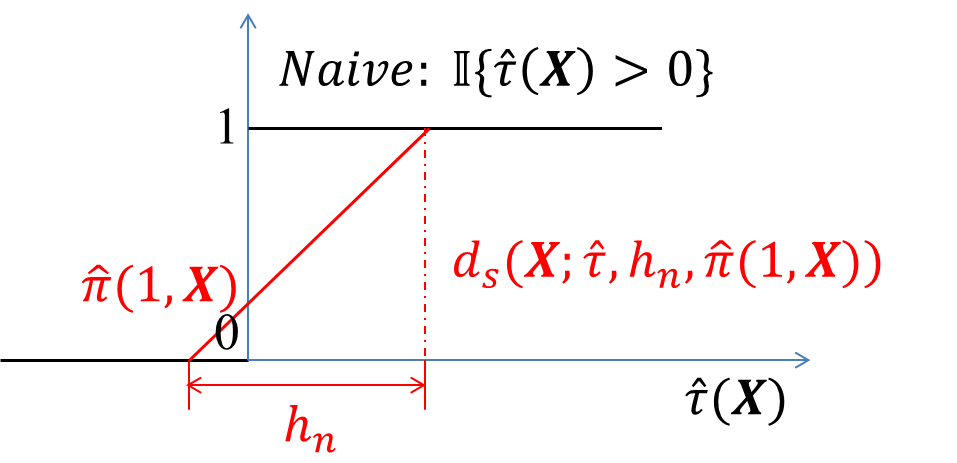}
\end{minipage}
\begin{minipage}[t]{0.45\textwidth}
\centering
\textbf{(b)}\par\smallskip
\includegraphics[width=4.3cm]{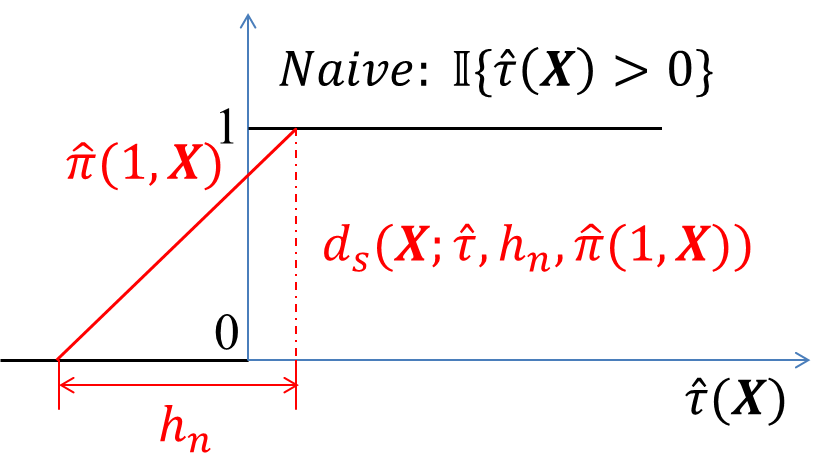}
\end{minipage}%
\EndAccSupp{}%
\caption{Illustration of the proposed adaptive smoothing decision function with (a) a smaller propensity score and (b) a larger propensity score.}
\label{adaillu}
\end{figure}

\begin{Assumption}[Homoscedasticity]\label{A8}
     $P\left(\operatorname{Var}(Y|\mathbf{X},A=1)=\operatorname{Var}(Y|\mathbf{X},A=0)\right)=1$. 
\end{Assumption}
Under Assumption \ref{A8}, {let $\sigma^2(\mathbf{X}):=\operatorname{Var}(Y|\mathbf{X},A=1)=\operatorname{Var}(Y|\mathbf{X},A=0)$ denote this common conditional variance, then} the nonregular part of the asymptotic variance of the smoothing estimator $\widehat{V}_s$ in Theorem \ref{thm1} is equal to $E u(X,t_0)$ where
\begin{equation}\label{quadraticminimizer}
   u(X,t_0)={\sigma^2(\mathbf{X})}\left(\frac{t_0^2}{\pi\left(1, \mathbf{X}\right)}+\frac{(1-t_0)^2}{\pi\left(0, \mathbf{X}\right)}\right) \mathbb{I}\left\{\tau\left(\mathbf{X}\right)=0\right\},
\end{equation}
and given the baseline covariate $\mathbf{X}$, the minimizer of Eq. \eqref{quadraticminimizer} w.r.t $t_0$ is 
\begin{align}
    \pi(1,\mathbf{X})=\arg\min_{t_0}\frac{t_0^2}{\pi\left(1, \mathbf{X}\right)}+\frac{(1-t_0)^2}{\pi\left(0, \mathbf{X}\right)};
\end{align}
i.e., propensity score on $\mathbf{X}$. Thus, a natural idea to improve efficiency of the smoothing estimator $\widehat{V}_{s}$ is to replace the fixed asymmetric tuning $t_0$ with the estimate of the propensity score $\widehat{\pi}_{\mathcal{I}_{3-j}}$ in the proposed smoothing decision function ${d_s}$ and consider an adaptive smoothing estimator
\begin{equation}
   \widehat{V}_{as}:= \frac{1}{4}\sum_{j=1,2}\sum_{k=1,2}\frac{1}{|\mathcal{I}_{j,k}|}\sum_{i\in\mathcal{I}_{j,k}}\psi_i(d_{s}(\mathbf{X}_i;\widehat{\tau}_{\mathcal{I}_{3-j}},h_{n,\mathcal{I}_{3-j}},\widehat{\pi}_{\mathcal{I}_{3-j}}(1,\mathbf{X}_i)),\widehat{\pi}_{\mathcal{I}_{j,k}^c},\widehat{\mu}_{\mathcal{I}_{j,k}^c}). 
\end{equation}
The intuitive explanation of why replacing the fixed $t_0$ with the estimated propensity score helps improve efficiency is that the effective sample size in the influence function $\psi$ depends on $Ad_{s}+(1-A)(1-d_{s})$ positively \citep{2024Effective}, so when the propensity score becomes larger, i.e., more patients receive the treatment, a left-shifted smoothing decision function $d_{s}$ can better utilize the treated patients and increase the effective sample size. The extent of the asymmetry should be proportional to the propensity score as illustrated in Figure \ref{adaillu}. Based on the adaptive smoothing estimator, the proposed method for inference on the mean outcome under OTR, $V_0$, is summarized in Algorithm \ref{alg1}. Note that compared with existing methods allowing nonregularity, such as $m$-out-of-$n$ bootstrap \citep{chakraborty2010inference,chakraborty2014inference}, online one step estimator \citep{luedtke2016statistical} and subbagging \citep{shi2020breaking}, Algorithm \ref{alg1} is far more efficient in terms of the computational time, as it does not require extensive resampling or refitting of the working models.

To justify the proposed method in Algorithm \ref{alg1}, we note that Theorem \ref{thm1} is not directly applicable. First, the propensity score $\pi(1,\mathbf{X})$ typically depends on $\mathbf{X}$, while Theorem \ref{thm1} is derived for a fixed $t_0$ only. Fortunately, we can show that Theorem \ref{thm1} holds uniformly w.r.t. $t_0\in[c,1-c]$ for any given constant $c>0$, which, together with the overlap condition for the propensity score in Assumption \ref{A3}, facilitates the justification of the proposed method.  Second,  because the propensity score $\pi(1,\mathbf{X})$ is usually unknown in practice, we use its estimated value; i.e., $\widehat{\pi}(1,\mathbf{X})$, instead in $\widehat{V}_{as}$, and the difference between $\pi(1,\mathbf{X})$ and $\widehat{\pi}(1,\mathbf{X})$ might be nonnegligible and break the asymptotic analysis in Theorem \ref{thm1}. Fortunately, because for the subjects whose $\tau(\mathbf{X})$ is around zero, $\widehat{\tau}(\mathbf{X})$ is also close to zero, the error induced by the estimation of the propensity score is controlled by $\widehat{\tau}(\mathbf{X})\cdot|\widehat{\pi}(1,\mathbf{X})-\pi(1,\mathbf{X})|$ and is negligible under Assumption \ref{A7}. Therefore, compared with Theorem \ref{thm1}, no additional assumptions are required to justify the proposed method. The establishment of the asymptotic normality and thus the validity of $\widehat{V}_{as}$ is stated in Theorem \ref{ef-opt}. The efficiency optimality of $\widehat{V}_{as}$ and the variance reduction over $\widehat{V}_{s}$ and the current state-of-the-art will be given in Section 3. 

\begin{algorithm}
\caption{The Adaptive Smoothing Estimator and Its Confidence Interval}
    \begin{algorithmic}[1]
        \State \textbf{Input}: Data set $\mathcal{O}_n$; Randomly picked subsets of $[n]$: $\mathcal{I}_1$ and $\mathcal{I}_2$, and randomly picked subsets $\mathcal{I}_{j,k}$ ($k=1,2$) of $\mathcal{I}_j$ ($j=1,2$), where all the $\mathcal{I}_{j,k}$ with the equal size; Parameters $h_{n,\mathcal{I}_j}$, $j=1,2$; Working models: $\widehat{\pi}$, $\widehat{\mu}$.
        \State \textbf{Construction of 
 The Adaptive Smoothing Estimator:} 
 \begin{align}
     \widehat{V}_{as}&:= \frac{1}{4}\sum_{j=1,2}\sum_{k=1,2}\frac{1}{|\mathcal{I}_{j,k}|}\sum_{i\in\mathcal{I}_{j,k}}\psi_i(d_{s}(\mathbf{X}_i;\widehat{\tau}_{\mathcal{I}_{3-j}},h_{n,\mathcal{I}_{3-j}},\widehat{\pi}_{\mathcal{I}_{3-j}}(1,\mathbf{X}_i)),\widehat{\pi}_{\mathcal{I}_{j,k}^c},\widehat{\mu}_{\mathcal{I}_{j,k}^c});\label{sin-opt-est}
 \end{align}
 \State \textbf{Construction of The Confidence Interval} The standard deviation estimate:
 \begin{align}
    \widehat{\sigma}_{as}^2&:=\frac{1}{n-1}\sum_{j=1,2}\sum_{k=1,2}\sum_{i\in\mathcal{I}_{j,k}}\left(\psi_i(d_{s}(\mathbf{X}_i;\widehat{\tau}_{\mathcal{I}_{3-j}},h_{n,\mathcal{I}_{3-j}},\widehat{\pi}_{\mathcal{I}_{3-j}}(1,\mathbf{X}_i)),\widehat{\pi}_{\mathcal{I}_{j,k}^c},\widehat{\mu}_{\mathcal{I}_{j,k}^c})-\widehat{V}_{as}\right)^2\label{sin-opt-std}
 \end{align}
 and thus the two-sided level-{$\eta$} confidence interval can be given as:
 \begin{align*}
     \left[\widehat{V}_{as}-{z_{\eta/2}}\cdot\frac{\widehat{\sigma}_{as}}{\sqrt{n}},\widehat{V}_{as}+{z_{\eta/2}}\cdot\frac{\widehat{\sigma}_{as}}{\sqrt{n}}\right].
 \end{align*}
    \end{algorithmic}
    \label{alg1}
\end{algorithm}
\begin{theorem}[Validity of $\widehat{V}_{as}$]\label{ef-opt}
Under the assumptions of Theorem \ref{thm1} and Assumption \ref{A8}, $\sqrt{n}(\widehat{V}_{as}-V_0)/\widehat{\sigma}_{as}\to_d N(0,1)$,  where $\widehat{\sigma}_{as}^2$ converges in probability to
 \begin{align*}
     &\underbrace{\operatorname{Var}\left\{\mu\left(d_0^{o p t}, \mathbf{X}\right)\right\} + E\left[{\sigma^2(\mathbf{X})}\left(\frac{\mathbb{I}\left\{\tau\left(\mathbf{X}\right)>0\right\}}{\pi\left(1, \mathbf{X}\right)}+\frac{\mathbb{I}\left\{\tau\left(\mathbf{X}\right)<0\right\}}{\pi\left(0, \mathbf{X}\right)}\right)\right]}_{\text{Induced by the regular part}}\\
    +& \underbrace{E\left[{\sigma^2(\mathbf{X})} \mathbb{I}\left\{\tau\left(\mathbf{X}\right)=0\right\}\right]}_{\text{Induced by the nonregular part}}.
 \end{align*}
\end{theorem}

\subsection{Tuning Selection and Practical Implementation}
In this subsection, we introduce a data-driven approach to select the smoothing tuning parameter $h_{n,\mathcal{I}_j}$ ($j=1,2$) in practice. The proposed method guarantees that the selected $h_{n,\mathcal{I}_j}$ falls into the theoretical range required in Assumptions \ref{A5} and \ref{A6} with probability tending to 1. The key idea of the proposed tuning approach is to construct a sample-split-based statistic, the  estimated approximation error (EAE) of $\widehat{\tau}_{\mathcal{I}_j}$, that approximates the order of the lower bound $n^{1/4}\left(E[|\widehat{\tau}_{\mathcal{I}_j}(\mathbf{X})-\tau(\mathbf{X})|^2]\right)^{3/4}$, after truncation and select $h_{n,\mathcal{I}_j}$ slightly larger than the truncated EAE but in a smaller order of the upper bound of $h_{n,\mathcal{I}_j}$, $n^{-\frac{1}{2(1+\alpha)}}$. Specifically, to prevent the selected $h_{n,\mathcal{I}_j}$ from being too small in finite sample, we consider a truncation {$(\log(n)/(C_0 n))$} for the estimated approximation error where {$C_0$} is a constant. Additional simulation results in Section A.11 of the supplementary material demonstrate that the tuning selection remains relatively robust across different choices of {$C_0$}, and for practical use, we suggest {$C_0$} is chosen between 0.01 and 0.05. Details of the tuning selection of $h_{n,\mathcal{I}_j}$ are summarized in Algorithm \ref{alg2}.

\begin{algorithm}
\caption{Selection of $h_{n,\mathcal{I}_j}$ ($j=1,2$)}
    \begin{algorithmic}[1]
        \State \textbf{Input}: Data set $\mathcal{O}_n$; Randomly picked subsets of $[n]$: $\mathcal{I}_1$ and $\mathcal{I}_2$ with equal size; Working models: $\widehat{\mu}$.
        \State \textbf{Estimated Approximation Error of $\widehat{\tau}_{\mathcal{I}_j}$ ($j=1,2$)}: 
 \begin{align*}
     EAE(\widehat{\tau}_{\mathcal{I}_j})=\frac{1}{|\mathcal{I}_{3-j}|}\left[\sum_{k=1,2}\sum_{i\in\mathcal{I}_{3-j,k}}\left(\widehat{\tau}_{\mathcal{I}_{3-j,3-k}}(\mathbf{X}_i)- \widehat{\tau}_{\mathcal{I}_{j}}(\mathbf{X}_i)\right)^2\right]
 \end{align*}
 \State \textbf{Selection of $h_{n,\mathcal{I}_j}$}: $h_{n,\mathcal{I}_j}={C_0}\cdot \log(n) \cdot n^{1/4}\cdot ((\log(n)/({C_0} n))\lor EAE(\widehat{\tau}_{\mathcal{I}_j}))^{3/4}$, where ${C_0}$ is a constant.
    \end{algorithmic}
    \label{alg2}
\end{algorithm}
Algorithm \ref{alg2} is justified under a rate requirement of $\widehat{\tau}$ in Assumption \ref{A9} which aims to ensure that there exists a reasonable gap between the lower bound, $n^{1/4}\left(E[|\widehat{\tau}_{\mathcal{I}_j}(\mathbf{X})-\tau(\mathbf{X})|^2]\right)^{3/4}$, and the upper bound, $n^{-\frac{1}{2(1+\alpha)}}$, of $h_{n,\mathcal{I}_j}$ in Assumptions \ref{A5} and \ref{A6}. Since $\beta\in(0,\alpha)$, the rate requirement in Assumption \ref{A9} is only slightly stronger than Eq. \eqref{sufficient-A6&7}. Proposition \ref{tuntheory} shows that under Assumption \ref{A9} and mild conditions on the contrast estimators as detailed in Section A.5 of the supplementary material, the selected tuning satisfies Assumptions \ref{A5} and \ref{A6} with probability tending to one. In other words, our tuning does not rely on knowing or estimating the convergence rate of $\widehat{\tau}$; the selected $h_{n,\mathcal{I}_j}$ automatically falls within the valid range required by Assumptions \ref{A5} and \ref{A6} with high probability as long as there exists a polynomial gap between Assumptions \ref{A5} and \ref{A6} implied by Assumption \ref{A9}.

Note that the goal of Algorithm \ref{alg2} is to ensure the selected $h_{n,\mathcal{I}_j}$ falls within the valid range for inference (Assumptions \ref{A5} and \ref{A6}), rather than minimize the MSE \citep{su2020adaptive}. The approximation of the MSE by EAE in Algorithm 2 is only to construct a statistic with the same order as the lower bound rather than minimize MSE. This is fundamentally different from typical nonparametric estimation problems where undersmoothing or bandwidth optimization is needed.

\begin{Assumption}[Rate Requirement for Tuning]\label{A9}
    $\exists$ $\beta\in(0,\alpha)$, s.t. $ E[|\widehat{\tau}_{\mathcal{I}_j}(\mathbf{X})-\tau(\mathbf{X})|^2]\prec_p n^{-\frac{2+\frac{2}{3}\beta}{2(1+\beta)}}$ for any $j$. 
\end{Assumption}
\begin{Proposition}[Validity of Tuning]\label{tuntheory}
Suppose Assumption \ref{A9} holds, and that the contrast estimators satisfy either Condition (C1) or both Conditions (C2) and (C3), stated in Section A.5 of the supplementary material. Then, the tuning $h_{n,\mathcal{I}_j}$ selected by Algorithm \ref{alg2} satisfies
$n^{1/4}\left(E[|\widehat{\tau}_{\mathcal{I}_j}(\mathbf{X})-\tau(\mathbf{X})|^2]\right)^{3/4}\prec_p h_{n,\mathcal{I}_j}\prec_p n^{-\frac{1}{2(1+\alpha)}}$, $\forall~j$.
\end{Proposition}
Repeated cross fitting and averaging (Algorithm~3 in the supplement) 
can improve finite sample performance without changing asymptotic 
efficiency (Corollary~A.1). However, simulations in Section~A.11 
show that Algorithm~\ref{alg1} already performs comparably while 
saving computational time.


\section{Efficiency Optimality}
In this section, we establish the efficiency optimality of the proposed adaptive smoothing approach in Algorithm \ref{alg1}. Specifically, we first introduce the challenge of developing an efficiency optimality theory for inference on the mean outcome under OTR, \(V_0\), when nonregularity is allowed. Then, we propose the Robust Asymptotically Linear Unbiased estimator class, and discuss the relevance and generalizablity of the proposed class. In the end, we demonstrate the efficiency optimality of our proposed adaptive smoothing method by deriving a lower bound on the asymptotic variance for our proposed estimator class and showing that our proposed adaptive smoothing estimator belongs to the proposed estimator class and achieves its asymptotic variance lower bound.

\subsection{Challenges of Deriving Optimality under Nonregularity}
Although semiparametric optimality has been well established for inference on the mean outcome under OTR under regularity, establishing efficiency optimality under nonregularity is well-known to be challenging due to the lack of path-wise differentiability \citep{van2011targeted,luedtke2016statistical}. This is because in semiparametric theory, the concept of a regular estimator is only well-defined when the target parameter is pathwise differentiable \citep{bickel1993efficient,van2000asymptotic}. Specifically, the semiparametric efficiency bound is established within the regular estimator for OTR which requires the map from the distribution of $(\mathbf{X},A, Y)$ to the quantity of interest $V_0$ is path-wise differentiable \citep{kennedy2017semiparametric,bickel1993efficient}. As shown in \citet{luedtke2016statistical}, the path-wise differentiability of the mean outcome under OTR, $V_0$, is indeed equivalent to
\begin{align}\label{eq-path-wise}
    P\left(\mathbf{X}: \tau(\mathbf{X})\neq 0 \lor \max_{A\in\{0,1\}}\operatorname{Var}(Y|A,\mathbf{X})=0\right)=1,
\end{align}
which is naturally satisfied under regularity; i.e., $P(\tau(\mathbf{X})=0)=0$. However, under nonregularity; i.e., $P(\tau(\mathbf{X})=0)>0$, the path-wise differentiability of $V_0$, or equivalently, Eq. \eqref{eq-path-wise}, is satisfied if and only if for $A=0,1$, 
$\operatorname{Var}(Y|A,\mathbf{X})=0\text{ a.e. for } \mathbf{X}\text{ where }\tau(\mathbf{X})=0,$ 
which requires given $A$ and $\mathbf{X}$, $Y$ is a constant for the patients in the nonregular set $\{\mathbf{X}:\tau(\mathbf{X})=0\}$ and is clearly an unrealistic condition. Therefore, under nonregularity that $P(\tau(\mathbf{X})=0)>0$, the regular estimator for $V_0$ cannot be well defined, not to mention deriving semiparametric optimality.


To establish efficiency optimality for inference on $V_0$ when nonregularity is allowed, we need to consider a new estimator class which bears the following properties: (1) its definition remains valid under nonregularity; (2) it should be general enough to include the proposed estimator and common estimators considered in both regular and nonregular case and (3) the variance bound exists and is meaningful. Considering all these criteria, one class naturally comes to mind is the asymptotically linear estimator,
as its definition only depends on linearity, and is free of path-wise differentiability and is valid under nonregularity. Moreover, the asymptotically linear estimator includes not only the proposed adaptive smoothing estimator $\widehat{V}_{as}$ but also efficient regular estimators \citep{bickel1993efficient}, the most efficient ones among regular estimators for $V_0$, which we denote by $\widehat{V}_{er}$. However, deriving a variance lower bound for asymptotically linear estimators is  well recognized as impossible because the class is too broad \citep{van2000asymptotic}. Therefore, appropriate constraints are needed for the class of  asymptotically linear estimators to derive efficiency optimality for inference on the mean outcome under OTR, $V_0$, when nonregularity is allowed. 


\subsection{Robust Asymptotically Linear Unbiased Estimator}
Motivated by the discussion of the asymptotically linear estimator in Section 3.1, we propose the robust asymptotically linear unbiased estimator class, $\mathcal{T}_{RL}$, for the mean outcome under OTR, $V_0$, as defined in Definition \ref{ralu}. Here, ``robust'' refers to double robustness and inclusion of $r_a$ is to ensure that the class $\mathcal{T}_{RL}$ is rich enough to accommodate heteroscedasticity case as detailed in  Sections A.12 and D of the supplementary material. In Definition \ref{ralu}, we impose no restrictions on the working models $\mu_a$, $\pi_1$ and $r_a$, which may or may not coincide with their population truth. The essential constraint is on the function $f$, which requires the doubly robust condition that $E_P[f]=V_0$ holds whenever either the outcome model or the propensity score is correctly specified. Such doubly robust constraints on the influence function are widely adopted in observational studies \citep{ding2023first}.

\begin{figure}
    \centering
    \BeginAccSupp{method=escape,Alt={\FigureThreeAlt},ActualText={\FigureThreeAlt}}%
    \includegraphics[width=6cm]{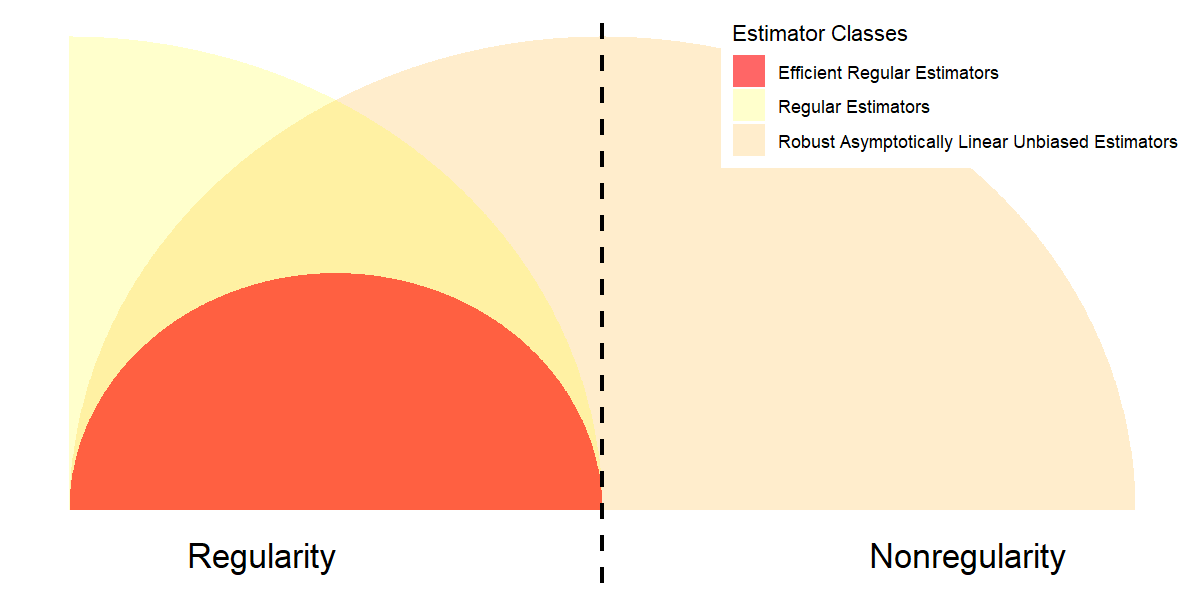}%
    \EndAccSupp{}%
    \caption{Relationships between efficient regular estimators, regular estimators and the proposed robust asymptotically linear unbiased estimators.}
    \label{estimator classes}
\end{figure}

\begin{definition}[Robust Asymptotically Linear Unbiased Estimator]\label{ralu}
    Let $c\in(0,1/2)$ be a fixed constant and let $\mathcal{P}$ denote the set of distributions satisfying Assumptions \ref{A1} to \ref{A3}. The class $\mathcal{T}_{RL}$ is the set of estimators $T_n$ for which there exists a measurable function
    $f:\mathcal{X}\times\{0,1\}\times\mathbb{R}^{3}\times(c,1-c)\times\{-1,0,1\}\times\mathbb{R}_{\geq 0}^{2}\to\mathbb{R}$ such that, for every $P\in\mathcal{P}$,
    \begin{equation}
    \begin{split}
    T_n(\mathcal{O}_n)=n^{-1}\sum_{i=1}^n
    f\bigl(\mathbf{X}_i,&A_i,Y_i,\mu_1(\mathbf{X}_i),\mu_0(\mathbf{X}_i),\pi_1(\mathbf{X}_i),\\[-2pt]
    &\operatorname{sgn}(\tau_P(\mathbf{X}_i)),r_1(\mathbf{X}_i),r_0(\mathbf{X}_i)\bigr)+o_p(n^{-1/2})
    \quad\text{under }P.
    \end{split}
    \end{equation}
    Here, $\tau_P(\mathbf{X})=E_P[Y\mid\mathbf{X},A=1]-E_P[Y\mid\mathbf{X},A=0]$, and $\mu_a$, $\pi_1$ and $r_a$ are the working models for  $E_P[Y\mid\mathbf{X},A=a]$, $E_P[A\mid\mathbf{X}]$ and $\operatorname{Var}_P(Y\mid\mathbf{X},A=a)$, respectively, for $a=0,1$. Moreover, for any $P\in\mathcal{P}$, $f$ satisfies
    \begin{align}\label{unbias}
        E_P[f] = E_P[E_P\{Y|\mathbf{X},A=d^{opt}(\mathbf{X})\}]
    \end{align}
    as long as one of the following equations holds $P$-a.s.:
    \begin{align}
       E_P[Y|\mathbf{X},A] =& A \mu_1(\mathbf{X})+(1-A)\mu_0(\mathbf{X}),\label{robust-1}\\ E_P[A|\mathbf{X}]=&\pi_1(\mathbf{X}).\label{robust-2}     
    \end{align}
\end{definition}
Due to the mild constraints imposed in $\mathcal{T}_{RL}$, the proposed robust asymptotically linear unbiased estimator class is general and includes or dominates most common estimators for the mean outcome under OTR, $V_0$, under both regularity and nonregularity. First, $\mathcal{T}_{RL}$ contains the proposed smoothing estimator $\widehat{V}_s$ in Section 2.3 and the adaptive smoothing estimator $\widehat{V}_{as}$ in Section 2.4. Second, $\mathcal{T}_{RL}$ contains the efficient regular estimator $\widehat{V}_{er}$ and thus the plug-in estimator under regularity. Third, $\mathcal{T}_{RL}$ contains a very broad and useful class of estimators $\mathcal{T}_{\mathcal{D}}:=\{\widehat{V}_{\mathcal{D}}^* = \frac{1}{n}\sum_{i=1}^n \psi(\mathcal{D}(\mathbf{X}_i),\pi,\mu)+o_p(n^{-1/2})\mid \mathcal{D}:\mathcal{X}\to[0,1]\land E[\psi(\mathcal{D}(\mathbf{X}),\pi,\mu)]=V_0\}$. Both the oracle estimator $\widehat{V}_{or}$ with true $d_{0}^{opt}$, $\pi$ and $\mu$ in $\psi_i(\cdot)$ and the current state-of-the-art estimator 
\begin{align*}
    \widehat{V}_{sb}=\frac{1}{n}\sum_{i=1}^n \psi_i(d_{s_n}(\mathbf{X}_i),\pi,\mu)+o_p(n^{-1/2}), \text{ where } d_{s_n}(\cdot)=E[\mathbb{I}\{\widehat{\tau}_{[s_n]}(\cdot)>0\}] \text{ and } s_n=o(n),
\end{align*}
as defined in \citet{shi2020breaking} are included in the class of $\widehat{V}_{\mathcal{D}}^*$ and thus $\mathcal{T}_{RL}$, and the latter implies that the other common estimators for $V_0$ allowing nonregularity are dominated by $\mathcal{T}_{RL}$. The above results are summarized in Corollary \ref{embedding} and illustrated in Figure \ref{estimator classes}. With small modifications to Definition \ref{ralu}, $\mathcal{T}_{RL}$ can further include other common estimators allowing nonregularity for $V_0$, such as the simple sample split estimator \citep{wager2018estimation} and the online one-step estimator \citep{luedtke2016statistical}. Details are provided in Section A.8 of the supplementary material.

\begin{Corollary}\label{embedding}
Under the same assumptions of Theorem \ref{ef-opt}, and assumptions of Theorem 3 in \citet{shi2020breaking}, we have
    (i) $\widehat{V}_{s}\in \mathcal{T}_{RL}$ for any $t_0\in[0,1]$; (ii) $\widehat{V}_{as}\in \mathcal{T}_{RL}$; (iii) $\widehat{V}_{er}\in \mathcal{T}_{RL}$;  (iv) $\widehat{V}_{or}, \widehat{V}_{sb}\in \mathcal{T}_{\mathcal{D}}\subset \mathcal{T}_{RL}$.
\end{Corollary}

\subsection{Efficiency Optimality}
Under the same assumptions of Theorem \ref{ef-opt} and some mild conditions, Theorem \ref{opt-ralu} derives the variance lower bound of the proposed robust asymptotically linear unbiased class $\mathcal{T}_{RL}$ and shows that the proposed adaptive smoothing estimator $\widehat{V}_{as}$ achieves the variance lower bound of $\mathcal{T}_{RL}$. The correct specification of the working models in Theorem \ref{opt-ralu} is analogous to the requirement in classical semiparametric efficiency bounds that all nuisance functions need to be correctly specified in the Convolution Theorem \citep{bickel1993efficient}. This is mainly because under misspecification, inference validity is not guaranteed in general and optimizing efficiency without validity is not meaningful \citep{tsiatis2006semiparametric}.

\begin{theorem}[Efficiency Optimality of $\widehat{V}_{as}$] \label{opt-ralu}
Under the same assumptions of Theorem \ref{ef-opt}, when the working models $\mu_1$, $\mu_0$ and $\pi_1$ in $T_n$ are correctly specified, and $T_n$ satisfies asymptotic uniform integrability, i.e., $ \lim _{M \rightarrow \infty} \sup _n E\left[n\left(T_n-V_0\right)^2 \cdot \mathbf{1}\left(n\left(T_n-V_0\right)^2>M\right)\right]=0$, we have
    \begin{align*}
        &\inf_{T_n\in\mathcal{T}_{RL}}\lim_{n\to\infty}\operatorname{Var}(\sqrt{n}T_n)= \lim_{n\to\infty}\operatorname{Var}(\sqrt{n}\widehat{V}_{as})\\
         =&\operatorname{Var}\left\{\mu\left(d_0^{o p t}, \mathbf{X}\right)\right\} + E\left[{\sigma^2(\mathbf{X})}\left(\frac{\mathbb{I}\left\{\tau\left(\mathbf{X}\right)>0\right\}}{\pi\left(1, \mathbf{X}\right)}+\frac{\mathbb{I}\left\{\tau\left(\mathbf{X}\right)<0\right\}}{\pi\left(0, \mathbf{X}\right)}\right)\right]\\
    +& E\left[{\sigma^2(\mathbf{X})} \mathbb{I}\left\{\tau\left(\mathbf{X}\right)=0\right\}\right].
    \end{align*}
\end{theorem}

The efficiency optimality of $\widehat{V}_{as}$ established in Theorem \ref{opt-ralu} within $\mathcal{T}_{RL}$ is due to the special hidden structure of $\mathcal{T}_{RL}$. Specifically, by variational principle, we can show that the estimator in $\mathcal{T}_{RL}$ can be decomposed into one part asymptotically equivalent to $\widehat{V}_{as}$ and the other part uncorrelated with $\widehat{V}_{as}$. Then, by variance decomposition, we can show that $\widehat{V}_{as}$ exactly attains the variance lower bound of $\mathcal{T}_{RL}$. This is substantially different from the convolution theorem of classical semiparametric efficiency optimality \citep{bickel1993efficient}.

Three implications from Theorem~\ref{opt-ralu} are worth noting, as all the estimators in Corollary \ref{embedding} have correctly specified working models of $\mu_1$, $\mu_0$ and $\pi_1$ under the assumptions therein in order to ensure their inference validity. First, because the current state-of-the-art estimator $\widehat{V}_{sb}$ belongs to $\mathcal{T}_{RL}$,  the proposed adaptive smoothing estimator $\widehat{V}_{as}$ dominates all existing methods allowing nonregularity, with a significant variance reduction over 
subbagging $\widehat{V}_{sb}$ stated in Corollary~\ref{var-reduce-sb}. Second, because the proposed smoothing estimator $\widehat{V}_s$ belongs to $\mathcal{T}_{RL}$,  the variance reduction of the proposed adaptive one $\widehat{V}_{as}$ over $\widehat{V}_{s}$ 
is always nonnegative no matter what $t_0$ is chosen and is indeed strict unless 
$P(\pi(1,\mathbf{X})=t_0 \lor \tau(\mathbf{X})\neq 0)=1$; 
the proof is in Section~A.10 of the supplementary material. Third, because the efficient regular estimator belongs to $\mathcal{T}_{RL}$, under regularity, $\widehat{V}_{as}$ is asymptotically equivalent to the efficient regular estimator and achieves the semiparametric efficiency bound, to which the derived variance bound in Theorem \ref{opt-ralu} indeed reduces. From this perspective, the optimality established by Definition~\ref{ralu} and Theorem~\ref{opt-ralu} encompasses classical semiparametric theories in inference on the mean outcome under OTR, and $\widehat{V}_{as}$ achieves optimal efficiency in 
both regular and nonregular cases.
\begin{Corollary}[Variance Reduction over Subbagging]\label{var-reduce-sb}
    Under the same assumptions of Theorem \ref{opt-ralu}, and assumptions of Theorem 3 in \citet{shi2020breaking}, we have
    \begin{align}\label{Vsb-Vas}
   &\lim_{n\to\infty}\left[ \operatorname{Var}(\sqrt{n}\widehat{V}_{sb}) - \operatorname{Var}(\sqrt{n}\widehat{V}_{as})\right]\notag\\
   =&E\left[\left(\frac{1}{4 \pi\left(1, \mathbf{X}\right)}+\frac{1}{4\pi\left(0, \mathbf{X}\right)}-1\right){\sigma^2(\mathbf{X})} \mathbb{I}\left\{\tau\left(\mathbf{X}\right)=0\right\}\right]\ge 0,
\end{align}
which is strict unless the study is a randomized and balanced design or regular; i.e., $P(\pi\left(1, \mathbf{X}\right)=0.5 \lor \tau(\mathbf{X})\neq 0)=1$.
\end{Corollary}

\section{Simulation Studies}
In this section, we study the finite sample performance of the proposed method via Monte Carlo simulation in the single time point case. To start with, we consider a data generating model similar to that in \citet{shi2020breaking}, i.e., $Y=\Phi\left(X_{1}, X_{2}\right)+A \tau\left(X_{1}, X_{2}\right)+e$, where the random error term $e\sim \operatorname{Normal}(0,0.5^2)$ is independent of $A$, $X_1$ and $X_2$. We vary the generation of $X_1$ and $X_{2}$, the baseline function $\Phi$, the contrast function $\tau$ and the propensity score $\pi$, and consider five different scenarios (A)-(E) including both regular cases and nonregular cases as summarized in Table \ref{sin-setting}.
In all scenarios, $X_{1}$ and $X_{2}$ are independent, and the mean outcome under OTR, $V_0$, can be explicitly calculated as shown in Table \ref{sin-setting}. In Scenarios (A), (C) and (E), the OTR is not uniquely defined; i.e., nonregular, since the contrast functions in these scenarios satisfy $P\left(\tau\left(X_{1}, X_{2}\right)=0\right)=P\left(X_{1}=0\right)$, which equals $0.6$, $0.7$ and $0.9$, respectively. On the contrary, in Scenarios (B) and (D), we have $P\left(\tau\left(X_{1}, X_{2}\right)=0\right)=0$; i.e., regular. For each scenario, we let the sample size $n=1000$, and comparison is made among the following methods:
\begin{itemize}
    \item Proposed: confidence interval obtained from adaptive smoothing combined with repeatedly cross fitting and averaging (details in Algorithm 3 in the supplementary material), and $h_{n,\mathcal{I}_j}$'s selected by Algorithm \ref{alg2}.
    \item Subbagging: confidence interval of subbagging in \citet{shi2020breaking}.
    \item SSS: confidence interval by simply splitting sample into two parts and using half to estimate the decision function, propensity score and the conditional mean functions, and the other half to evaluate the plug-in estimator.
\end{itemize}

In Scenarios (A) and (B), we apply nonparametric maximum likelihood estimators to estimate the propensity score $\pi$ and the conditional mean function $h$. In Scenarios (C)-(E), we apply the cubic spline to estimate $\pi$ and $h$, and use the same truncation on the estimation of $\pi$ as that in \citet{shi2020breaking} (with the lower bound of 0.05 and upper bound 0.95). In this simulation, we let {$C_0=0.05$} for the tuning selection in Algorithm \ref{alg2} and subsample size $K_0=5$ in subbaging. The resampling times $B$ for all the methods are 1000, and the observed results including Empirical Coverage Probability (ECP) and Average Length of confidence intervals (AL) are aggregated over 500 Monte Carlo samples.

From Table \ref{Single-sim}, we see that the proposed method achieves the nominal coverage probability; i.e., 95\%, well in both regular, Scenarios (B) and (D), and nonregular cases, Scenarios (A), (C) and (E), which is in accordance with the asymptotic theory in Theorem \ref{ef-opt}. The results in Table \ref{Single-sim} demonstrate that in all the scenarios regardless of whether it is regular or not, the proposed method has shortest confidence interval among all the competing methods. In particular, in the nonregular cases, Scenarios (A), (C) and (E), the efficiency gain of the proposed method over the current state-of-the-art method, subbagging, is significant with about 10\%-25\% reduction of average length from the proposed method over that from subbaging, which is consistent with the efficiency optimality in Theorem \ref{opt-ralu}. 

We also conduct additional simulations to demonstrate (1) the robustness w.r.t. the constant {$C_0$} in Algorithm \ref{alg2}; (2) the robustness w.r.t. the number of the split in the proposed method and (3) the computational benefit of the proposed method in Section A.11, and (4) the robustness of the proposed method w.r.t. sample size under heteroscedasticity and more challenging settings, e.g., $\alpha<1$ and multi-dimensional covariates, in Section A.12 of the supplementary material. Simulation results demonstrating the merit of the proposed method in multiple time point cases are also included in Section B.4 and Section D of the supplementary material. 

\section{Real Data Analysis}
In this section, we apply the proposed method to the data from AIDS Clinical Trials Group Protocol 175 (ACTG175) which is a randomized clinical trial comparing the treatment effects of two monotherapies, zidovudine (ZDV) and didanosine (ddI), and two combination therapies, ZDV+ddI and ZDV+zalcitabine (zal), for patients infected with the human immunodeficiency virus (HIV) and of CD4 T cell counts between 200 and 500 per cubic millimeter \citep{hammer1996trial}. Here, we focus on a subset of 1046 patients  who receive the treatment, ZDV + ddI, denoted by $A=1$ or the control, ZDV + zal, denoted by $A=0$. We choose age as the covariate $\mathbf{X}$ as suggested in \citet{fan2017concordance}, consider the CD4 T cell count at 20+/-5 weeks as the response $Y$, and compare the performance of the proposed method in Algorithm 3 with {$C_0=0.01$} and the subbagging proposed in \citet{shi2020breaking} with $K_0=4$.

From Table \ref{real}, we see that the length of the 95\% confidence interval from the proposed method is shorter than that from the subbagging, which is consistent to Theorem \ref{opt-ralu} and the simulation results. The length reduction is about 5\%, and such a significant reduction implies that nonregularity might exist in this study, which is in accordance with previous findings in \citet{fan2017concordance} and \citet{shi2020breaking}. Here, the length reduction is smaller than that in simulation, which is due to the relatively balanced design in ACTG175 \citep{hammer1996trial} as discussed in Eq. \eqref{Vsb-Vas}. The efficiency gain with the proposed method also helps generate statistical evidence in a more trustworthy and powerful way. Consider a toy example where we test the null hypothesis that the average difference of the CD4 T cell count under OTR is 387. Then, the proposed method can reject the null hypothesis while the subbagging can not under the significance level 0.05. 

\section{Discussion}
Statistical inference on the mean outcome under OTR plays an essential role in precision medicine, and how to address challenges brought by potential nonregularity in the inference is a critical question in drug development and regulatory decision-making. In this paper, we propose an adaptive smoothing approach to infer the mean outcome under OTR regardless of whether the regularity assumption is satisfied or not. The proposed method is valid, easy-to-compute and efficiency-optimal within the robust asymptotically linear unbiased estimators. Revisiting the ACTG 175 trial, we not only show that the proposed method can lead to more trustworthy statistical findings in a more efficient way, but also demonstrate that before any statistical inference is made on the mean outcome under OTR, we must appropriately account for potential nonregularity. 

Theoretically, this paper paves a novel way to establish efficiency optimality theories for OTR in the general scenario allowing nonregularity via variational principle. In particular, we derive a variance lower bound for the class of robust asymptotically linear unbiased estimators which is free of path-wise differentiability required in classical semiparametric theories and can be well defined under nonregularity in OTR. Under regularity, the proposed class includes the efficient regular estimators and the derived variance lower bound reduces to the classical semiparametric efficiency bound in inference on the mean outcome under OTR. 

In this paper, we focus on the binary OTR without cost constraints. However, several more complex scenarios of OTR are also of interest in practice. First, the OTR might be defined with cost constraints as in precision medicine, we sometimes need to strike a balance between the benefit and the cost \citep{illenberger2023identifying}. Second, the OTR might be considered when the treatment is of continuous values \citep{ai2024data}. Third, the OTR may be restricted to a parametric class, such as linear index rules, that might be misspecified and induce additional bias \citep{wu2021resampling}. When nonregularity is allowed, how to extend the proposed adaptive smoothing approach and efficiency optimality theories to infer the mean outcome under constrained OTR, continuous OTR and parametric OTR requires further investigation.

\section*{Data Availability}
The ACTG 175 data analysed in this article are publicly available as the
\texttt{ACTG175} data set in the R package \texttt{speff2trial}, version
1.0.5, through the Comprehensive R Archive Network at
\url{https://CRAN.R-project.org/package=speff2trial}.

\section*{Funding}
This work was supported by the Research Grants Council of the Hong Kong Special Administrative Region, China [HKUST 26308323, HKUST 16310125]; the Seed Fund of the Big Data for Bio-Intelligence Laboratory [Z0428]; and the Hong Kong University of Science and Technology [L0438].

\section*{Conflict of Interest}
The authors declare no conflict of interest.

\begin{table}[H]
    \centering
    \scriptsize
     \caption{Simulation Settings}
\begin{tabular}{cccccc}
\hline \hline & $(\mathrm{A})$ & $(\mathrm{B})$ & $(\mathrm{C})$ & $(\mathrm{D})$ & (E)  \\
\hline $X_1$ & $\operatorname{Bernoulli}(0.4)$ & $\operatorname{Bernoulli}(0.5)$ & $\operatorname{Bernoulli}(0.3)$ & $\operatorname{Bernoulli}(0.5)$ & $\operatorname{Bernoulli}(0.1)$\\
\hline $X_2$ & $\operatorname{Bernoulli}(0.5)$ & $\operatorname{Bernoulli}(0.5)$ & $\operatorname{Uniform}[-2,2]$ & $\operatorname{Uniform}[-2,2]$ & $\operatorname{Uniform}[-2,2]$ \\
\hline$\Phi\left(x_1, x_2\right)$ & 0.3 & 0.3 & $(x_2/4)^2$ & $x_2^2$ & $(x_2/4)^2$  \\
\hline$\tau\left(x_1, x_2\right)$ & $0.4 x_1$ & 0.4 & $x_1 (x_2/4)^2$ & $x_2^2-4 / 3$ & $ x_1 \cos \left(\pi x_2 / 4\right)/8$ \\
\hline $\pi(1,x_1,x_2)$ & $0.7+0.1x_1$ & $0.5+0.3x_1$ & 0.8 & $0.5+0.3x_1$ & 0.8 \\
\hline$V_0$ & 0.46 & 0.7 & 13/120 & 1.85 & $1/(40\pi) + 1/12$ \\
\hline & nonregular & regular & nonregular & regular & nonregular
\\\hline
\end{tabular}
\label{sin-setting}
\end{table}
\begin{table}[H]
\centering
\scriptsize
    \caption{ECP and AL of the CIs with standard error smaller than 0.01.}
\begin{tabular}{|c|c|c|c|c|c|c|}
\hline \multirow{2}{*}{Settings}  & \multicolumn{2}{c|}{ Proposed} & \multicolumn{2}{c|}{ Subbagging } & \multicolumn{2}{c|}{ SSS }\\
\cline{2-7}   & $\operatorname{ECP}(\%)$ & AL*100 & $\operatorname{ECP}(\%)$ & AL*100 & $\operatorname{ECP}(\%)$ & AL*100 \\
\hline (A)  & $95.0$ & $7.1$ & $94.4$ & $8.2$ & $95.4$ & $12.6$ \\
\hline (B)  & $95.6$ & $8.0$ & $94.2$ & $8.0$ & $95.6$ & $11.2$ \\
\hline (C)  & $94.4$ & $7.2$ & $94.2$ & $9.4$ & $94.0$ & $15.2$ \\
\hline (D)  & $95.6$ & $26.2$ & $95.0$ & $26.2$ & $94.8$ & $37.0$ \\
\hline (E) & $94.4$ & $7.2$ & $91.6$ & $9.6$ & $93.6$ & $15.6$\\
\hline
\end{tabular}
    \label{Single-sim}
\end{table}
\begin{table}[H]
    \centering
    \scriptsize
        \caption{Estimated value functions and confidence intervals}
    \label{real}
   \begin{tabular}{cccc}
\hline & Estimated value function & $95 \%$ CI & Length of CI \\
\hline Proposed & 398.8 & {$[387.7,409.9]$} & 22.3 \\
\hline Subbagging & 398.4 & {$[386.7,410.2]$} & 23.5 \\
\hline
\end{tabular}
\end{table}

\bibliographystyle{apalike}
\bibliography{reference}

\clearpage
\begin{center}
{\LARGE\bfseries SUPPLEMENTARY MATERIAL}
\end{center}
In Appendix \ref{appA}, we present additional details of the single time point case, including the proofs of all the theoretical results in the single time point case, the sensitivity 
analysis of tuning selection, and the extensions of the methodological and theoretical results to the case of heteroscedasticity together with 
additional simulation studies examining heteroscedasticity, $\alpha < 1$ and multi-dimensional covariates. In Appendix \ref{appB}, 
we present the setting, methodology, theoretical results and simulation studies in the multiple time point case. In Appendix \ref{appC}, we 
present proofs of all the theoretical results in the multiple time point case. In Appendix \ref{appD}, we extend the methodological and 
theoretical results in the multiple time point case to the case of heteroscedasticity and provide additional simulation studies.

\appendix
\counterwithin{theorem}{section}
\setcounter{equation}{25}
\section{Additional Details and Proofs of Theoretical Results in Single Time Point Case}\label{appA}
In this section, we focus on the theoretical results in the single time point case.
\subsection{Asymptotic Normality for Plug-in Estimator in Regular Case}
{For the inverse-weighted terms below, we also use a fixed fitted-overlap condition \citep{chernozhukov2018double,shi2020breaking}, consistent with the common practice of bounding or truncating fitted propensity scores to control extreme inverse weights \citep{kang2007demystifying,tan2010bounded,ju2019adaptive}: for some constant $c_{\mathrm f}\in(0,1/2)$, with probability tending to one over the training data, $\widehat{\pi}_{\mathcal I}(a,\mathbf X)\in(c_{\mathrm f},1-c_{\mathrm f})$ $P_{\mathbf X}$-a.s. for $a=0,1$ and all fitted folds $\mathcal I$. The corresponding condition is used for the stage-specific fitted propensity scores in the multiple time point case, and $c_{\mathrm f}$ need not be known.}
\begin{proof}
The justification is mainly based on Assumption 1 to 4 in the main text {and the fitted-overlap condition above}. In specific, recall that
    \begin{align*}
          \widehat{V}_0=\frac{1}{n}\sum_{i\in\mathcal{I}} \psi_i\left(\widehat{d}_{\mathcal{I}^c}, \widehat{\pi}_{\mathcal{I}^c}, \widehat{\mu}_{\mathcal{I}^c}\right)+\frac{1}{n}\sum_{i\in\mathcal{I}^c} \psi_i\left(\widehat{d}_{\mathcal{I}}, \widehat{\pi}_{\mathcal{I}}, \widehat{\mu}_{\mathcal{I}}\right),\quad \mathcal{I}\subset[n] \text{ and } |\mathcal{I}|=\lfloor n/2\rfloor.
    \end{align*}
    To show that $\sqrt{n}(\widehat{V}_0-V_0)$ is asymptotically normal with zero mean in the regular case when $P(\tau(\mathbf{X})=0)=0$, it suffice to show that
    \begin{align}\label{regular-1}
        \left|\widehat{V}_0 - \frac{1}{n}\sum_{i\in\mathcal{I}} \psi_i\left(\widehat{d}_{\mathcal{I}^c}, \pi, \mu\right)-\frac{1}{n}\sum_{i\in\mathcal{I}^c} \psi_i\left(\widehat{d}_{\mathcal{I}}, \pi, \mu\right)\right|=o_p(n^{-1/2})
    \end{align}
    and
    \begin{align}\label{regular-2}
    \begin{aligned}
          \left|\frac{1}{n}\sum_{i\in\mathcal{I}} \psi_i\left(\widehat{d}_{\mathcal{I}^c}, \pi, \mu\right)+\frac{1}{n}\sum_{i\in\mathcal{I}^c} \psi_i\left(\widehat{d}_{\mathcal{I}}, \pi, \mu\right)\right.\\
        \left.-\underbrace{\left(\frac{1}{n}\sum_{i\in\mathcal{I}} \psi_i\left(d_0^{opt}, \pi, \mu\right)+\frac{1}{n}\sum_{i\in\mathcal{I}^c} \psi_i\left(d_0^{opt}, \pi, \mu\right)\right)}_{\widehat{V}_{or}}\right|=o_p(n^{-1/2}).
    \end{aligned}
    \end{align}
    To see Eq. \eqref{regular-1}, we notice the following decomposition
    \begin{align}
        &\frac{1}{n}\sum_{i\in\mathcal{I}} \psi_i\left(\widehat{d}_{\mathcal{I}^c}, \widehat{\pi}_{\mathcal{I}^c}, \widehat{\mu}_{\mathcal{I}^c}\right) - \frac{1}{n}\sum_{i\in\mathcal{I}} \psi_i\left(\widehat{d}_{\mathcal{I}^c}, \pi, \mu\right)\label{regular-1.0}\\
        =&\frac{1}{n}\sum_{i\in\mathcal{I}}\left(\widehat{\mu}_{\mathcal{I}^c}(1,\mathbf{X}_i)-\mu(1,\mathbf{X}_i)\right)\left(1-\frac{A_i}{\pi(1,\mathbf{X}_i)}\right)\widehat{d}_{\mathcal{I}^c}(\mathbf{X}_i)\label{regular-1.1}\\
        +&\frac{1}{n}\sum_{i\in\mathcal{I}}A_i\left(Y_i-\mu(1,\mathbf{X}_i)\right)\left(\frac{1}{\widehat{\pi}_{\mathcal{I}^c}(1,\mathbf{X}_i)}-\frac{1}{\pi(1,\mathbf{X}_i)}\right)\widehat{d}_{\mathcal{I}^c}(\mathbf{X}_i)\label{regular-1.2}\\
        +&\frac{1}{n}\sum_{i\in\mathcal{I}}A_i\left(\widehat{\mu}_{\mathcal{I}^c}(1,\mathbf{X}_i)-\mu(1,\mathbf{X}_i)\right)\left(\frac{1}{\widehat{\pi}_{\mathcal{I}^c}(1,\mathbf{X}_i)}-\frac{1}{\pi(1,\mathbf{X}_i)}\right)\widehat{d}_{\mathcal{I}^c}(\mathbf{X}_i)\label{regular-1.3}\\
        +&\frac{1}{n}\sum_{i\in\mathcal{I}}\left(\widehat{\mu}_{0,\mathcal{I}^c}(\mathbf{X}_i)-\mu(0,\mathbf{X}_i)\right)\left(1-\frac{1-A_i}{\pi(0,\mathbf{X}_i)}\right)\left(1-\widehat{d}_{\mathcal{I}^c}(\mathbf{X}_i)\right)\label{regular-1.4}\\
        +&\frac{1}{n}\sum_{i\in\mathcal{I}}(1-A_i)\left(Y_i-\mu(0,\mathbf{X}_i)\right)\left(\frac{1}{\widehat{\pi}_{\mathcal{I}^c}(0,\mathbf{X}_i)}-\frac{1}{\pi(0,\mathbf{X}_i)}\right)\left(1-\widehat{d}_{\mathcal{I}^c}(\mathbf{X}_i)\right)\label{regular-1.5}\\
        +&\frac{1}{n}\sum_{i\in\mathcal{I}}(1-A_i)\left(\widehat{\mu}_{\mathcal{I}^c}(0,\mathbf{X}_i)-\mu(0,\mathbf{X}_i)\right)\left(\frac{1}{\widehat{\pi}_{\mathcal{I}^c}(0,\mathbf{X}_i)}-\frac{1}{\pi(0,\mathbf{X}_i)}\right)\left(1-\widehat{d}_{\mathcal{I}^c}(\mathbf{X}_i)\right).\label{regular-1.6}
    \end{align}
    For Eq. \eqref{regular-1.1}, we have
    \begin{align*}
       & E[\text{Eq. }\eqref{regular-1.1}^2|\widehat{d}_{\mathcal{I}^c}, \widehat{\pi}_{\mathcal{I}^c}, \widehat{\mu}_{\mathcal{I}^c}]\\
       =&\frac{|\mathcal{I}|}{n^2} E\left[\left|\left(\widehat{\mu}_{\mathcal{I}^c}(1,\mathbf{X}_i)-\mu(1,\mathbf{X}_i)\right)\left(1-\frac{A_i}{\pi(1,\mathbf{X}_i)}\right)\widehat{d}_{\mathcal{I}^c}(\mathbf{X}_i)\right|^2\bigg|\widehat{d}_{\mathcal{I}^c}, \widehat{\pi}_{\mathcal{I}^c}, \widehat{\mu}_{\mathcal{I}^c}\right]\\
       +&\frac{1}{n^2}\sum_{i,j\in\mathcal{I},i\neq j}E\left[\left(\widehat{\mu}_{\mathcal{I}^c}(1,\mathbf{X}_i)-\mu(1,\mathbf{X}_i)\right)\left(1-\frac{A_i}{\pi(1,\mathbf{X}_i)}\right)\widehat{d}_{\mathcal{I}^c}(\mathbf{X}_i)\bigg|\widehat{d}_{\mathcal{I}^c}, \widehat{\pi}_{\mathcal{I}^c}, \widehat{\mu}_{\mathcal{I}^c}\right]\\
       &\cdot E\left[\left(\widehat{\mu}_{\mathcal{I}^c}(1,\mathbf{X}_j)-\mu(1,\mathbf{X}_j)\right)\left(1-\frac{A_j}{\pi(1,\mathbf{X}_j)}\right)\widehat{d}_{\mathcal{I}^c}(\mathbf{X}_j)\bigg|\widehat{d}_{\mathcal{I}^c}, \widehat{\pi}_{\mathcal{I}^c}, \widehat{\mu}_{\mathcal{I}^c}\right]\\
       =&\frac{|\mathcal{I}|}{n^2} E\left[\left|\left(\widehat{\mu}_{\mathcal{I}^c}(1,\mathbf{X}_i)-\mu(1,\mathbf{X}_i)\right)\left(1-\frac{A_i}{\pi(1,\mathbf{X}_i)}\right)\widehat{d}_{\mathcal{I}^c}(\mathbf{X}_i)\right|^2\bigg|\widehat{d}_{\mathcal{I}^c}, \widehat{\pi}_{\mathcal{I}^c}, \widehat{\mu}_{\mathcal{I}^c}\right]\\
       =&\frac{|\mathcal{I}|}{n^2} O\left(E\left[\left(\widehat{\mu}_{\mathcal{I}^c}(1,\mathbf{X}_i)-\mu(1,\mathbf{X}_i)\right)^2\bigg|\widehat{d}_{\mathcal{I}^c}, \widehat{\pi}_{\mathcal{I}^c}, \widehat{\mu}_{\mathcal{I}^c}\right]\right)=o_p(n^{-1})
    \end{align*}
    where the last two "=" is due to the overlap assumption and the consistency of $\widehat{\mu}$ in Assumption 4. As a result, Eq. \eqref{regular-1.1} is $o_p(n^{-1/2})$. For Eq. \eqref{regular-1.2}, we have
    \begin{align*}
        & E[\text{Eq. }\eqref{regular-1.2}^2|\widehat{d}_{\mathcal{I}^c}, \widehat{\pi}_{\mathcal{I}^c}, \widehat{\mu}_{\mathcal{I}^c}]\\
        =&\frac{|\mathcal{I}|}{n^2} E\left[\left|A_i\left(Y_i-\mu(1,\mathbf{X}_i)\right)\left(\frac{1}{\widehat{\pi}_{\mathcal{I}^c}(1,\mathbf{X}_i)}-\frac{1}{\pi(1,\mathbf{X}_i)}\right)\widehat{d}_{\mathcal{I}^c}(\mathbf{X}_i)\right|^2\bigg|\widehat{d}_{\mathcal{I}^c}, \widehat{\pi}_{\mathcal{I}^c}, \widehat{\mu}_{\mathcal{I}^c}\right]\\
       +&\frac{1}{n^2}\sum_{i,j\in\mathcal{I},i\neq j}E\left[A_i\left(Y_i-\mu(1,\mathbf{X}_i)\right)\left(\frac{1}{\widehat{\pi}_{\mathcal{I}^c}(1,\mathbf{X}_i)}-\frac{1}{\pi(1,\mathbf{X}_i)}\right)\widehat{d}_{\mathcal{I}^c}(\mathbf{X}_i)\bigg|\widehat{d}_{\mathcal{I}^c}, \widehat{\pi}_{\mathcal{I}^c}, \widehat{\mu}_{\mathcal{I}^c}\right]\\
       &\cdot E\left[A_j\left(Y_j-\mu(1,\mathbf{X}_j)\right)\left(\frac{1}{\widehat{\pi}_{\mathcal{I}^c}(1,\mathbf{X}_j)}-\frac{1}{\pi(1,\mathbf{X}_j)}\right)\widehat{d}_{\mathcal{I}^c}(\mathbf{X}_j)\bigg|\widehat{d}_{\mathcal{I}^c}, \widehat{\pi}_{\mathcal{I}^c}, \widehat{\mu}_{\mathcal{I}^c}\right]\\
       =&\frac{|\mathcal{I}|}{n^2} E\left[\left|A_i\left(Y_i-\mu(1,\mathbf{X}_i)\right)\left(\frac{1}{\widehat{\pi}_{\mathcal{I}^c}(1,\mathbf{X}_i)}-\frac{1}{\pi(1,\mathbf{X}_i)}\right)\widehat{d}_{\mathcal{I}^c}(\mathbf{X}_i)\right|^2\bigg|\widehat{d}_{\mathcal{I}^c}, \widehat{\pi}_{\mathcal{I}^c}, \widehat{\mu}_{\mathcal{I}^c}\right]\\
       =&\frac{|\mathcal{I}|}{n^2} O\left(E\left[\left(\frac{1}{\widehat{\pi}_{\mathcal{I}^c}(1,\mathbf{X}_i)}-\frac{1}{\pi(1,\mathbf{X}_i)}\right)^2\bigg|\widehat{d}_{\mathcal{I}^c}, \widehat{\pi}_{\mathcal{I}^c}, \widehat{\mu}_{\mathcal{I}^c}\right]\right)=o_p(n^{-1})
    \end{align*}
    where the last two "=" is due to the boundedness of second moment of $Y$ and the consistency of $\widehat{\pi}$ in Assumption 4 {together with the fitted-overlap condition above}.  As a result, Eq. \eqref{regular-1.1} is $o_p(n^{-1/2})$. For Eq. \eqref{regular-1.3}, we have
    \begin{align*}
       &|\eqref{regular-1.3}|
       \leq \sqrt{\frac{1}{n}\sum_{i\in\mathcal{I}}\left(\widehat{\mu}_{\mathcal{I}^c}(1,\mathbf{X}_i)-\mu(1,\mathbf{X}_i)\right)^2}\cdot\sqrt{\frac{1}{n}\sum_{i\in\mathcal{I}}\left(\frac{1}{\widehat{\pi}_{\mathcal{I}^c}(1,\mathbf{X}_i)}-\frac{1}{\pi(1,\mathbf{X}_i)}\right)^2}=o_p(n^{-1/2})
    \end{align*}
    where the last "=" is due to Assumption 4 {and the fitted-overlap condition above}. Similarly, we can show that Eq. \eqref{regular-1.4}, Eq. \eqref{regular-1.5} and Eq. \eqref{regular-1.6} are all $o_p(n^{-1/2})$.

    Now for term Eq. \eqref{regular-2}, we observe that
    \begin{align}
        &\frac{1}{n}\sum_{i\in\mathcal{I}} \psi_i\left(\widehat{d}_{\mathcal{I}^c}, \pi, \mu\right) - \frac{1}{n}\sum_{i\in\mathcal{I}} \psi_i\left(d_0^{opt}, \pi, \mu\right)\label{regular-2.0}\\
        =&\frac{1}{n}\sum_{i\in\mathcal{I}}\frac{(2A_i-1)(\widehat{d}_{\mathcal{I}^c}(\mathbf{X}_i)-d_0^{opt}(\mathbf{X}_i))}{\pi(A_i,\mathbf{X}_i)}\{Y_i-\mu(A_i,\mathbf{X}_i)\}\label{regular-2.1}\\
         +&\frac{1}{n}\sum_{i\in\mathcal{I}}\tau(\mathbf{X}_i)(\widehat{d}_{\mathcal{I}^c}(\mathbf{X}_i)-d_0^{opt}(\mathbf{X}_i)).\label{regular-2.2}
    \end{align}
    For Eq. \eqref{regular-2.1}, similar to what we do for Eq. \eqref{regular-1.1} and Eq. \eqref{regular-1.3}, we can show that $E[\eqref{regular-2.1}^2|\widehat{d}_{\mathcal{I}^c}]=o_p(n^{-1})$ and thus Eq. \eqref{regular-2.1} is $o_p(n^{-1/2})$. For term Eq. \eqref{regular-2.2}, we use the margin assumption, the lower and upper bounds on the smoothing scale, and the moment-comparability condition in Eq. (10) of the main text. In particular, the two bounds on the smoothing scale give an $h_{n,\mathcal{I}^c}$ such that
    \begin{align}\label{Eq.10 bound}
        n^{1/4}\left(E[|\widehat{\tau}_{\mathcal{I}^c}(\mathbf{X})-\tau(\mathbf{X})|^2]\right)^{3/4}\prec_p h_{n,\mathcal{I}^c}\prec n^{-\frac{1}{2(1+\alpha)}}.
    \end{align}
    Then we have
    \begin{align}
        \text{Eq. }\eqref{regular-2.2}=&\frac{1}{n}\sum_{i\in\mathcal{I}}\tau(\mathbf{X}_i)(\mathbb{I}\{\widehat{\tau}_{\mathcal{I}^c}(\mathbf{X}_i)>0\}-\mathbb{I}\{\tau(\mathbf{X}_i)>0\})\notag\\
        =&\frac{1}{n}\sum_{i\in\mathcal{I},|\tau(\mathbf{X}_i)|\leq 2 h_{n,\mathcal{I}^c}}\tau(\mathbf{X}_i)(\mathbb{I}\{\widehat{\tau}_{\mathcal{I}^c}(\mathbf{X}_i)>0\}-\mathbb{I}\{\tau(\mathbf{X}_i)>0\})\label{regular-2.2.1}\\
        +&\frac{1}{n}\sum_{i\in\mathcal{I},|\tau(\mathbf{X}_i)|\ge 2 h_{n,\mathcal{I}^c}, |\tau(\mathbf{X}_i)-\widehat{\tau}_{\mathcal{I}^c}(\mathbf{X}_i)|\leq h_{n,\mathcal{I}^c}}\tau(\mathbf{X}_i)(\mathbb{I}\{\widehat{\tau}_{\mathcal{I}^c}(\mathbf{X}_i)>0\}-\mathbb{I}\{\tau(\mathbf{X}_i)>0\})\label{regular-2.2.2}\\
        +&\frac{1}{n}\sum_{i\in\mathcal{I},|\tau(\mathbf{X}_i)|\ge 2 h_{n,\mathcal{I}^c}, |\tau(\mathbf{X}_i)-\widehat{\tau}_{\mathcal{I}^c}(\mathbf{X}_i)|\ge h_{n,\mathcal{I}^c}}\tau(\mathbf{X}_i)(\mathbb{I}\{\widehat{\tau}_{\mathcal{I}^c}(\mathbf{X}_i)>0\}-\mathbb{I}\{\tau(\mathbf{X}_i)>0\})\label{regular-2.2.3}
    \end{align}
    For Eq. \eqref{regular-2.2.1}, by Assumption 5 and the upper bound in Eq. \eqref{Eq.10 bound} we have
    \begin{align*}
        |\text{Eq. }\eqref{regular-2.2.1}|=o_p(n^{-1/2})+O\left(E[\tau(\mathbf{X})\mathbb{I}\{|\tau(\mathbf{X})|\leq h_{n,\mathcal{I}^c}\}]\right)=o_p(n^{-1/2}).
    \end{align*}
    It is clear that Eq. \eqref{regular-2.2.2} is 0, and for Eq. \eqref{regular-2.2.3}, we have
    \begin{align*}
        \text{Eq. }\eqref{regular-2.2.3}=&\frac{1}{n}\sum_{i\in\mathcal{I},|\tau(\mathbf{X}_i)|\ge 2 h_{n,\mathcal{I}^c}, |\tau(\mathbf{X}_i)-\widehat{\tau}_{\mathcal{I}^c}(\mathbf{X}_i)|\ge h_{n,\mathcal{I}^c}}\widehat{\tau}_{\mathcal{I}^c}(\mathbf{X}_i) \mathbb{I}\{\widehat{\tau}_{\mathcal{I}^c}(\mathbf{X}_i)>0\} - \tau(\mathbf{X}_i)\mathbb{I}\{\tau(\mathbf{X}_i)>0\}\\
        +&\frac{1}{n}\sum_{i\in\mathcal{I},|\tau(\mathbf{X}_i)|\ge 2 h_{n,\mathcal{I}^c}, |\tau(\mathbf{X}_i)-\widehat{\tau}_{\mathcal{I}^c}(\mathbf{X}_i)|\ge h_{n,\mathcal{I}^c}}\left(\tau(\mathbf{X}_i)-\widehat{\tau}_{\mathcal{I}^c}(\mathbf{X}_i)\right)\mathbb{I}\{\widehat{\tau}_{\mathcal{I}^c}(\mathbf{X}_i)>0\}.
    \end{align*}
    Since for any $x_1$, $x_2\in\mathbb{R}$
    \begin{align*}
       | x_1\mathbb{I}\{x_1>0\}-x_2\mathbb{I}\{x_2>0\}|\leq |x_1-x_2|,
    \end{align*}
    we have
    \begin{align*}
        \text{Eq. }\eqref{regular-2.2.3}\leq& \frac{1}{n}\sum_{i\in\mathcal{I}}|\tau(\mathbf{X}_i)-\widehat{\tau}_{\mathcal{I}^c}(\mathbf{X}_i)|\mathbb{I}\{|\tau(\mathbf{X}_i)-\widehat{\tau}_{\mathcal{I}^c}(\mathbf{X}_i)|\ge h_{n,\mathcal{I}^c}\}\\
        =&o_p(n^{-1/2})+E\left[|\tau(\mathbf{X})-\widehat{\tau}_{\mathcal{I}^c}(\mathbf{X})|\mathbb{I}\{|\tau(\mathbf{X})-\widehat{\tau}_{\mathcal{I}^c}(\mathbf{X})|\ge h_{n,\mathcal{I}^c}\}\right]\\
        \leq&o_p(n^{-1/2})+\frac{E|\tau(\mathbf{X})-\widehat{\tau}_{\mathcal{I}^c}(\mathbf{X})|^3}{(h_{n,\mathcal{I}^c})^2}\\
        =&o_p(n^{-1/2})+O_p\left(\frac{\{E|\tau(\mathbf{X})-\widehat{\tau}_{\mathcal{I}^c}(\mathbf{X})|^2\}^{3/2}}{(h_{n,\mathcal{I}^c})^2}\right)=o_p(n^{-1/2}).
    \end{align*}
    Here $|u|\mathbb{I}\{|u|\ge h\}\le |u|^3/h^2$, the penultimate equality uses Eq. (10) at $q=3$, and the last equality follows from Eq. \eqref{Eq.10 bound}. In contrast, \citet{shi2020breaking} use the second-moment Markov route through Theorem 3.1 of \citet{qian2011performance}; the third-moment route used here trades the additional moment-comparability condition for a weaker rate requirement. Therefore, Eq. \eqref{regular-2.2} is $o_p(n^{-1/2})$.
\end{proof}
However, in the nonregular case, the naive cross-fitted estimator does not work due to the inconsistency of the estimated decision rule, as is illustrated in Example \ref{eg1} below.
\begin{example}\label{eg1}
Assume there are 400 patients with 200 males and 200 females randomly assigned with treatment and control. The potential outcomes are defined as follows, (1) outcomes of female patients under treatment follow $N(1,1)$; (2) outcomes of the others follow $N(0,1)$. Individualized treatment effect is estimated by difference in means estimator, i.e.,
    \begin{align}\label{eg1-did}
        \widehat{\tau}(G)=\frac{\sum_{i=1}^n Y_i\mathbb{I}\{\mathbf{X}_i=G,A=1\}}{\sum_{i=1}^n \mathbb{I}\{\mathbf{X}_i=G,A=1\}} - \frac{\sum_{i=1}^n Y_i\mathbb{I}\{\mathbf{X}_i=G,A=0\}}{\sum_{i=1}^n \mathbb{I}\{\mathbf{X}_i=G,A=0\}}, \quad G\in\{\text{male},\text{female}\}.
    \end{align}
     It is clear that there exists nonregularity in this toy example as for male patients, the treatment and the control have no differences. Figure \ref{toy} illustrates the bias issue under nonregularity.      Figure \ref{fig-eg1}  shows that how the consistency of the estimated decision function is broken under nonregularity and fixed by smoothing.
     \begin{figure}[H]
    \centering
    \includegraphics[width=6cm]{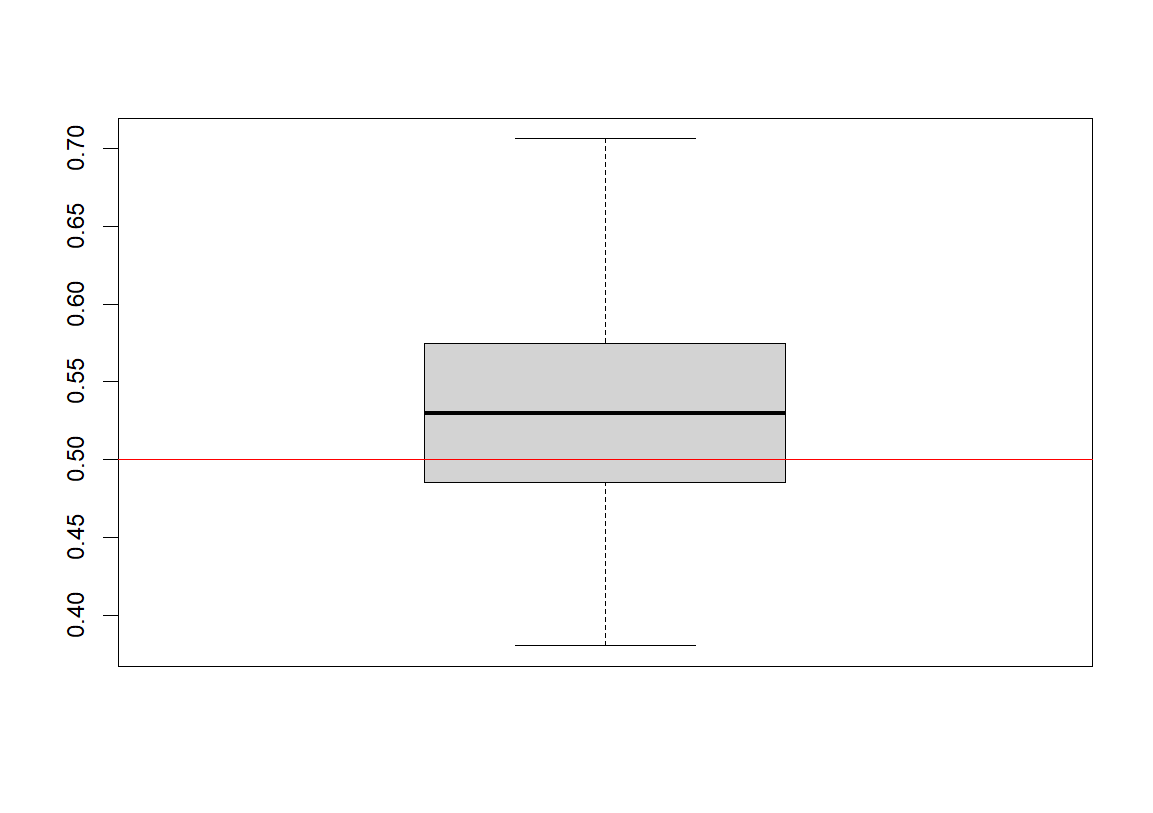}
    \caption{The boxplot of the plug-in estimator in Eq. (4) of the mean outcome under OTR in Example \ref{eg1} where the red line is the mean outcome under the true OTR.}
    \label{toy}
\end{figure}

\begin{figure}[H]
\centering
\begin{minipage}[t]{0.3\textwidth}
\centering
\includegraphics[width=4cm]{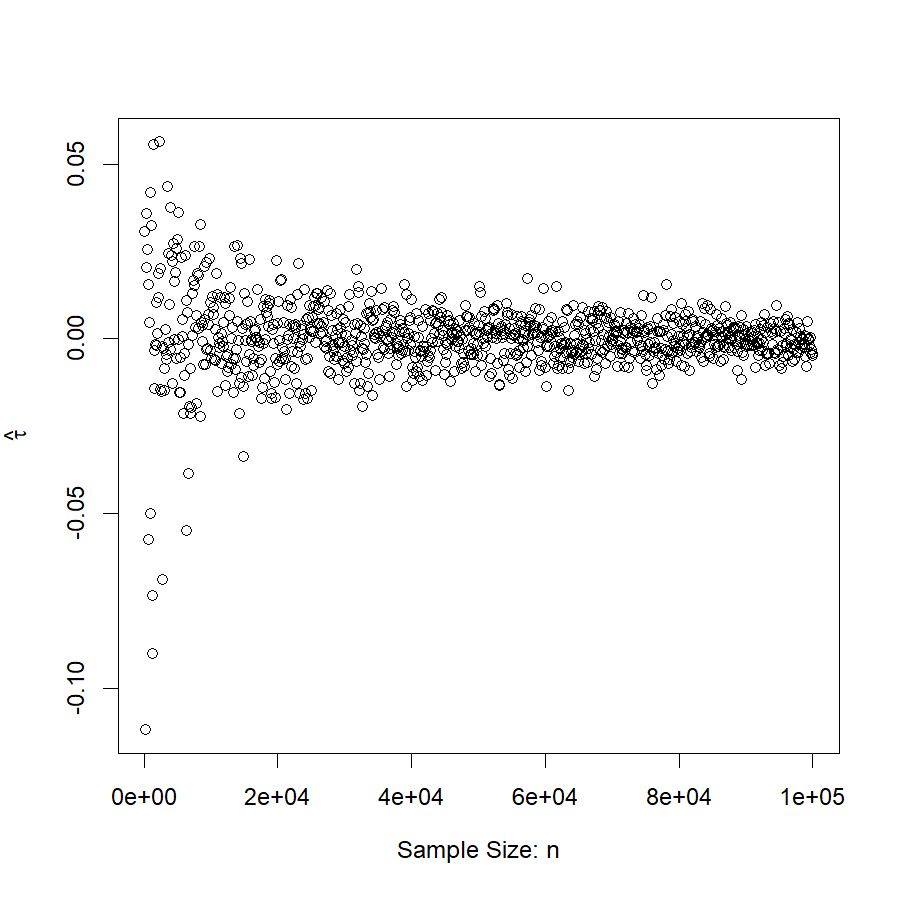}
\end{minipage}
\begin{minipage}[t]{0.3\textwidth}
\centering
\includegraphics[width=4cm]{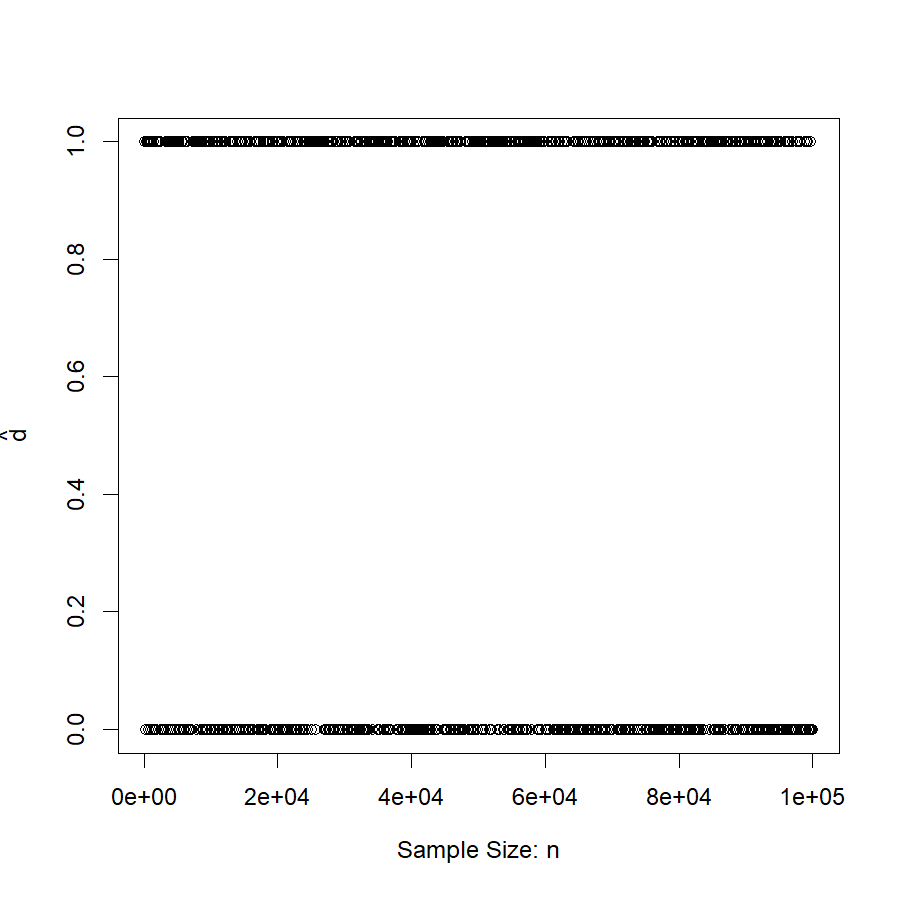}
\end{minipage}
\begin{minipage}[t]{0.3\textwidth}
\centering
\includegraphics[width=4cm]{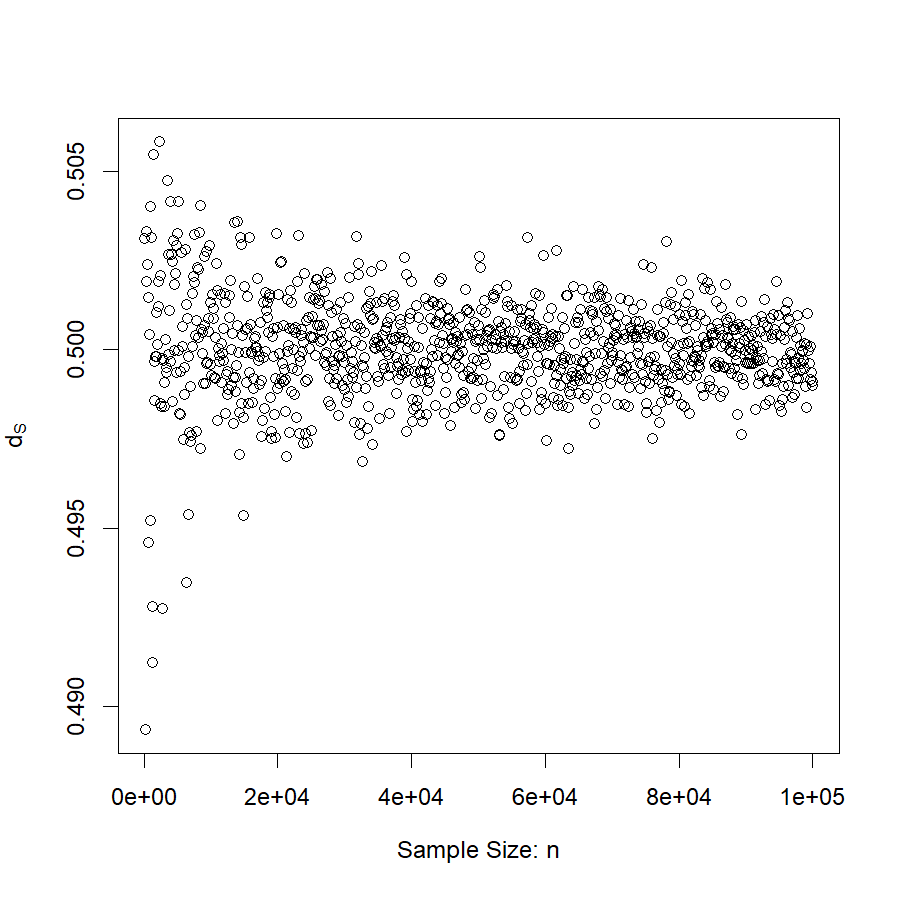}
\end{minipage}
\caption{The left panel is $\widehat{\tau}(\text{male})$, the middle panel is $\widehat{d}(\text{male})$, and the right panel is $d_{s}(\text{male})$ as sample size increases in Example \ref{eg1}.}
\label{fig-eg1}
\end{figure}
\end{example}
\subsection{Lemma \ref{single-lem} and Its Proof}
Define the following class of decision functions:
\begin{align*}
    \mathbb{D}(V_0):=\{\mathcal{D}|\mathcal{D}:\tau(\mathcal{X})\to[0,1]\land E(\psi(\mathcal{D}(\tau(\mathbf{X}_0)),\pi,h))=V_0\}.
\end{align*}
where the expectation is equal to $V_0$. For the purpose of generalization and to cover as many situations as possible, we consider a sequence of decision function  $\{\mathcal{D}_n\}_{n\in\mathbb{N}_+}$ $(\mathcal{X}\times\mathbb{R}\to[0,1])$ with the following properties.
\setcounter{assumption}{9}
\begin{assumption}[Uniform continuity]\label{A10}
    $\exists$ $\delta_n$ s.t. $P_{\mathbf{X}}(|\widehat{\tau}_{\mathcal{I}}(\mathbf{X})-\tau(\mathbf{X})|\ge \delta_n)=o_p(1)$ and $$\sup_{\mathbf{X}\in\mathcal{X},|s-t|\leq \delta_n}|\mathcal{D}_n(\mathbf{X},s)-\mathcal{D}_n(\mathbf{X},t)|\to 0$$ as $n\to\infty$.
\end{assumption}
\begin{assumption}[Consistency]\label{A11}
    $\mathcal{D}(\mathbf{X}):=\displaystyle{\lim_{n\to\infty}}\mathcal{D}_n(\mathbf{X},\tau(\mathbf{X}))\in\mathbb{D}(V_0)$.
\end{assumption}
\begin{assumption}[Concentration]\label{A12}
    $\exists$ $\delta_n'$ s.t. $\delta_n\lor n^{1/4}\left(E[|\widehat{\tau}_{\mathcal{I}_j}(\mathbf{X})-\tau(\mathbf{X})|^2]\right)^{3/4} \prec_p\delta_n'\prec n^{-\frac{1}{2(1+\alpha)}}$ ($\delta_n$ is the same as that in Assumption \ref{A10}) s.t. 
\begin{align}
    |\mathcal{D}_n(\mathbf{X},s)-\mathbb{I}\{s>0\}|=o\left(\frac{1}{{|s|}\sqrt{n}}\right)\label{A10-1}
\end{align}
when $|s|>\delta_n'$.
\end{assumption}
All the above assumptions are set to guarantee the sequence of decision function $\mathcal{D}_n(\cdot, \widehat{\tau}_{\mathcal{I}}(\cdot))$ ($\mathcal{I}$ is certain subset of $[n]$) converges to $\mathcal{D}(\cdot)\in\mathbb{D}(V_0)$ in an uniform and fast enough manner.  Assumption \ref{A10} requires the smoothness of $\mathcal{D}_n$, ensuring the sequence of decision function $\mathcal{D}_n(\cdot, \widehat{\tau}_{\mathcal{I}}(\cdot))$ being convergent. Assumption \ref{A11} indicates that the sequence $\mathcal{D}_n$ is meaningful for the task of inferring $V_0$ as the limit is OTR. Assumption \ref{A12} suggests that $\mathcal{D}_n(\cdot, s)$ should converge to $\mathbb{I}\{s>0\}$ in a faster enough rate when $s$ is away from zero in the same side, thus ensures the convergence rate from $V(\mathcal{D}_n(\mathbf{X}, \widehat{\tau}_{\mathcal{I}}(\mathbf{X})))$ to $V_0$ is $o(n^{-1/2})$. Assumptions \ref{A10} to \ref{A12} can be implied by Assumption 6 and 7 when we justify Theorems 1 and 2.

To facilitate our analysis, we define the following estimators for $V_0$ based on the sequence of decision function, $\mathcal{D}_n$, or its limit,
\begin{align}
\widehat{V}_{\mathcal{D}}^*=&\frac{1}{n}\sum_{i=1}^n \psi_i(\mathcal{D}(\mathbf{X}_i),\pi,\mu),\text{ ($\tau$, $\pi$ and $\mu$ are all true)}\\
\widehat{V}_{\mathcal{D}_n}^*=&\frac{1}{n}\sum_{i=1}^n \psi_i(\mathcal{D}_n(\mathbf{X}_i,\tau(\mathbf{X}_i)),\pi,\mu),\text{ ($\tau$, $\pi$ and $\mu$ are all true)}\label{VD_oracle}\\
    \widehat{V}_{\mathcal{D}_n}=&\frac{1}{4}\sum_{j=1,2}\sum_{k=1,2}\frac{1}{|\mathcal{I}_{j,k}|}\sum_{i\in\mathcal{I}_{j,k}}\psi_i(\mathcal{D}_n(\mathbf{X}_i,\widehat{\tau}_{\mathcal{I}_{3-j}}(\mathbf{X}_i)),\widehat{\pi}_{\mathcal{I}_{j,k}^c},\widehat{\mu}_{\mathcal{I}_{j,k}^c})\label{VD},
\end{align}
and the estimator of standard deviation
\begin{align*}
    \widehat{\sigma}_{\mathcal{D}_n}:=\left(\frac{1}{n-1}\sum_{j=1,2}\sum_{k=1,2}\sum_{i\in\mathcal{I}_{j,k}}\left(\psi_i(\mathcal{D}_n(\mathbf{X}_i,\widehat{\tau}_{\mathcal{I}_{3-j}}(\mathbf{X}_i)),\widehat{\pi}_{\mathcal{I}_{j,k}^c},\widehat{\mu}_{\mathcal{I}_{j,k}^c})- \widehat{V}_{\mathcal{D}_n}\right)^2\right)^{1/2}.
\end{align*}
Lemma \ref{single-lem} shows that the difference between $\widehat{V}_{\mathcal{D}_n}$, the cross-fitting estimator based on the estimated $\tau$, $\pi$ and $\mu$, and $\widehat{V}_{\mathcal{D}_n}^*$, the estimator based on the true $\tau$, $\pi$ and $\mu$, is negligible for inference, and establishes asymptotic normality of $\widehat{V}_{\mathcal{D}_n}$ centered at $V_0$. 

\begin{lemma}\label{single-lem}
 Under Assumptions 1 to 5 and \ref{A10} to \ref{A12}, the moment-comparability condition in Eq. (10) of the main text with $q=3$, {and the fitted-overlap condition above,} Eq. \eqref{VD_oracle} and Eq. \eqref{VD} are asymptotically equivalent: $|\widehat{V}_{\mathcal{D}_n}^* - \widehat{V}_{\mathcal{D}_n}|=o_p(n^{-1/2})$, and $\sqrt{n}(\widehat{V}_{\mathcal{D}_n}-V_0)/\widehat{\sigma}_{\mathcal{D}_n}$ converges weakly to a standard normal distribution and $(\widehat{\sigma}_{\mathcal{D}_n})^2$ converges in probability to
    \begin{align*}
          \operatorname{Var}\left\{\mu\left(d_0^{o p t}, \mathbf{X}\right)\right\}+ \mathrm{E}\left(\frac{\mathbb{I}\left\{\tau\left(\mathbf{X}\right)>0\right\}}{\pi\left(1, \mathbf{X}\right)}\operatorname{Var}(Y|\mathbf{X},A=1)+\frac{\mathbb{I}\left\{\tau\left(\mathbf{X}\right)<0\right\}}{\pi\left(0, \mathbf{X}\right)}\operatorname{Var}(Y|\mathbf{X},A=0)\right)\\
    +\underbrace{\mathrm{E}\left[\left(\frac{(\mathcal{D}(\mathbf{X}))^2}{\pi\left(1, \mathbf{X}\right)}\operatorname{Var}(Y|\mathbf{X},A=1)+\frac{(1-\mathcal{D}(\mathbf{X}))^2}{\pi\left(0, \mathbf{X}\right)}\operatorname{Var}(Y|\mathbf{X},A=0)\right) \mathbb{I}\left\{\tau\left(\mathbf{X}\right)=0\right\}\right]}_{\text{Varying with $\mathcal{D}$}}.
    \end{align*}
\end{lemma}
\begin{proof}
     The key idea is to show that $|\widehat{V}_{\mathcal{D}_n}^* - \widehat{V}_{\mathcal{D}}^*|=o_p(n^{-1/2})$ and $|\widehat{V}_{\mathcal{D}_n} - \widehat{V}_{\mathcal{D}}^*|=o_p(n^{-1/2})$, then the result naturally follows by applying central limit theorem to $\widehat{V}_{\mathcal{D}}^*$.  Firstly, we show that for $j=1,2,$ and $k=1,2$,
    \begin{align}
       &\frac{1}{|\mathcal{I}_{j,k}|}\sum_{i\in\mathcal{I}_{j,k}}\psi_i(\mathcal{D}_n(\mathbf{X}_i,\widehat{\tau}_{\mathcal{I}_{3-j}}(\mathbf{X}_i)),\widehat{\pi}_{\mathcal{I}_{j,k}^c},\widehat{\mu}_{\mathcal{I}_{j,k}^c})\notag\\
       &-\frac{1}{|\mathcal{I}_{j,k}|}\sum_{i\in\mathcal{I}_{j,k}}\psi_i(\mathcal{D}_n(\mathbf{X}_i,\widehat{\tau}_{\mathcal{I}_{3-j}}(\mathbf{X}_i)),\pi,\mu)=o_p(n^{-1/2});\label{thm2_1}\\
        &\frac{1}{|\mathcal{I}_{j,k}|}\sum_{i\in\mathcal{I}_{j,k}}\psi_i(\mathcal{D}_n(\mathbf{X}_i,\widehat{\tau}_{\mathcal{I}_{3-j}}(\mathbf{X}_i)),\pi,\mu)\notag\\
        &-\frac{1}{|\mathcal{I}_{j,k}|}\sum_{i\in\mathcal{I}_{j,k}}\psi_i(\mathcal{D}_n(\mathbf{X}_i,\tau(\mathbf{X}_i)),\pi,\mu)=o_p(n^{-1/2}).\label{thm2_2}
    \end{align}
    The term Eq. \eqref{thm2_1} can be decomposed as follows:
    \begin{align}
        &\text{Eq. }\eqref{thm2_1}\notag\\
        =&\frac{1}{|\mathcal{I}_{j,k}|}\sum_{i\in\mathcal{I}_{j,k}}\left(\widehat{\mu}_{\mathcal{I}_{j,k}^c}(1,\mathbf{X}_i)-\mu(1,\mathbf{X}_i)\right)\left(1-\frac{A_i}{\pi(1,\mathbf{X}_i)}\right)\cdot \mathcal{D}_n(\mathbf{X}_i,\widehat{\tau}_{\mathcal{I}_{3-j}}(\mathbf{X}_i))\label{thm2_1_1}\\
        +&\frac{1}{|\mathcal{I}_{j,k}|}\sum_{i\in\mathcal{I}_{j,k}} A_i (Y_i-\mu(1,\mathbf{X}_i))\left(\frac{1}{\widehat{\pi}_{\mathcal{I}_{j,k}^c}(1,\mathbf{X}_i)}-\frac{1}{\pi(1,\mathbf{X}_i)}\right)\cdot \mathcal{D}_n(\mathbf{X}_i,\widehat{\tau}_{\mathcal{I}_{3-j}}(\mathbf{X}_i))\label{thm2_1_2}\\
        +&\frac{1}{|\mathcal{I}_{j,k}|}\sum_{i\in\mathcal{I}_{j,k}} A_i \left(\widehat{\mu}_{\mathcal{I}_{j,k}^c}(1,\mathbf{X}_i)-\mu(1,\mathbf{X}_i)\right)\left(\frac{1}{\widehat{\pi}_{\mathcal{I}_{j,k}^c}(1,\mathbf{X}_i)}-\frac{1}{\pi(1,\mathbf{X}_i)}\right)\notag\\
        &\cdot \mathcal{D}_n(\mathbf{X}_i,\widehat{\tau}_{\mathcal{I}_{3-j}}(\mathbf{X}_i))\label{thm2_1_3}\\
        +&\frac{1}{|\mathcal{I}_{j,k}|}\sum_{i\in\mathcal{I}_{j,k}}\left(\widehat{\mu}_{\mathcal{I}_{j,k}^c}(0,\mathbf{X}_i)-\mu(0,\mathbf{X}_i)\right)\left(1-\frac{1-A_i}{\pi(0,\mathbf{X}_i)}\right)\cdot (1-\mathcal{D}_n(\mathbf{X}_i,\widehat{\tau}_{\mathcal{I}_{3-j}}(\mathbf{X}_i)))\label{thm2_1_4}\\
        +&\frac{1}{|\mathcal{I}_{j,k}|}\sum_{i\in\mathcal{I}_{j,k}} (1-A_i) (Y_i-\mu(0,\mathbf{X}_i))\left(\frac{1}{\widehat{\pi}_{\mathcal{I}_{j,k}^c}(0,\mathbf{X}_i)}-\frac{1}{\pi(0,\mathbf{X}_i)}\right)\notag\\
        &\cdot (1-\mathcal{D}_n(\mathbf{X}_i,\widehat{\tau}_{\mathcal{I}_{3-j}}(\mathbf{X}_i)))\label{thm2_1_5}\\
        +&\frac{1}{|\mathcal{I}_{j,k}|}\sum_{i\in\mathcal{I}_{j,k}} (1-A_i) \left(\widehat{\mu}_{\mathcal{I}_{j,k}^c}(0,\mathbf{X}_i)-\mu(0,\mathbf{X}_i)\right)\left(\frac{1}{\widehat{\pi}_{\mathcal{I}_{j,k}^c}(0,\mathbf{X}_i)}-\frac{1}{\pi(0,\mathbf{X}_i)}\right)\notag\\
        &\cdot (1-\mathcal{D}_n(\mathbf{X}_i,\widehat{\tau}_{\mathcal{I}_{3-j}}(\mathbf{X}_i)))\label{thm2_1_6}
    \end{align}
   With similar techniques in justifying the double robustness, it is not hard to show that all of Eq. \eqref{thm2_1_1} to Eq. \eqref{thm2_1_6} are $o_p(n^{-1/2})$ under Assumption 4 {and the fitted-overlap condition above}. Specifically, for term Eq. \eqref{thm2_1_1}
    \begin{align*}
        &n\cdot E[\text{Eq. }\eqref{thm2_1_1}^2]\\
        =&E_{\widehat{\mu}_{\mathcal{I}_{j,k}^c}}\left[ \frac{n}{|\mathcal{I}_{j,k}|}E_{\mathbf{X}_i,A_i}\left[\left|\left(\widehat{\mu}_{\mathcal{I}_{j,k}^c}(1,\mathbf{X}_i)-\mu(1,\mathbf{X}_i)\right)\left(1-\frac{A_i}{\pi(1,\mathbf{X}_i)}\right)\cdot \mathcal{D}_n(\mathbf{X}_i,\widehat{\tau}_{\mathcal{I}_{3-j}}(\mathbf{X}_i))\right|^2\right]\right]\\
        \leq & E_{\widehat{\mu}_{\mathcal{I}_{j,k}^c}}\left[ \frac{n}{|\mathcal{I}_{j,k}|}E_{\mathbf{X}_i,A_i}\left[\left|\widehat{\mu}_{\mathcal{I}_{j,k}^c}(1,\mathbf{X}_i)-\mu(1,\mathbf{X}_i)\right|^2\right]\cdot\frac{
        1-c
        }{c}\right]=o(1), \text{ (by Assumption 4)}
    \end{align*}
    for Eq. \eqref{thm2_1_2}, thanks to the overlap, Assumption 3, {and the fitted-overlap condition above,} similarly we have $ n\cdot E[\text{Eq. }\eqref{thm2_1_2}^2]=o(1)$.
    For Eq. \eqref{thm2_1_3} we have,
    \begin{align*}
        |\text{Eq. }\eqref{thm2_1_3}|\leq&\sqrt{\frac{1}{|\mathcal{I}_{j,k}|}\sum_{i\in\mathcal{I}_{j,k}}A_i^2 \left(\widehat{\mu}_{\mathcal{I}_{j,k}^c}(1,\mathbf{X}_i)-\mu(1,\mathbf{X}_i)\right)^2}\\
        &\cdot\sqrt{\frac{1}{|\mathcal{I}_{j,k}|}\sum_{i\in\mathcal{I}_{j,k}}\left(\frac{1}{\widehat{\pi}_{\mathcal{I}_{j,k}^c}(1,\mathbf{X}_i)}-\frac{1}{\pi(1,\mathbf{X}_i)}\right)^2\cdot (\mathcal{D}_n(\mathbf{X}_i,\widehat{\tau}_{\mathcal{I}_{3-j}}(\mathbf{X}_i)))^2}\\
        \leq & \sqrt{\frac{1}{|\mathcal{I}_{j,k}|}\sum_{i\in\mathcal{I}_{j,k}} \left(\widehat{\mu}_{\mathcal{I}_{j,k}^c}(1,\mathbf{X}_i)-\mu(1,\mathbf{X}_i)\right)^2}\\
        &\cdot\sqrt{\frac{1}{|\mathcal{I}_{j,k}|}\sum_{i\in\mathcal{I}_{j,k}}\left(\frac{1}{\widehat{\pi}_{\mathcal{I}_{j,k}^c}(1,\mathbf{X}_i)}-\frac{1}{\pi(1,\mathbf{X}_i)}\right)^2}\\
        &=o_p(n^{-1/2}), \text{ (by Assumption 4)}
    \end{align*}
    for Eq. \eqref{thm2_1_4} we have
    \begin{align*}
        &n\cdot E[\text{Eq. }\eqref{thm2_1_4}^2]\\
        =&E_{\widehat{\mu}_{\mathcal{I}_{j,k}^c}}\left[\frac{n}{|\mathcal{I}_{j,k}|}E_{\mathbf{X}_i,A_i}\left[\left|\left(\widehat{\mu}_{\mathcal{I}_{j,k}^c}(0,\mathbf{X}_i)-\mu(0,\mathbf{X}_i)\right)\left(1-\frac{1-A_i}{\pi(0,\mathbf{X}_i)}\right)\right.\right.\right.\\
        &\left.\left.\left.\cdot (1-\mathcal{D}_n(\mathbf{X}_i,\widehat{\tau}_{\mathcal{I}_{3-j}}(\mathbf{X}_i)))\right|^2 \right]\right]\\
        \leq & E_{\widehat{\mu}_{\mathcal{I}_{j,k}^c}}\left[\frac{n}{|\mathcal{I}_{j,k}|}E_{\mathbf{X}_i,A_i}\left[\left|\widehat{\mu}_{\mathcal{I}_{j,k}^c}(0,\mathbf{X}_i)-\mu(0,\mathbf{X}_i)\right|^2\right]\right]\max\{1,1-1/(1-c)\}\\
        =&o(1), \text{ (by Assumption 4)}
    \end{align*}
    for Eq. \eqref{thm2_1_5}, still by Assumption 3 {and the fitted-overlap condition above} we have $n\cdot E[\text{Eq. }\eqref{thm2_1_5}^2]=o(1)$.
    For Eq. \eqref{thm2_1_6} we have
     \begin{align*}
        |\text{Eq. }\eqref{thm2_1_6}|\leq&\sqrt{\frac{1}{|\mathcal{I}_{j,k}|}\sum_{i\in\mathcal{I}_{j,k}}(1-A_i)^2 \left(\widehat{\mu}_{\mathcal{I}_{j,k}^c}(0,\mathbf{X}_i)-\mu(0,\mathbf{X}_i)\right)^2}\\
        &\cdot\left|\frac{1}{|\mathcal{I}_{j,k}|}\sum_{i\in\mathcal{I}_{j,k}}\left(\frac{1}{\widehat{\pi}_{\mathcal{I}_{j,k}^c}(0,\mathbf{X}_i)}-\frac{1}{\pi(0,\mathbf{X}_i)}\right)^2\cdot (1-\mathcal{D}_n(\mathbf{X}_i,\widehat{\tau}_{\mathcal{I}_{3-j}}(\mathbf{X}_i)))^2\right|^{1/2}\\
        \leq & \sqrt{\frac{1}{|\mathcal{I}_{j,k}|}\sum_{i\in\mathcal{I}_{j,k}} \left(\widehat{\mu}_{\mathcal{I}_{j,k}^c}(0,\mathbf{X}_i)-\mu(0,\mathbf{X}_i)\right)^2}\\
        &\cdot\sqrt{\frac{1}{|\mathcal{I}_{j,k}|}\sum_{i\in\mathcal{I}_{j,k}}\left(\frac{1}{\widehat{\pi}_{\mathcal{I}_{j,k}^c}(0,\mathbf{X}_i)}-\frac{1}{\pi(0,\mathbf{X}_i)}\right)^2}\\
        &=o_p(n^{-1/2}), \text{ (by Assumption 4)}
    \end{align*}
    and therefore, Eq. \eqref{thm2_1} is $o_p(n^{-1/2})$.
    
    The Eq. \eqref{thm2_2} can be decomposed as follows
    \begin{align}
        &\text{Eq. }\eqref{thm2_2}\notag\\
        =&\frac{1}{|\mathcal{I}_{j,k}|}\sum_{i\in\mathcal{I}_{j,k}}\frac{(2A_i-1)(\mathcal{D}_n(\mathbf{X}_i,\widehat{\tau}_{\mathcal{I}_{3-j}}(\mathbf{X}_i))-\mathcal{D}_n(\mathbf{X}_i,\tau(\mathbf{X}_i)))}{\pi(A_i,\mathbf{X}_i)}\{Y_i-\mu(A_i,\mathbf{X}_i)\}\label{thm2_2_1}\\
         +&\frac{1}{|\mathcal{I}_{j,k}|}\sum_{i\in\mathcal{I}_{j,k}}\tau(\mathbf{X}_i)(\mathcal{D}_n(\mathbf{X}_i,\widehat{\tau}_{\mathcal{I}_{3-j}}(\mathbf{X}_i))-\mathcal{D}_n(\mathbf{X}_i,\tau(\mathbf{X}_i))).\label{thm2_2_2}
    \end{align}
    Since 
    \begin{align*}
        \frac{(2A_i-1)(\mathcal{D}_n(\mathbf{X}_i,\widehat{\tau}_{\mathcal{I}_{3-j}}(\mathbf{X}_i))-\mathcal{D}_n(\mathbf{X}_i,\tau(\mathbf{X}_i)))}{\pi(A_i,\mathbf{X}_i)}\{Y_i-\mu(A_i,\mathbf{X}_i)\}, \quad i\in\mathcal{I}_{j,k}
    \end{align*}
 are i.i.d. terms, 
    \begin{align*}
        E\left[\frac{(2A_i-1)(\mathcal{D}_n(\mathbf{X}_i,\widehat{\tau}_{\mathcal{I}_{3-j}}(\mathbf{X}_i))-\mathcal{D}_n(\mathbf{X}_i,\tau(\mathbf{X}_i)))}{\pi(A_i,\mathbf{X}_i)}\{Y_i-\mu(A_i,\mathbf{X}_i)\}\bigg|\mathbf{X},A,\mathcal{I}_{j,k}\right]=0,
    \end{align*}
    and by Assumption \ref{A10}, $E[Y^2]<\infty$ and consistency of $\widehat{\tau}$ required by Assumption \ref{A12}, we have
    \begin{align*}
        \operatorname{Var}\left(\frac{(2A_i-1)(\mathcal{D}_n(\mathbf{X}_i,\widehat{\tau}_{\mathcal{I}_{3-j}}(\mathbf{X}_i))-\mathcal{D}_n(\mathbf{X}_i,\tau(\mathbf{X}_i)))}{\pi(A_i,\mathbf{X}_i)}\{Y_i-\mu(A_i,\mathbf{X}_i)\}\bigg|\mathbf{X},A,\mathcal{I}_{j,k}\right)=o_p(1),
    \end{align*}
    we can conclude that Eq. \eqref{thm2_2_1} is $o_p(n^{-1/2})$. For Eq. \eqref{thm2_2_2}, we have
    \begin{align}
        &\text{Eq. }\eqref{thm2_2_2}\notag\\
        =&\frac{1}{|\mathcal{I}_{j,k}|}\sum_{i\in\mathcal{I}_{j,k},|\tau(\mathbf{X}_i)|<2\delta_n'}\tau(\mathbf{X}_i)(\mathcal{D}_n(\mathbf{X}_i,\widehat{\tau}_{\mathcal{I}_{3-j}}(\mathbf{X}_i))-\mathcal{D}_n(\mathbf{X}_i,\tau(\mathbf{X}_i)))\label{thm2_2_3}\\
        +&\frac{1}{|\mathcal{I}_{j,k}|}\sum_{i\in\mathcal{I}_{j,k},|\tau(\mathbf{X}_i)|\ge2\delta_n'}\tau(\mathbf{X}_i)(\mathcal{D}_n(\mathbf{X}_i,\widehat{\tau}_{\mathcal{I}_{3-j}}(\mathbf{X}_i))-\mathcal{D}_n(\mathbf{X}_i,\tau(\mathbf{X}_i))).\label{thm2_2_4}
    \end{align}
     By Assumptions 5 and \ref{A11}, it is not hard to verify that
    \begin{align*}
        |\text{Eq. }\eqref{thm2_2_3}|\leq \frac{2}{|\mathcal{I}_{j,k}|}\sum_{i\in\mathcal{I}_{j,k},|\tau(\mathbf{X}_i)|<2\delta_n'}|\tau(\mathbf{X}_i)|=O_p((2\delta_n')^{1+\alpha})+o_p(n^{-1/2})=o_p(n^{-1/2})
    \end{align*}
    and by Assumption \ref{A12}
    \begin{align*}
        &\text{Eq. }\eqref{thm2_2_4}\\
        =& \frac{1}{|\mathcal{I}_{j,k}|}\sum_{i\in\mathcal{I}_{j,k},|\tau(\mathbf{X}_i)|\ge2\delta_n'}\tau(\mathbf{X}_i)(\mathcal{D}_n(\mathbf{X}_i,\widehat{\tau}_{\mathcal{I}_{3-j}}(\mathbf{X}_i))-\mathcal{D}_n(\mathbf{X}_i,\tau(\mathbf{X}_i)))\\
        =&\frac{1}{|\mathcal{I}_{j,k}|}\sum_{i\in\mathcal{I}_{j,k},|\tau(\mathbf{X}_i)|\ge2\delta_n',|\tau(\mathbf{X}_i)-\widehat{\tau}_{\mathcal{I}_{3-j}}(\mathbf{X}_i)|>\delta_n'}\tau(\mathbf{X}_i)(\mathcal{D}_n(\mathbf{X}_i,\widehat{\tau}_{\mathcal{I}_{3-j}}(\mathbf{X}_i))-\mathcal{D}_n(\mathbf{X}_i,\tau(\mathbf{X}_i)))\\
        +&\underbrace{\frac{1}{|\mathcal{I}_{j,k}|}\sum_{i\in\mathcal{I}_{j,k},|\tau(\mathbf{X}_i)|\ge2\delta_n',|\tau(\mathbf{X}_i)-\widehat{\tau}_{\mathcal{I}_{3-j}}(\mathbf{X}_i)|\leq\delta_n'}\tau(\mathbf{X}_i)(\mathcal{D}_n(\mathbf{X}_i,\widehat{\tau}_{\mathcal{I}_{3-j}}(\mathbf{X}_i))-\mathcal{D}_n(\mathbf{X}_i,\tau(\mathbf{X}_i)))}_{{\text{The two arguments have the same sign and magnitude at least $\delta_n'$; thus, by Assumption \ref{A12}, this term is $o(n^{-1/2})$}}}\\
        =&\frac{1}{|\mathcal{I}_{j,k}|}\sum_{i\in\mathcal{I}_{j,k},|\tau(\mathbf{X}_i)|\ge2\delta_n',|\tau(\mathbf{X}_i)-\widehat{\tau}_{\mathcal{I}_{3-j}}(\mathbf{X}_i)|>\delta_n'}\widehat{\tau}_{\mathcal{I}_{3-j}}(\mathbf{X}_i)\mathcal{D}_n(\mathbf{X}_i,\widehat{\tau}_{\mathcal{I}_{3-j}}(\mathbf{X}_i)) -\tau(\mathbf{X}_i)\mathcal{D}_n(\mathbf{X}_i,\tau(\mathbf{X}_i))\\
        +&\frac{1}{|\mathcal{I}_{j,k}|}\sum_{i\in\mathcal{I}_{j,k},|\tau(\mathbf{X}_i)|\ge2\delta_n',|\tau(\mathbf{X}_i)-\widehat{\tau}_{\mathcal{I}_{3-j}}(\mathbf{X}_i)|>\delta_n'}(\tau(\mathbf{X}_i)-\widehat{\tau}_{\mathcal{I}_{3-j}}(\mathbf{X}_i))\mathcal{D}_n(\mathbf{X}_i,\widehat{\tau}_{\mathcal{I}_{3-j}}(\mathbf{X}_i))+o(n^{-1/2})
    \end{align*}
By the definition of $\mathcal{D}_n$, when $|\tau(\mathbf{X}_i)|\ge2\delta_n'$ and $|\tau(\mathbf{X}_i)-\widehat{\tau}_{\mathcal{I}_{3-j}}(\mathbf{X}_i)|>\delta_n'$, we have
\begin{align}
    \left|\widehat{\tau}_{\mathcal{I}_{3-j}}(\mathbf{X}_i)\mathcal{D}_n(\mathbf{X}_i,\widehat{\tau}_{\mathcal{I}_{3-j}}(\mathbf{X}_i)) -\tau(\mathbf{X}_i)\mathcal{D}_n(\mathbf{X}_i,\tau(\mathbf{X}_i))\right|= O\left(|\tau(\mathbf{X}_i)-\widehat{\tau}_{\mathcal{I}_{3-j}}(\mathbf{X}_i)|\right)+o(n^{-1/2}),\label{Lip1}\\
    \left|(\tau(\mathbf{X}_i)-\widehat{\tau}_{\mathcal{I}_{3-j}}(\mathbf{X}_i))\mathcal{D}_n(\mathbf{X}_i,\widehat{\tau}_{\mathcal{I}_{3-j}}(\mathbf{X}_i))\right|\leq  |\tau(\mathbf{X}_i)-\widehat{\tau}_{\mathcal{I}_{3-j}}(\mathbf{X}_i)|.\label{Lip2}
\end{align}
Eq. \eqref{Lip2} is clear as $\mathcal{D}_n$ taking value in $[0,1]$. To justify Eq. \eqref{Lip1}, we discuss different cases. Firstly, when $|\tau(\mathbf{X}_i)|\ge 2\delta_n'$ and $|\widehat{\tau}_{\mathcal{I}_{3-j}}(\mathbf{X}_i)|\ge \delta_n'$, by Assumption \ref{A12} we have
\begin{align*}
    \left|\tau(\mathbf{X}_i)\mathcal{D}_n(\mathbf{X}_i,\tau(\mathbf{X}_i)) - \tau(\mathbf{X}_i)\mathbb{I}\{\tau(\mathbf{X}_i)>0\}\right|=o(n^{-1/2}),\\
    \left|\widehat{\tau}_{\mathcal{I}_{3-j}}(\mathbf{X}_i)\mathcal{D}_n(\mathbf{X}_i,\widehat{\tau}_{\mathcal{I}_{3-j}}(\mathbf{X}_i)) - \widehat{\tau}_{\mathcal{I}_{3-j}}(\mathbf{X}_i)\mathbb{I}\{\widehat{\tau}_{\mathcal{I}_{3-j}}(\mathbf{X}_i)>0\}\right|=o(n^{-1/2}),
\end{align*}
and thus
\begin{align*}
     &\left|\widehat{\tau}_{\mathcal{I}_{3-j}}(\mathbf{X}_i)\mathcal{D}_n(\mathbf{X}_i,\widehat{\tau}_{\mathcal{I}_{3-j}}(\mathbf{X}_i)) -\tau(\mathbf{X}_i)\mathcal{D}_n(\mathbf{X}_i,\tau(\mathbf{X}_i))\right|\\
     \leq&o(n^{-1/2})+\left|\tau(\mathbf{X}_i)\mathbb{I}\{\tau(\mathbf{X}_i)>0\} - \widehat{\tau}_{\mathcal{I}_{3-j}}(\mathbf{X}_i)\mathbb{I}\{\widehat{\tau}_{\mathcal{I}_{3-j}}(\mathbf{X}_i)>0\}\right|\\
     \leq&o(n^{-1/2})+O(|\tau(\mathbf{X}_i)-\widehat{\tau}_{\mathcal{I}_{3-j}}(\mathbf{X}_i)|).
\end{align*}
Second, when $|\tau(\mathbf{X}_i)|\ge 2\delta_n'$ and $|\widehat{\tau}_{\mathcal{I}_{3-j}}(\mathbf{X}_i)|< \delta_n'$, we have
\begin{align*}
    |\widehat{\tau}_{\mathcal{I}_{3-j}}(\mathbf{X}_i)|\leq |\widehat{\tau}_{\mathcal{I}_{3-j}}(\mathbf{X}_i)-\tau(\mathbf{X}_i)|
\end{align*}
and
\begin{align*}
    &\left|\widehat{\tau}_{\mathcal{I}_{3-j}}(\mathbf{X}_i)\mathcal{D}_n(\mathbf{X}_i,\widehat{\tau}_{\mathcal{I}_{3-j}}(\mathbf{X}_i)) -\tau(\mathbf{X}_i)\mathcal{D}_n(\mathbf{X}_i,\tau(\mathbf{X}_i))\right|\\
    \leq&|\tau(\mathbf{X}_i)|+|\widehat{\tau}_{\mathcal{I}_{3-j}}(\mathbf{X}_i)|\\
    \leq& |\tau(\mathbf{X}_i)-\widehat{\tau}_{\mathcal{I}_{3-j}}(\mathbf{X}_i)|+2|\widehat{\tau}_{\mathcal{I}_{3-j}}(\mathbf{X}_i)|\\
    \leq& 3|\tau(\mathbf{X}_i)-\widehat{\tau}_{\mathcal{I}_{3-j}}(\mathbf{X}_i)|.
\end{align*}
Hence we have justified Eq. \eqref{Lip1}. Put
$e_{3-j}(\mathbf{X})=\widehat{\tau}_{\mathcal{I}_{3-j}}(\mathbf{X})-\tau(\mathbf{X})$.
Both terms in Eqs.~\eqref{Lip1}--\eqref{Lip2}, on $|e_{3-j}|>\delta_n'$, are bounded by a constant multiple of $|e_{3-j}|$. Directly using Eq. (10) at $q=3$,
\begin{align*}
E[|e_{3-j}(\mathbf{X})|\mathbb{I}\{|e_{3-j}(\mathbf{X})|>\delta_n'\}]
&\leq \frac{E|e_{3-j}(\mathbf{X})|^3}{(\delta_n')^2}\\
&=O_p\left(\frac{\{E|e_{3-j}(\mathbf{X})|^2\}^{3/2}}{(\delta_n')^2}\right)=o_p(n^{-1/2}),
\end{align*}
because $n^{1/4}\{E|e_{3-j}(\mathbf{X})|^2\}^{3/4}\prec_p\delta_n'$.
Therefore, 
    \begin{align*}
        E\bigg|\frac{1}{|\mathcal{I}_{j,k}|}\sum_{i\in\mathcal{I}_{j,k},|\tau(\mathbf{X}_i)|\ge2\delta_n',|\tau(\mathbf{X}_i)-\widehat{\tau}_{\mathcal{I}_{3-j}}(\mathbf{X}_i)|>\delta_n'}\widehat{\tau}_{\mathcal{I}_{3-j}}(\mathbf{X}_i)\mathcal{D}_n(\mathbf{X}_i,\widehat{\tau}_{\mathcal{I}_{3-j}}(\mathbf{X}_i)) \\-\tau(\mathbf{X}_i)\mathcal{D}_n(\mathbf{X}_i,\tau(\mathbf{X}_i))\bigg|=o(n^{-1/2})
    \end{align*}
    and
    \begin{align*}
        E\bigg|\frac{1}{|\mathcal{I}_{j,k}|}\sum_{i\in\mathcal{I}_{j,k},|\tau(\mathbf{X}_i)|\ge2\delta_n',|\tau(\mathbf{X}_i)-\widehat{\tau}_{\mathcal{I}_{3-j}}(\mathbf{X}_i)|>\delta_n'}(\tau(\mathbf{X}_i)-\widehat{\tau}_{\mathcal{I}_{3-j}}(\mathbf{X}_i))\\\cdot\mathcal{D}_n(\mathbf{X}_i,\widehat{\tau}_{\mathcal{I}_{3-j}}(\mathbf{X}_i))\bigg|=o(n^{-1/2}).
    \end{align*}
These above results together imply $\text{Eq. }\eqref{thm2_2_4}=o_p(n^{-1/2})$.
Hence we have shown that both Eq. \eqref{thm2_1} and Eq. \eqref{thm2_2} are $o_p(n^{-1/2})$. Then we observe that
    \begin{align}
        &\frac{1}{|\mathcal{I}_{j,k}|}\sum_{i\in\mathcal{I}_{j,k}}\psi_i(\mathcal{D}_n(\mathbf{X}_i,\tau(\mathbf{X}_i)),\pi,\mu)-\frac{1}{|\mathcal{I}_{j,k}|}\sum_{i\in\mathcal{I}_{j,k}}\psi_i(\mathcal{D}(\mathbf{X}_i),\pi,\mu)\label{thm2_3}\\
        =&\frac{1}{|\mathcal{I}_{j,k}|}\sum_{i\in\mathcal{I}_{j,k}}\frac{(2A_i-1)(\mathcal{D}_n(\mathbf{X}_i,\tau(\mathbf{X}_i))-\mathcal{D}(\mathbf{X}_i))}{\pi(A_i,\mathbf{X}_i)}\{Y_i-\mu(A_i,\mathbf{X}_i)\}\label{thm2_2_7}\\
        +&\frac{1}{|\mathcal{I}_{j,k}|}\sum_{i\in\mathcal{I}_{j,k}}\tau(\mathbf{X}_i)(\mathcal{D}_n(\mathbf{X}_i,\tau(\mathbf{X}_i))-\mathcal{D}(\mathbf{X}_i))\label{thm2_2_8}
    \end{align}
    where Eq. \eqref{thm2_2_7} can be shown to be $o_p(n^{-1/2})$ similarly to Eq. \eqref{thm2_2_1}. For Eq. \eqref{thm2_2_8} we have 
    \begin{align}
        &\text{Eq. }\eqref{thm2_2_8}\notag\\
        =&\frac{1}{|\mathcal{I}_{j,k}|}\sum_{i\in\mathcal{I}_{j,k},|\tau(\mathbf{X}_i)|<\delta_n'}\tau(\mathbf{X}_i)(\mathcal{D}_n(\mathbf{X}_i,\tau(\mathbf{X}_i))-\mathcal{D}(\mathbf{X}_i))\label{thm2_2_9}\\
        +&\frac{1}{|\mathcal{I}_{j,k}|}\sum_{i\in\mathcal{I}_{j,k},|\tau(\mathbf{X}_i)|\ge\delta_n'}\tau(\mathbf{X}_i)(\mathcal{D}_n(\mathbf{X}_i,\tau(\mathbf{X}_i))-\mathcal{D}(\mathbf{X}_i)).\label{thm2_2_10}
    \end{align}
      By Assumptions \ref{A10} to \ref{A12}, similar to the arguments for Eq.\eqref{thm2_2_3} and Eq. \eqref{thm2_2_4} we can verify that
    \begin{align*}
         |\text{Eq. }\eqref{thm2_2_9}|\leq \frac{2}{|\mathcal{I}_{j,k}|}\sum_{i\in\mathcal{I}_{j,k},|\tau(\mathbf{X}_i)|<\delta_n'}|\tau(\mathbf{X}_i)|=O_p((\delta_n')^{1+\alpha})+o_p(n^{-1/2})=o_p(n^{-1/2})
    \end{align*}
    and $\text{Eq. }\eqref{thm2_2_10}=o_p(n^{-1/2})$. Therefore, we have shown that
    \begin{align*}
        |\widehat{V}_{\mathcal{D}}^*-\widehat{V}_{\mathcal{D}_n}^*|=o_p(n^{-1/2}).
    \end{align*}
    Combined with Eq. \eqref{thm2_1} and Eq. \eqref{thm2_2}, we have
    \begin{align*}
        |\widehat{V}_{\mathcal{D}}^*-\widehat{V}_{\mathcal{D}_n}|=o_p(n^{-1/2}).
    \end{align*}
    Thus for the variance we have
    \begin{align*}
        &n\operatorname{Var}\left(\widehat{V}_{\mathcal{D}_n}\right)-o_p(1)=n\operatorname{Var}\left(\widehat{V}_{\mathcal{D}}^*\right)\\
          =& \operatorname{Var}\left\{\mu\left(d_0^{o p t}, \mathbf{X}\right)\right\}+ \mathrm{E}\left(\frac{\mathbb{I}\left\{\tau\left(\mathbf{X}\right)>0\right\}}{\pi\left(1, \mathbf{X}\right)}\operatorname{Var}(Y|\mathbf{X},A=1)+\frac{\mathbb{I}\left\{\tau\left(\mathbf{X}\right)<0\right\}}{\pi\left(0, \mathbf{X}\right)}\operatorname{Var}(Y|\mathbf{X},A=0)\right)\\
    +&\underbrace{\mathrm{E}\left[\left(\frac{(\mathcal{D}(\mathbf{X}))^2}{\pi\left(1, \mathbf{X}\right)}\operatorname{Var}(Y|\mathbf{X},A=1)+\frac{(1-\mathcal{D}(\mathbf{X}))^2}{\pi\left(0, \mathbf{X}\right)}\operatorname{Var}(Y|\mathbf{X},A=0)\right) \mathbb{I}\left\{\tau\left(\mathbf{X}\right)=0\right\}\right]}_{\text{Varying with $\mathcal{D}$}}.
    \end{align*}
    Align with the basic identification Assumptions 1 to 3, we have $E[\widehat{V}_{\mathcal{D}}^*] = V_0$ which implies the mean zero asymptotic normality of $\sqrt{n}(\widehat{V}_{\mathcal{D}}^* - V_0)$. Combined with $ |\widehat{V}_{\mathcal{D}}^*-\widehat{V}_{\mathcal{D}_n}|=o_p(n^{-1/2})$, our desired results can be proved.
\end{proof}
\subsection{Proof of Theorem 1}
\begin{proof}
    We verify that the smoothing decision function $d_s$ in 
    Eq.~(6) of the main text satisfies Assumptions~\ref{A10} 
    to~\ref{A12} by Assumptions 6 and 7, so that Lemma~\ref{single-lem} can be applied.

    Let $\mathcal{D}_n(\mathbf{X}_i,s)
    =d_{s}(\mathbf{X}_i;s,h_{n,\mathcal{I}_{j(i)}^c},t_0)$, 
    where $j(i)$ is the index of subset that the $i$-th sample 
    belongs to (for example, if $i\in\mathcal{I}_1$ then 
    $j(i)=1$). {For each $j=1,2$, define}
    \begin{align*}
        {\delta_{n,j}}
        &={\sqrt{h_{n,\mathcal{I}_{j}^c}\cdot n^{1/4}
        \left(E\!\left[|\widehat{\tau}_{\mathcal{I}_{j}^c}
        (\mathbf{X})-\tau(\mathbf{X})|^2\right]\right)^{3/4}}},\\
        \delta_n'&=2\max_j h_{n,\mathcal{I}_{j}^c}.
    \end{align*}

    We first verify Assumption~\ref{A11} (Consistency). 
    By the definition of $d_s$ in Eq.~(6) of the main text 
    and the fact that $h_{n,\mathcal{I}_j^c}\to 0$ by 
    Assumption~7 (upper bound of $h_{n,\mathcal{I}_j}$), 
    for any fixed $\mathbf{X}$: if $\tau(\mathbf{X})>0$, 
    then for all sufficiently large $n$, 
    $\tau(\mathbf{X})\ge (1-t_0)h_{n,\mathcal{I}_j^c}$, 
    so $d_s(\mathbf{X};\tau(\mathbf{X}),
    h_{n,\mathcal{I}_j^c},t_0)=1$; 
    if $\tau(\mathbf{X})<0$, then 
    $\tau(\mathbf{X})<-t_0 h_{n,\mathcal{I}_j^c}$ 
    for large $n$, so 
    $d_s(\mathbf{X};\tau(\mathbf{X}),
    h_{n,\mathcal{I}_j^c},t_0)=0$; 
    if $\tau(\mathbf{X})=0$, then 
    $\tau(\mathbf{X})\in(-t_0 h_{n,\mathcal{I}_j^c},
    (1-t_0)h_{n,\mathcal{I}_j^c})$ and 
    $d_s(\mathbf{X};\tau(\mathbf{X}),
    h_{n,\mathcal{I}_j^c},t_0)
    =\tau(\mathbf{X})/h_{n,\mathcal{I}_j^c}+t_0=t_0$. 
    Therefore, 
    $\mathcal{D}(\mathbf{X})
    :=\lim_{n\to\infty}\mathcal{D}_n(\mathbf{X},\tau(\mathbf{X}))
    =t_0\cdot\mathbb{I}\{\tau(\mathbf{X})=0\}
    +\mathbb{I}\{\tau(\mathbf{X})>0\}$. 
    Since the value of $\mathcal{D}$ on 
    $\{\tau(\mathbf{X})=0\}$ does not affect the mean outcome 
    (because $\tau(\mathbf{X})\cdot\mathcal{D}(\mathbf{X})
    =\max\{0,\tau(\mathbf{X})\}$ a.s.), we have 
    $E[\psi(\mathcal{D}(\mathbf{X}),\pi,\mu)]=V_0$, 
    so $\mathcal{D}\in\mathbb{D}(V_0)$ and 
    Assumption~\ref{A11} is satisfied.

    We next verify Assumption~\ref{A10} (Uniform continuity). 
    By the piecewise linear structure of $d_s$ in Eq.~(6) 
    of the main text, the Lipschitz constant of 
    $s\mapsto d_s(\mathbf{X};s,h_{n,\mathcal{I}_j^c},t_0)$ 
    is at most $1/h_{n,\mathcal{I}_j^c}$. Therefore, for 
    $|s-t|\leq {\delta_{n,j}}$,
    \begin{align*}
        \sup_{\mathbf{X}\in\mathcal{X}}
        |\mathcal{D}_n(\mathbf{X},s)-\mathcal{D}_n(\mathbf{X},t)|
        \leq \frac{{\delta_{n,j}}}{h_{n,\mathcal{I}_j^c}}
        =\sqrt{\frac{n^{1/4}
        \left(E[|\widehat{\tau}_{\mathcal{I}_j^c}(\mathbf{X})
        -\tau(\mathbf{X})|^2]\right)^{3/4}}
        {h_{n,\mathcal{I}_j^c}}}
        =o_p(1)
    \end{align*}
    where the last step follows from Assumption~6, which 
    requires $n^{1/4}(E[|\widehat{\tau}_{{\mathcal{I}_j^c}}
    (\mathbf{X})-\tau(\mathbf{X})|^2])^{3/4}
    \prec_p h_{n,{\mathcal{I}_j^c}}$. It remains to show 
    $P_{\mathbf{X}}(|\widehat{\tau}_{\mathcal{I}_j^c}
    (\mathbf{X})-\tau(\mathbf{X})|\ge {\delta_{n,j}})=o_p(1)$. 
    By Chebyshev's inequality and the definition of ${\delta_{n,j}}$,
    \begin{align*}
        P_{\mathbf{X}}\!\left(|\widehat{\tau}_{\mathcal{I}_j^c}
        (\mathbf{X})-\tau(\mathbf{X})|\ge {\delta_{n,j}}\right)
        \leq \frac{E[|\widehat{\tau}_{\mathcal{I}_j^c}
        (\mathbf{X})-\tau(\mathbf{X})|^2]}{{\delta_{n,j}^2}}
        =\frac{(E[|\widehat{\tau}_{\mathcal{I}_j^c}(\mathbf{X})
        -\tau(\mathbf{X})|^2])^{1/4}}
        {h_{n,\mathcal{I}_j^c}\cdot n^{1/4}}
        =o_p(1)
    \end{align*}
    again by Assumption~6. Hence Assumption~\ref{A10} 
    is satisfied.

    Finally, we verify Assumption~\ref{A12} (Concentration). 
    For the rate condition on $\delta_n'$, by the definition 
    of $\delta_n'$ and Assumption~6, we have 
    $n^{1/4}(E[|\widehat{\tau}_{{\mathcal{I}_j^c}}(\mathbf{X})
    -\tau(\mathbf{X})|^2])^{3/4}\prec_p h_{n,{\mathcal{I}_j^c}}
    \leq \delta_n'/2$, {and hence this term is $\prec_p\delta_n'$;} by Assumption~7, 
    $\delta_n'=2\max_j h_{n,\mathcal{I}_j^c}
    \prec n^{-\frac{1}{2(1+\alpha)}}$. Moreover, 
    ${\delta_{n,j}\prec_p\delta_n'}$ since 
    ${\delta_{n,j}^2}=h_{n,\mathcal{I}_j^c}\cdot n^{1/4}
    (E[|\widehat{\tau}_{\mathcal{I}_j^c}(\mathbf{X})
    -\tau(\mathbf{X})|^2])^{3/4}
    \prec_p h_{n,\mathcal{I}_j^c}^2
    \leq (\delta_n'/2)^2$ by Assumption~6. 
    For the approximation condition Eq.~\eqref{A10-1}, 
    when $|s|>\delta_n'=2\max_j h_{n,\mathcal{I}_j^c}$, 
    we have $|s|>2 h_{n,\mathcal{I}_j^c}$ for all $j$. 
    If $s>\delta_n'$, then 
    $s>2h_{n,\mathcal{I}_j^c}>(1-t_0)h_{n,\mathcal{I}_j^c}$ 
    for $t_0\in(0,1)$, so $\mathcal{D}_n(\mathbf{X},s)=1
    =\mathbb{I}\{s>0\}$; if $s<-\delta_n'$, then 
    $s<-2h_{n,\mathcal{I}_j^c}<-t_0 h_{n,\mathcal{I}_j^c}$, 
    so $\mathcal{D}_n(\mathbf{X},s)=0=\mathbb{I}\{s>0\}$. 
    In both cases, $|\mathcal{D}_n(\mathbf{X},s)
    -\mathbb{I}\{s>0\}|=0=o(1/({|s|}\sqrt{n}))$, so 
    Eq.~\eqref{A10-1} holds trivially and 
    Assumption~\ref{A12} is satisfied.

    Since Assumptions~\ref{A10} to~\ref{A12} are verified 
    and Assumptions~1 to 5 from the main text are 
    assumed, all conditions of Lemma~\ref{single-lem} are 
    satisfied, and therefore Theorem~1 holds.
\end{proof}

\subsection{Proof of Theorem 2}
\begin{proof}
    We verify that the adaptive smoothing decision function 
    with $t_0$ replaced by $\widehat{\pi}_{\mathcal{I}_{j}^c}
    (1,\mathbf{X})$ satisfies Assumptions~\ref{A10} 
    to~\ref{A12} by Assumptions 6 and 7, so that Lemma~\ref{single-lem} can be applied.

    Let $\mathcal{D}_n(\mathbf{X}_i,s)
    =d_{s}(\mathbf{X}_i;s,h_{n,\mathcal{I}_{j(i)}^c},
    \widehat{\pi}_{\mathcal{I}_{j(i)}^c}(1,\mathbf{X}_i))$, 
    where $j(i)$ is the index of subset that the $i$-th sample 
    belongs to. {For each $j$, use $\delta_{n,j}$ as defined for that fold and the same $\delta_n'$} 
    in the proof of Theorem~1. The only difference from the 
    proof of Theorem~1 is that the fixed asymmetric tuning 
    parameter $t_0$ is replaced by the estimated propensity 
    score $\widehat{\pi}_{\mathcal{I}_j^c}(1,\mathbf{X})$, 
    which takes values in $({c_{\mathrm f}},1-{c_{\mathrm f}})$ by the {fitted-overlap condition above}. We now verify each assumption.

    We first verify Assumption~\ref{A11} (Consistency). 
    By the same case analysis as in the proof of Theorem~1, 
    using $h_{n,\mathcal{I}_j^c}\to 0$ from Assumption~7: 
    if $\tau(\mathbf{X})>0$, then for large $n$, 
    $\tau(\mathbf{X})\ge (1-\widehat{\pi}_{\mathcal{I}_j^c}
    (1,\mathbf{X}))h_{n,\mathcal{I}_j^c}$ since 
    $(1-\widehat{\pi}_{\mathcal{I}_j^c}(1,\mathbf{X}))$ 
    is bounded above by $1-{c_{\mathrm f}}$ by {the fitted-overlap condition above} and 
    $h_{n,\mathcal{I}_j^c}\to 0$, so 
    $\mathcal{D}_n(\mathbf{X},\tau(\mathbf{X}))=1$; 
    if $\tau(\mathbf{X})<0$, then 
    $\tau(\mathbf{X})<-\widehat{\pi}_{\mathcal{I}_j^c}
    (1,\mathbf{X})h_{n,\mathcal{I}_j^c}$ since 
    $\widehat{\pi}_{\mathcal{I}_j^c}(1,\mathbf{X})>{c_{\mathrm f}}>0$, 
    so $\mathcal{D}_n(\mathbf{X},\tau(\mathbf{X}))=0$; 
    if $\tau(\mathbf{X})=0$, then 
    $\mathcal{D}_n(\mathbf{X},\tau(\mathbf{X}))
    =0/h_{n,\mathcal{I}_j^c}
    +\widehat{\pi}_{\mathcal{I}_j^c}(1,\mathbf{X})
    =\widehat{\pi}_{\mathcal{I}_j^c}(1,\mathbf{X})
    \to_p \pi(1,\mathbf{X})$ by the consistency of 
    $\widehat{\pi}$ in Assumption~4. Therefore, 
    $\mathcal{D}(\mathbf{X})
    :=\lim_{n\to\infty}\mathcal{D}_n(\mathbf{X},\tau(\mathbf{X}))
    =\pi(1,\mathbf{X})\cdot\mathbb{I}\{\tau(\mathbf{X})=0\}
    +\mathbb{I}\{\tau(\mathbf{X})>0\}$. 
    Since $\tau(\mathbf{X})\cdot\mathcal{D}(\mathbf{X})
    =\max\{0,\tau(\mathbf{X})\}$ a.s., we have 
    $E[\psi(\mathcal{D}(\mathbf{X}),\pi,\mu)]=V_0$, 
    so $\mathcal{D}\in\mathbb{D}(V_0)$ and 
    Assumption~\ref{A11} is satisfied.

    We next verify Assumption~\ref{A10} (Uniform continuity). 
    The piecewise linear structure of $d_s$ is unchanged; 
    for any fixed $\mathbf{X}$, the Lipschitz constant of 
    $s\mapsto d_s(\mathbf{X};s,h_{n,\mathcal{I}_j^c},
    \widehat{\pi}_{\mathcal{I}_j^c}(1,\mathbf{X}))$ 
    is still at most $1/h_{n,\mathcal{I}_j^c}$, regardless 
    of the value of $\widehat{\pi}_{\mathcal{I}_j^c}
    (1,\mathbf{X})\in({c_{\mathrm f}},1-{c_{\mathrm f}})$. Therefore, by the identical 
    argument as in the proof of Theorem~1,
    \begin{align*}
        \sup_{\mathbf{X}\in\mathcal{X}}
        |\mathcal{D}_n(\mathbf{X},s)-\mathcal{D}_n(\mathbf{X},t)|
        \leq \frac{{\delta_{n,j}}}{h_{n,\mathcal{I}_j^c}}=o_p(1)
        \quad\text{for }|s-t|\leq{\delta_{n,j}}
    \end{align*}
    by Assumption~6, and 
    $P_{\mathbf{X}}(|\widehat{\tau}_{\mathcal{I}_j^c}(\mathbf{X})
    -\tau(\mathbf{X})|\ge{\delta_{n,j}})=o_p(1)$ 
    by Chebyshev's inequality and Assumption~6. Hence 
    Assumption~\ref{A10} is satisfied.

    Finally, we verify Assumption~\ref{A12} (Concentration). 
    The rate condition on $\delta_n'$ is identical to that 
    in the proof of Theorem~1: 
    ${\delta_{n,j}}\lor n^{1/4}(E[|\widehat{\tau}_{{\mathcal{I}_j^c}}
    (\mathbf{X})-\tau(\mathbf{X})|^2])^{3/4}
    \prec_p \delta_n'$ by Assumption~6 and 
    $\delta_n'\prec n^{-\frac{1}{2(1+\alpha)}}$ 
    by Assumption~7. For the approximation condition 
    Eq.~\eqref{A10-1}, when 
    $|s|>\delta_n'=2\max_j h_{n,\mathcal{I}_j^c}$, we have 
    $|s|>2 h_{n,\mathcal{I}_j^c}$ for all $j$. Since 
    $\widehat{\pi}_{\mathcal{I}_j^c}(1,\mathbf{X})\in({c_{\mathrm f}},1-{c_{\mathrm f}})$ 
    by {the fitted-overlap condition above}, both 
    $(1-\widehat{\pi}_{\mathcal{I}_j^c}(1,\mathbf{X}))
    h_{n,\mathcal{I}_j^c}\leq (1-{c_{\mathrm f}})h_{n,\mathcal{I}_j^c}
    <h_{n,\mathcal{I}_j^c}<|s|/2$ and 
    $\widehat{\pi}_{\mathcal{I}_j^c}(1,\mathbf{X})
    h_{n,\mathcal{I}_j^c}\leq (1-{c_{\mathrm f}})h_{n,\mathcal{I}_j^c}
    <|s|/2$. Therefore, if $s>\delta_n'$, then 
    $s>(1-\widehat{\pi}_{\mathcal{I}_j^c}(1,\mathbf{X}))
    h_{n,\mathcal{I}_j^c}$, so 
    $\mathcal{D}_n(\mathbf{X},s)=1=\mathbb{I}\{s>0\}$; 
    if $s<-\delta_n'$, then 
    $s<-\widehat{\pi}_{\mathcal{I}_j^c}(1,\mathbf{X})
    h_{n,\mathcal{I}_j^c}$, so 
    $\mathcal{D}_n(\mathbf{X},s)=0=\mathbb{I}\{s>0\}$. 
    In both cases, $|\mathcal{D}_n(\mathbf{X},s)
    -\mathbb{I}\{s>0\}|=0=o(1/({|s|}\sqrt{n}))$, 
    so Eq.~\eqref{A10-1} holds trivially. Note that 
    unlike the proof of Theorem~1 where $t_0$ is fixed, 
    here the thresholds 
    $-\widehat{\pi}_{\mathcal{I}_j^c}(1,\mathbf{X})
    h_{n,\mathcal{I}_j^c}$ and 
    $(1-\widehat{\pi}_{\mathcal{I}_j^c}(1,\mathbf{X}))
    h_{n,\mathcal{I}_j^c}$ are random, but the {fitted-overlap condition above} ensures they remain uniformly bounded by 
    $(1-{c_{\mathrm f}})h_{n,\mathcal{I}_j^c}<\delta_n'/2$, which 
    guarantees the same conclusion. Hence 
    Assumption~\ref{A12} is satisfied.

    Since Assumptions~\ref{A10} to~\ref{A12} are verified 
    and Assumptions~1 to~5 from the main text are 
    assumed, all conditions of Lemma~\ref{single-lem} are 
    satisfied. Therefore, by Lemma~\ref{single-lem}, 
    $|\widehat{V}_{as}-\widehat{V}_{\mathcal{D}}^*|
    =o_p(n^{-1/2})$ where 
    $\mathcal{D}(\mathbf{X})=\pi(1,\mathbf{X})\cdot
    \mathbb{I}\{\tau(\mathbf{X})=0\}
    +\mathbb{I}\{\tau(\mathbf{X})>0\}$, 
    and the asymptotic normality and variance formula in 
    Theorem~2 follow from applying 
    Lemma~\ref{single-lem} with this specific $\mathcal{D}$ 
    and Assumption~8 (homoscedasticity) to simplify the 
    nonregular part of the variance.
\end{proof}

\subsection{Additional Conditions and Proof of Proposition 1}
Throughout this subsection, $E_{\mathbf X}$ denotes expectation over a fresh draw of $\mathbf X$, with the training data fixed. A fold $\mathcal I$ is one of the half or quarter folds used by Algorithm 2, and every such fold has size of order $n$.

\medskip\noindent\textbf{Condition (C1) (near-parametric rate).}
$E_{\mathbf X}|\widehat\tau_{\mathcal I}(\mathbf X)-\tau(\mathbf X)|^2=O_p(\log n/n)$ for every such fold.

\medskip\noindent\textbf{Condition (C2) (stochastic boundedness).}
$\|\tau\|_\infty<\infty$ and $\|\widehat\tau_{\mathcal I}\|_\infty=O_p(1)$ for every such fold.

\medskip\noindent\textbf{Condition (C3) (variability among splits).}
For any two disjoint such folds $\mathcal I,\mathcal I'$ with $|\mathcal I|\asymp|\mathcal I'|\asymp n$,
\[
E_{\mathbf X}|\widehat\tau_{\mathcal I}(\mathbf X)-\widehat\tau_{\mathcal I'}(\mathbf X)|^2
\asymp_p E_{\mathbf X}|\widehat\tau_{\mathcal I}(\mathbf X)-\tau(\mathbf X)|^2.
\]

Condition (C1) requires the near-parametric rate $E_{\bX}[(\tauh_{\cI}(\bX)-\tau(\bX))^2]=O_p(\log n/n)$, which correctly specified parametric working models attain. When the working model satisfies the rate requirement in Condition (C1), the truncation term $\log(n)/({C_0}n)$ of Algorithm 2 becomes the leading term. Therefore, the desired range of the tuning parameter is satisfied automatically by Algorithm 2 because the truncation term already dominates the estimation error and is of order sufficiently smaller than that of the desired upper bound by Assumption 9.

Condition (C2) requires the estimators to be stochastically bounded, which is a standard assumption in the causal inference literature. In practice, Condition (C2) can also be enforced by truncating $\widehat\tau_{\mathcal I}$ at a known bound. Under Conditions (C2) and (C3), Condition (C2) provides conditional concentration of the two validation-fold averages entering the $EAE$, and Condition (C3) links their population counterparts to the estimation error. Under Condition (C1), the truncation term $\log(n)/({C_0}n)$ yields the comparison in Eq.~\eqref{eae-error-path} directly.

Condition (C3) requires the average squared difference between two independently trained estimators, which is what the $EAE$ estimates, to be of the same order as the estimation error. Condition (C3) is mild, and for two folds of the same size, averaging over the training data of both sides of Condition (C3), we can show that Condition (C3) basically asks the squared bias of the estimator not to dominate the order of its variance on average
\[
E_{\mathbf X}\left[\left(E_{\widehat\tau_{\mathcal I}}\{\widehat\tau_{\mathcal I}(\mathbf X)-\tau(\mathbf X)\}\right)^2\right]
\lesssim
E_{\mathbf X}\left[\operatorname{Var}_{\widehat\tau_{\mathcal I}}\{\widehat\tau_{\mathcal I}(\mathbf X)\}\right]
\]
as the other direction $\gtrsim$ holds trivially. Lemma~\ref{parametric-tuning-conditions} covers correctly specified parametric working models. Lemma~\ref{linear-smoother-c3} gives high-level sufficient conditions for undersmoothed estimators, and Examples~\ref{example:undersmoothed-spline} and~\ref{example:undersmoothed-kernel} verify those conditions for explicit spline and kernel estimators. Section A.11 illustrates the condition numerically.

\begin{proof}[Proof of Proposition 1]
Because the random sample split is independent of the observations, we
condition on the realized split and regard the fold index sets as fixed
throughout the proof. The bounds below depend on the split only through
the fixed fold sizes and therefore also hold without conditioning on the
split.

We first prove, simultaneously for $j=1,2$, the two comparisons
\begin{align}
&
\frac{
EAE(\widehat\tau_{\mathcal I_j})\vee\{\log n/({C_0}n)\}
}{
E_{\mathbf X}
|\widehat\tau_{\mathcal I_j}(\mathbf X)-\tau(\mathbf X)|^2
\vee\{\log n/({C_0}n)\}
}
=O_p(1),
\label{eae-error-path}
\\
&
\frac{
E_{\mathbf X}
|\widehat\tau_{\mathcal I_j}(\mathbf X)-\tau(\mathbf X)|^2
\vee\{\log n/({C_0}n)\}
}{
EAE(\widehat\tau_{\mathcal I_j})\vee\{\log n/({C_0}n)\}
}
=O_p(1).
\label{error-to-eae}
\end{align}

For $j,\ell\in\{1,2\}$, the observations indexed by
$\mathcal I_{3-j,\ell}$ are independent of the two fitted contrast
estimators
$\widehat\tau_{\mathcal I_j}$ and
$\widehat\tau_{\mathcal I_{3-j,3-\ell}}$. Conditional on these two
estimators, the observations
$\{\mathbf X_i:i\in\mathcal I_{3-j,\ell}\}$ remain independent and
identically distributed. Hence
\begin{align}
&E\left[
\left.
\frac{1}{|\mathcal I_{3-j,\ell}|}
\sum_{i\in\mathcal I_{3-j,\ell}}
\left|
\widehat\tau_{\mathcal I_{3-j,3-\ell}}(\mathbf X_i)
-\widehat\tau_{\mathcal I_j}(\mathbf X_i)
\right|^2
\,\right|\,
\widehat\tau_{\mathcal I_{3-j,3-\ell}},
\widehat\tau_{\mathcal I_j}
\right]
\notag\\
&\qquad=
E_{\mathbf X}\left|
\widehat\tau_{\mathcal I_{3-j,3-\ell}}(\mathbf X)
-\widehat\tau_{\mathcal I_j}(\mathbf X)
\right|^2,
\label{conditional-eae-mean}
\end{align}
where $E_{\mathbf X}$ is taken over an independent copy $\mathbf X$ of
the covariate, holding the two fitted contrast estimators fixed.

Because the two validation subfolds have equal sizes, the definition of
the estimated approximation error gives the exact identity
\begin{align}
EAE(\widehat\tau_{\mathcal I_j})
=
\frac12\sum_{\ell=1}^2
\frac{1}{|\mathcal I_{3-j,\ell}|}
\sum_{i\in\mathcal I_{3-j,\ell}}
\left|
\widehat\tau_{\mathcal I_{3-j,3-\ell}}(\mathbf X_i)
-\widehat\tau_{\mathcal I_j}(\mathbf X_i)
\right|^2.
\label{exact-eae-expression}
\end{align}

For every fixed $L>0$, conditional Markov's inequality and
Eq.~\eqref{conditional-eae-mean} give
\begin{align}
&P\Bigg[
\frac{1}{|\mathcal I_{3-j,\ell}|}
\sum_{i\in\mathcal I_{3-j,\ell}}
\left|
\widehat\tau_{\mathcal I_{3-j,3-\ell}}(\mathbf X_i)
-\widehat\tau_{\mathcal I_j}(\mathbf X_i)
\right|^2
\notag\\
&\hspace{40mm}>
L\left\{
E_{\mathbf X}\left|
\widehat\tau_{\mathcal I_{3-j,3-\ell}}(\mathbf X)
-\widehat\tau_{\mathcal I_j}(\mathbf X)
\right|^2
\vee\frac{\log n}{{C_0}n}
\right\}
\,\Bigg|\,
\widehat\tau_{\mathcal I_{3-j,3-\ell}},
\widehat\tau_{\mathcal I_j}
\Bigg]
\notag\\
&\qquad\le
\frac{
E_{\mathbf X}\left|
\widehat\tau_{\mathcal I_{3-j,3-\ell}}(\mathbf X)
-\widehat\tau_{\mathcal I_j}(\mathbf X)
\right|^2
}{
L\left\{
E_{\mathbf X}\left|
\widehat\tau_{\mathcal I_{3-j,3-\ell}}(\mathbf X)
-\widehat\tau_{\mathcal I_j}(\mathbf X)
\right|^2
\vee\frac{\log n}{{C_0}n}
\right\}
}
\le\frac1L.
\label{conditional-markov}
\end{align}
Taking expectations in Eq.~\eqref{conditional-markov} shows that, for
each $j$ and $\ell$,
\begin{align}
&
\frac{
\left[
\displaystyle
\frac{1}{|\mathcal I_{3-j,\ell}|}
\sum_{i\in\mathcal I_{3-j,\ell}}
\left|
\widehat\tau_{\mathcal I_{3-j,3-\ell}}(\mathbf X_i)
-\widehat\tau_{\mathcal I_j}(\mathbf X_i)
\right|^2
\right]\vee\{\log n/({C_0}n)\}
}{
E_{\mathbf X}\left|
\widehat\tau_{\mathcal I_{3-j,3-\ell}}(\mathbf X)
-\widehat\tau_{\mathcal I_j}(\mathbf X)
\right|^2
\vee\{\log n/({C_0}n)\}
}
=O_p(1).
\label{empirical-to-conditional}
\end{align}

First suppose Condition (C1) holds. For each $\ell$, the squared
triangle inequality gives
\begin{align*}
&E_{\mathbf X}\left|
\widehat\tau_{\mathcal I_{3-j,3-\ell}}(\mathbf X)
-\widehat\tau_{\mathcal I_j}(\mathbf X)
\right|^2\\
&\quad\le
2E_{\mathbf X}\left|
\widehat\tau_{\mathcal I_j}(\mathbf X)-\tau(\mathbf X)
\right|^2
+
2E_{\mathbf X}\left|
\widehat\tau_{\mathcal I_{3-j,3-\ell}}(\mathbf X)-\tau(\mathbf X)
\right|^2\\
&\quad=
O_p(\log n/n),
\end{align*}
where the last equality follows from Condition (C1), applied to the
estimators trained on $\mathcal I_j$ and
$\mathcal I_{3-j,3-\ell}$. Both training-fold sizes are of order $n$.

Equation~\eqref{empirical-to-conditional} therefore implies, for each
$\ell$,
\[
\frac{1}{|\mathcal I_{3-j,\ell}|}
\sum_{i\in\mathcal I_{3-j,\ell}}
\left|
\widehat\tau_{\mathcal I_{3-j,3-\ell}}(\mathbf X_i)
-\widehat\tau_{\mathcal I_j}(\mathbf X_i)
\right|^2
=
O_p\{\log n/({C_0}n)\}.
\]
It follows from Eq.~\eqref{exact-eae-expression} that
\[
EAE(\widehat\tau_{\mathcal I_j})
=
O_p\{\log n/({C_0}n)\}.
\]
Condition (C1), applied to $\widehat\tau_{\mathcal I_j}$, also gives
\[
E_{\mathbf X}
|\widehat\tau_{\mathcal I_j}(\mathbf X)-\tau(\mathbf X)|^2
=
O_p\{\log n/({C_0}n)\}.
\]
Both the numerator and denominator in each of
Eqs.~\eqref{eae-error-path} and \eqref{error-to-eae} are therefore
bounded below by $\log n/({C_0}n)$ and are
$O_p\{\log n/({C_0}n)\}$. This proves both comparisons under Condition
(C1). Neither Condition (C2) nor a lower concentration inequality is
needed in this case.

Now suppose Conditions (C2) and (C3) hold. Since there are only two
values of $\ell$, Eqs.~\eqref{exact-eae-expression} and
\eqref{empirical-to-conditional} imply
\begin{align}
&
\frac{
EAE(\widehat\tau_{\mathcal I_j})\vee\{\log n/({C_0}n)\}
}{
\left\{
\displaystyle
\frac12\sum_{\ell=1}^2
E_{\mathbf X}\left|
\widehat\tau_{\mathcal I_{3-j,3-\ell}}(\mathbf X)
-\widehat\tau_{\mathcal I_j}(\mathbf X)
\right|^2
\right\}
\vee\{\log n/({C_0}n)\}
}
=O_p(1).
\label{markov-direction}
\end{align}

We next establish the reverse comparison. Fix $\varepsilon>0$. By
Condition (C2), applied to all relevant half-fold and quarter-fold
estimators, and a union bound over the four pairs $(j,\ell)$, there
exists a constant $B<\infty$ such that
\begin{align}
P\Bigg[
\bigcap_{j,\ell=1}^2
\left\{
\|\widehat\tau_{\mathcal I_j}\|_\infty\le B,\quad
\|\widehat\tau_{\mathcal I_{3-j,3-\ell}}\|_\infty\le B
\right\}
\Bigg]
\ge1-\varepsilon
\label{common-boundedness}
\end{align}
for all sufficiently large $n$.

Fix $j$ and $\ell$. On the boundedness event appearing in
Eq.~\eqref{common-boundedness}, every summand in the corresponding
validation-fold average is between zero and $4B^2$. Moreover,
\begin{align*}
&\operatorname{Var}\left(
\left.
\left|
\widehat\tau_{\mathcal I_{3-j,3-\ell}}(\mathbf X_i)
-\widehat\tau_{\mathcal I_j}(\mathbf X_i)
\right|^2
\,\right|\,
\widehat\tau_{\mathcal I_{3-j,3-\ell}},
\widehat\tau_{\mathcal I_j}
\right)\\
&\quad\le
E\left(
\left.
\left|
\widehat\tau_{\mathcal I_{3-j,3-\ell}}(\mathbf X_i)
-\widehat\tau_{\mathcal I_j}(\mathbf X_i)
\right|^4
\,\right|\,
\widehat\tau_{\mathcal I_{3-j,3-\ell}},
\widehat\tau_{\mathcal I_j}
\right)\\
&\quad\le
4B^2
E_{\mathbf X}\left|
\widehat\tau_{\mathcal I_{3-j,3-\ell}}(\mathbf X)
-\widehat\tau_{\mathcal I_j}(\mathbf X)
\right|^2.
\end{align*}
Consequently, whenever
\[
E_{\mathbf X}\left|
\widehat\tau_{\mathcal I_{3-j,3-\ell}}(\mathbf X)
-\widehat\tau_{\mathcal I_j}(\mathbf X)
\right|^2>0,
\]
the one-sided Bernstein inequality gives
\begin{align}
&P\Bigg[
\frac{1}{|\mathcal I_{3-j,\ell}|}
\sum_{i\in\mathcal I_{3-j,\ell}}
\left|
\widehat\tau_{\mathcal I_{3-j,3-\ell}}(\mathbf X_i)
-\widehat\tau_{\mathcal I_j}(\mathbf X_i)
\right|^2
\notag\\
&\hspace{27mm}\le
\frac12
E_{\mathbf X}\left|
\widehat\tau_{\mathcal I_{3-j,3-\ell}}(\mathbf X)
-\widehat\tau_{\mathcal I_j}(\mathbf X)
\right|^2
\,\Bigg|\,
\widehat\tau_{\mathcal I_{3-j,3-\ell}},
\widehat\tau_{\mathcal I_j}
\Bigg]
\notag\\
&\qquad\le
\exp\left\{
-\frac{
3|\mathcal I_{3-j,\ell}|
E_{\mathbf X}\left|
\widehat\tau_{\mathcal I_{3-j,3-\ell}}(\mathbf X)
-\widehat\tau_{\mathcal I_j}(\mathbf X)
\right|^2
}{
112B^2
}
\right\}.
\label{conditional-bernstein}
\end{align}

Because $|\mathcal I_{3-j,\ell}|\asymp n$, a constant $K_0>0$ can be
chosen sufficiently large, depending only on $B$ and the fixed fold
proportions, so that the tower property and
Eq.~\eqref{conditional-bernstein} give
\begin{align*}
&P\Bigg[
\left\{
\frac{1}{|\mathcal I_{3-j,\ell}|}
\sum_{i\in\mathcal I_{3-j,\ell}}
\left|
\widehat\tau_{\mathcal I_{3-j,3-\ell}}(\mathbf X_i)
-\widehat\tau_{\mathcal I_j}(\mathbf X_i)
\right|^2
\right.\\
&\hspace{28mm}\left.
\le
\frac12
E_{\mathbf X}\left|
\widehat\tau_{\mathcal I_{3-j,3-\ell}}(\mathbf X)
-\widehat\tau_{\mathcal I_j}(\mathbf X)
\right|^2
\right\}\\
&\hspace{7mm}\cap
\left\{
E_{\mathbf X}\left|
\widehat\tau_{\mathcal I_{3-j,3-\ell}}(\mathbf X)
-\widehat\tau_{\mathcal I_j}(\mathbf X)
\right|^2
\ge K_0\log n/n
\right\}\\
&\hspace{7mm}\cap
\left\{
\|\widehat\tau_{\mathcal I_j}\|_\infty\le B,\quad
\|\widehat\tau_{\mathcal I_{3-j,3-\ell}}\|_\infty\le B
\right\}
\Bigg]
\le n^{-3}.
\end{align*}
Indeed, the left side equals the expectation of the indicator of the
last two events multiplied by the conditional probability in
Eq.~\eqref{conditional-bernstein}; on those events, the conditional
probability is at most $n^{-3}$ by the choice of $K_0$.

Combining the preceding probability bound with
Eq.~\eqref{common-boundedness} and taking a union bound over the four
pairs gives
\begin{align}
P\Bigg[
\bigcap_{j,\ell=1}^2
\Bigg(
&
\left\{
E_{\mathbf X}\left|
\widehat\tau_{\mathcal I_{3-j,3-\ell}}(\mathbf X)
-\widehat\tau_{\mathcal I_j}(\mathbf X)
\right|^2
<K_0\log n/n
\right\}
\notag\\
&\quad\cup
\left\{
\frac{1}{|\mathcal I_{3-j,\ell}|}
\sum_{i\in\mathcal I_{3-j,\ell}}
\left|
\widehat\tau_{\mathcal I_{3-j,3-\ell}}(\mathbf X_i)
-\widehat\tau_{\mathcal I_j}(\mathbf X_i)
\right|^2
\right.
\notag\\
&\hspace{31mm}\left.
>
\frac12
E_{\mathbf X}\left|
\widehat\tau_{\mathcal I_{3-j,3-\ell}}(\mathbf X)
-\widehat\tau_{\mathcal I_j}(\mathbf X)
\right|^2
\right\}
\Bigg)
\Bigg]
\ge1-\varepsilon-4n^{-3}.
\label{simultaneous-bernstein}
\end{align}

On the event in Eq.~\eqref{simultaneous-bernstein}, fix any
$(j,\ell)$. If
\[
E_{\mathbf X}\left|
\widehat\tau_{\mathcal I_{3-j,3-\ell}}(\mathbf X)
-\widehat\tau_{\mathcal I_j}(\mathbf X)
\right|^2
\ge K_0\log n/n,
\]
then Eq.~\eqref{exact-eae-expression} and the nonnegativity of both
validation-fold averages give
\begin{align*}
EAE(\widehat\tau_{\mathcal I_j})
&\ge
\frac12
\frac{1}{|\mathcal I_{3-j,\ell}|}
\sum_{i\in\mathcal I_{3-j,\ell}}
\left|
\widehat\tau_{\mathcal I_{3-j,3-\ell}}(\mathbf X_i)
-\widehat\tau_{\mathcal I_j}(\mathbf X_i)
\right|^2\\
&>
\frac14
E_{\mathbf X}\left|
\widehat\tau_{\mathcal I_{3-j,3-\ell}}(\mathbf X)
-\widehat\tau_{\mathcal I_j}(\mathbf X)
\right|^2.
\end{align*}
If instead
\[
E_{\mathbf X}\left|
\widehat\tau_{\mathcal I_{3-j,3-\ell}}(\mathbf X)
-\widehat\tau_{\mathcal I_j}(\mathbf X)
\right|^2
<K_0\log n/n
=
K_0{C_0}\{\log n/({C_0}n)\},
\]
then the common truncation level controls this conditional squared
difference. Therefore, in either case,
\begin{align}
&
E_{\mathbf X}\left|
\widehat\tau_{\mathcal I_{3-j,3-\ell}}(\mathbf X)
-\widehat\tau_{\mathcal I_j}(\mathbf X)
\right|^2
\vee\{\log n/({C_0}n)\}
\notag\\
&\qquad\le
\max\{4,K_0{C_0},1\}
\left[
EAE(\widehat\tau_{\mathcal I_j})
\vee\{\log n/({C_0}n)\}
\right].
\label{deterministic-reverse-comparison}
\end{align}
For each $\varepsilon>0$, the constant in
Eq.~\eqref{deterministic-reverse-comparison} is fixed, and the
inequality holds simultaneously over the four pairs with probability
at least $1-\varepsilon-4n^{-3}$. Since $\varepsilon$ is arbitrary,
Eq.~\eqref{deterministic-reverse-comparison} implies
\begin{align}
&
\frac{
\left\{
\displaystyle
\frac12\sum_{\ell=1}^2
E_{\mathbf X}\left|
\widehat\tau_{\mathcal I_{3-j,3-\ell}}(\mathbf X)
-\widehat\tau_{\mathcal I_j}(\mathbf X)
\right|^2
\right\}
\vee\{\log n/({C_0}n)\}
}{
EAE(\widehat\tau_{\mathcal I_j})\vee\{\log n/({C_0}n)\}
}
=O_p(1).
\label{bernstein-direction}
\end{align}

For each $\ell$, the folds $\mathcal I_j$ and
$\mathcal I_{3-j,3-\ell}$ are disjoint and both have sizes of order
$n$. Condition (C3), applied with $\mathcal I_j$ as the first fold and
$\mathcal I_{3-j,3-\ell}$ as the second fold, gives, after applying the
same truncation to numerator and denominator,
\begin{align}
&
\frac{
E_{\mathbf X}\left|
\widehat\tau_{\mathcal I_{3-j,3-\ell}}(\mathbf X)
-\widehat\tau_{\mathcal I_j}(\mathbf X)
\right|^2
\vee\{\log n/({C_0}n)\}
}{
E_{\mathbf X}
|\widehat\tau_{\mathcal I_j}(\mathbf X)-\tau(\mathbf X)|^2
\vee\{\log n/({C_0}n)\}
}
=O_p(1),
\label{c3-first-direction}
\\
&
\frac{
E_{\mathbf X}
|\widehat\tau_{\mathcal I_j}(\mathbf X)-\tau(\mathbf X)|^2
\vee\{\log n/({C_0}n)\}
}{
E_{\mathbf X}\left|
\widehat\tau_{\mathcal I_{3-j,3-\ell}}(\mathbf X)
-\widehat\tau_{\mathcal I_j}(\mathbf X)
\right|^2
\vee\{\log n/({C_0}n)\}
}
=O_p(1).
\label{c3-second-direction}
\end{align}
Because there are only two values of $\ell$,
Eqs.~\eqref{c3-first-direction}--\eqref{c3-second-direction} imply
\begin{align}
&
\frac{
\left\{
\displaystyle
\frac12\sum_{\ell=1}^2
E_{\mathbf X}\left|
\widehat\tau_{\mathcal I_{3-j,3-\ell}}(\mathbf X)
-\widehat\tau_{\mathcal I_j}(\mathbf X)
\right|^2
\right\}
\vee\{\log n/({C_0}n)\}
}{
E_{\mathbf X}
|\widehat\tau_{\mathcal I_j}(\mathbf X)-\tau(\mathbf X)|^2
\vee\{\log n/({C_0}n)\}
}
=O_p(1)
\end{align}
and the same bound with the numerator and denominator interchanged.
Combining these two bounds with
Eqs.~\eqref{markov-direction} and \eqref{bernstein-direction} proves
Eqs.~\eqref{eae-error-path} and \eqref{error-to-eae} under Conditions
(C2) and (C3). Thus, the two comparisons hold under either alternative
in Proposition 1.

By Algorithm 2,
\[
h_{n,\mathcal I_j}
=
{C_0}\log n\,n^{1/4}
\left\{
EAE(\widehat\tau_{\mathcal I_j})
\vee\frac{\log n}{{C_0}n}
\right\}^{3/4}.
\]
Consequently,
\begin{align*}
&\frac{
n^{1/4}
\left\{
E_{\mathbf X}
|\widehat\tau_{\mathcal I_j}(\mathbf X)-\tau(\mathbf X)|^2
\right\}^{3/4}
}{
h_{n,\mathcal I_j}
}\\
&\qquad=
\frac1{{C_0}\log n}
\left\{
\frac{
E_{\mathbf X}
|\widehat\tau_{\mathcal I_j}(\mathbf X)-\tau(\mathbf X)|^2
}{
EAE(\widehat\tau_{\mathcal I_j})\vee\{\log n/({C_0}n)\}
}
\right\}^{3/4}\\
&\qquad\le
\frac1{{C_0}\log n}
\left\{
\frac{
E_{\mathbf X}
|\widehat\tau_{\mathcal I_j}(\mathbf X)-\tau(\mathbf X)|^2
\vee\{\log n/({C_0}n)\}
}{
EAE(\widehat\tau_{\mathcal I_j})\vee\{\log n/({C_0}n)\}
}
\right\}^{3/4}\\
&\qquad=
O_p(1/\log n)
=
o_p(1),
\end{align*}
where the last equality follows from Eq.~\eqref{error-to-eae}. Hence
\[
n^{1/4}
\left\{
E_{\mathbf X}
|\widehat\tau_{\mathcal I_j}(\mathbf X)-\tau(\mathbf X)|^2
\right\}^{3/4}
\prec_p
h_{n,\mathcal I_j}.
\]

Finally, Assumption 9 gives some $\beta\in(0,\alpha)$ such that
\[
E_{\mathbf X}
|\widehat\tau_{\mathcal I_j}(\mathbf X)-\tau(\mathbf X)|^2
=
o_p\left(
n^{-\frac{2+2\beta/3}{2(1+\beta)}}
\right).
\]
Because $\beta>0$,
\[
\frac{2+2\beta/3}{2(1+\beta)}
=
\frac{1+\beta/3}{1+\beta}
<1,
\]
and therefore
\[
\frac{\log n}{{C_0}n}
=
o\left(
n^{-\frac{2+2\beta/3}{2(1+\beta)}}
\right).
\]
Equation~\eqref{eae-error-path} now implies
\[
EAE(\widehat\tau_{\mathcal I_j})
\vee\frac{\log n}{{C_0}n}
=
o_p\left(
n^{-\frac{2+2\beta/3}{2(1+\beta)}}
\right).
\]
Substitution into the formula for $h_{n,\mathcal I_j}$ gives
\[
h_{n,\mathcal I_j}
=
o_p\left\{
\log n\,n^{-1/\{2(1+\beta)\}}
\right\},
\]
where
\[
n^{1/4}
\left\{
n^{-\frac{2+2\beta/3}{2(1+\beta)}}
\right\}^{3/4}
=
n^{-1/\{2(1+\beta)\}}.
\]
Since $\beta<\alpha$,
\[
\frac{
\log n\,n^{-1/\{2(1+\beta)\}}
}{
n^{-1/\{2(1+\alpha)\}}
}
=
\log n\,
n^{-\frac{\alpha-\beta}
{2(1+\alpha)(1+\beta)}}
\longrightarrow0.
\]
It follows that
\[
h_{n,\mathcal I_j}
\prec_p
n^{-1/\{2(1+\alpha)\}}.
\]
All preceding bounds hold simultaneously for $j=1,2$, completing the
proof.
\end{proof}

Here and below, a relevant training fold means one of the half folds
$\mathcal I_j$ or quarter folds $\mathcal I_{j,k}$ used by Algorithm 2.
\begin{lemma}[Eq.~(10) and Condition (C1) for correctly specified parametric working models]\label{parametric-tuning-conditions}
Fix $q\ge2$. Suppose that, for every relevant fold $\mathcal I$,
\[
\widehat\tau_{\mathcal I}(x)-\tau(x)
=(\widehat\theta_{\mathcal I}-\theta)^\top\varphi(x),
\]
where $\varphi(x)\in\mathbb R^d$ has fixed finite dimension,
$\Sigma=E_{\mathbf X}\{\varphi(\mathbf X)\varphi(\mathbf X)^\top\}$ is positive definite,
$E_{\mathbf X}\|\varphi(\mathbf X)\|^q<\infty$, and
$\|\widehat\theta_{\mathcal I}-\theta\|=O_p(|\mathcal I|^{-1/2})$. Then, for every relevant fold, the following conclusions hold:
\begin{enumerate}
\item[\textup{(a)}] The moment-comparability condition in Eq. (10) of the main text holds at $q$:
\[
E_{\mathbf X}|\widehat\tau_{\mathcal I}(\mathbf X)-\tau(\mathbf X)|^q
=O_p\left[
\{E_{\mathbf X}|\widehat\tau_{\mathcal I}(\mathbf X)-\tau(\mathbf X)|^2\}^{q/2}
\right];
\]
\item[\textup{(b)}] Condition (C1) holds. In fact,
\[
E_{\mathbf X}|\widehat\tau_{\mathcal I}(\mathbf X)-\tau(\mathbf X)|^2=O_p(n^{-1}).
\]
\end{enumerate}
\end{lemma}
\begin{proof}
Fix a relevant fold $\mathcal I$ and write
$\delta_{\mathcal I}=\widehat\theta_{\mathcal I}-\theta$.
For conclusion (a), Cauchy--Schwarz and the positive definiteness of $\Sigma$ give the continuous chain
\begin{align*}
E_{\mathbf X}|\widehat\tau_{\mathcal I}(\mathbf X)-\tau(\mathbf X)|^q
&=E_{\mathbf X}|\delta_{\mathcal I}^\top\varphi(\mathbf X)|^q\\
&\le \|\delta_{\mathcal I}\|^q E_{\mathbf X}\|\varphi(\mathbf X)\|^q\\
&\le
\frac{E_{\mathbf X}\|\varphi(\mathbf X)\|^q}
{\lambda_{\min}(\Sigma)^{q/2}}
\{\delta_{\mathcal I}^\top\Sigma\delta_{\mathcal I}\}^{q/2}\\
&=
\frac{E_{\mathbf X}\|\varphi(\mathbf X)\|^q}
{\lambda_{\min}(\Sigma)^{q/2}}
\{E_{\mathbf X}|\widehat\tau_{\mathcal I}(\mathbf X)-\tau(\mathbf X)|^2\}^{q/2}.
\end{align*}
The multiplicative constant is finite and does not depend on $n$, $\mathcal I$, or the realized training sample. This proves Eq. (10), including the case $\delta_{\mathcal I}=0$.

For conclusion (b),
\begin{align*}
E_{\mathbf X}|\widehat\tau_{\mathcal I}(\mathbf X)-\tau(\mathbf X)|^2
&=\delta_{\mathcal I}^\top\Sigma\delta_{\mathcal I}\\
&\le\lambda_{\max}(\Sigma)\|\delta_{\mathcal I}\|^2
=O_p(|\mathcal I|^{-1})
=O_p(n^{-1}),
\end{align*}
because $|\mathcal I|\asymp n$. This is stronger than $O_p(\log n/n)$ and therefore proves Condition (C1).
\end{proof}

\begin{lemma}[Eq.~(10) and Conditions (C2)--(C3) for undersmoothed estimators]\label{linear-smoother-c3}
Fix $q\ge2$. In this lemma, a relevant fold means any half-sample or
quarter-sample training fold used in Algorithm~2. For every relevant fold
$\mathcal I$, regard
$\widehat\tau_{\mathcal I}$ as a random function through the training
observations in $\mathcal I$, and let
$E_{\widehat\tau_{\mathcal I}}$ denote expectation with respect to its
sampling distribution. Thus,
$E_{\widehat\tau_{\mathcal I}}\{\widehat\tau_{\mathcal I}\}$ denotes the
function
$x\mapsto E_{\widehat\tau_{\mathcal I}}
\{\widehat\tau_{\mathcal I}(x)\}$.
Let $E_{\mathbf X}$ denote expectation over an independent covariate draw,
and define the integrated variance
\[
v_{\mathcal I}
=
E_{\widehat\tau_{\mathcal I}}\left[
E_{\mathbf X}
\left|
\widehat\tau_{\mathcal I}(\mathbf X)
-E_{\widehat\tau_{\mathcal I}}
\{\widehat\tau_{\mathcal I}(\mathbf X)\}
\right|^2
\right].
\]
The quantity $v_{\mathcal I}$ is deterministic, and all probability limits
and stochastic-order statements below are with respect to the training
observations. Suppose
$v_{\mathcal I}>0$ and that there exist fixed constants
$0<c_v\le C_v<\infty$ such that, for every pair of relevant folds
$\mathcal I,\mathcal I'$ and all sufficiently large $n$,
\[
c_v
\le \frac{v_{\mathcal I'}}{v_{\mathcal I}}
\le C_v.
\]
Assume further that the following requirements hold uniformly over the
relevant folds:
\begin{enumerate}
\item[\textup{(U1)}] \textit{Strict undersmoothing:}
\[
\left\|
E_{\widehat\tau_{\mathcal I}}\{\widehat\tau_{\mathcal I}\}
-\tau
\right\|_\infty
=o(v_{\mathcal I}^{1/2}).
\]

\item[\textup{(U2)}] \textit{Quadratic concentration and cross-fold
orthogonality:}
\[
\frac{
E_{\mathbf X}
\left|
\widehat\tau_{\mathcal I}(\mathbf X)
-E_{\widehat\tau_{\mathcal I}}
\{\widehat\tau_{\mathcal I}(\mathbf X)\}
\right|^2
}{v_{\mathcal I}}
\longrightarrow_p1,
\]
and, for every pair of disjoint relevant folds $\mathcal I,\mathcal I'$,
\[
\frac{
E_{\mathbf X}\left[
\left\{
\widehat\tau_{\mathcal I}(\mathbf X)
-E_{\widehat\tau_{\mathcal I}}
\{\widehat\tau_{\mathcal I}(\mathbf X)\}
\right\}
\left\{
\widehat\tau_{\mathcal I'}(\mathbf X)
-E_{\widehat\tau_{\mathcal I'}}
\{\widehat\tau_{\mathcal I'}(\mathbf X)\}
\right\}
\right]
}{(v_{\mathcal I}v_{\mathcal I'})^{1/2}}
\longrightarrow_p0.
\]

\item[\textup{(U3)}] \textit{Integrated $q$th-moment bound:}
\[
E_{\mathbf X}
\left|
\widehat\tau_{\mathcal I}(\mathbf X)
-E_{\widehat\tau_{\mathcal I}}
\{\widehat\tau_{\mathcal I}(\mathbf X)\}
\right|^q
=O_p(v_{\mathcal I}^{q/2}).
\]

\item[\textup{(U4)}] \textit{Stochastic sup-norm boundedness:}
$\|\tau\|_\infty<\infty$ and
\[
\|\widehat\tau_{\mathcal I}-\tau\|_\infty=O_p(1).
\]
\end{enumerate}
Then the following conclusions hold:
\begin{enumerate}
\item[\textup{(a)}] The moment-comparability condition in Eq.~(10) of the
main text holds at $q$ for every relevant fold:
\[
E_{\mathbf X}
|\widehat\tau_{\mathcal I}(\mathbf X)-\tau(\mathbf X)|^q
=O_p\left[
\left\{
E_{\mathbf X}
|\widehat\tau_{\mathcal I}(\mathbf X)-\tau(\mathbf X)|^2
\right\}^{q/2}
\right].
\]

\item[\textup{(b)}] Condition (C2) holds for every relevant fold:
\[
\|\widehat\tau_{\mathcal I}\|_\infty=O_p(1).
\]

\item[\textup{(c)}] Condition (C3) holds for every pair of disjoint relevant
folds $\mathcal I,\mathcal I'$:
\[
E_{\mathbf X}
|\widehat\tau_{\mathcal I}(\mathbf X)
-\widehat\tau_{\mathcal I'}(\mathbf X)|^2
\asymp_p
E_{\mathbf X}
|\widehat\tau_{\mathcal I}(\mathbf X)-\tau(\mathbf X)|^2.
\]
\end{enumerate}
\end{lemma}

\begin{proof}
For $r\ge1$, write
$\|f\|_r=\{E_{\mathbf X}|f(\mathbf X)|^r\}^{1/r}$, and define
\[
b_{\mathcal I}
=E_{\widehat\tau_{\mathcal I}}\{\widehat\tau_{\mathcal I}\}-\tau,
\qquad
u_{\mathcal I}
=\widehat\tau_{\mathcal I}
-E_{\widehat\tau_{\mathcal I}}\{\widehat\tau_{\mathcal I}\}.
\]
Thus,
$\widehat\tau_{\mathcal I}-\tau=u_{\mathcal I}+b_{\mathcal I}$.
By (U1),
\[
\|b_{\mathcal I}\|_2
\le \|b_{\mathcal I}\|_\infty
=o(v_{\mathcal I}^{1/2}),
\]
whereas the first part of (U2) gives
\[
\|u_{\mathcal I}\|_2^2
=v_{\mathcal I}\{1+o_p(1)\},
\qquad
\|u_{\mathcal I}\|_2
=v_{\mathcal I}^{1/2}\{1+o_p(1)\}.
\]
The reverse triangle inequality therefore gives
\begin{align*}
\left|
\|\widehat\tau_{\mathcal I}-\tau\|_2
-\|u_{\mathcal I}\|_2
\right|
&=
\left|
\|u_{\mathcal I}+b_{\mathcal I}\|_2
-\|u_{\mathcal I}\|_2
\right|\\
&\le \|b_{\mathcal I}\|_2
=o(v_{\mathcal I}^{1/2}).
\end{align*}
Consequently,
\begin{equation}\label{undersmooth-error-scale}
E_{\mathbf X}
|\widehat\tau_{\mathcal I}(\mathbf X)-\tau(\mathbf X)|^2
=\|\widehat\tau_{\mathcal I}-\tau\|_2^2
=v_{\mathcal I}\{1+o_p(1)\}.
\end{equation}

We first prove conclusion (a). By (U3),
\[
\|u_{\mathcal I}\|_q
=O_p(v_{\mathcal I}^{1/2}),
\]
and by (U1),
\[
\|b_{\mathcal I}\|_q
\le \|b_{\mathcal I}\|_\infty
=o(v_{\mathcal I}^{1/2}).
\]
Minkowski's inequality hence yields
\[
\|\widehat\tau_{\mathcal I}-\tau\|_q
=\|u_{\mathcal I}+b_{\mathcal I}\|_q
\le \|u_{\mathcal I}\|_q+\|b_{\mathcal I}\|_q
=O_p(v_{\mathcal I}^{1/2}).
\]
Raising both sides to the $q$th power and using
Eq.~\eqref{undersmooth-error-scale}, we obtain
\begin{align*}
E_{\mathbf X}
|\widehat\tau_{\mathcal I}(\mathbf X)-\tau(\mathbf X)|^q
&=O_p(v_{\mathcal I}^{q/2})\\
&=O_p\left[
\left\{
E_{\mathbf X}
|\widehat\tau_{\mathcal I}(\mathbf X)-\tau(\mathbf X)|^2
\right\}^{q/2}
\right].
\end{align*}
This is the moment-comparability condition in Eq.~(10) at $q$.

For conclusion (b), (U4) and the triangle inequality give
\[
\|\widehat\tau_{\mathcal I}\|_\infty
\le
\|\tau\|_\infty
+\|\widehat\tau_{\mathcal I}-\tau\|_\infty
=O_p(1),
\]
which is Condition (C2).

It remains to prove conclusion (c). Let $\mathcal I$ and $\mathcal I'$ be
disjoint relevant folds. The two parts of (U2) imply
\begin{align*}
\|u_{\mathcal I}\|_2^2
&=v_{\mathcal I}\{1+o_p(1)\},\\
\|u_{\mathcal I'}\|_2^2
&=v_{\mathcal I'}\{1+o_p(1)\},\\
E_{\mathbf X}
\{u_{\mathcal I}(\mathbf X)u_{\mathcal I'}(\mathbf X)\}
&=o_p\{(v_{\mathcal I}v_{\mathcal I'})^{1/2}\}.
\end{align*}
Because
$(v_{\mathcal I}v_{\mathcal I'})^{1/2}
\le (v_{\mathcal I}+v_{\mathcal I'})/2$, it follows that
\begin{align}
\|u_{\mathcal I}-u_{\mathcal I'}\|_2^2
&=\|u_{\mathcal I}\|_2^2
+\|u_{\mathcal I'}\|_2^2
-2E_{\mathbf X}
\{u_{\mathcal I}(\mathbf X)u_{\mathcal I'}(\mathbf X)\}
\notag\\
&=(v_{\mathcal I}+v_{\mathcal I'})\{1+o_p(1)\}.
\label{centered-split-scale}
\end{align}
Moreover, (U1) gives
\begin{align*}
\|b_{\mathcal I}-b_{\mathcal I'}\|_2
&\le
\|b_{\mathcal I}\|_\infty
+\|b_{\mathcal I'}\|_\infty\\
&=o(v_{\mathcal I}^{1/2})
+o(v_{\mathcal I'}^{1/2})\\
&=o\{(v_{\mathcal I}+v_{\mathcal I'})^{1/2}\}.
\end{align*}
Applying the reverse triangle inequality once more and using
Eq.~\eqref{centered-split-scale},
\begin{align*}
&\left|
\|\widehat\tau_{\mathcal I}-\widehat\tau_{\mathcal I'}\|_2
-\|u_{\mathcal I}-u_{\mathcal I'}\|_2
\right|\\
&\qquad\le
\|b_{\mathcal I}-b_{\mathcal I'}\|_2
=o\{(v_{\mathcal I}+v_{\mathcal I'})^{1/2}\}.
\end{align*}
Therefore,
\begin{equation}\label{split-discrepancy-scale}
E_{\mathbf X}
|\widehat\tau_{\mathcal I}(\mathbf X)
-\widehat\tau_{\mathcal I'}(\mathbf X)|^2
=(v_{\mathcal I}+v_{\mathcal I'})\{1+o_p(1)\}.
\end{equation}
Combining Eqs.~\eqref{undersmooth-error-scale} and
\eqref{split-discrepancy-scale} gives
\[
\frac{
E_{\mathbf X}
|\widehat\tau_{\mathcal I}(\mathbf X)
-\widehat\tau_{\mathcal I'}(\mathbf X)|^2
}{
E_{\mathbf X}
|\widehat\tau_{\mathcal I}(\mathbf X)-\tau(\mathbf X)|^2
}
=
\frac{v_{\mathcal I}+v_{\mathcal I'}}{v_{\mathcal I}}
\{1+o_p(1)\}.
\]
Since
$c_v\le v_{\mathcal I'}/v_{\mathcal I}\le C_v$, the preceding ratio is
bounded above and away from zero by fixed constants with probability
tending to one. Hence
\[
E_{\mathbf X}
|\widehat\tau_{\mathcal I}(\mathbf X)
-\widehat\tau_{\mathcal I'}(\mathbf X)|^2
\asymp_p
E_{\mathbf X}
|\widehat\tau_{\mathcal I}(\mathbf X)-\tau(\mathbf X)|^2,
\]
which is precisely Condition (C3) under the definition of $\asymp_p$ in
the main text.
\end{proof}

The next two examples define concrete spline and kernel contrast estimators and verify the variance-comparability premise and (U1)--(U4) separately. Their assumptions are deliberately explicit sufficient conditions rather than claims about every possible implementation of the two smoothers.

\begin{example}[Undersmoothed spline least squares]
\label{example:undersmoothed-spline}
Consider i.i.d. observations $O_i=(\mathbf X_i,A_i,Y_i)$ satisfying
\[
Y_i=\mu_{A_i}(\mathbf X_i)+\varepsilon_i,
\qquad
E(\varepsilon_i\mid \mathbf X_i,A_i)=0,
\qquad
\tau=\mu_1-\mu_0.
\]
Suppose the following conditions hold.
\begin{enumerate}
\item[\textup{(S1)}] The support of $\mathbf X$ is $[0,1]^p$, its density is bounded above and away from zero, and, for $a=0,1$,
\[
c_\pi\le P(A=a\mid\mathbf X=x)\le1-c_\pi.
\]
Conditional on $(\mathbf X,A)$, $\varepsilon$ has a uniformly bounded sub-Gaussian norm and
\[
0<c_\varepsilon
\le E(\varepsilon^2\mid\mathbf X,A)
\le C_\varepsilon<\infty.
\]
\item[\textup{(S2)}] For $a=0,1$, $\mu_a$ belongs to a H\"older ball of smoothness $s>p/2$. For a relevant fold $\mathcal I$, set $N_{\mathcal I}=|\mathcal I|$. Let $\mathcal S_{\mathcal I}$ be a tensor-product spline space of order $r>s$ with quasi-uniform knots and
\[
K_{\mathcal I}=\dim(\mathcal S_{\mathcal I})
\asymp N_{\mathcal I}^{\kappa},
\qquad
\frac{p}{2s+p}<\kappa<\frac12.
\]
The knots and $K_{\mathcal I}$ depend only on $N_{\mathcal I}$.
\item[\textup{(S3)}] Let $\mathbf B_{\mathcal I}(x)\in\mathbb R^{K_{\mathcal I}}$ be a stably normalized tensor-product B-spline basis. Uniformly over the relevant fold sizes,
\[
\sup_x\|\mathbf B_{\mathcal I}(x)\|
\lesssim K_{\mathcal I}^{1/2},
\]
and the eigenvalues of
\[
G_{\mathcal I}
=E_{\mathbf X}\{\mathbf B_{\mathcal I}(\mathbf X)
\mathbf B_{\mathcal I}(\mathbf X)^\top\}
\]
are bounded above and away from zero.
Write
\[
Q_{a,\mathcal I}
=E\left[\mathbb I(A=a)\mathbf B_{\mathcal I}(\mathbf X)
\mathbf B_{\mathcal I}(\mathbf X)^\top\right]
\]
and define the corresponding weighted population projection by
\[
(P_{a,\mathcal I}f)(x)
=\mathbf B_{\mathcal I}(x)^\top Q_{a,\mathcal I}^{-1}
E\{\mathbb I(A=a)\mathbf B_{\mathcal I}(\mathbf X)f(\mathbf X)\}.
\]
Assume that these projectors are
uniformly stable in sup norm,
\[
\sup_{a,\mathcal I}\|P_{a,\mathcal I}\|_{\infty\to\infty}<\infty,
\]
and that the analogous empirical projectors are uniformly sup-norm stable
whenever their empirical Gram matrices have minimum eigenvalue at least one
half of the uniform population lower bound.
\end{enumerate}

For $a=0,1$, define the within-arm least-squares estimator by
\begin{align*}
\widehat Q_{a,\mathcal I}
&=\frac1{N_{\mathcal I}}
\sum_{i\in\mathcal I}\mathbb I(A_i=a)
\mathbf B_{\mathcal I}(\mathbf X_i)
\mathbf B_{\mathcal I}(\mathbf X_i)^\top,\\
\widehat g_{a,\mathcal I}
&=\frac1{N_{\mathcal I}}
\sum_{i\in\mathcal I}\mathbb I(A_i=a)
\mathbf B_{\mathcal I}(\mathbf X_i)Y_i.
\end{align*}
Let
\[
\underline c_Q=\inf_{a,\mathcal I}\lambda_{\min}(Q_{a,\mathcal I})>0,
\qquad
\mathcal E_{a,\mathcal I}
=\{\lambda_{\min}(\widehat Q_{a,\mathcal I})\ge\underline c_Q/2\},
\]
and use the stabilized coefficient
\[
\widehat\gamma_{a,\mathcal I}
=
\begin{cases}
\widehat Q_{a,\mathcal I}^{-1}\widehat g_{a,\mathcal I},
&\mathcal E_{a,\mathcal I}\text{ holds},\\[2mm]
0,&\mathcal E_{a,\mathcal I}^{c}\text{ holds}.
\end{cases}
\]
Thus the usual within-arm least-squares fit is used whenever its Gram matrix
is well conditioned, with a bounded fallback on an event of exponentially
small probability. Finally, set
\begin{align*}
\widehat\mu_{a,\mathcal I}(x)
&=\mathbf B_{\mathcal I}(x)^\top\widehat\gamma_{a,\mathcal I},
\qquad
\widehat\tau_{\mathcal I}
=\widehat\mu_{1,\mathcal I}-\widehat\mu_{0,\mathcal I}.
\end{align*}
Then the variance-comparability premise and (U1)--(U4) of
Lemma~\ref{linear-smoother-c3} hold for every fixed $q\ge2$. Consequently:
\begin{enumerate}
\item[\textup{(a)}] Eq. (10) of the main text holds at $q$;
\item[\textup{(b)}] Condition (C2) holds;
\item[\textup{(c)}] Condition (C3) holds for every pair of disjoint relevant folds.
\end{enumerate}
Moreover,
\[
E_{\mathbf X}|\widehat\tau_{\mathcal I}(\mathbf X)-\tau(\mathbf X)|^2
\asymp_p \frac{K_{\mathcal I}}{N_{\mathcal I}}
\asymp N_{\mathcal I}^{\kappa-1}.
\]
\end{example}

\begin{proof}[Verification]
For each $a$, let $\mu_{a,\mathcal I}^{*}
=\mathbf B_{\mathcal I}^{\top}\gamma_{a,\mathcal I}^{*}$ be the corresponding population weighted least-squares projection of $\mu_a$ onto $\mathcal S_{\mathcal I}$. The design and overlap conditions imply that the eigenvalues of $Q_{a,\mathcal I}$ are bounded above and away from zero. Tensor-product spline approximation gives
\begin{equation}
\|\mu_{a,\mathcal I}^{*}-\mu_a\|_\infty
\le\{1+\|P_{a,\mathcal I}\|_{\infty\to\infty}\}
\inf_{g\in\mathcal S_{\mathcal I}}\|g-\mu_a\|_\infty
\lesssim K_{\mathcal I}^{-s/p}.
\label{spline-projection-bias}
\end{equation}
Write $r_{a,\mathcal I}=\mu_a-\mu_{a,\mathcal I}^{*}$. The population normal equations give
\[
E\left[\mathbb I(A=a)\mathbf B_{\mathcal I}(\mathbf X)
r_{a,\mathcal I}(\mathbf X)\right]=0.
\]
Matrix Bernstein's inequality, local support, and
$K_{\mathcal I}^2\log N_{\mathcal I}/N_{\mathcal I}\to0$ give
\[
P(\mathcal E_{a,\mathcal I}^{c})
\le C K_{\mathcal I}\exp(-cN_{\mathcal I}/K_{\mathcal I}).
\]
By the uniform sub-Gaussian bound and the bounded fallback, this event is
negligible for every unconditional moment used below. On
$\mathcal E_{a,\mathcal I}$, the sample normal
equations give the exact representation
\begin{align}
\widehat\gamma_{a,\mathcal I}-\gamma_{a,\mathcal I}^{*}
=\widehat Q_{a,\mathcal I}^{-1}
\frac1{N_{\mathcal I}}
\sum_{i\in\mathcal I}\mathbb I(A_i=a)
\mathbf B_{\mathcal I}(\mathbf X_i)
\{\varepsilon_i+r_{a,\mathcal I}(\mathbf X_i)\}.
\label{spline-normal-equation}
\end{align}
In particular,
$\|\widehat Q_{a,\mathcal I}-Q_{a,\mathcal I}\|_{\mathrm{op}}=o_p(1)$
uniformly over the finitely many relevant folds. Write
\[
u_{\mathcal I}
=\widehat\tau_{\mathcal I}
-E_{\widehat\tau_{\mathcal I}}
\{\widehat\tau_{\mathcal I}\}.
\]

\medskip\noindent\textit{Verification of (U1).}
Equation~\eqref{spline-normal-equation}, conditional mean zero of $\varepsilon$, the exponentially small fallback event, and Eq.~\eqref{spline-projection-bias} give
\[
\left\|
E_{\widehat\tau_{\mathcal I}}
\{\widehat\mu_{a,\mathcal I}\}-\mu_a
\right\|_\infty
=O(K_{\mathcal I}^{-s/p})
+O\left(K_{\mathcal I}^{-s/p}
\frac{K_{\mathcal I}}{N_{\mathcal I}^{1/2}}\right).
\]
For the second term, the population normal equations imply
\[
E\left\|
\frac1{N_{\mathcal I}}
\sum_{i\in\mathcal I}\mathbb I(A_i=a)
\mathbf B_{\mathcal I}(\mathbf X_i)
r_{a,\mathcal I}(\mathbf X_i)
\right\|^2
\lesssim K_{\mathcal I}^{-2s/p}
\frac{K_{\mathcal I}}{N_{\mathcal I}},
\]
and $\sup_x\|\mathbf B_{\mathcal I}(x)\|\lesssim K_{\mathcal I}^{1/2}$. Since $s>p/2$,
the second bias term is $o\{(K_{\mathcal I}/N_{\mathcal I})^{1/2}\}$. The leading bias satisfies
\[
\frac{K_{\mathcal I}^{-s/p}}
{(K_{\mathcal I}/N_{\mathcal I})^{1/2}}
\asymp
N_{\mathcal I}^{1/2-\kappa(s/p+1/2)}
\longrightarrow0
\]
because $\kappa>p/(2s+p)$. Hence
\begin{equation}
\left\|
E_{\widehat\tau_{\mathcal I}}
\{\widehat\tau_{\mathcal I}\}-\tau
\right\|_\infty
=o\{(K_{\mathcal I}/N_{\mathcal I})^{1/2}\}.
\label{spline-strict-undersmoothing}
\end{equation}

\medskip\noindent\textit{Verification of (U2).}
Let $s_1=1$, $s_0=-1$, and define the population-Gram leading term
\[
\xi_{\mathcal I}(x)
=\frac1{N_{\mathcal I}}
\sum_{i\in\mathcal I}\sum_{a=0}^1
s_a\,\mathbf B_{\mathcal I}(x)^\top Q_{a,\mathcal I}^{-1}
\mathbb I(A_i=a)\mathbf B_{\mathcal I}(\mathbf X_i)\varepsilon_i.
\]
Define
\[
\Omega_{a,\mathcal I}
=E\left[\mathbb I(A=a)\varepsilon^2
\mathbf B_{\mathcal I}(\mathbf X)
\mathbf B_{\mathcal I}(\mathbf X)^\top\right].
\]
This leading term has integrated variance
\begin{align}
v_{\mathcal I}^{*}
&=E\|\xi_{\mathcal I}\|_2^2
=\frac1{N_{\mathcal I}}
\sum_{a=0}^1\operatorname{tr}\left(
G_{\mathcal I}Q_{a,\mathcal I}^{-1}
\Omega_{a,\mathcal I}Q_{a,\mathcal I}^{-1}\right)
\asymp\frac{K_{\mathcal I}}{N_{\mathcal I}}.
\label{spline-integrated-variance}
\end{align}
The lower conditional-variance bound in (S1) gives the lower inequality. The empirical-projection term involving $r_{a,\mathcal I}$ has integrated squared norm
\[
O_p\left(K_{\mathcal I}^{-2s/p}
\frac{K_{\mathcal I}}{N_{\mathcal I}}\right)
=o_p(K_{\mathcal I}/N_{\mathcal I}),
\]
and replacing $Q_{a,\mathcal I}^{-1}$ with
$\widehat Q_{a,\mathcal I}^{-1}$ contributes another smaller term. More
precisely, Eq.~\eqref{spline-normal-equation}, the Gram bound, and the
exponentially small fallback event give
\[
E\|u_{\mathcal I}-\xi_{\mathcal I}\|_2^2
=O\left(
K_{\mathcal I}^{-2s/p}\frac{K_{\mathcal I}}{N_{\mathcal I}}
+\frac{K_{\mathcal I}^2\log N_{\mathcal I}}{N_{\mathcal I}^2}
\right)
=o(K_{\mathcal I}/N_{\mathcal I}).
\]
Therefore
\[
v_{\mathcal I}
=E_{\widehat\tau_{\mathcal I}}\left[
E_{\mathbf X}|u_{\mathcal I}(\mathbf X)|^2
\right]
=E\|\xi_{\mathcal I}\|_2^2\{1+o(1)\}
=v_{\mathcal I}^{*}\{1+o(1)\}
\asymp\frac{K_{\mathcal I}}{N_{\mathcal I}}.
\]
The corresponding quadratic-form calculation for the leading term gives
\[
\operatorname{Var}\{\|\xi_{\mathcal I}\|_2^2\}
=O\left(\frac{K_{\mathcal I}}{N_{\mathcal I}^2}
+\frac{K_{\mathcal I}^2}{N_{\mathcal I}^3}\right).
\]
After division by $v_{\mathcal I}^2\asymp K_{\mathcal I}^2/N_{\mathcal I}^2$, the relative variance is
$O(K_{\mathcal I}^{-1}+N_{\mathcal I}^{-1})=o(1)$. Together with the
mean-square remainder bound, this implies
\[
\frac{E_{\mathbf X}|u_{\mathcal I}(\mathbf X)|^2}{v_{\mathcal I}}
\longrightarrow_p1.
\]
For disjoint folds $\mathcal I$ and $\mathcal I'$, the leading terms are
independent. If $\Gamma_{\mathcal I}$ is the covariance operator of
$\xi_{\mathcal I}$, then
$\operatorname{rank}(\Gamma_{\mathcal I})\le K_{\mathcal I}$ and
$\|\Gamma_{\mathcal I}\|_{\mathrm{op}}=O(N_{\mathcal I}^{-1})$.
Consequently,
\[
E\left[E_{\mathbf X}\{\xi_{\mathcal I}(\mathbf X)
\xi_{\mathcal I'}(\mathbf X)\}\right]^2
=\operatorname{tr}(\Gamma_{\mathcal I}\Gamma_{\mathcal I'})
=O\left(\frac{K_{\mathcal I}\wedge K_{\mathcal I'}}
{N_{\mathcal I}N_{\mathcal I'}}\right).
\]
Since $v_{\mathcal I}v_{\mathcal I'}\asymp
K_{\mathcal I}K_{\mathcal I'}/(N_{\mathcal I}N_{\mathcal I'})$ and the two dimensions are comparable, Chebyshev's inequality gives
\[
\frac{E_{\mathbf X}\{\xi_{\mathcal I}(\mathbf X)\xi_{\mathcal I'}(\mathbf X)\}}
{(v_{\mathcal I}v_{\mathcal I'})^{1/2}}
\longrightarrow_p0.
\]
The mean-square remainder bound and Cauchy--Schwarz allow
$\xi_{\mathcal I},\xi_{\mathcal I'}$ to be replaced by
$u_{\mathcal I},u_{\mathcal I'}$, proving the cross-fold part of (U2).

\medskip\noindent\textit{Verification of (U3).}
The stabilized normal-equation decomposition, empirical-projector stability,
conditional sub-Gaussianity, Rosenthal's inequality, and local support give,
for every fixed $q\ge2$,
\begin{align*}
E_{\widehat\tau_{\mathcal I}}\left[
E_{\mathbf X}|u_{\mathcal I}(\mathbf X)|^q
\right]
&\lesssim
\left(\frac{K_{\mathcal I}}{N_{\mathcal I}}\right)^{q/2}
+\left(\frac{K_{\mathcal I}}{N_{\mathcal I}}\right)^{q-1}\\
&=O\left\{(K_{\mathcal I}/N_{\mathcal I})^{q/2}\right\}.
\end{align*}
The term involving $r_{a,\mathcal I}$ is smaller by the factor
$K_{\mathcal I}^{-sq/p}$. Markov's inequality therefore yields
\[
E_{\mathbf X}|u_{\mathcal I}(\mathbf X)|^q
=O_p\{(K_{\mathcal I}/N_{\mathcal I})^{q/2}\}
=O_p(v_{\mathcal I}^{q/2}).
\]

\medskip\noindent\textit{Verification of (U4).}
The maximal inequality for a local B-spline basis and
Eq.~\eqref{spline-normal-equation} give
\[
\|\widehat\tau_{\mathcal I}-\tau\|_\infty
=O_p\left\{
K_{\mathcal I}^{-s/p}
+\left(\frac{K_{\mathcal I}\log N_{\mathcal I}}{N_{\mathcal I}}\right)^{1/2}
+K_{\mathcal I}^{-s/p}\frac{K_{\mathcal I}}{N_{\mathcal I}^{1/2}}
\right\}=o_p(1).
\]
Thus (U4), and in fact sup-norm consistency, holds.

Finally, because the knots and $K_{\mathcal I}$ depend on a fold only
through $N_{\mathcal I}$, $v_{\mathcal I}$ is the same for all folds of
the same size. The half- and quarter-fold sizes are fixed fractions of $n$,
so
\[
\frac{K_{\mathcal I}}{N_{\mathcal I}}
\asymp N_{\mathcal I}^{\kappa-1}
\asymp N_{\mathcal I'}^{\kappa-1}
\asymp\frac{K_{\mathcal I'}}{N_{\mathcal I'}}.
\]
Together with
$v_{\mathcal I}\asymp K_{\mathcal I}/N_{\mathcal I}$, this implies that
there exist fixed constants $0<c_v\le C_v<\infty$ such that
\[
c_v
\le \frac{v_{\mathcal I'}}{v_{\mathcal I}}
\le C_v
\]
for every pair of relevant folds and all sufficiently large $n$. This
verifies the variance-comparability premise of
Lemma~\ref{linear-smoother-c3}. The lemma now gives conclusions (a)--(c).
\end{proof}

The MSE-optimal spline dimension balances
$K_{\mathcal I}^{-2s/p}$ and $K_{\mathcal I}/N_{\mathcal I}$ and therefore has exponent $p/(2s+p)$. The strict inequality
$\kappa>p/(2s+p)$ in Example~\ref{example:undersmoothed-spline} is precisely the undersmoothing requirement; the upper bound $\kappa<1/2$ is a transparent sufficient condition for the empirical-Gram calculations above.

\begin{example}[Undersmoothed kernel contrast]
\label{example:undersmoothed-kernel}
Let $\mathbb T^p=[0,1]^p$ be the $p$-dimensional torus. Suppose the observations in every training fold are i.i.d. copies of $(\mathbf X,Z)$ satisfying
\[
\mathbf X\sim\operatorname{Uniform}(\mathbb T^p),
\qquad
E(Z\mid\mathbf X=x)=\tau(x),
\qquad
|Z|\le M
\]
almost surely, and
\[
\inf_{x\in\mathbb T^p}\operatorname{Var}(Z\mid\mathbf X=x)>0.
\]
For example, with bounded outcomes, let
$\pi(x)=P(A=1\mid\mathbf X=x)$,
$\mu_a(x)=E(Y\mid A=a,\mathbf X=x)$, and
$\tau=\mu_1-\mu_0$. If the known propensity score is bounded away from zero and one, then
\[
Z=\frac{AY}{\pi(\mathbf X)}-
\frac{(1-A)Y}{1-\pi(\mathbf X)}
\]
is a bounded contrast response satisfying $E(Z\mid\mathbf X)=\tau(\mathbf X)$.

Suppose $\tau$ belongs to the periodic H\"older class
$C^{s-1,1}(\mathbb T^p)$ for an integer $s\ge1$. Let
$K:\mathbb R^p\to\mathbb R$ be bounded, compactly supported, and Lipschitz, with
\[
\int K(u)\,du=1,
\qquad
0<\int K(u)^2\,du<\infty,
\]
and
\[
\int u^{\boldsymbol a}K(u)\,du=0
\quad(1\le|\boldsymbol a|\le s-1).
\]
For $b>0$, define the periodic rescaling
\[
K_b^{\mathrm{per}}(u)
=\sum_{k\in\mathbb Z^p}b^{-p}K\{(u+k)/b\}.
\]
For a relevant fold $\mathcal I$, set $N_{\mathcal I}=|\mathcal I|$ and choose
\[
b_{\mathcal I}=N_{\mathcal I}^{-\gamma},
\qquad
\frac1{2s+p}<\gamma<\frac1p.
\]
Define
\[
\widehat\tau_{\mathcal I}(x)
=\frac1{N_{\mathcal I}}
\sum_{i\in\mathcal I}K_{b_{\mathcal I}}^{\mathrm{per}}(x-\mathbf X_i)Z_i.
\]
Then the variance-comparability premise and (U1)--(U4) of
Lemma~\ref{linear-smoother-c3} hold for every fixed $q\ge2$. Consequently:
\begin{enumerate}
\item[\textup{(a)}] Eq. (10) of the main text holds at $q$;
\item[\textup{(b)}] Condition (C2) holds;
\item[\textup{(c)}] Condition (C3) holds for every pair of disjoint relevant folds.
\end{enumerate}
Moreover,
\[
E_{\mathbf X}|\widehat\tau_{\mathcal I}(\mathbf X)-\tau(\mathbf X)|^2
\asymp_p N_{\mathcal I}^{-(1-\gamma p)}.
\]
\end{example}

\begin{proof}[Verification]
Write $N=N_{\mathcal I}$ and $b=b_{\mathcal I}$ only within this verification, and set
\[
\bar\tau_{\mathcal I}
=E_{\widehat\tau_{\mathcal I}}
\{\widehat\tau_{\mathcal I}\},
\qquad
u_{\mathcal I}=\widehat\tau_{\mathcal I}-\bar\tau_{\mathcal I}.
\]

\medskip\noindent\textit{Verification of (U1).}
The uniform design and kernel moment conditions give
\[
\bar\tau_{\mathcal I}=K_b^{\mathrm{per}}*\tau,
\qquad
\|\bar\tau_{\mathcal I}-\tau\|_\infty=O(b^s).
\]
For all sufficiently small $b$, direct integration gives
\begin{align}
v_{\mathcal I}
&=E_{\widehat\tau_{\mathcal I}}
\{\|u_{\mathcal I}\|_2^2\}\notag\\
&=\frac1N\left[
b^{-p}\|K\|_2^2E(Z^2)
-\|K_b^{\mathrm{per}}*\tau\|_2^2
\right]
\asymp\frac1{Nb^p}.
\label{kernel-variance-rate}
\end{align}
The conditional-variance lower bound gives the lower inequality. Therefore
\[
\frac{\|\bar\tau_{\mathcal I}-\tau\|_\infty}{v_{\mathcal I}^{1/2}}
=O\{(Nb^{2s+p})^{1/2}\}
=O\{N^{[1-\gamma(2s+p)]/2}\}=o(1),
\]
which is (U1). Equation~\eqref{kernel-variance-rate} also gives
$v_{\mathcal I}\asymp N_{\mathcal I}^{-(1-\gamma p)}$. Because the bandwidth
$b_{\mathcal I}$ depends on a fold only through $N_{\mathcal I}$ and the
half- and quarter-fold sizes are fixed fractions of $n$, there therefore exist
fixed constants $0<c_v\le C_v<\infty$ such that
\[
c_v
\le \frac{v_{\mathcal I'}}{v_{\mathcal I}}
\le C_v
\]
for every pair of relevant folds and all sufficiently large $n$. This verifies
the variance-comparability premise of
Lemma~\ref{linear-smoother-c3}.

\medskip\noindent\textit{Verification of (U2).}
Define the mean-zero random element of $L^2(\mathbb T^p)$
\[
\xi_{i,\mathcal I}(\cdot)
=K_b^{\mathrm{per}}(\cdot-\mathbf X_i)Z_i
-\bar\tau_{\mathcal I}(\cdot),
\qquad
u_{\mathcal I}=\frac1N\sum_{i\in\mathcal I}\xi_{i,\mathcal I}.
\]
Let $\mathcal C_{\mathcal I}$ be the covariance operator of
$\xi_{i,\mathcal I}$. Boundedness of $Z$ and compact support of $K$ give
\[
\operatorname{tr}(\mathcal C_{\mathcal I})\asymp b^{-p},
\qquad
E\|\xi_{i,\mathcal I}\|_2^4=O(b^{-2p}),
\qquad
\operatorname{tr}(\mathcal C_{\mathcal I}^2)=O(b^{-p}).
\]
To see the last order, write the covariance kernel as
\[
C_{\mathcal I}(x,y)
=A_b(x,y)-\bar\tau_{\mathcal I}(x)\bar\tau_{\mathcal I}(y),
\]
where
\[
A_b(x,y)
=E\left[K_b^{\mathrm{per}}(x-\mathbf X)
K_b^{\mathrm{per}}(y-\mathbf X)Z^2\right].
\]
The kernel $A_b$ is $O(b^{-p})$ and vanishes unless $x$ and $y$ are
within periodic distance $O(b)$, so
\[
\int_{\mathbb T^p}\int_{\mathbb T^p}A_b(x,y)^2\,dx\,dy
=O(b^{-p}).
\]
Moreover, $\|\bar\tau_{\mathcal I}\|_\infty=O(1)$, so the rank-one
kernel $\bar\tau_{\mathcal I}(x)\bar\tau_{\mathcal I}(y)$ has squared
$L^2$ norm $O(1)$. Hence
\[
\operatorname{tr}(\mathcal C_{\mathcal I}^2)
=\int_{\mathbb T^p}\int_{\mathbb T^p}C_{\mathcal I}(x,y)^2\,dx\,dy
=O(b^{-p}).
\]
The fourth-moment formula for an average of independent mean-zero Hilbert-space random elements then gives
\begin{align*}
\frac{\operatorname{Var}\{\|u_{\mathcal I}\|_2^2\}}{v_{\mathcal I}^2}
&\lesssim
\frac{N^{-3}b^{-2p}+N^{-2}b^{-p}}
{N^{-2}b^{-2p}}\\
&=O(N^{-1}+b^p)=o(1).
\end{align*}
Thus $\|u_{\mathcal I}\|_2^2/v_{\mathcal I}\to_p1$.

For disjoint folds $\mathcal I,\mathcal I'$, the centered fits are independent. The Hilbert--Schmidt inequality and comparable bandwidths give
\begin{align*}
E\left[
\frac{\langle u_{\mathcal I},u_{\mathcal I'}\rangle_2}
{(v_{\mathcal I}v_{\mathcal I'})^{1/2}}
\right]^2
&=\frac{\operatorname{tr}(\mathcal C_{\mathcal I}\mathcal C_{\mathcal I'})}
{N_{\mathcal I}N_{\mathcal I'}v_{\mathcal I}v_{\mathcal I'}}\\
&=O\{(b_{\mathcal I}b_{\mathcal I'})^{p/2}\}=o(1).
\end{align*}
This proves the cross-fold part of (U2).

\medskip\noindent\textit{Verification of (U3).}
For every fixed $q\ge2$, the pointwise Rosenthal inequality followed by integration gives
\begin{align*}
E_{\widehat\tau_{\mathcal I}}
\{\|u_{\mathcal I}\|_q^q\}
&\le C_q\left\{
N^{-q/2}b^{-pq/2}
+N^{-(q-1)}b^{-p(q-1)}\right\}\\
&=O\{(Nb^p)^{-q/2}\},
\end{align*}
because $Nb^p\to\infty$. Since
$v_{\mathcal I}\asymp(Nb^p)^{-1}$, Markov's inequality gives
\[
E_{\mathbf X}|u_{\mathcal I}(\mathbf X)|^q
=O_p(v_{\mathcal I}^{q/2}).
\]

\medskip\noindent\textit{Verification of (U4).}
The translation class generated by a bounded Lipschitz kernel has polynomial covering numbers, envelope $O(b^{-p})$, and pointwise variance $O(b^{-p})$. The Bernstein maximal inequality gives
\[
\|u_{\mathcal I}\|_\infty
=O_p\left\{
\sqrt{\frac{\log N}{Nb^p}}
+\frac{\log N}{Nb^p}\right\}=o_p(1),
\]
because $\gamma<1/p$ implies $Nb^p/\log N\to\infty$. Together with the bias bound,
\[
\|\widehat\tau_{\mathcal I}-\tau\|_\infty
\le\|u_{\mathcal I}\|_\infty
+\|\bar\tau_{\mathcal I}-\tau\|_\infty=o_p(1).
\]
Since a periodic H\"older function is bounded, (U4) holds. Lemma~\ref{linear-smoother-c3} now gives conclusions (a)--(c).
\end{proof}

For the kernel estimator, the MSE-optimal exponent is
$1/(2s+p)$. The strict inequality $\gamma>1/(2s+p)$ is the undersmoothing requirement, while $\gamma<1/p$ ensures that the effective number of observations in each kernel neighborhood, $N_{\mathcal I}b_{\mathcal I}^p$, diverges.

\subsection{Algorithm \ref{alg3} and Its Property}
Algorithm \ref{alg3} presents a refinement of the proposed adaptive smoothing method in the manuscript by repeatedly cross fitting to improve the finite sample performance. Corollary \ref{CV-eq} justifies Algorithm \ref{alg3} and suggests the asymptotic variance remains unchanged.
\setcounter{algorithm}{2}
\begin{algorithm}
\caption{Finite Sample Refinement: Repeatedly Cross Fitting and  Averaging}
    \begin{algorithmic}[1]
        \State \textbf{Input}: Data set $\mathcal{O}_n$; 
$B$ Random bipartitions of $[n]$: $\mathcal{I}_{b,1}$ and $\mathcal{I}_{b,2}$ ($b=1,\dots,B$), and randomly picked subsets $\mathcal{I}_{b,j,k}$ ($k=1,2$) of $\mathcal{I}_{b,j}$ ($j=1,2$), where all the $\mathcal{I}_{b,j,k}$ with the equal size; Parameters $h_{n,\mathcal{I}_{b,j}}$, $j=1,2$, and $t_0$; Working models: $\widehat{\pi}$, $\widehat{\mu}$, $\widehat{D}$.
        \State \textbf{Construction of 
 The Adaptive Smoothing Estimator:} 
 \begin{align}
     \widehat{V}_{as,B}&:= \frac{1}{4B}\sum_{b=1}^B \sum_{j=1,2}\sum_{k=1,2}\frac{1}{|\mathcal{I}_{b,j,k}|}\sum_{i\in\mathcal{I}_{b,j,k}}\psi_i(d_{s}(\mathbf{X}_i;\widehat{\tau}_{\mathcal{I}_{b,3-j}},\notag\\&h_{n,\mathcal{I}_{b,3-j}},\widehat{\pi}_{\mathcal{I}_{b,3-j}}(1,\mathbf{X}_i)),\widehat{\pi}_{\mathcal{I}_{b,j,k}^c},\widehat{\mu}_{\mathcal{I}_{b,j,k}^c})\label{sin-opt-est-cv}
 \end{align}
 \State \textbf{Construction of The Confidence Interval} The standard deviation estimate:
 \begin{align}
     \widehat{\sigma}_{as,B}:=&\left(\frac{1}{B\cdot(n-1)}\sum_{b=1}^B\sum_{j=1,2}\sum_{k=1,2}\sum_{i\in\mathcal{I}_{b,j,k}}\right.\notag\\
    & \left.\left(\psi_i(d_{s}(\mathbf{X}_i;\widehat{\tau}_{\mathcal{I}_{b,3-j}},h_{n,\mathcal{I}_{b,3-j}},\widehat{\pi}_{\mathcal{I}_{b,3-j}}(1,\mathbf{X}_i)),\widehat{\pi}_{\mathcal{I}_{b,j,k}^c},\widehat{\mu}_{\mathcal{I}_{b,j,k}^c})-\widehat{V}_{as,B}\right)^2\right)^{1/2}\label{sin-opt-std-cv}
 \end{align}
 and thus the two-sided level-{$\eta$} confidence interval can be given as:
\begin{align}\label{sin-CI-opt-cv}
     \left[\widehat{V}_{as,B}-{z_{\eta/2}}\cdot\frac{\widehat{\sigma}_{as,B}}{\sqrt{n}},\widehat{V}_{as,B}+{z_{\eta/2}}\cdot\frac{\widehat{\sigma}_{as,B}}{\sqrt{n}}\right]
 \end{align}
    \end{algorithmic}
    \label{alg3}
\end{algorithm}
\begin{corollary}\label{CV-eq}
    Under Assumptions 1 to 7, $\sqrt{n}\left(\widehat{V}_{as}-\widehat{V}_{as,B}\right)=o_p(1)$ as $B\to\infty$.
\end{corollary}
\begin{proof}
    Let $\mathcal{D}_n(\mathbf{X}_i,s)=d_{s}(\mathbf{X}_i;s,h_{n,\mathcal{I}_{j(i)}^c},\widehat{\pi}_{\mathcal{I}_{j(i)}^c}(1,\mathbf{X}_i))$, where $j(i)$ is the index of subset that $i$-th sample in (for example, if $i\in\mathcal{I}_1$ then $j(i)=1$). Known from Lemma \ref{single-lem} that $|\widehat{V}_{\mathcal{D}_n}^* - \widehat{V}_{\mathcal{D}_n}|=o_p(n^{-1/2})$ holds regardless of the specific choice of $\mathcal{I}_1$ and $\mathcal{I}_2$. It is not hard to verify that the decision function in $\widehat{V}_{as}$ also satisfy the conditions required for $\mathcal{D}_n$ in Theorem 2, therefore, 
    \begin{align*}
        \left|\widehat{V}_{as}-\frac{1}{n}\sum_{i=1}^n \psi_i(\mathcal{D}(\mathbf{X}_i),\pi,\mu)\right|=o_p(n^{-1/2})
    \end{align*}
    where $\mathcal{D}(\mathbf{X})=\pi(1,\mathbf{X})\cdot\mathbb{I}\{\tau(\mathbf{X})=0\}+\mathbb{I}\{\tau(\mathbf{X})>0\}$ and consequently, 
    \begin{align*}
        \widehat{V}_{as,B}-\frac{1}{n}\sum_{i=1}^n \psi_i(\mathcal{D}(\mathbf{X}_i),\pi,\mu)=o_p(n^{-1/2}).
    \end{align*}
    Therefore, the asymptotic variance of the proposed estimator with Repeatedly Cross Fitting and Averaging (RCFA) is the same as before (without RCFA), but the finite sample performance would be improved, which is also suggested by \citet{buhlmann2002analyzing}.
\end{proof}
\subsection{Proof of Corollary 1}
Firstly we start with showing (v). By the definition of $\widehat{V}_{\mathcal{D}}^*$, the corresponding influence function has the following form:
\begin{align*}
    &f(\mathbf{X},A,Y,\mu_1(\mathbf{X}),\mu_0(\mathbf{X}),\pi_1(\mathbf{X}),\dots)\\
    =&\frac{A \mathcal{D}(\mathbf{X})+(1-A)(1-\mathcal{D}(\mathbf{X}))}{A\pi_1(\mathbf{X})+(1-A)(1-\pi_1(\mathbf{X}))}(Y-A \mu_1(\mathbf{X})-(1-A)\mu_0(\mathbf{X}))\\
    &+\mathcal{D}(\mathbf{X})\mu_1(\mathbf{X})+(1-\mathcal{D}(\mathbf{X}))\mu_0(\mathbf{X})\\
    =&\frac{A \mathcal{D}(\mathbf{X})+(1-A)(1-\mathcal{D}(\mathbf{X}))}{A\pi_1(\mathbf{X})+(1-A)(1-\pi_1(\mathbf{X}))}Y\\&+\mu_1(\mathbf{X})\mathcal{D}(\mathbf{X})\left(1-\frac{A}{\pi_1(\mathbf{X})}\right)+\mu_0(\mathbf{X})(1-\mathcal{D}(\mathbf{X}))\left(1-\frac{1-A}{1-\pi_1(\mathbf{X})}\right).
\end{align*}
If $\mu(A,\mathbf{X}) = A\mu_1(\mathbf{X})+(1-A)\mu_0(\mathbf{X})$, then 
\begin{align*}
    &E\left[\frac{A \mathcal{D}(\mathbf{X})+(1-A)(1-\mathcal{D}(\mathbf{X}))}{A\pi_1(\mathbf{X})+(1-A)(1-\pi_1(\mathbf{X}))}(Y-A \mu_1(\mathbf{X})-(1-A)\mu_0(\mathbf{X}))\right]\\
    =&E\left[E\left[\frac{A \mathcal{D}(\mathbf{X})+(1-A)(1-\mathcal{D}(\mathbf{X}))}{A\pi_1(\mathbf{X})+(1-A)(1-\pi_1(\mathbf{X}))}(Y-A \mu_1(\mathbf{X})-(1-A)\mu_0(\mathbf{X}))\bigg|\mathbf{X},A\right]\right]\\
    =&E\left[\frac{A \mathcal{D}(\mathbf{X})+(1-A)(1-\mathcal{D}(\mathbf{X}))}{A\pi_1(\mathbf{X})+(1-A)(1-\pi_1(\mathbf{X}))}\underbrace{E\left[(Y-A \mu_1(\mathbf{X})-(1-A)\mu_0(\mathbf{X}))\bigg|\mathbf{X},A\right]}_{=0}\right]\\
    =&0.
\end{align*}
and 
\begin{align*}
    &E\left[f(\mathbf{X},A,Y,\mu_1(\mathbf{X}),\mu_0(\mathbf{X}),\pi_1(\mathbf{X}),\dots)\right]\\
    =&E\left[E\left[\mathcal{D}(\mathbf{X})\mu_1(\mathbf{X})+(1-\mathcal{D}(\mathbf{X}))\mu_0(\mathbf{X})|\mathbf{X},A\right]\right]=V_0.
\end{align*}
If $\pi_1(\mathbf{X}) = \pi(1,\mathbf{X})$, we have
\begin{align*}
    E\left[E\left[\mu_1(\mathbf{X})\mathcal{D}(\mathbf{X})\left(1-\frac{A}{\pi_1(\mathbf{X})}\right)+\mu_0(\mathbf{X})(1-\mathcal{D}(\mathbf{X}))\left(1-\frac{1-A}{1-\pi_1(\mathbf{X})}\right)\bigg|\mathbf{X}\right]\right]=0
\end{align*}
and 
\begin{align*}
    E\left[E\left[\frac{A \mathcal{D}(\mathbf{X})+(1-A)(1-\mathcal{D}(\mathbf{X}))}{A\pi_1(\mathbf{X})+(1-A)(1-\pi_1(\mathbf{X}))}Y\bigg|\mathbf{X},A\right]\right]=V_0.
\end{align*}
For (i) to (iv), we have
\begin{align}
    &\widehat{V}_s = \widehat{V}_{\mathcal{D}}^*+o_p(n^{-1/2}), \text{ s.t. }\mathcal{D}(\mathbf{X}) = t_0\cdot\mathbb{I}\{\tau(\mathbf{X})=0\}+\mathbb{I}\{\tau(\mathbf{X})>0\};\\
     &\widehat{V}_{as} = \widehat{V}_{\mathcal{D}}^*+o_p(n^{-1/2}), \text{ s.t. }\mathcal{D}(\mathbf{X}) = \pi(1,\mathbf{X})\cdot\mathbb{I}\{\tau(\mathbf{X})=0\}+\mathbb{I}\{\tau(\mathbf{X})>0\};\\
     & \widehat{V}_{or} = \widehat{V}_{\mathcal{D}}^*+o_p(n^{-1/2}), \text{ s.t. }\mathcal{D}(\mathbf{X}) = \mathbb{I}\{\tau(\mathbf{X})\ge 0\};\\
     & \widehat{V}_{sb} = \widehat{V}_{\mathcal{D}}^*+o_p(n^{-1/2}), \text{ s.t. }\mathcal{D}(\mathbf{X}) = 0.5\cdot\mathbb{I}\{\tau(\mathbf{X})=0\}+\mathbb{I}\{\tau(\mathbf{X})>0\}.\label{Vsb-embed}
\end{align}
Note that Eq. \eqref{Vsb-embed} is derived under Assumption 7 and Eq. (16) in \citet{shi2020breaking}. Hence we have shown all results from (i) to (v).

\subsection{Extensions of $\mathcal{T}_{RL}$}
With some slight modifications to replace the linear i.i.d. summation with linear martingale summation, we can extend $\mathcal{T}_{RL}$, the proposed robust asymptotically linear unbiased estimator class, to a more general version containing simple sample split estimator \citep{belloni2014inference,wager2018estimation} and online one step estimator \citep{luedtke2016statistical} as shown in Definition \ref{M-ralu}. 
\setcounter{definition}{1}
\begin{definition}[Martingale Robust Asymptotically Linear Unbiased Estimator]\label{M-ralu}
    The class $\mathcal{T}_{MRL}$ is defined as the set of the 
    estimator $T_m$ $(m\leq n)$ such that there exists a 
    measurable function 
    $f(\cdot|\mathcal{O}_{i-1}): \mathcal{X} \times\{0,1\} 
    \times \mathbb{R}^3 \times(c, 1-c) \times\{-1,0,1\} 
    \times \mathbb{R}_{\geq 0}^2$ and for all $P\in\mathcal{P}$ 
    which is the set of distributions satisfying 
    Assumptions 1 to 3,
    \begin{align*}
        T_m(\mathcal{O}_m)=m^{-1}\sum_{i=1}^m 
        f(\mathbf{X}_i,A_i,Y_i,\mu_1(\mathbf{X}_i),
        \mu_0(\mathbf{X}_i),\pi_1(\mathbf{X}_i),
        \operatorname{sgn}\left(\tau_P\left(\mathbf{X}_i
        \right)\right),\\
        r_1(\mathbf{X}_i),r_0(\mathbf{X}_i);
        \mathcal{O}_{i-1})+o_p(m^{-1/2})
    \end{align*}
    where $\tau_P(\mathbf{X})=E_P[Y|\mathbf{X},A=1]
    -E_P[Y|\mathbf{X},A=0]$, and $\mu_a$, $\pi_1$ and $r_a$ 
    are the working models for $E_P[Y|\mathbf{X},A=a]$, 
    $E_P[A|\mathbf{X}]$ and $\operatorname{Var}_P(Y|\mathbf{X},A=a)$, 
    respectively, for $a=0,1$. Moreover, for any 
    $P\in\mathcal{P}$, $f$ satisfies
    \begin{align}\label{M-unbias}
        E_P[f|\mathcal{O}_{i-1}] = 
        E_P[E_P\{Y|\mathbf{X},A=d^{opt}(\mathbf{X})\}]
    \end{align}
    as long as one of the following equations holds $P$-a.s.:
    \begin{align}\label{M-robust}
    \begin{aligned}
       E_P[Y|\mathbf{X},A] =& A \mu_1(\mathbf{X})
       +(1-A)\mu_0(\mathbf{X}),~
        E_P[A|\mathbf{X}]=\pi_1(\mathbf{X}).        
    \end{aligned}
    \end{align}
\end{definition}
Simple sample split estimator belongs to $\mathcal{T}_{MRL}$ when $m=|\mathcal{I}|$, as i.i.d. is a special case of such a martingale sequence. The proof of Theorem 2 in \citet{luedtke2016statistical} has already implied that the online one step estimator belongs to $\mathcal{T}_{MRL}$. Note that the variance lower bound of $\mathcal{T}_{MRL}$ is the same as that of $\mathcal{T}_{RL}$, which can be shown directly by replacing the $f(\cdot)$ in the proof of Theorem \ref{single-eff-opt-heter} with $f(\cdot;\mathcal{O}_{i-1})$ in $\mathcal{T}_{MRL}$.
\subsection{Proof of Theorem 3}
\begin{proof}
    By Theorem \ref{single-eff-opt-heter}, under Assumption 8, the decision function defined in Eq. \eqref{single-optimal-decision-heter} would reduce to $\mathcal{D}(\mathbf{X})=\pi(1,\mathbf{X})\cdot\mathbb{I}\{\tau(\mathbf{X})=0\}+\mathbb{I}\{\tau(\mathbf{X})>0\}$, and since   
    \begin{align*}
        \left|\widehat{V}_{as}-\frac{1}{n}\sum_{i=1}^n \psi_i(\mathcal{D}(\mathbf{X}_i),\pi,\mu)\right|=o_p(n^{-1/2})
    \end{align*}
    where $\mathcal{D}(\mathbf{X})=\pi(1,\mathbf{X})\cdot\mathbb{I}\{\tau(\mathbf{X})=0\}+\mathbb{I}\{\tau(\mathbf{X})>0\}$, the results in Theorem \ref{single-eff-opt-heter} imply Theorem 3.
\end{proof}
\subsection{Proof of Corollary 2 and Variance Reduction}
\begin{proof}[Proof of Corollary 2]
As is stated in Equation (16) in Section 3 of \citet{shi2020breaking}, when undersmoothing is employed for $\widehat{V}_{sb}$, we have 
\begin{align*}
    P\left(\widehat{\tau}_{\mathcal{I}}(\mathbf{X})>0\right) \rightarrow \frac{1}{2}, \quad \forall \mathbf{X}\text { with } \tau(\mathbf{X})=0.
\end{align*}
In this case, the function $p_0(\mathbf{X})$ in Assumption (A7) in \citet{shi2020breaking} becomes a constant $1/2$. Therefore, by the Proof of Theorem 5 in \citet{shi2020breaking}, the asymptotic variance of $\widehat{V}_{sb}$ can be calculated as follows: 
\begin{align*}
    &\lim_{n\to\infty}\operatorname{Var}(\sqrt{n}\widehat{V}_{sb})\\
    =&\operatorname{Var}\left\{h\left(d_0^{o p t}, \mathbf{X}\right)\right\} + E\left(\frac{\mathbb{I}\left\{\tau\left(\mathbf{X}\right)>0\right\}}{\pi\left(1, \mathbf{X}\right)}\operatorname{Var}(Y|\mathbf{X},A=1)+\frac{\mathbb{I}\left\{\tau\left(\mathbf{X}\right)<0\right\}}{\pi\left(0, \mathbf{X}\right)}\operatorname{Var}(Y|\mathbf{X},A=0)\right)\\
    +& E\left[\left(\frac{1}{4 \pi\left(1, \mathbf{X}\right)}\operatorname{Var}(Y|\mathbf{X},A=1)+\frac{1}{4\pi\left(0, \mathbf{X}\right)}\operatorname{Var}(Y|\mathbf{X},A=0)\right) \mathbb{I}\left\{\tau\left(\mathbf{X}\right)=0\right\}\right],
\end{align*}
which is actually the same with that of $\widehat{V}_{s}$ when $t_0=1/2$. Then the variance reduction can be directly calculated as
\begin{align*}
   &\lim_{n\to\infty}\left[ \operatorname{Var}(\sqrt{n}\widehat{V}_{sb}) - \operatorname{Var}(\sqrt{n}\widehat{V}_{as})\right]\notag\\
    =&\operatorname{Var}\left\{h\left(d_0^{o p t}, \mathbf{X}\right)\right\} + E\left(\frac{\mathbb{I}\left\{\tau\left(\mathbf{X}\right)>0\right\}}{\pi\left(1, \mathbf{X}\right)}\operatorname{Var}(Y|\mathbf{X},A=1)+\frac{\mathbb{I}\left\{\tau\left(\mathbf{X}\right)<0\right\}}{\pi\left(0, \mathbf{X}\right)}\operatorname{Var}(Y|\mathbf{X},A=0)\right)\\
    +& E\left[\left(\frac{1}{4 \pi\left(1, \mathbf{X}\right)}\operatorname{Var}(Y|\mathbf{X},A=1)+\frac{1}{4\pi\left(0, \mathbf{X}\right)}\operatorname{Var}(Y|\mathbf{X},A=0)\right) \mathbb{I}\left\{\tau\left(\mathbf{X}\right)=0\right\}\right]\\
    -&\operatorname{Var}\left\{h\left(d_0^{o p t}, \mathbf{X}\right)\right\} - E\left[{\sigma^2(\mathbf{X})}\left(\frac{\mathbb{I}\left\{\tau\left(\mathbf{X}\right)>0\right\}}{\pi\left(1, \mathbf{X}\right)}+\frac{\mathbb{I}\left\{\tau\left(\mathbf{X}\right)<0\right\}}{\pi\left(0, \mathbf{X}\right)}\right)\right]\\
    -& E\left[{\sigma^2(\mathbf{X})} \mathbb{I}\left\{\tau\left(\mathbf{X}\right)=0\right\}\right]\\
   =&E\left[\left(\frac{1}{4 \pi\left(1, \mathbf{X}\right)}+\frac{1}{4\pi\left(0, \mathbf{X}\right)}-1\right){\sigma^2(\mathbf{X})} \mathbb{I}\left\{\tau\left(\mathbf{X}\right)=0\right\}\right]\ge 0
\end{align*}
which is larger than 0 unless the study is a randomized and completely balanced design or regular; i.e. $P(\pi\left(1, \mathbf{X}\right)=0.5\lor \tau(\mathbf{X})\neq 0)=1$. 
\end{proof}
For the variance deduction from $\widehat{V}_s$ to $\widehat{V}_{as}$, directly from the results in Theorem 1 and 2 in the main text,  we have 
\begin{align*}
  &\lim_{n\to\infty}\left[ \operatorname{Var}(\sqrt{n}\widehat{V}_{s}) - \operatorname{Var}(\sqrt{n}\widehat{V}_{as})\right]\notag\\
   =&E\left[\left(\frac{t_0^2}{\pi\left(1, \mathbf{X}\right)}+\frac{(1-t_0)^2}{\pi\left(0, \mathbf{X}\right)}-1\right){\sigma^2(\mathbf{X})} \mathbb{I}\left\{\tau\left(\mathbf{X}\right)=0\right\}\right]\ge 0
\end{align*}
which is larger than 0 unless the study is a randomized design with $t_0\equiv\pi(1,\mathbf{X})$ 
or regular; i.e. $P(\pi(1,\mathbf{X})\equiv t_0 \lor \tau(\mathbf{X})\neq 0)=1$. 

\subsection{Sensitivity Analysis of Tuning}
Here, we provide simulation results to demonstrate the robustness of the algorithms w.r.t. the constant ${C_0}$ in the proposed tuning selection method in Algorithm 2 and the number of the split; i.e. Algorithm 1 (one split) and Algorithm \ref{alg3} (repeated split), and illustrate the computational benefit of the proposed method in Algorithm 1 (one split). The simulation settings are the same as that in Section 4 of the main text. 

By comparing the results in Table \ref{tab4} and Table \ref{tab5}, we see that (1) the ECP and AL for both Algorithm 1 and Algorithm \ref{alg3} are robust to the choice of the constant ${C_0}$; (2) the performances in terms of ECP and AL of Algorithm 1 (one split) and Algorithm \ref{alg3} (repeated split) are comparable and almost the same which indicates the proposed method is robust to the number of the split; (3) the computational time of Algorithm 1 is much shorter than that of Algorithm \ref{alg3} and that of subbagging in \citet{shi2020breaking}, the current state-of-the-art method,  in Table \ref{tab6}, as Algorithm 1 does not require repeatedly sample splitting, thus avoiding multiple training of the working models. 
\setcounter{table}{3}
\begin{table}[H]
\centering
\scriptsize
\begin{tabular}{|c|c|c|c|c|c|c|}
\hline
Value of ${C_0}$        & Settings      & (A)  & (B)  & (C)   & (D)   & (E)   \\ \hline
\multirow{3}{*}{0.01} & ECP(\%)       & 95.6 & 95.6 & 94.2  & 95.8  & 93.6  \\ \cline{2-7} 
                      & AL$\times$100 & 7.2  & 8.0  & 7.8  & 26.2  & 7.8  \\ \cline{2-7} 
                      & Time (sec)    & 6.66 & 3.28 & 26.38 & 21.62 & 26.60 \\ \hline
\multirow{3}{*}{0.02} & ECP(\%)       & 95.4 & 95.6 & 94.4  & 95.8  & 93.6  \\ \cline{2-7} 
                      & AL$\times$100 & 7.1  & 8.0  & 7.5  & 26.2  & 7.5  \\ \cline{2-7} 
                      & Time (sec)    & 6.49 & 2.97 & 26.49 & 21.84 & 25.88 \\ \hline
\multirow{3}{*}{0.03} & ECP(\%)       & 95.0 & 95.6 & 94.4  & 95.6  & 94.0  \\ \cline{2-7} 
                      & AL$\times$100 & 7.1  & 8.0  & 7.4  & 26.2  & 7.4  \\ \cline{2-7} 
                      & Time (sec)    & 6.72 & 3.26 & 25.95 & 20.23 & 26.63 \\ \hline
\multirow{3}{*}{0.04} & ECP(\%)       & 95.0 & 95.6 & 94.4  & 95.6  & 94.2  \\ \cline{2-7} 
                      & AL$\times$100 & 7.1  & 8.0  & 7.3  & 26.2  & 7.3  \\ \cline{2-7} 
                      & Time (sec)    & 6.28 & 2.95 & 26.44 & 21.87 & 27.92 \\ \hline
\multirow{3}{*}{0.05} & ECP(\%)       & 95.0 & 95.6 & 94.4  & 95.6  & 94.4  \\ \cline{2-7} 
                      & AL$\times$100 & 7.1  & 8.0  & 7.2  & 26.2  & 7.2  \\ \cline{2-7} 
                      & Time (sec)    & 6.01 & 2.94 & 26.53 & 23.08 & 27.29 \\ \hline
\end{tabular}
\caption{Performance of Algorithm 1 (simple splitting) under different values of ${C_0}$. The device: CPU Intel(R) Xeon(R) CPU E5-4650 v4 @ 2.20GHz. The implementation is without parallel computing.}
\label{tab4}
\end{table}

\begin{table}[H]
\centering
\scriptsize
\begin{tabular}{|c|c|c|c|c|c|c|}
\hline
Value of ${C_0}$        & Settings      & (A)  & (B)  & (C)   & (D)   & (E)   \\ \hline
\multirow{3}{*}{0.01} & ECP(\%)       & 95.2 & 95.6 & 93.6  & 96.0  & 93.8  \\ \cline{2-7} 
                      & AL$\times$100 & 7.1  & 8.0  & 7.1  & 26.2  & 7.0  \\ \cline{2-7} 
                      & Time (min)    & 7.68 & 4.27 & 25.25 & 17.96 & 24.83 \\ \hline
\multirow{3}{*}{0.02} & ECP(\%)       & 95.4 & 95.6 & 94.2  & 96.0  & 94.2  \\ \cline{2-7} 
                      & AL$\times$100 & 7.0 & 8.0  & 7.0  & 26.2  & 6.9  \\ \cline{2-7} 
                      & Time (min)    & 7.58 & 4.13 & 23.90 & 18.26 & 25.55 \\ \hline
\multirow{3}{*}{0.03} & ECP(\%)       & 95.6 & 95.6 & 94.2  & 96.0  & 94.2  \\ \cline{2-7} 
                      & AL$\times$100 & 7.0  & 8.0  & 6.9  & 26.1  & 6.8  \\ \cline{2-7} 
                      & Time (min)    & 7.98 & 4.10 & 24.63 & 18.05 & 25.18 \\ \hline
\multirow{3}{*}{0.04} & ECP(\%)       & 95.4 & 95.4 & 94.4  & 96.0  & 94.2  \\ \cline{2-7} 
                      & AL$\times$100 & 7.0  & 8.0  & 6.9  & 26.1  & 6.8  \\ \cline{2-7} 
                      & Time (min)    & 8.42 & 4.08 & 24.23 & 18.18 & 25.62 \\ \hline
\multirow{3}{*}{0.05} & ECP(\%)       & 95.6 & 95.4 & 94.2  & 95.8  & 94.6  \\ \cline{2-7} 
                      & AL$\times$100 & 7.0  & 8.0  & 6.8  & 26.1  & 6.7  \\ \cline{2-7} 
                      & Time (min)    & 8.02 & 4.21 & 24.40 & 18.34 & 24.95 \\ \hline
\end{tabular}
\caption{Performance of Algorithm \ref{alg3} (repeatedly splitting and averaging) under different values of ${C_0}$. The device: CPU Intel(R) Xeon(R) CPU E5-4650 v4 @ 2.20GHz. The implementation is under parallel computing of 56 cores.}
\label{tab5}
\end{table}
\begin{table}[H]
\centering
\scriptsize
\begin{tabular}{|c|c|c|c|c|c|}
\hline
Subbagging & (A)  & (B)  & (C)  & (D)  & (E)  \\ \hline
Time (min) & 4.27 & 3.20 & 8.45 & 6.59 & 9.23 \\ \hline
\end{tabular}
\caption{Computational time of subbagging in different cases. The device: CPU Intel(R) Xeon(R) CPU E5-4650 v4 @ 2.20GHz. The implementation is under parallel computing of 56 cores.}
\label{tab6}
\end{table}
\medskip\noindent\textbf{Numerical check of the tuning conditions.}
We conduct simulation and report two quantities, the variability ratio and the truncated tracking ratio, to numerically verify the commonly used parametric and undersmoothing estimators satisfy the desired tuning selection condition and property as summarized in Table \ref{tab:numeric-c1-c3}. Specifically, the variability ratio illustrates Condition (C3), and the truncated tracking ratio illustrates the approximation of the order of the estimated contrast function by $EAE$ after truncation. We consider the correctly specified parametric working model and undersmoothed kernel and spline estimators of the same types used in the simulation and application of the manuscript. The results support that the additional conditions are mild in these settings and illustrate the finite-sample behavior of Algorithm 2. 

Specifically, we use
$\mathbf X\sim\operatorname{Uniform}(0,1)$, $A\sim\operatorname{Bernoulli}(1/2)$, and
$Y=\mu_0(\mathbf X)+A\tau(\mathbf X)+\varepsilon$, where
$\mu_0(x)=\sin(2x)+x^2$ and $\varepsilon\sim N(0,0.25)$. We take
$n\in\{500,2000,8000\}$, with 300 replications for the first two sample sizes and 150 for the last. The parametric design uses
$\tau(x)=1-2x$ and a correctly specified linear interaction model. The nonparametric design uses
$\tau(x)=x\sin(3x)-0.3$ and separate within-arm regressions: quadratic splines with
$\lfloor1.2|\mathcal I|^{1/3}\rfloor$ interior knots on a truncated-power basis, undersmoothed relative to the $|\mathcal I|^{1/5}$-optimal choice, or a Nadaraya--Watson estimator with a Gaussian kernel and bandwidth
$0.8|\mathcal I|^{-1/3}$, which is also undersmoothed.

The sample is split as in Algorithms 1 and 2 into half folds $\mathcal I_1,\mathcal I_2$ and quarter folds $\mathcal I_{j,k}$, $j,k\in\{1,2\}$, with $|\mathcal I_j|=n/2$. For each $k$, the half-fold fit trained on $\mathcal I_j$ is compared with the quarter-fold fit trained on $\mathcal I_{3-j,3-k}$, and their squared difference is evaluated on $\mathcal I_{3-j,k}$. Population $L^2(P_{\mathbf X})$ norms are approximated on a grid of 1{,}500 points. The table reports two ratios for each estimator and each $n$:
\begin{itemize}
\item The variability ratio is
\[
\frac{E_{\mathbf X}|\widehat\tau_{\mathcal I_{3-j,1}}(\mathbf X)-\tau(\mathbf X)|^2}
{E_{\mathbf X}|\widehat\tau_{\mathcal I_{3-j,1}}(\mathbf X)-\widehat\tau_{\mathcal I_{3-j,2}}(\mathbf X)|^2}.
\]
It is bounded above and below exactly when the two sides of Condition (C3) are of the same order on average, and it equals $1/2$ when the error has no bias.
\item The truncated tracking ratio is
\[
\frac{EAE\vee(\log n/n)}{E_{\mathbf X}|\widehat\tau_{\mathcal I_j}(\mathbf X)-\tau(\mathbf X)|^2\vee(\log n/n)}.
\]
This ratio is the finite-sample analogue of the comparison used in the proof, and under the stated sufficient conditions, it should remain bounded away from $0$ and $\infty$.
\end{itemize}
\begin{table}[H]
\centering
\small
\begin{tabular}{lccc}
\hline
Estimator & $n$ & variability ratio & truncated tracking ratio \\
\hline
parametric (OLS) & 500  & 0.46 & 1.51 [1.00, 3.62] \\
                 & 2000 & 0.46 & 1.19 [1.00, 2.85] \\
                 & 8000 & 0.49 & 1.22 [1.00, 2.94] \\
spline (undersmoothed) & 500  & 0.46 & 3.06 [1.77, 5.70] \\
                       & 2000 & 0.49 & 2.86 [1.79, 4.53] \\
                       & 8000 & 0.51 & 2.66 [1.92, 3.84] \\
kernel (undersmoothed) & 500  & 0.59 & 1.85 [0.85, 3.61] \\
                       & 2000 & 0.64 & 1.93 [1.00, 3.39] \\
                       & 8000 & 0.59 & 2.40 [1.36, 3.84] \\
\hline
\end{tabular}
\caption{Monte Carlo check of Condition (C3) and of the desired property of Proposition 1. Entries are medians; brackets contain the 10th and 90th percentiles. There are 300 replications for $n=500,2000$ and 150 for $n=8000$.}
\label{tab:numeric-c1-c3}
\end{table}
The result in Table \ref{tab:numeric-c1-c3} is consistent with the theory. For the two undersmoothed nonparametric rows, the variability ratio stays near $1/2$ and the truncated tracking ratio stays bounded away from $0$ and $\infty$. For the parametric row, Condition (C1) applies and the truncated tracking ratio stays bounded away from $0$ and $\infty$. In summary, the results illustrate that Condition (C3) can hold for common undersmoothed nonparametric estimators and that the $EAE$ in Algorithm 2 can track the order of the squared estimation error after truncation in the settings considered.
\subsection{Extensions to Heteroscedasticity in Single Time Point Case}
In the single time point case allowing heteroscedasticity, we can show that the optimal efficiency can be achieved by $\widehat{V}_{\mathcal{D}}$ where
\begin{align}\label{single-optimal-decision-heter}
    \mathcal{D}(\mathbf{X})=\mathbb{I}\{\tau(\mathbf{X})>0\}+\frac{\operatorname{Var}(Y|\mathbf{X},A=0)\pi(1,\mathbf{X})}{\operatorname{Var}(Y|\mathbf{X},A=1)\pi(0,\mathbf{X})+\operatorname{Var}(Y|\mathbf{X},A=0)\pi(1,\mathbf{X})}\mathbb{I}\{\tau(\mathbf{X})=0\}.
\end{align}
The results are summarized in Theorem \ref{single-eff-opt-heter}.
\begin{theorem}\label{single-eff-opt-heter}
         Under Assumptions 1 to 3, when the working models $\mu_1$, $\mu_0$ and $\pi_1$ in $T_n$ are correctly specified, and $T_n$ satisfies asymptotic uniform integrability, i.e.,
    \[
        \lim _{M \rightarrow \infty} \sup _n E\left[n\left(T_n-V_0\right)^2 \cdot \mathbf{1}\left(n\left(T_n-V_0\right)^2>M\right)\right]=0,
    \]
    we have
    \begin{align*}
        \inf_{T_n\in\mathcal{T}_{RL}}\lim_{n\to\infty}\operatorname{Var}(\sqrt{n}T_n)= \lim_{n\to\infty}\operatorname{Var}(\sqrt{n}\widehat{V}_{\mathcal{D}}^{*})
    \end{align*}
    where $\mathcal{D}$ is defined as in Eq. \eqref{single-optimal-decision-heter}.
\end{theorem}
Before we present the proof of Theorem \ref{single-eff-opt-heter}, we firstly state and prove some technical lemmas.
\begin{lemma}\label{lem: quad est}
Let $\mu \in \mathbb{R}$ and $\sigma^2 > 0$ be fixed constants. If $E[f(U)]$ is a constant $C$ for any $U$ s.t. $E[U] = \mu$ and $\text{Var}[U] = \sigma^2$, then $f(u)$ is a quadratic polynomial in $u$. 
\end{lemma}
\begin{proof}
Consider a discrete random variable $U$  s.t. $E[U] = \mu$ and $\operatorname{Var}[U] = \sigma^2$ supported on a finite set of $n$ distinct points $\{u_1, u_2, \dots, u_n\} \subset \mathbb{R}$ (where $n > 3$). Let $p_i = P(U = u_i)$. The conditions that $E[U] = \mu$ and $\operatorname{Var}[U] = \sigma^2$ impose the following constraints on the probability vector $\mathbf{p} = [p_1, \dots, p_n]^\top$:

\begin{enumerate}
    \item Normalization: $\sum_{i=1}^n p_i = 1$, $p_i\ge0$, $i=1,\dots,n$
    \item Mean: $\sum_{i=1}^n p_i u_i = \mu$
    \item Variance: $\sum_{i=1}^n p_i (u_i - \mu)^2 = \sigma^2$, which is equivalent to $\sum_{i=1}^n p_i u_i^2 = \sigma^2 + \mu^2$
\end{enumerate}
We can express these constraints as a linear system $A\mathbf{p} = \mathbf{b}$, where:
$$
A = \begin{bmatrix}
1 & 1 & \dots & 1 \\
u_1 & u_2 & \dots & u_n \\
u_1^2 & u_2^2 & \dots & u_n^2
\end{bmatrix}, \quad
\mathbf{b} = \begin{bmatrix}
1 \\
\mu \\
\sigma^2 + \mu^2
\end{bmatrix}
$$
To make sure this linear system (1) has a all positive solution ($p_i>0$ for all $i$) and (2) the solution is nonunique, we consider the following construction of $u_i$, $i=1,\dots,n$. We let $n=5$ and $\{u_1,\dots,u_5\} = \{\mu-4\sigma,\mu-2\sigma,\mu,\mu+2\sigma,\mu+4\sigma\}$. In this case, one all positive solution is $(p_1,\dots,p_5) = (1/64,1/16,27/32,1/16,1/64)$. The solution is also nonunique, as there is another all positive solution $(p_1,\dots,p_5) = (1/48,1/24,7/8,1/24,1/48)$. This shows that we are able to choose $\{u_i\}$ such that a solution exists with strictly positive probabilities, i.e., $\{\mathbf{p} \mid A \mathbf{p}=\mathbf{b}, \mathbf{p} > 0\}$ is not $\varnothing$ or singleton.

The expectation of $f(U)$ is given by the linear form:
$$ E[f(U)] = \sum_{i=1}^n p_i f(u_i) = \mathbf{f}^\top \mathbf{p} $$
where $\mathbf{f} = [f(u_1), f(u_2), \dots, f(u_n)]^\top$.
Note that $E[f(Y)]\equiv C$ for any  $\mathbf{p}\in\{\mathbf{p} \mid A \mathbf{p}=\mathbf{b}, \mathbf{p} > 0\}$ is equivalent to
\begin{align}
    \mathbf{f}^\top \mathbf{p}\equiv C,~\forall \mathbf{p}\in\{\mathbf{p} \mid A \mathbf{p}=\mathbf{b}, \mathbf{p} > 0\}.
\end{align}
(actually we have $E[f(U)]\equiv C$ for any  $\mathbf{p}\in\{\mathbf{p} \mid A \mathbf{p}=\mathbf{b}, \mathbf{p} \ge 0\}$ but here it is enough to only focus on $\mathbf{p} > 0$.) Let $\operatorname{Ker}(A):=\{\mathbf{h}: A\mathbf{h}=\mathbf{0}\}$. Since $\{\mathbf{p} \mid A \mathbf{p}=\mathbf{b}, \mathbf{p} > 0\}$ is not $\varnothing$ or singleton, $\operatorname{Ker}(A)$ contains nonzero element. Then for any $\mathbf{p}\in\{\mathbf{p} \mid A \mathbf{p}=\mathbf{b}, \mathbf{p} > 0\} $ and any nonzero (zero case is trivial) $\mathbf{h}\in \operatorname{Ker}(A)$, since $\{\mathbf{p}:\mathbf{p}>0\}$ is an open set of $\mathbf{p}$, $\{\mathbf{p} \mid A \mathbf{p}=\mathbf{b}, \mathbf{p} > 0\}$ is also an open set in the subspace topology in $\{\mathbf{p} \mid A \mathbf{p}=\mathbf{b}\}$, then there exists a positive $\delta(\mathbf{p},\mathbf{h})$ s.t. 
\begin{align}
    \{\mathbf{p}+\varepsilon\mathbf{h}:0\leq\varepsilon<\delta(\mathbf{p},\mathbf{h})\}\subset\{\mathbf{p} \mid A \mathbf{p}=\mathbf{b}, \mathbf{p} > 0\}
\end{align}
Therefore, 
\begin{align}
    \mathbf{f}^\top (\mathbf{p}+\varepsilon\mathbf{h})\equiv C, ~\forall\varepsilon\in[0,\delta(\mathbf{p},\mathbf{h})),
\end{align}
which suggests $\mathbf{f}^\top\mathbf{h}=0$ for any  $\mathbf{h}\in \operatorname{Ker}(A)$. Denote $\operatorname{Ker}(\mathbf{f}^\top):=\{\mathbf{h}: \mathbf{f}^\top\mathbf{h}=\mathbf{0}\}$, and we have $\operatorname{Ker}(A)\subset\operatorname{Ker}(\mathbf{f}^\top)$. Therefore, $\operatorname{Im}(\mathbf{f}^\top)\subset \operatorname{Im}(A)$, that is,  $\mathbf{f}^\top$ should be linear representable by the rows of $A$. Thus, there exist scalars $\lambda_0, \lambda_1, \lambda_2$ (corresponding to the three rows of $A$) such that:
$$ \mathbf{f}^\top = \lambda_0 A_{1\cdot} + \lambda_1 A_{2\cdot} + \lambda_2 A_{3\cdot} $$
where $A_{j\cdot}$, $j=1,2,3$ denotes the $j$-th row of $A$. Writing this component-wise for $u_i$, $i=1,\dots n$ ($n=5$ in our construction):
$$ f(u_i) = \lambda_0 + \lambda_1 u_i + \lambda_2 u_i^2 $$
For any $u^*$ other than $\{\mu-4\sigma,\mu-2\sigma,\mu,\mu+2\sigma,\mu+4\sigma\}$, we expand the construction to be $n=7$:
\begin{align*}
    \{u_1,\dots,u_7\} = \{\mu-4\sigma,\mu-2\sigma,\mu,\mu+2\sigma,\mu+4\sigma,u^*,2\mu-u^*\}
\end{align*}
In this case, denote $A$ in the $n=5$ example extended by two additional rows $(1,u^*,(u^*)^2)^\top$ and $(1,2\mu-u^*,(2\mu-u^*)^2)^\top$ as
\begin{align*}
    \widetilde{A} = \begin{bmatrix}
1 & 1 & \dots & 1 & 1 & 1 \\
u_1 & u_2 & \dots & u_n & u^* & 2\mu-u^* \\
u_1^2 & u_2^2 & \dots & u_n^2 & (u^*)^2 & (2\mu-u^*)^2.
\end{bmatrix}
\end{align*}
In this case, for any $u^*$ other than $\{\mu-4\sigma,\mu-2\sigma,\mu,\mu+2\sigma,\mu+4\sigma\}$, $\{\mathbf{p} \mid \widetilde{A}\mathbf{p}=\mathbf{b}, \mathbf{p} > 0\}$ is still not $\varnothing$ or singleton. To see this, we discuss two cases. First, when $|u^* - \mu|\ge \sigma$, we can verify that both 
\begin{align*}
    p_6=p_7 = \frac{1}{4}\frac{\sigma^2}{|\mu-u^*|^2}, p_1 = p_5 = 1/128, p_2 = p_4 = 1/32,\\ p_3 = 1-\frac{5}{64} - \frac{1}{2}\frac{\sigma^2}{|\mu-u^*|^2}\ge \frac{27}{64}
\end{align*}
and
\begin{align*}
      p_6=p_7 = \frac{1}{4}\frac{\sigma^2}{|\mu-u^*|^2}, p_1 = p_5 = 1/96, p_2 = p_4 = 1/48,\\ p_3 = 1-\frac{1}{16} - \frac{1}{2}\frac{\sigma^2}{|\mu-u^*|^2}\ge \frac{7}{16}
\end{align*}
are different positive solutions in $\{\mathbf{p} \mid \widetilde{A}\mathbf{p}=\mathbf{b}, \mathbf{p} > 0\}$.

When $|u^* - \mu|< \sigma$, we can verify that both 
\begin{align*}
    p_6=p_7 = \frac{1}{4}, p_1 = p_5 = \frac{1}{64}\left(1-\frac{1}{2}\frac{|\mu-u^*|^2}{\sigma^2}\right), p_2 = p_4 = \frac{1}{16}\left(1-\frac{1}{2}\frac{|\mu-u^*|^2}{\sigma^2}\right),\\ \frac{27}{64}>p_3 = \frac{11}{32} + \frac{5}{64}\frac{|\mu-u^*|^2}{\sigma^2}>\frac{11}{32}
\end{align*}
and 
\begin{align*}
    p_6=p_7 = \frac{1}{4}, p_1 = p_5 = \frac{1}{48}\left(1-\frac{1}{2}\frac{|\mu-u^*|^2}{\sigma^2}\right), p_2 = p_4 = \frac{1}{24}\left(1-\frac{1}{2}\frac{|\mu-u^*|^2}{\sigma^2}\right),\\ \frac{7}{16}>p_3 = \frac{3}{8} + \frac{1}{16}\frac{|\mu-u^*|^2}{\sigma^2}> \frac{3}{8}
\end{align*}
are different positive solutions in $\{\mathbf{p} \mid \widetilde{A}\mathbf{p}=\mathbf{b}, \mathbf{p} > 0\}$. Denote $\widetilde{\mathbf{f}}^\top=[f(u_1), f(u_2), \dots, f(u_7)]$, by the same derivation as in the $n=5$ case, we have $\widetilde{\mathbf{f}}^\top$ can also be linear representable by the rows of $\widetilde{A}$. This is to say that, there exists $c_0$, $c_1$ and $c_2$ s.t. 
$$ f(u_i) = c_0 + c_1 u_i + c_2 u_i^2 $$
for $i=1,\dots,7$. However, for $i=1,\dots,5$, we have
\begin{align*}
    f(u_i) = c_0 + c_1 u_i + c_2 u_i^2 = \lambda_0 +\lambda_1 u_i + \lambda_2 u_i^2.
\end{align*}
Since three distinct points can determine a quadratic polynomial, and $5\ge 3$, so here we must have
\begin{align*}
    c_0 = \lambda_0, c_1 = \lambda_1, c_2 = \lambda_2.
\end{align*}
Therefore, for $u_6=u^*$, we have 
\begin{align*}
    f(u^*) = \lambda_0+\lambda_1 u^* + \lambda_2(u^*)^2.
\end{align*}
Since $u^*$ can be any points other than $\{\mu-4\sigma,\mu-2\sigma,\mu,\mu+2\sigma,\mu+4\sigma\}$, we have shown that for any $u$, $f(u) =\lambda_0+\lambda_1 u + \lambda_2 u^2$, that is, $f(u)$ is a quadratic polynomial in $u$.
\end{proof}

\begin{lemma}\label{lem: linear constraint}
    If $E[f(U,\operatorname{Var} (U))] = 0$ for any random variable $U$ s.t. $E[U]> \mu_0$, then $f(u,v) = 0$ for any $u>\mu_0$ and $v\ge 0$.
\end{lemma}
\begin{proof}[Proof of Lemma \ref{lem: linear constraint}]
    Consider a distribution of $U$ defined as 
    \begin{align*}
        P(U = \mu+\alpha) = \frac{\beta}{\alpha+\beta},~P(U = \mu-\beta) = \frac{\alpha}{\alpha+\beta},~\mu>\mu_0,~\alpha,\beta>0.
    \end{align*}
    For any $\alpha,\beta>0$, we have
    \begin{align*}
        E[U] = \mu,~\operatorname{Var}(U)=\alpha\beta.
    \end{align*}
    Since $E[f(U,\operatorname{Var} (U))] = 0$ for any $U$, for any $\alpha,\beta>0$, we have
    \begin{align}\label{ab-constraint}
       \forall\mu>\mu_0,~ \beta f(\mu+\alpha,\alpha\beta)+\alpha f(\mu-\beta,\alpha\beta)=0.
    \end{align}
    Firstly, we let $\alpha=\beta=\sqrt{v}$, then Eq. \eqref{ab-constraint} becomes
    \begin{align}
          \forall\mu>\mu_0,~  f(\mu+\sqrt{v},v)+f(\mu-\sqrt{v},v)=0
    \end{align}
    and this induces
    \begin{align}\label{ab-constraint-1}
          \forall\mu>\mu_0-\sqrt{v},~ f(\mu+2\sqrt{v},v) = -f(\mu,v).
    \end{align}
    Then, let $\alpha = 2\sqrt{v}$, $\beta=\sqrt{v}/2$, then Eq. \eqref{ab-constraint} becomes
    \begin{align}\label{ab-constraint-2}
        \forall\mu>\mu_0,~ \frac{\sqrt{v}}{2}f(\mu+2\sqrt{v},v)+2\sqrt{v}f(\mu - \frac{\sqrt{v}}{2},v)=0
    \end{align}
    and by Eq. \eqref{ab-constraint-1}, Eq. \eqref{ab-constraint-2} becomes
    \begin{align}
         \forall\mu>\mu_0,~ -\frac{\sqrt{v}}{2}f(\mu,v)+2\sqrt{v}f(\mu - \frac{\sqrt{v}}{2},v)=0
    \end{align}
    which is equivalent to
    \begin{align}\label{ab-constraint-3}
       \forall\mu>\mu_0-\frac{\sqrt{v}}{2},~ f(\mu+\frac{\sqrt{v}}{2},v) = 4 f(\mu,v).
    \end{align}
    Now, we have
    \begin{align}\label{ab-constraint-4}
        \forall\mu>\mu_0,~f(\mu+\sqrt{v},v) =& f(\mu+\frac{\sqrt{v}}{2}+\frac{\sqrt{v}}{2},v)\notag\\
        \text{by Eq. \eqref{ab-constraint-3}}=&4f(\mu+\frac{\sqrt{v}}{2},v)\notag\\
        \text{by Eq. \eqref{ab-constraint-3}}=&16f(\mu,v)
    \end{align}
    and Eq. \eqref{ab-constraint-4} again suggests
    \begin{align}
         \forall\mu>\mu_0,~f(\mu+2\sqrt{v},v) = 16 f(\mu+\sqrt{v},v) = 16^2 f(\mu,v).
    \end{align}
    However, combined with Eq. \eqref{ab-constraint-1}, we have
    \begin{align*}
        \forall\mu>\mu_0,~f(\mu+2\sqrt{v},v) = 16^2 f(\mu,v) = -f(\mu,v).
    \end{align*}
    This suggests that $f(\mu,v)\equiv 0$ for any $\mu>\mu_0$ and $v>0$. When $v=\operatorname{Var}(U)=0$, $U\equiv \mu$ almost surely, where $\mu = E(U)>\mu_0$. The condition $E[f(U,\operatorname{Var}(U))] = 0$ then gives $f(\mu,0) = 0$ for $\mu>\mu_0$. Therefore, we have proved that $f(u,v)=0$ for any $u>\mu_0$ and $v\ge 0$.
\end{proof}
\begin{lemma}\label{lem: linear constraint neg}
    If $E[f(U,\operatorname{Var} (U))] = 0$ for any random variable $U$ s.t. $E[U]< \mu_0$, then $f(u,v) = 0$ for any $u<\mu_0$ and $v\ge 0$.
\end{lemma}
\begin{proof}
    Apply the Lemma \ref{lem: linear constraint} to $-U$ and $-\mu_0$ can lead to the result.
\end{proof}
\begin{proof}[Proof of Theorem \ref{single-eff-opt-heter}]
Here, we show the results in a more general case allowing heteroscedasticity; i.e. without Assumption 8. The sketch of this proof is as follows: (1) show that the influence function $f$ is also unbiased conditional on $\mathbf{X}$; (2) by variational principle, show that the influence function $f$ must follow a standard form of being a special linear function w.r.t. $Y$ and (3) by variance decomposition of the standard form, show that $\widehat{V}_{as}$ is optimal.

Firstly, we show that the unbiasedness required in Definition 1; i.e. $E[f] = E[E[Y|\mathbf{X},A=d^{opt}(\mathbf{X})]]$ for any distribution of $(\mathbf{X},A,Y)$, implies conditional unbiasedness 
    \begin{align*}
        E[f|\mathbf{X}] = E[Y|\mathbf{X},A=d^{opt}(\mathbf{X})]
    \end{align*}
    with probability 1 for any distribution of $(\mathbf{X},A,Y)$. We show this by contradiction. Assume that for joint distribution $(\mathbf{X},A,Y)$, we have
    \begin{align*}
          E[f|\mathbf{X}] \neq E[Y|\mathbf{X},A=d^{opt}(\mathbf{X})]
    \end{align*}
    in a positive probability. Then, one of the following two sets:
    \begin{align*}
        \mathcal{X}_>:=&\{\mathbf{X}: E[f|\mathbf{X}] > E[Y|\mathbf{X},A=d^{opt}(\mathbf{X})]\};\\
         \mathcal{X}_<:=&\{\mathbf{X}: E[f|\mathbf{X}] < E[Y|\mathbf{X},A=d^{opt}(\mathbf{X})]\}
    \end{align*}
    should have positive probability. WLOG, we let $P(\mathbf{X}\in\mathcal{X}_>)>0$. Then for jont distribution $(\mathbf{X},A,Y)|\mathbf{X}\in\mathcal{X}_>$, let $p_{\mathbf{X}}(\cdot)$ be the probability density of $\mathbf{X}$, and we have
    \begin{align*}
        &E_{(\mathbf{X},A,Y)|\mathbf{X}\in\mathcal{X}_>}[f-E[Y|\mathbf{X},A=d^{opt}(\mathbf{X})]]\\
        =&\int_{x\in\mathcal{X}_>}\frac{p_{\mathbf{X}}(x)}{P(\mathbf{X}\in\mathcal{X}_>)}\left( E[f|\mathbf{X}=x] - E[Y|\mathbf{X}=x,A=d^{opt}(x)]\right) dx>0,
    \end{align*}
    which contradicts the fact that $f$ is unbiased for any joint distribution of $(\mathbf{X},A,Y)$.

    Then, we show that conditional on $\mathbf{X}$ and $A$, the influence function $f$ should be a linear function of $Y$. We start by the $\mathbf{X}$ s.t. $\tau(\mathbf{X}) = 0$. In this case, by the identification Assumptions 1 to 3, we have
    \begin{align}\label{thm3.2_1}
        E[Y|\mathbf{X}] = E[Y|\mathbf{X},A=1] = E[Y|\mathbf{X},A=0] = E[Y|\mathbf{X},A=d^{opt}(\mathbf{X})].
    \end{align}
When the working model for propensity equals the population truth, i.e., $\pi_1(\mathbf{X}) = \pi(1,\mathbf{X})$, almost surely we have
\begin{align}\label{single-unbias-eq-tau=0}
   \pi(1,\mathbf{X})E_{Y|A=1,\mathbf{X}}[ f(\mathbf{X},1,Y,\dots)] + \pi(0,\mathbf{X}) E_{Y|A=0,\mathbf{X}}[ f(\mathbf{X},0,Y,\dots)] = E[Y|\mathbf{X}]
\end{align}
holds for any model satisfying $E[Y|A=1,\mathbf{X}] = E[Y|A=0,\mathbf{X}]=E[Y|\mathbf{X}]$. Now we fix $E[Y|\mathbf{X}]=\mu_0^*$ and $\pi(1,\mathbf{X})=p$ and treat them as constants, and fix the distribution of $Y|\mathbf{X},A=0$, and then we observe that for any given $\widetilde{\sigma}_1>0$ and any random variable $U_1$ satisfying $E[U_1]=\mu_0^*$ and $\operatorname{Var}(U_1)=\widetilde{\sigma}_1^2$, under the correct specification of $\pi_1$, $f$ defined in Definition 1 satisfies
\begin{align*}
    E_{U_1}[f(\mathbf{X},1,U_1,\mu_1(\mathbf{X}),\mu_0(\mathbf{X}),p,0,r_1(\mathbf{X}),r_0(\mathbf{X}))] \\= \frac{\mu_0^*-(1-p)E_{Y|\mathbf{X},A=0}[f(\mathbf{X},0,Y,\dots)]}{p}\equiv \text{const}.
\end{align*}
By the definition of $f$ and Lemma \ref{lem: quad est}, $f(\mathbf{X},1,U_1,\dots)$ is a quadratic polynomial with all coefficients being function of $p$, $\mu_a(\mathbf{X})$ and $r_a(\mathbf{X})$, $a=0,1$. That is, 
\begin{align}\label{quad-single-1}
    &f(\mathbf{X},1,U_1,\mu_1(\mathbf{X}),\mu_0(\mathbf{X}),p,0,r_1(\mathbf{X}),r_0(\mathbf{X}))\notag\\ =& a_1(p,\mu_1(\mathbf{X}),\mu_0(\mathbf{X}),r_1(\mathbf{X}),r_0(\mathbf{X})) U_1^2 + b_1(p,\mu_1(\mathbf{X}),\mu_0(\mathbf{X}),r_1(\mathbf{X}),r_0(\mathbf{X})) U_1 \notag\\+ &c_1(p,\mu_1(\mathbf{X}),\mu_0(\mathbf{X}),r_1(\mathbf{X}),r_0(\mathbf{X}))
\end{align}
where $a_1$, $b_1$ and $c_1$ are functions of $p,\mu_1(\mathbf{X}),\mu_0(\mathbf{X}),r_1(\mathbf{X}),r_0(\mathbf{X})$. Similarly, take any random variable $U_0$ satisfying $E[U_0]=\mu_0^*$ and $\operatorname{Var}(U_0)=\widetilde{\sigma}_0^2$ and fix $Y|\mathbf{X},A=1$, we have
\begin{align}\label{quad-single-0}
     &f(\mathbf{X},0,U_0,\mu_1(\mathbf{X}),\mu_0(\mathbf{X}),p,0,r_1(\mathbf{X}),r_0(\mathbf{X})) \notag\\
     =& a_0(p,\mu_1(\mathbf{X}),\mu_0(\mathbf{X}),r_1(\mathbf{X}),r_0(\mathbf{X})) U_0^2 + b_0(p,\mu_1(\mathbf{X}),\mu_0(\mathbf{X}),r_1(\mathbf{X}),r_0(\mathbf{X})) U_0 \notag\\+& c_0(p,\mu_1(\mathbf{X}),\mu_0(\mathbf{X}),r_1(\mathbf{X}),r_0(\mathbf{X})).
\end{align}
If we set $\widetilde{Y}$ s.t. $E[\widetilde{Y}|\mathbf{X},A=1] = E[\widetilde{Y}|\mathbf{X},A=0]=\mu_0^*$, $\widetilde{\sigma}_1^2:=\operatorname{Var}(\widetilde{Y}|\mathbf{X},A=1)$ and $\widetilde{\sigma}_0^2:=\operatorname{Var}(\widetilde{Y}|\mathbf{X},A=0)$, by Eqs. \eqref{quad-single-1} and \eqref{quad-single-0}, $f$ in Definition 1 has the following constrained form 
\begin{align*}
    &f(\mathbf{X},1,\widetilde{Y},\mu_1(\mathbf{X}),\mu_0(\mathbf{X}),p,0,r_1(\mathbf{X}),r_0(\mathbf{X}))\\
     = &a_1(p,\mu_1(\mathbf{X}),\mu_0(\mathbf{X}),r_1(\mathbf{X}),r_0(\mathbf{X})) \widetilde{Y}^2 + b_1(p,\mu_1(\mathbf{X}),\mu_0(\mathbf{X}),r_1(\mathbf{X}),r_0(\mathbf{X})) \widetilde{Y}\\ +& c_1(p,\mu_1(\mathbf{X}),\mu_0(\mathbf{X}),r_1(\mathbf{X}),r_0(\mathbf{X})).
\end{align*}
and
\begin{align*}
    &f(\mathbf{X},0,\widetilde{Y},\mu_1(\mathbf{X}),\mu_0(\mathbf{X}),p,0,r_1(\mathbf{X}),r_0(\mathbf{X}))\\
     = &a_0(p,\mu_1(\mathbf{X}),\mu_0(\mathbf{X}),r_1(\mathbf{X}),r_0(\mathbf{X})) \widetilde{Y}^2 + b_0(p,\mu_1(\mathbf{X}),\mu_0(\mathbf{X}),r_1(\mathbf{X}),r_0(\mathbf{X})) \widetilde{Y}\\ +& c_0(p,\mu_1(\mathbf{X}),\mu_0(\mathbf{X}),r_1(\mathbf{X}),r_0(\mathbf{X})).
\end{align*}

Then, Eq. \eqref{single-unbias-eq-tau=0} becomes
\begin{align*}
   & p a_1(p,\mu_1(\mathbf{X}),\mu_0(\mathbf{X}),r_1(\mathbf{X}),r_0(\mathbf{X})) ((\mu_0^*)^2+\widetilde{\sigma}_1^2) + p b_1(p,\mu_1(\mathbf{X}),\mu_0(\mathbf{X}),r_1(\mathbf{X}),r_0(\mathbf{X})) \mu_0^*\\+ &pc_1(p,\mu_1(\mathbf{X}),\mu_0(\mathbf{X}),r_1(\mathbf{X}),r_0(\mathbf{X}))
    +(1-p) a_0(p,\mu_1(\mathbf{X}),\mu_0(\mathbf{X}),r_1(\mathbf{X}),r_0(\mathbf{X}))(\mu_0^*)^2+\widetilde{\sigma}_0^2) \\+& (1-p) b_0(p,\mu_1(\mathbf{X}),\mu_0(\mathbf{X}),r_1(\mathbf{X}),r_0(\mathbf{X})) \mu_0^*+(1-p) c_0(p,\mu_1(\mathbf{X}),\mu_0(\mathbf{X}),r_1(\mathbf{X}),r_0(\mathbf{X}))\\=&\mu_0^*.
\end{align*}
Taking derivative w.r.t. $\widetilde{\sigma}_1^2$ and $\widetilde{\sigma}_0^2$ on both sides (it is differentiable as any $\widetilde{\sigma}_1$ and $\widetilde{\sigma}_1$ are in the interior of the positive real line), we can see that
\begin{align*}
    a_1(p,\mu_1(\mathbf{X}),\mu_0(\mathbf{X}),r_1(\mathbf{X}),r_0(\mathbf{X}))\equiv0,~a_0(p,\mu_1(\mathbf{X}),\mu_0(\mathbf{X}),r_1(\mathbf{X}),r_0(\mathbf{X}))\equiv 0.
\end{align*}
Therefore, 
\begin{align*}
   & f(\mathbf{X},A,Y,\dots)\\ =  &[Ab_1(p,\mu_1(\mathbf{X}),\mu_0(\mathbf{X}),r_1(\mathbf{X}),r_0(\mathbf{X})) + (1-A)b_0(p,\mu_1(\mathbf{X}),\mu_0(\mathbf{X}),r_1(\mathbf{X}),r_0(\mathbf{X}))]Y \\+& Ac_1(p,\mu_1(\mathbf{X}),\mu_0(\mathbf{X}),r_1(\mathbf{X}),r_0(\mathbf{X}))+(1-A) c_0(p,\mu_1(\mathbf{X}),\mu_0(\mathbf{X}),r_1(\mathbf{X}),r_0(\mathbf{X}))
\end{align*}
Taking derivative on both sides of 
\begin{align*}
   & p b_1(p,\mu_1(\mathbf{X}),\mu_0(\mathbf{X}),r_1(\mathbf{X}),r_0(\mathbf{X})) \mu_0^*+ pc_1(p,\mu_1(\mathbf{X}),\mu_0(\mathbf{X}),r_1(\mathbf{X}),r_0(\mathbf{X}))
   \\+& (1-p) b_0(p,\mu_1(\mathbf{X}),\mu_0(\mathbf{X}),r_1(\mathbf{X}),r_0(\mathbf{X})) \mu_0^*+(1-p) c_0(p,\mu_1(\mathbf{X}),\mu_0(\mathbf{X}),r_1(\mathbf{X}),r_0(\mathbf{X}))\\=&\mu_0^*
\end{align*}
w.r.t. $\mu_0^*$, we have
\begin{align}
     p b_1(p,\mu_1(\mathbf{X}),\mu_0(\mathbf{X}),r_1(\mathbf{X}),r_0(\mathbf{X}))+ (1-p) b_0(p,\mu_1(\mathbf{X}),\mu_0(\mathbf{X}),r_1(\mathbf{X}),r_0(\mathbf{X}))=1.\label{slop-tau-0-constraint}
\end{align}
Next, we derive the variance lower bound as
    \begin{align}\label{single-var-lb-tau0}
    \begin{aligned}
      &  \operatorname{Var}(f|\mathbf{X})\\
      =&\operatorname{Var}\left([Ab_1(p,\mu_1(\mathbf{X}),\mu_0(\mathbf{X}),r_1(\mathbf{X}),r_0(\mathbf{X})) + (1-A)b_0(p,\mu_1(\mathbf{X}),\mu_0(\mathbf{X}),r_1(\mathbf{X}),r_0(\mathbf{X}))]\mu_0^*\right.\\&+\left.f(\mathbf{X},A,0,\dots)\bigg|\mathbf{X}\right)\\
        +&E\left([Ab_1(p,\mu_1(\mathbf{X}),\mu_0(\mathbf{X}),r_1(\mathbf{X}),r_0(\mathbf{X}))\right.\\
        &\left.+ (1-A)b_0(p,\mu_1(\mathbf{X}),\mu_0(\mathbf{X}),r_1(\mathbf{X}),r_0(\mathbf{X}))]^2\operatorname{Var}(Y|\mathbf{X},A)\bigg|\mathbf{X}\right)\\
        \ge & E\left([Ab_1(p,\mu_1(\mathbf{X}),\mu_0(\mathbf{X}),r_1(\mathbf{X}),r_0(\mathbf{X}))\right.\\ &\left.+ (1-A)b_0(p,\mu_1(\mathbf{X}),\mu_0(\mathbf{X}),r_1(\mathbf{X}),r_0(\mathbf{X}))]^2\operatorname{Var}(Y|\mathbf{X},A)\bigg|\mathbf{X}\right)\\
        =&E\left(A b_1^2 \operatorname{Var}(Y|\mathbf{X},A) + (1-A) b_0^2 \operatorname{Var}(Y|\mathbf{X},A)\mid\mathbf{X}\right)\\
        =& p \operatorname{Var}(Y|\mathbf{X},A=1)b_1^2 + (1-p)\operatorname{Var}(Y|\mathbf{X},A=0)b_0^2
        \end{aligned}
    \end{align}
    which achieved the extreme value as function of $b_1$ (by Eq. \eqref{slop-tau-0-constraint}, $b_0=\frac{1-pb_1}{1-p}$) at
    \begin{align*}
    &\frac{d}{d b_1}\left(p \operatorname{Var}(Y|\mathbf{X},A=1)b_1^2 + (1-p)\operatorname{Var}(Y|\mathbf{X},A=0)b_0^2\right)\\
    =&\frac{d}{d b_1}\left(p \operatorname{Var}(Y|\mathbf{X},A=1)b_1^2 + (1-p)\operatorname{Var}(Y|\mathbf{X},A=0)\left(\frac{1-pb_1}{1-p}\right)^2\right)
    \\
      =&  2p\operatorname{Var}(Y|\mathbf{X},A=1)b_1 +2(1-p)\operatorname{Var}(Y|\mathbf{X},A=0)\frac{1-p b_1}{1-p}\cdot\frac{-p}{1-p}=0
    \end{align*}
    which suggest the minimal value of $p \operatorname{Var}(Y|\mathbf{X},A=1)b_1^2 + (1-p)\operatorname{Var}(Y|\mathbf{X},A=0)b_0^2$ is taken at
    \begin{align*}
        b_1 = \frac{\operatorname{Var}(Y|\mathbf{X},A=0)}{\operatorname{Var}(Y|\mathbf{X},A=1)(1-p)+\operatorname{Var}(Y|\mathbf{X},A=0)p}
    \end{align*}
    and consequently by the constraint in Eq. \eqref{slop-tau-0-constraint}, we have
    \begin{align*}
         b_1 = \frac{\operatorname{Var}(Y|\mathbf{X},A=1)}{\operatorname{Var}(Y|\mathbf{X},A=1)(1-p)+\operatorname{Var}(Y|\mathbf{X},A=0)p}.
    \end{align*}
Such $b_1$ and $b_0$ are achievable when
\begin{align*}
        &b_1(p,\mu_1(\mathbf{X}),\mu_0(\mathbf{X}),r_1(\mathbf{X}),r_0(\mathbf{X})) =  \frac{r_0(\mathbf{X})}{r_1(\mathbf{X}) (1-p)+r_0(\mathbf{X}) p}\\
        \text{ and }&b_0(p,\mu_1(\mathbf{X}),\mu_0(\mathbf{X}),r_1(\mathbf{X}),r_0(\mathbf{X})) =  \frac{r_1(\mathbf{X})}{r_1(\mathbf{X}) (1-p)+r_0(\mathbf{X}) p}\\
        &\text{ where }r_a(\mathbf{X}) = \operatorname{Var}(Y|\mathbf{X},A=a),~a=0,1.
\end{align*}
Next, we discuss the case when $\tau_P(\mathbf{X})>0$, i.e., $\operatorname{sgn}\left(\tau_P\left(\mathbf{X}\right)\right)=1$. When the working model for propensity is correctly specified, i.e., equals the population truth, i.e., $\pi_1(\mathbf{X}) = \pi(1,\mathbf{X})$, based on the identification condition, Assumptions 1 to 3, and similar to the derivation of Eq. \eqref{single-unbias-eq-tau=0}, we have almost surely 
\begin{align}\label{single-unbias-eq-tau>0}
   \pi(1,\mathbf{X})E_{Y|A=1,\mathbf{X}}[ f(\mathbf{X},1,Y,\dots)] + \pi(0,\mathbf{X}) E_{Y|A=0,\mathbf{X}}[ f(\mathbf{X},0,Y,\dots)] = E[Y|\mathbf{X},A=1]
\end{align}
holds for any model satisfying $E[Y|A=1,\mathbf{X}] > E[Y|A=0,\mathbf{X}]$. Fix $p:= \pi(1,\mathbf{X})$ and the model of $Y|\mathbf{X},A=0$, we get
\begin{align}
    E_{Y|\mathbf{X},A=1}\left[f(\mathbf{X},1,Y,\dots,)-\frac{Y}{p} + \frac{1-p}{p}E_{Y|A=0,\mathbf{X}}[ f(\mathbf{X},0,Y,\dots)]\right]=0
\end{align}
holds for any  $Y|\mathbf{X},A=1$ s.t. $E[Y|A=1,\mathbf{X}] > E[Y|A=0,\mathbf{X}]$. By Lemma \ref{lem: linear constraint} we can view $Y|\mathbf{X},A=1$ as $U$,  and since we can choose $E[Y|\mathbf{X},A=0]$ to be arbitrarily small, $E[U]$ can also be arbitrarily small, and thus the conclusion of Lemma \ref{lem: linear constraint} can be extended to the entire real line (i.e., $f(u,v)=0$ for any $u$ and $v\ge 0$), we have
\begin{align*}
    f(\mathbf{X},1,Y,\dots,) = \frac{Y}{p} + c_{1,1}(p,\mu_1(\mathbf{X}),\mu_0(\mathbf{X}),r_1(\mathbf{X}),r_0(\mathbf{X}))
\end{align*}
where $c_{1,1}$ is a function of $p,\mu_1(\mathbf{X}),\mu_0(\mathbf{X}),r_1(\mathbf{X}),r_0(\mathbf{X})$.

Similarly, fix $Y|\mathbf{X},A=1$, we get
\begin{align}
    E_{Y|\mathbf{X},A=0}\left[f(\mathbf{X},0,Y,\dots) + \frac{E_{Y|A=1,\mathbf{X}}[ p f(\mathbf{X},1,Y,\dots)-Y]}{1-p}\right]
\end{align}
holds for any $Y|\mathbf{X},A=0$ s.t.  $E[Y|A=1,\mathbf{X}] > E[Y|A=0,\mathbf{X}]$. By Lemma \ref{lem: linear constraint neg}, viewing $Y|\mathbf{X},A=0$ as $U$, we have
    \begin{align*}
        f(\mathbf{X},0,Y,\dots,) =  c_{1,0}(p,\mu_1(\mathbf{X}),\mu_0(\mathbf{X}),r_1(\mathbf{X}),r_0(\mathbf{X}))
    \end{align*}
where $c_{1,0}$ is a function of $p,\mu_1(\mathbf{X}),\mu_0(\mathbf{X}),r_1(\mathbf{X}),r_0(\mathbf{X})$.

Thus
\begin{align*}
    f(\mathbf{X},A,Y,\dots) = \frac{AY}{p} + A c_{1,1}(p,\mu_1(\mathbf{X}),\mu_0(\mathbf{X}),r_1(\mathbf{X}),r_0(\mathbf{X}))\\+ (1-A) c_{1,0}(p,\mu_1(\mathbf{X}),\mu_0(\mathbf{X}),r_1(\mathbf{X}),r_0(\mathbf{X})).\\
    =\frac{AY}{p} + c_1(A,p,\mu_1(\mathbf{X}),\mu_0(\mathbf{X}),r_1(\mathbf{X}),r_0(\mathbf{X}))
\end{align*}
As for the variance lower bound, we have
    \begin{align}\label{single-var-lb-tau+}
    \begin{aligned}
       & \operatorname{Var}(f|\mathbf{X})=\operatorname{Var}\left(\frac{AY}{p}+c_1(A,p,\mu_1(\mathbf{X}),\mu_0(\mathbf{X}),r_1(\mathbf{X}),r_0(\mathbf{X}))\bigg|\mathbf{X}\right)\\
        =&\operatorname{Var}\left(E\left[\frac{AY}{p}+c_1(A,p,\mu_1(\mathbf{X}),\mu_0(\mathbf{X}),r_1(\mathbf{X}),r_0(\mathbf{X}))\bigg|\mathbf{X},A\right]\bigg|\mathbf{X}\right)\\
        +&E\left(\operatorname{Var}\left[\frac{AY}{p}+c_1(A,p,\mu_1(\mathbf{X}),\mu_0(\mathbf{X}),r_1(\mathbf{X}),r_0(\mathbf{X}))\bigg|\mathbf{X},A\right]\bigg|\mathbf{X}\right)\\
        =&\operatorname{Var}\left(\frac{A}{p}E[Y|\mathbf{X},A]+c_1(A,p,\mu_1(\mathbf{X}),\mu_0(\mathbf{X}),r_1(\mathbf{X}),r_0(\mathbf{X}))\bigg|\mathbf{X}\right)\\
        +& E\left(\frac{A}{p^2}\operatorname{Var}(Y|\mathbf{X},A)\bigg|\mathbf{X}\right)\\
        \ge &  E\left(\frac{A}{p^2}\operatorname{Var}(Y|\mathbf{X},A)\bigg|\mathbf{X}\right) = \frac{\sigma_1^2}{p}
        \end{aligned}
    \end{align}
 Next, we discuss the case when $\tau_P(\mathbf{X})<0$, i.e., $\operatorname{sgn}\left(\tau_P\left(\mathbf{X}\right)\right)=-1$. When the working model for propensity equals the population truth, i.e., $\pi_1(\mathbf{X}) = \pi(1,\mathbf{X})$, based on the identification Assumptions 1 to 3, almost surely we have
\begin{align}\label{single-unbias-eq-tau<0}
   \pi(1,\mathbf{X})E_{Y|A=1,\mathbf{X}}[ f(\mathbf{X},1,Y,\dots)] + \pi(0,\mathbf{X}) E_{Y|A=0,\mathbf{X}}[ f(\mathbf{X},0,Y,\dots)] = E[Y|\mathbf{X},A=0]
\end{align}
holds for any model preserve $E[Y|A=1,\mathbf{X}] < E[Y|A=0,\mathbf{X}]$. Fix $p:= \pi(1,\mathbf{X})$ and the model of $Y|\mathbf{X},A=1$, we get
\begin{align}
    E_{Y|\mathbf{X},A=0}\left[f(\mathbf{X},0,Y,\dots,)-\frac{Y}{1-p} + \frac{p}{1-p}E_{Y|A=1,\mathbf{X}}[ f(\mathbf{X},1,Y,\dots)]\right]=0
\end{align}
holds for any  $Y|\mathbf{X},A=0$ s.t. $E[Y|A=1,\mathbf{X}] < E[Y|A=0,\mathbf{X}]$. By Lemma \ref{lem: linear constraint}, and since we can choose $E[Y|\mathbf{X},A=0]$ to be arbitrarily small,  we have
\begin{align*}
    f(\mathbf{X},0,Y,\dots,) = \frac{Y}{1-p} + c_{0,0}(p,\mu_1(\mathbf{X}),\mu_0(\mathbf{X}),r_1(\mathbf{X}),r_0(\mathbf{X})).
\end{align*}
Similarly, fix $Y|\mathbf{X},A=0$, we get
\begin{align}
    E_{Y|\mathbf{X},A=1}\left[f(\mathbf{X},1,Y,\dots) + \frac{E_{Y|A=0,\mathbf{X}}[ (1-p) f(\mathbf{X},1,Y,\dots)-Y]}{p}\right]
\end{align}
holds for any $Y|\mathbf{X},A=1$ s.t.  $E[Y|A=1,\mathbf{X}] < E[Y|A=0,\mathbf{X}]$. By Lemma \ref{lem: linear constraint neg}, we have
    \begin{align*}
        f(\mathbf{X},1,Y,\dots,) =  c_{0,1}(p,\mu_1(\mathbf{X}),\mu_0(\mathbf{X}),r_1(\mathbf{X}),r_0(\mathbf{X})).
    \end{align*}
Thus
\begin{align*}
    f(\mathbf{X},A,Y,\dots) = \frac{(1-A)Y}{1-p} + (1-A) c_{0,0}(p,\mu_1(\mathbf{X}),\mu_0(\mathbf{X}),r_1(\mathbf{X}),r_0(\mathbf{X}))\\+ A c_{0,1}(p,\mu_1(\mathbf{X}),\mu_0(\mathbf{X}),r_1(\mathbf{X}),r_0(\mathbf{X})).\\
    =\frac{(1-A)Y}{1-p} + c_0(A,p,\mu_1(\mathbf{X}),\mu_0(\mathbf{X}),r_1(\mathbf{X}),r_0(\mathbf{X}))
\end{align*}
As for the variance lower bound, we have
    \begin{align}\label{single-var-lb-tau-}
    \begin{aligned}
       & \operatorname{Var}(f|\mathbf{X})=\operatorname{Var}\left(\frac{(1-A)Y}{1-p}+c_0(A,p,\mu_1(\mathbf{X}),\mu_0(\mathbf{X}),r_1(\mathbf{X}),r_0(\mathbf{X}))\bigg|\mathbf{X}\right)\\
        =&\operatorname{Var}\left(E\left[\frac{(1-A)Y}{1-p}+c_0(A,p,\mu_1(\mathbf{X}),\mu_0(\mathbf{X}),r_1(\mathbf{X}),r_0(\mathbf{X}))\bigg|\mathbf{X},A\right]\bigg|\mathbf{X}\right)\\
        +&E\left(\operatorname{Var}\left[\frac{(1-A)Y}{1-p}+c_0(A,p,\mu_1(\mathbf{X}),\mu_0(\mathbf{X}),r_1(\mathbf{X}),r_0(\mathbf{X}))\bigg|\mathbf{X},A\right]\bigg|\mathbf{X}\right)\\
        =&\operatorname{Var}\left(\frac{1-A}{1-p}E[Y|\mathbf{X},A]+c_0(A,p,\mu_1(\mathbf{X}),\mu_0(\mathbf{X}),r_1(\mathbf{X}),r_0(\mathbf{X}))\bigg|\mathbf{X}\right)\\
        +& E\left(\frac{1-A}{(1-p)^2}\operatorname{Var}(Y|\mathbf{X},A)\bigg|\mathbf{X}\right)\\
        =& (1-p)\left(\frac{E[Y|\mathbf{X},A=0]}{1-p}+c_0(0,p,\mu_1(\mathbf{X}),\mu_0(\mathbf{X}),r_1(\mathbf{X}),r_0(\mathbf{X}))-E[Y|\mathbf{X},A=0]\right)^2\\
        +&p\left(c_1(1,p,\mu_1(\mathbf{X}),\mu_0(\mathbf{X}),r_1(\mathbf{X}),r_0(\mathbf{X}))-E[Y|\mathbf{X},A=0]\right)^2+\frac{\sigma_0^2}{1-p}\\
        \ge & \frac{\sigma_0^2}{1-p}
        \end{aligned}
    \end{align}   
Therefore, to sum up, the variance lower bound of $f$ that we derive can be summarized as follows
\begin{align*}
    &\operatorname{Var}(f)=\operatorname{Var}\left[E\left(f\mid \mathbf{X}\right)\right]+E\left[\operatorname{Var}\left(f\mid \mathbf{X}\right)\right] \\
    =&\operatorname{Var}[\max\{E[Y|\mathbf{X},A=1],E[Y|\mathbf{X},A=0]\}]+E\left[(\mathbb{I}\{\tau(\mathbf{X})=0\}+\mathbb{I}\{\tau(\mathbf{X})\neq0\})\operatorname{Var}\left(f\mid \mathbf{X}\right)\right]\\
    \ge&\operatorname{Var}[\max\{E[Y|\mathbf{X},A=1],E[Y|\mathbf{X},A=0]\}]\\
    +&E\left[\mathbb{I}\{\tau(\mathbf{X})=0\}\underbrace{\frac{\pi(1,\mathbf{X})\operatorname{Var}(Y|\mathbf{X},A=1)[\operatorname{Var}(Y|\mathbf{X},A=0)]^2}{[\operatorname{Var}(Y|\mathbf{X},A=1)\pi(0,\mathbf{X})+\operatorname{Var}(Y|\mathbf{X},A=0)\pi(1,\mathbf{X})]^2}}_{\text{from Eq. \eqref{single-var-lb-tau0}}}\right]\\
    +&E\left[\mathbb{I}\{\tau(\mathbf{X})=0\}\underbrace{\frac{\pi(0,\mathbf{X})\operatorname{Var}(Y|\mathbf{X},A=0)[\operatorname{Var}(Y|\mathbf{X},A=1)]^2}{[\operatorname{Var}(Y|\mathbf{X},A=1)\pi(0,\mathbf{X})+\operatorname{Var}(Y|\mathbf{X},A=0)\pi(1,\mathbf{X})]^2}}_{\text{from Eq. \eqref{single-var-lb-tau0}}}\right]\\
    +&E\left[\mathbb{I}\{\tau(\mathbf{X})>0\}\underbrace{\frac{\operatorname{Var}(Y|\mathbf{X},A=1)}{\pi(1,\mathbf{X})}}_{\text{from Eq. \eqref{single-var-lb-tau+}}}+\mathbb{I}\{\tau(\mathbf{X})<0\}\underbrace{\frac{\operatorname{Var}(Y|\mathbf{X},A=0)}{\pi(0,\mathbf{X})}}_{\text{from Eq. \eqref{single-var-lb-tau-}}}\right],
\end{align*}
and the variance lower bound can be reached by $\widehat{V}_{\mathcal{D}}^*$ s.t. 
    \begin{align*}
        \mathcal{D}(\mathbf{X})=\mathbb{I}\{\tau(\mathbf{X})>0\}+  \frac{\operatorname{Var}(Y|\mathbf{X},A=0)\pi(1,\mathbf{X})}{\operatorname{Var}(Y|\mathbf{X},A=1)\pi(0,\mathbf{X})+\operatorname{Var}(Y|\mathbf{X},A=0)\pi(1,\mathbf{X})}\mathbb{I}\{\tau(\mathbf{X})=0\}.
    \end{align*}
We check that the equality of Eqs. \eqref{single-var-lb-tau0} \eqref{single-var-lb-tau+} and \eqref{single-var-lb-tau-} can be caused by $f=\psi(\mathcal{D}(\mathbf{X}),\pi,\mu)$ where
\begin{align*}
        \mathcal{D}(\mathbf{X})=\mathbb{I}\{\tau(\mathbf{X})>0\}+  \frac{\operatorname{Var}(Y|\mathbf{X},A=0)\pi(1,\mathbf{X})}{\operatorname{Var}(Y|\mathbf{X},A=1)\pi(0,\mathbf{X})+\operatorname{Var}(Y|\mathbf{X},A=0)\pi(1,\mathbf{X})}\mathbb{I}\{\tau(\mathbf{X})=0\}.
    \end{align*}
First, for Eq. \eqref{single-var-lb-tau0}, which means when $\tau(\mathbf{X})=0$, $r_a(\mathbf{X}) = \operatorname{Var}(Y|\mathbf{X},A=a)$, we have
\begin{align*}
      &b_1(\pi(1,\mathbf{X}),\mu_1(\mathbf{X}),\mu_0(\mathbf{X}),r_1(\mathbf{X}),r_0(\mathbf{X})) \\=&  \frac{\operatorname{Var}(Y|\mathbf{X},A=0)}{\operatorname{Var}(Y|\mathbf{X},A=1)\pi(0,\mathbf{X})+\operatorname{Var}(Y|\mathbf{X},A=0)\pi(1,\mathbf{X})},\\
        &b_0(\pi(1,\mathbf{X}),\mu_1(\mathbf{X}),\mu_0(\mathbf{X}),r_1(\mathbf{X}),r_0(\mathbf{X}))\\ = & \frac{\operatorname{Var}(Y|\mathbf{X},A=1)}{\operatorname{Var}(Y|\mathbf{X},A=1)\pi(0,\mathbf{X})+\operatorname{Var}(Y|\mathbf{X},A=0)\pi(1,\mathbf{X})},\\
        &f(\mathbf{X},A,0,\dots)\\
        =&\frac{A \mathcal{D}(\mathbf{X})+(1-A)(1-\mathcal{D}(\mathbf{X}))}{\pi(A,\mathbf{X})}[0-\mu(A,\mathbf{X})]+\mathcal{D}(\mathbf{X})\mu(1,\mathbf{X})+(1-\mathcal{D}(\mathbf{X}))\mu(0,\mathbf{X})\\
        =&\mathcal{D}(\mathbf{X})\mu(1,\mathbf{X})\left(1-\frac{A}{\pi(1,\mathbf{X})}\right)+(1-\mathcal{D}(\mathbf{X}))\mu(0,\mathbf{X})\left(1-\frac{1-A}{\pi(0,\mathbf{X})}\right)\\
        &\text{(Recall that $\mu_0^* = \mu(1,\mathbf{X}) = \mu(0,\mathbf{X})$)}\\
        =&\mu_0^* - \mu_0^*[Ab_0(\pi(1,\mathbf{X}),\mu_1(\mathbf{X}),\mu_0(\mathbf{X}),r_1(\mathbf{X}),r_0(\mathbf{X}))\\
        &-(1-A)b_0(\pi(1,\mathbf{X}),\mu_1(\mathbf{X}),\mu_0(\mathbf{X}),r_1(\mathbf{X}),r_0(\mathbf{X}))],
\end{align*}
which means, 
\begin{align*}
    &\operatorname{Var}\left([Ab_1(p,\mu_1(\mathbf{X}),\mu_0(\mathbf{X}),r_1(\mathbf{X}),r_0(\mathbf{X})) + (1-A)b_0(p,\mu_1(\mathbf{X}),\mu_0(\mathbf{X}),r_1(\mathbf{X}),r_0(\mathbf{X}))]\mu_0^*\right.\\&+\left.f(\mathbf{X},A,0,\dots)\bigg|\mathbf{X}\right)=\operatorname{Var}(\mu_0^*\mid\mathbf{X})=0
\end{align*}
and the final equality in Eq. \eqref{single-var-lb-tau0} holds.

Then, for Eq. \eqref{single-var-lb-tau+}, that is when $\tau(\mathbf{X})>0$,  we  have
\begin{align*}
    &f = \psi(\mathcal{D}(\mathbf{X}),\pi,\mu)\\
    =&\frac{A}{\pi(1,\mathbf{X})}(Y - \mu(1,\mathbf{X})) + \mu(1,\mathbf{X}).
\end{align*}
In other words, when $\mu_a$ is correctly specified, i.e., $\mu_a(\mathbf{X}) = \mu(a,\mathbf{X})$,
\begin{align*}
    c_1(A,p,\mu_1(\mathbf{X}),\mu_0(\mathbf{X}),r_1(\mathbf{X}),r_0(\mathbf{X})) = \left(1-\frac{A}{\pi(1,\mathbf{X})}\right)\mu(1,\mathbf{X})
\end{align*}
and thus
\begin{align*}
    \operatorname{Var}\left(\frac{A}{p}E[Y|\mathbf{X},A]+c_1(A,p,\mu_1(\mathbf{X}),\mu_0(\mathbf{X}),r_1(\mathbf{X}),r_0(\mathbf{X}))\bigg|\mathbf{X}\right) = \operatorname{Var}(\mu(1,\mathbf{X})\mid \mathbf{X}) = 0.
\end{align*}
Finally, for Eq. \eqref{single-var-lb-tau-}, that is when $\tau(\mathbf{X})<0$,  we  have
\begin{align*}
    &f = \psi(\mathcal{D}(\mathbf{X}),\pi,\mu)\\
    =&\frac{1-A}{\pi(0,\mathbf{X})}(Y - \mu(0,\mathbf{X})) + \mu(0,\mathbf{X}).
\end{align*}
In other words, when $\mu_a(\mathbf{X})$ is correctly specified, i.e., $\mu_a(\mathbf{X}) = \mu(a,\mathbf{X})$,
\begin{align*}
    c_0(A,p,\mu_1(\mathbf{X}),\mu_0(\mathbf{X}),r_1(\mathbf{X}),r_0(\mathbf{X})) = \left(1-\frac{1-A}{\pi(0,\mathbf{X})}\right)\mu(0,\mathbf{X})
\end{align*}
and thus
\begin{align*}
    \operatorname{Var}\left(\frac{1-A}{1-p}E[Y|\mathbf{X},A]+c_0(A,p,\mu_1(\mathbf{X}),\mu_0(\mathbf{X}),r_1(\mathbf{X}),r_0(\mathbf{X}))\bigg|\mathbf{X}\right) = \operatorname{Var}(\mu(0,\mathbf{X})\mid \mathbf{X}) = 0.
\end{align*}
Therefore, we have shown that
\begin{align*}
    &\inf_{T_n\in\mathcal{T}_{RL}}\operatorname{Var}(f) 
    =\operatorname{Var}\left[\psi(\mathcal{D},\pi,\mu)\right]
\end{align*}
where $\mathcal{D}$ is defined in Eq. \eqref{single-optimal-decision-heter}. When $T_n$ has asymptotic uniform integrability, i.e., 
    \begin{align}\label{AUI}
        \lim _{M \rightarrow \infty} \sup _n E\left[n\left(T_n-V_0\right)^2 \cdot \mathbf{1}\left(n\left(T_n-V_0\right)^2>M\right)\right]=0,
    \end{align}
we have
\begin{align*}
    \inf_{T_n\in\mathcal{T}_{RL}}\lim_{n\to\infty}\operatorname{Var}(\sqrt{n}T_n)= \inf_{T_n\in\mathcal{T}_{RL}}\operatorname{Var}(f) 
\end{align*}
and since $\widehat{V}_{\mathcal{D}}^{*}\in\mathcal{T}_{RL}$,
\begin{align*}
    \operatorname{Var}\left[\psi(\mathcal{D},\pi,\mu)\right] = \lim_{n\to\infty}\operatorname{Var}(\sqrt{n}\widehat{V}_{\mathcal{D}}^{*})
\end{align*}
where $\mathcal{D}$ is defined in Eq. \eqref{single-optimal-decision-heter}. Hence we have shown the general case under heteroscedasticity.

In the case of homoscedasticity, it just reduces to the optimal adaptive smoothing estimator we proposed, i.e., $\widehat{V}_{\mathcal{D}}^*$ with $\mathcal{D}(\mathbf{X})=\mathbb{I}\{\tau(\mathbf{X})>0\}+\pi(1,\mathbf{X})\mathbb{I}\{\tau(\mathbf{X})=0\}$. 
\end{proof}
Then, in terms of the data driven implementation of achieving the above stated optimal efficiency, under the same choice $h_{n, \mathcal{I}_j}$'s and and working models of $h$ and $\pi$, replacing the $t_0$ in $\widehat{V}_s$ with any consistent estimator of 
\begin{align}\label{single-heter-t0}
    \frac{\operatorname{Var}(Y|\mathbf{X},A=0)\pi(1,\mathbf{X})}{\operatorname{Var}(Y|\mathbf{X},A=1)\pi(0,\mathbf{X})+\operatorname{Var}(Y|\mathbf{X},A=0)\pi(1,\mathbf{X})}
\end{align}
would lead to an estimator which is asymptotically equivalent with $\widehat{V}_{\mathcal{D}}$ where
\begin{align*}
    \mathcal{D}(\mathbf{X})=\mathbb{I}\{\tau(\mathbf{X})>0\}+\frac{\operatorname{Var}(Y|\mathbf{X},A=0)\pi(1,\mathbf{X})}{\operatorname{Var}(Y|\mathbf{X},A=1)\pi(0,\mathbf{X})+\operatorname{Var}(Y|\mathbf{X},A=0)\pi(1,\mathbf{X})}\mathbb{I}\{\tau(\mathbf{X})=0\},
\end{align*}
and hence the efficiency bound of $\mathcal{T}_{RL}$ can be achieved by a practically meaningful estimator in the case of heteroscedasticity. Here we emphasize that the rate requirement of estimating Eq. \eqref{single-heter-t0} is so mild that any consistent estimator would be satisfactory, which is exactly the meaning of Assumption \ref{A11}. Since we can already consistently estimate $\pi$ by Assumption 4, it suffice to obtain a consistent estimator for $\operatorname{Var}(Y|\mathbf{X},A)$ to achieve the efficiency optimality as described in \ref{var-const}. There exists several model free nonparametric estimators for such individualized variance, see an overview at \citet{chu2020standard}. Moreover, the additional condition of correct specification of conditional variance is required only for efficiency optimality under heteroscedasticity. Even if it is violated, inference validity under heteroscedasticity is still preserved, as the asymmetric extent $t_0$ does not influence the validity. In the heteroscedasticity case, the general validity of the proposed estimator allowing misspecification of conditional variances can be implied by Lemma \ref{single-lem}.  We summarize the algorithm as follows. 
\begin{algorithm}[H]
\caption{The Optimal Estimator Allowing Heteroscedasticity}
    \begin{algorithmic}[1]
        \State \textbf{Input}: Data set $\mathcal{O}_n$; Randomly picked subsets of $[n]$: $\mathcal{I}_1$ and $\mathcal{I}_2$, and randomly picked subsets $\mathcal{I}_{j,k}$ ($k=1,2$) of $\mathcal{I}_j$ ($j=1,2$), where all the $\mathcal{I}_{j,k}$ with the equal size; Parameters $h_{n,\mathcal{I}_j}$, $j=1,2$; Working models: $\widehat{\pi}$, $\widehat{\mu}$ and $\widehat{\operatorname{Var}}(Y|\mathbf{X},A)$.
        \State \textbf{Construction of 
 The Optimal Estimator Allowing Heteroscedasticity:} 
 \begin{align}
     \widehat{V}_{as,heter}&:= \frac{1}{4}\sum_{j=1,2}\sum_{k=1,2}\frac{1}{|\mathcal{I}_{j,k}|}\sum_{i\in\mathcal{I}_{j,k}}\psi_i\left(d_{s}\left(\mathbf{X}_i;\widehat{\tau}_{\mathcal{I}_{3-j}},h_{n,\mathcal{I}_{3-j}},\right.\right.\notag\\&\left.\left.  \frac{\widehat{\operatorname{Var}}_{\mathcal{I}_{3-j}}(Y|\mathbf{X},0)\widehat{\pi}_{\mathcal{I}_{3-j}}(1,\mathbf{X}_i)}{\widehat{\operatorname{Var}}_{\mathcal{I}_{3-j}}(Y|\mathbf{X},1)\widehat{\pi}_{\mathcal{I}_{3-j}}(0,\mathbf{X}_i)+\widehat{\operatorname{Var}}_{\mathcal{I}_{3-j}}(Y|\mathbf{X},0)\widehat{\pi}_{\mathcal{I}_{3-j}}(1,\mathbf{X}_i)}\right),\widehat{\pi}_{\mathcal{I}_{j,k}^c},\widehat{\mu}_{\mathcal{I}_{j,k}^c}\right);\label{sin-opt-est-heter}
 \end{align}
 \State \textbf{Construction of The Confidence Interval} The standard deviation estimate:
 \begin{align}\label{sin-opt-std-heter}
    &\widehat{\sigma}_{as,heter}^2:=\frac{1}{n-1}\sum_{j=1,2}\sum_{k=1,2}\sum_{i\in\mathcal{I}_{j,k}}\left(\psi_i\left(d_{s}\left(\mathbf{X}_i;\widehat{\tau}_{\mathcal{I}_{3-j}},h_{n,\mathcal{I}_{3-j}},\right.\right.\right.\notag\\&\left.\left.\left.\frac{\widehat{\operatorname{Var}}_{\mathcal{I}_{3-j}}(Y|\mathbf{X},0)\widehat{\pi}_{\mathcal{I}_{3-j}}(1,\mathbf{X}_i)}{\widehat{\operatorname{Var}}_{\mathcal{I}_{3-j}}(Y|\mathbf{X},1)\widehat{\pi}_{\mathcal{I}_{3-j}}(0,\mathbf{X}_i)+\widehat{\operatorname{Var}}_{\mathcal{I}_{3-j}}(Y|\mathbf{X},0)\widehat{\pi}_{\mathcal{I}_{3-j}}(1,\mathbf{X}_i)}\right),\widehat{\pi}_{\mathcal{I}_{j,k}^c},\widehat{\mu}_{\mathcal{I}_{j,k}^c}\right)-\widehat{V}_{as}\right)^2
 \end{align}
 and thus the two-sided level-{$\eta$} confidence interval can be given as:
 \begin{align*}
     \left[\widehat{V}_{as,heter}-{z_{\eta/2}}\cdot\frac{\widehat{\sigma}_{as,heter}}{\sqrt{n}},\widehat{V}_{as,heter}+{z_{\eta/2}}\cdot\frac{\widehat{\sigma}_{as,heter}}{\sqrt{n}}\right].
 \end{align*}
    \end{algorithmic}
    \label{alg-single-heter}
\end{algorithm}
The theoretical property of $\widehat{V}_{as,heter}$ is summarized in Theorem \ref{thm: single-opt-est-heter}.
\begin{theorem}\label{thm: single-opt-est-heter}
    Under Assumptions from 1 to 7, and if the working estimator for conditional variance satisfies
    \begin{align}\label{var-const}
        \left|\widehat{\operatorname{Var}}(Y|\mathbf{X},A) - \operatorname{Var}(Y|\mathbf{X},A)\right|\to_p0
    \end{align}
   almost surely for $\mathbf{X}$ and $A$, we have
    \begin{align}
        \widehat{V}_{as,heter} - \widehat{V}_{\mathcal{D}}^*=o_p(n^{-1/2})
    \end{align}
    where 
    \begin{align}
    \mathcal{D}(\mathbf{X})=\mathbb{I}\{\tau(\mathbf{X})>0\}+\frac{\operatorname{Var}(Y|\mathbf{X},A=0)\pi(1,\mathbf{X})}{\operatorname{Var}(Y|\mathbf{X},A=1)\pi(0,\mathbf{X})+\operatorname{Var}(Y|\mathbf{X},A=0)\pi(1,\mathbf{X})}\mathbb{I}\{\tau(\mathbf{X})=0\}.
\end{align}
\end{theorem}
\begin{proof}[Proof of Theorem \ref{thm: single-opt-est-heter}]
    We verify that the decision function $\mathcal{D}_n$ 
    with $t_0$ replaced by the estimated variance-weighted 
    propensity score satisfies Assumptions~\ref{A10} 
    to~\ref{A12} (in probability) by Assumptions~6, 7 and 
    the variance consistency condition 
    Eq.~\eqref{var-const}, so that 
    Lemma~\ref{single-lem} can be applied.

    Let 
    \begin{align*}
        \widehat{t}_0(\mathbf{X})
        :=\frac{\widehat{\operatorname{Var}}_{\mathcal{I}_{j}^c}
        (Y|\mathbf{X},A\!=\!0)\,
        \widehat{\pi}_{\mathcal{I}_{j}^c}(1,\mathbf{X})}
        {\widehat{\operatorname{Var}}_{\mathcal{I}_{j}^c}
        (Y|\mathbf{X},A\!=\!1)\,
        \widehat{\pi}_{\mathcal{I}_{j}^c}(0,\mathbf{X})
        +\widehat{\operatorname{Var}}_{\mathcal{I}_{j}^c}
        (Y|\mathbf{X},A\!=\!0)\,
        \widehat{\pi}_{\mathcal{I}_{j}^c}(1,\mathbf{X})}
    \end{align*}
    and define 
    $\mathcal{D}_n(\mathbf{X}_i,s)
    =d_{s}(\mathbf{X}_i;s,h_{n,\mathcal{I}_{j(i)}^c},
    \widehat{t}_0(\mathbf{X}_i))$, where $j(i)$ is 
    the index of the subset containing the $i$-th sample. 
    Define {$\delta_{n,j}$ for each $j$} and $\delta_n'$:
    \begin{align*}
        {\delta_{n,j}}
        &=\sqrt{h_{n,\mathcal{I}_{j}^c}\cdot n^{1/4}
        \left(E\!\left[|\widehat{\tau}_{\mathcal{I}_{j}^c}
        (\mathbf{X})-\tau(\mathbf{X})|^2\right]\right)^{3/4}},
        \qquad
        \delta_n'=2\max_j h_{n,\mathcal{I}_{j}^c}.
    \end{align*}
    Denote the population-level counterpart of 
    $\widehat{t}_0$ by
    \begin{align*}
        t_0^*(\mathbf{X})
        :=\frac{\operatorname{Var}(Y|\mathbf{X},A\!=\!0)\,
        \pi(1,\mathbf{X})}
        {\operatorname{Var}(Y|\mathbf{X},A\!=\!1)\,
        \pi(0,\mathbf{X})
        +\operatorname{Var}(Y|\mathbf{X},A\!=\!0)\,
        \pi(1,\mathbf{X})}.
    \end{align*}
    By the overlap Assumption~3, 
    $\pi(1,\mathbf{X})\in(c,1-c)$, so 
    $t_0^*(\mathbf{X})\in(0,1)$ almost surely. 
    Moreover, by the consistency of $\widehat{\pi}$ in 
    Assumption~4 and the variance consistency 
    Eq.~\eqref{var-const}, we have 
    $\widehat{t}_0(\mathbf{X})\to_p t_0^*(\mathbf{X})$ 
    almost surely for $\mathbf{X}$.

    We first verify Assumption~\ref{A11} (Consistency). 
    By $h_{n,\mathcal{I}_j^c}\to 0$ from Assumption~7 
    and $\widehat{t}_0(\mathbf{X})\in(0,1)$ with 
    probability tending to~1: if $\tau(\mathbf{X})>0$, 
    then for all sufficiently large $n$, 
    $\tau(\mathbf{X})
    \ge(1-\widehat{t}_0(\mathbf{X}))h_{n,\mathcal{I}_j^c}$ 
    since $(1-\widehat{t}_0(\mathbf{X}))$ is bounded 
    and $h_{n,\mathcal{I}_j^c}\to 0$, so 
    $\mathcal{D}_n(\mathbf{X},\tau(\mathbf{X}))=1$; 
    if $\tau(\mathbf{X})<0$, then 
    $\tau(\mathbf{X})
    <-\widehat{t}_0(\mathbf{X})h_{n,\mathcal{I}_j^c}$ 
    for large $n$ since $\widehat{t}_0(\mathbf{X})>0$ 
    w.p.t.1, so 
    $\mathcal{D}_n(\mathbf{X},\tau(\mathbf{X}))=0$; 
    if $\tau(\mathbf{X})=0$, then 
    $\mathcal{D}_n(\mathbf{X},\tau(\mathbf{X}))
    =\widehat{t}_0(\mathbf{X})
    \to_p t_0^*(\mathbf{X})$. 
    Therefore, 
    $\mathcal{D}(\mathbf{X})
    :=\lim_{n\to\infty}
    \mathcal{D}_n(\mathbf{X},\tau(\mathbf{X}))
    =\mathbb{I}\{\tau(\mathbf{X})>0\}
    +t_0^*(\mathbf{X})\cdot
    \mathbb{I}\{\tau(\mathbf{X})=0\}$. 
    Since $\tau(\mathbf{X})\cdot\mathcal{D}(\mathbf{X})
    =\max\{0,\tau(\mathbf{X})\}$ a.s., we have 
    $E[\psi(\mathcal{D}(\mathbf{X}),\pi,\mu)]=V_0$, 
    so $\mathcal{D}\in\mathbb{D}(V_0)$ and 
    Assumption~\ref{A11} is satisfied.

    We next verify Assumption~\ref{A10} 
    (Uniform continuity). The piecewise linear 
    structure of $d_s$ is unchanged; for any fixed 
    $\mathbf{X}$, the Lipschitz constant of 
    $s\mapsto d_s(\mathbf{X};s,h_{n,\mathcal{I}_j^c},
    \widehat{t}_0(\mathbf{X}))$ is at most 
    $1/h_{n,\mathcal{I}_j^c}$, regardless of the 
    value of $\widehat{t}_0(\mathbf{X})\in(0,1)$. 
    Therefore, by the identical argument as in the 
    proof of Theorem~1,
    \begin{align*}
        \sup_{\mathbf{X}\in\mathcal{X}}
        |\mathcal{D}_n(\mathbf{X},s)
        -\mathcal{D}_n(\mathbf{X},t)|
        \leq \frac{{\delta_{n,j}}}{h_{n,\mathcal{I}_j^c}}
        =o_p(1)
        \quad\text{for }|s-t|\leq{\delta_{n,j}}
    \end{align*}
    by Assumption~6, and 
    $P_{\mathbf{X}}(|\widehat{\tau}_{\mathcal{I}_j^c}
    (\mathbf{X})-\tau(\mathbf{X})|\ge{\delta_{n,j}})=o_p(1)$ 
    by Chebyshev's inequality and Assumption~6. Hence 
    Assumption~\ref{A10} is satisfied.

    Finally, we verify Assumption~\ref{A12} 
    (Concentration). The rate condition on $\delta_n'$ 
    follows identically to the proof of Theorem~1: 
    ${\delta_{n,j}}\lor n^{1/4}(E[|\widehat{\tau}_{{\mathcal{I}_j^c}}
    (\mathbf{X})-\tau(\mathbf{X})|^2])^{3/4}
    \prec_p\delta_n'$ by Assumption~6 and 
    $\delta_n'\prec n^{-\frac{1}{2(1+\alpha)}}$ by 
    Assumption~7. For the approximation condition 
    Eq.~\eqref{A10-1}, when 
    $|s|>\delta_n'=2\max_j h_{n,\mathcal{I}_j^c}$, 
    we have $|s|>2h_{n,\mathcal{I}_j^c}$ for all $j$. 
    Since $\widehat{t}_0(\mathbf{X})\in(0,1)$ with 
    probability tending to~1, both 
    $(1-\widehat{t}_0(\mathbf{X}))
    h_{n,\mathcal{I}_j^c}
    <h_{n,\mathcal{I}_j^c}<|s|/2$ and 
    $\widehat{t}_0(\mathbf{X})
    h_{n,\mathcal{I}_j^c}
    <h_{n,\mathcal{I}_j^c}<|s|/2$. Therefore, if 
    $s>\delta_n'$, then 
    $s>(1-\widehat{t}_0(\mathbf{X}))
    h_{n,\mathcal{I}_j^c}$, so 
    $\mathcal{D}_n(\mathbf{X},s)=1
    =\mathbb{I}\{s>0\}$; if $s<-\delta_n'$, then 
    $s<-\widehat{t}_0(\mathbf{X})
    h_{n,\mathcal{I}_j^c}$, so 
    $\mathcal{D}_n(\mathbf{X},s)=0
    =\mathbb{I}\{s>0\}$. In both cases, 
    $|\mathcal{D}_n(\mathbf{X},s)
    -\mathbb{I}\{s>0\}|=0=o(1/({|s|}\sqrt{n}))$, 
    so Eq.~\eqref{A10-1} holds trivially with 
    probability tending to~1. Note that unlike the 
    proof of Theorem~2 where 
    {$\widehat{\pi}(1,\mathbf{X})\in(c_{\mathrm f},1-c_{\mathrm f})$ follows from the fitted-overlap condition above}, here 
    $\widehat{t}_0(\mathbf{X})\in(0,1)$ holds only 
    with probability tending to~1 via the consistency 
    of $\widehat{\pi}$ and 
    $\widehat{\operatorname{Var}}$, but this suffices 
    for all the above arguments to hold in probability. 
    Hence Assumption~\ref{A12} is satisfied in 
    probability.

    Since Assumptions~\ref{A10} to~\ref{A12} are 
    verified (in probability) and Assumptions~1 to~7 
    from the main text are assumed, all conditions of 
    Lemma~\ref{single-lem} are satisfied. Therefore, by 
    Lemma~\ref{single-lem}, 
    $|\widehat{V}_{as,heter}
    -\widehat{V}_{\mathcal{D}}^*|
    =o_p(n^{-1/2})$ where 
    $\mathcal{D}(\mathbf{X})
    =\mathbb{I}\{\tau(\mathbf{X})>0\}
    +t_0^*(\mathbf{X})\cdot
    \mathbb{I}\{\tau(\mathbf{X})=0\}$, and 
    Theorem~\ref{thm: single-opt-est-heter} holds.
\end{proof}
To provide more numerical evidence, we present seven additional simulation settings in the single time point case that
examine the proposed method across different combinations of
heteroscedasticity, $\alpha$, and covariate dimension. All seven settings,
summarized in Table~\ref{heter-setting}, share the data generating model
$$
Y = \Phi(\mathbf{X}) + A\,\tau(X_1, X_2) + e,
$$
where the random error $e$ is independent of $(A, \mathbf{X})$ but its
variance may depend on $A$ and $\mathbf{X}$, introducing
heteroscedasticity. In all settings, the treatment effect $\tau$ depends
only on $(X_1, X_2)$ with $P(\tau(X_1, X_2) = 0) = P(X_1 = 0) > 0$,
so nonregularity is present.

Settings~(I) and~(II) feature heteroscedasticity with $\alpha < 1$ and
$X_2 \sim \operatorname{Uniform}[-2,2]$. In Setting~(I),
heteroscedasticity arises in the nonregular subgroup: for patients with
$X_1 = 0$ where $\tau = 0$, the conditional variance under treatment
 is three times that under control, while for patients with $X_1 = 1$ where
$\tau > 0$, the error variances are equal. In Setting~(II),
heteroscedasticity enters through the continuous covariate~$X_2$:
$\operatorname{Var}(Y \mid \mathbf{X}, A=1) = 0.2 + 0.3|X_2|$ increases
with $|X_2|$, while
$\operatorname{Var}(Y \mid \mathbf{X}, A=0) = 0.2$ remains constant.

Setting~(III) combines heteroscedasticity with discrete covariates and
$\alpha = \infty$. Its structure is similar to the main text Setting~A,
but the variance under control is over five
times that under treatment. Because
$\tau(x_1,x_2) = 0.4\,x_1$ takes values in $\{0, 0.4\}$, the treatment
effect is either exactly zero or bounded away from zero, yielding
$\alpha = \infty$.

Settings~(IV) and~(V) share the treatment effect
$\tau(x_1, x_2) = 0.5\,x_1\,x_2$ with
$X_2 \sim \operatorname{Beta}(0.5, 2)$, which induces $\alpha = 1/2$ due
to the unbounded density of the Beta distribution near zero. Setting~(IV)
is homoscedastic with
$\operatorname{Var}(Y \mid \mathbf{X}, A=a) = 0.25$ for both treatment
groups, while Setting~(V) introduces heteroscedasticity.

Settings~(VI) and~(VII) extend Settings~(I) and~(II) to a higher-dimensional covariate space with $p = 5$. Three additional covariates $X_3 \sim N(0,1)$, $X_4 \sim \operatorname{Uniform}(0,1)$, and $X_5 \sim \operatorname{Bernoulli}(0.5)$ enter the baseline outcome $\Phi$ and the propensity score $\pi$, but do not affect the treatment effect $\tau$, which remains a function of $(X_1, X_2)$ only. The margin exponent, heteroscedasticity structure, and nonregularity are identical to those in Settings~(I) and~(II), respectively. These settings examine whether the proposed method remains effective when the nuisance functions involve a larger number of covariates.

Across the seven settings, the margin exponent in Assumption~5 takes three
distinct values: $\alpha = 1/2$ in Settings~(I), (IV), (V), and~(VI),
$\alpha = 2/3$ in Settings~(II) and~(VII), and $\alpha = \infty$ in Setting~(III).
All settings with $\alpha < \infty$ correspond to the more challenging
case where the density of $|\tau(\mathbf{X})|$ near zero is unbounded, as
the functions $x_2^2/4$ and $|x_2|^{3/2}/4$ vanish at $x_2 = 0$ in
Settings~(I), (II), (VI), and~(VII), and $0.5\,x_2$ with
$x_2 \sim \operatorname{Beta}(0.5,2)$ concentrates mass near zero in
Settings~(IV) and~(V). This contrasts with the main text simulations
where $\alpha \geq 1$.

For each setting, we let the sample size $n \in \{500, 1000, 2000\}$. We compare the following four methods:
\begin{itemize}
    \item Proposed (Heter): adaptive smoothing with the efficiency-optimal
    asymmetric tuning under heteroscedasticity as derived in Section~D of
    the supplementary material, where $t_0$ is set to
    $\widehat{r}_0(\mathbf{X})\widehat{\pi}(1,\mathbf{X}) /
    \{\widehat{r}_1(\mathbf{X})(1-\widehat{\pi}(1,\mathbf{X})) +
    \widehat{r}_0(\mathbf{X})\widehat{\pi}(1,\mathbf{X})\}$ with
    $\widehat{r}_a(\mathbf{X})$ denoting the estimated conditional
    variance $\widehat{\operatorname{Var}}(Y|\mathbf{X},A=a)$.
    \item Proposed (Homo): adaptive smoothing with the efficiency-optimal
    asymmetric tuning under homoscedasticity as in Algorithm~1, where
    $t_0 = \widehat{\pi}(1,\mathbf{X})$.
    \item Subbagging: confidence interval of subbagging in
    \citet{shi2020breaking}.
    \item SSS: confidence interval by simply splitting the sample into two
    halves, using one half to estimate the decision function, propensity
    score, and conditional mean functions, and the other half to evaluate
    the plug-in estimator.
\end{itemize}
We
set ${C_0} = 0.05$ for the tuning selection in Algorithm~2 and subsample size
$K_0 = 5$ in subbagging. All results are aggregated over 500 Monte Carlo
replications.

In Settings~(I), (II), (VI) and~(VII), since $X_1$ is discrete, we stratify the outcome model by $X_1$ and estimate the conditional mean function $\mu(a, \mathbf{X})$ by ordinary least squares (OLS) with correctly specified basis functions of $X_2$ (e.g., $X_2^2$ in Settings~(I) and~(VI), and $|X_2|^{3/2}$ in Settings~(II) and~(VII)) within each $(X_1, A)$ stratum. In Settings~(VI) and~(VII), the additional covariates $X_3$, $X_4$ and $X_5$ enter the outcome model through their correctly specified linear terms. The propensity score is estimated by OLS regression of $A$ on the relevant covariates. In Setting~(III), where both covariates are discrete, the propensity score and the conditional mean functions are estimated by the nonparametric maximum likelihood estimator, i.e., sample means within strata, the same approach as in the main text Settings~(A) and~(B). In Settings~(IV) and~(V), where $X_2\sim\operatorname{Beta}(0.5,2)$ is continuous, the conditional mean functions are estimated using natural cubic splines with 4 degrees of freedom for $X_2$, interacted with the discrete covariate $X_1$. In all settings, the propensity score is truncated to $[0.05,0.95]$, the contrast function is estimated as $\widehat{\tau}(\mathbf{X})=\widehat{\mu}(1,\mathbf{X})-\widehat{\mu}(0,\mathbf{X})$, and the bandwidth is selected by Algorithm~2 with ${C_0}=0.05$. For the Proposed~(Heter) method, the conditional variance $\operatorname{Var}(Y\mid\mathbf{X},A=a)$ is estimated by a residual-based approach: after fitting the outcome model $\widehat{\mu}(a,\mathbf{X})$, we compute the squared residuals $\widehat{e}_i^2 = (Y_i - \widehat{\mu}(A_i,\mathbf{X}_i))^2$ for subjects with $A_i=a$ and then either take the stratum mean (when the true variance is constant within the stratum, e.g., Settings~(I), (III) and~(VI)) or regress $\widehat{e}_i^2$ on the relevant covariates (e.g., $|X_2|$ in Settings~(II) and~(VII)), with a lower truncation of $0.05$ to ensure numerical stability. The standard deviation of $\widehat{V}_{as}$ is computed as the sample standard deviation of the influence function values $\psi_i$ across all four folds, as in Eq.~\eqref{sin-opt-std-heter}. Both proposed methods are implemented via Algorithm~1 (single cross-fitting) and the results are aggregated over 500 Monte Carlo replications.

The results in Table~\ref{heter-sim} reveal several findings. First, in
all heteroscedastic settings~(I)--(III), (V), (VI), and~(VII), Proposed~(Heter) achieves
the shortest confidence intervals at every sample size, confirming the
efficiency gain from correctly incorporating the variance structure into
the asymmetric tuning. In the homoscedastic Setting~(IV),
Proposed~(Heter) and Proposed~(Homo) produce identical results, which is
expected: when
$\operatorname{Var}(Y \mid \mathbf{X}, A=1) =
\operatorname{Var}(Y \mid \mathbf{X}, A=0)$, the variance-weighted tuning
$t_0$ reduces exactly to the propensity-based tuning, and this empirical
equivalence validates the theoretical derivation in Section~A.12.

Second, the magnitude of the efficiency gain from Proposed~(Heter) over
Proposed~(Homo) increases with the severity of heteroscedasticity. In
Settings~(III) and~(V), where the variance ratio between treatment groups
exceeds four, the average length reduction is approximately 13\%--16\%. In
Settings~(I), (II), (VI), and~(VII), where heteroscedasticity is more moderate, the
reduction is approximately 3\%--9\%.

Third, comparing Settings~(VI) and~(VII) ($p = 5$ covariates) with their counterparts Settings~(I) and~(II) ($p = 2$), the proposed methods maintain valid coverage and comparable confidence interval lengths. The additional covariates enter the nuisance estimation but do not materially affect the efficiency of the value estimator, confirming the robustness of our data-driven tuning (Algorithm~2) to the covariate dimension.

Finally, across all the settings and all the sample sizes tested, the proposed method under both homoscedastic and heteroscedastic cases always achieves nominal coverage and substantially outperforms the methods for comparison. In
Settings~(IV) and~(V) with $\alpha = 1/2$ and
$\operatorname{Beta}$-distributed $X_2$, subbagging exhibits pronounced
undercoverage, with empirical coverage ranging from 83\% to 87\% across
sample sizes, well below the nominal 95\% level. In these settings,
subbagging's comparable or shorter confidence interval length is
misleading, as the intervals fail to achieve the desired coverage. In
contrast, both proposed methods maintain valid coverage around 93\%--96\%
throughout. SSS consistently achieves nominal coverage but suffer from efficiency loss due to the sample split.

\begin{table}[H]
    \centering
    \scriptsize
    \caption{Additional simulation settings. All settings are nonregular with $P\{\tau(X_1,X_2)=0\}>0$. Settings~(VI) and~(VII) include three additional covariates $X_3 \sim N(0,1)$, $X_4 \sim \operatorname{Uniform}(0,1)$, $X_5 \sim \operatorname{Bernoulli}(0.5)$ that enter $\Phi$ and $\pi$ but not $\tau$.}
    \resizebox{\textwidth}{!}{
    \begin{tabular}{cccccccc}
        \hline \hline
        & (I) & (II) & (III) & (IV) & (V) & (VI) & (VII) \\
        \hline
        $p$ & 2 & 2 & 2 & 2 & 2 & 5 & 5 \\
        \hline
        $X_1$ & $\operatorname{Ber}(0.3)$ & $\operatorname{Ber}(0.2)$ & $\operatorname{Ber}(0.4)$ & $\operatorname{Ber}(0.3)$ & $\operatorname{Ber}(0.3)$ & $\operatorname{Ber}(0.3)$ & $\operatorname{Ber}(0.2)$ \\
        \hline
        $X_2$ & $\operatorname{U}[-2,2]$ & $\operatorname{U}[-2,2]$ & $\operatorname{Ber}(0.5)$ & $\operatorname{Beta}(0.5,2)$ & $\operatorname{Beta}(0.5,2)$ & $\operatorname{U}[-2,2]$ & $\operatorname{U}[-2,2]$ \\
        \hline
        $\Phi(\mathbf{x})$ & $0.5$ & $(x_2/2)^2$ & $0.3$ & $0.2$ & $0.2$ & \makecell{$0.5 + 0.3 x_3$ \\ ${}+ 0.2 x_4 + 0.1 x_5$} & \makecell{$(x_2/2)^2 + 0.3 x_3$ \\ ${}+ 0.2 x_4 + 0.1 x_5$} \\
        \hline
        $\tau(x_1, x_2)$ & $x_1 x_2^2 / 4$ & $x_1 |x_2|^{3/2}/4$ & $0.4\, x_1$ & $0.5\, x_1 x_2$ & $0.5\, x_1 x_2$ & $x_1 x_2^2 / 4$ & $x_1 |x_2|^{3/2}/4$ \\
        \hline
        $\pi(1, \mathbf{x})$ & $0.6 + 0.2 x_1$ & $0.7 + 0.15 x_1$ & $0.7 + 0.1 x_1$ & $0.75$ & $0.6 + 0.2 x_1$ & \makecell{$0.6 + 0.2 x_1$ \\ ${}+ 0.05 x_3$} & \makecell{$0.7 + 0.15 x_1$ \\ ${}+ 0.05 x_4$} \\
        \hline
        $\operatorname{Var}(Y \mid \mathbf{X}, A\!=\!1)$ & $0.25 + 0.5(1\!-\!x_1)$ & $0.2 + 0.3|x_2|$ & $0.09$ & $0.25$ & $0.09$ & $0.25 + 0.5(1\!-\!x_1)$ & $0.2 + 0.3|x_2|$ \\
        \hline
        $\operatorname{Var}(Y \mid \mathbf{X}, A\!=\!0)$ & $0.25$ & $0.2$ & $0.49$ & $0.25$ & $0.36$ & $0.25$ & $0.2$ \\
        \hline
        Heteroscedastic & Yes & Yes & Yes & No & Yes & Yes & Yes \\
        \hline
        $\alpha$ & $1/2$ & $2/3$ & $\infty$ & $1/2$ & $1/2$ & $1/2$ & $2/3$ \\
        \hline
        $V_0$ & $0.60$ & $0.39$ & $0.46$ & $0.23$ & $0.23$ & $0.75$ & $0.54$ \\
        \hline
    \end{tabular}
    }
    \label{heter-setting}
\end{table}
\begin{table}[H]
    \centering
    \scriptsize
    \caption{ECP and AL of the CIs for additional settings with standard error smaller than 0.01.}
    \begin{tabular}{|c|c|c|c|c|c|c|c|c|c|}
        \hline
        \multirow{2}{*}{Setting} & \multirow{2}{*}{$n$} & \multicolumn{2}{c|}{Proposed (Heter)} & \multicolumn{2}{c|}{Proposed (Homo)} & \multicolumn{2}{c|}{Subbagging} & \multicolumn{2}{c|}{SSS} \\
        \cline{3-10}
        & & ECP(\%) & AL*100 & ECP(\%) & AL*100 & ECP(\%) & AL*100 & ECP(\%) & AL*100 \\
        \hline
        \multirow{3}{*}{(I)}
        & 500  & $94.8$ & $12.2$ & $95.6$ & $13.2$ & $89.4$ & $14.4$ & $95.4$ & $23.0$ \\
        & 1000 & $95.4$ & $8.5$  & $94.0$ & $9.3$  & $92.8$ & $10.0$ & $95.8$ & $16.1$ \\
        & 2000 & $94.2$ & $5.9$  & $93.6$ & $6.5$  & $90.8$ & $7.0$  & $94.6$ & $11.2$ \\
        \hline
        \multirow{3}{*}{(II)}
        & 500  & $93.6$ & $13.4$ & $94.2$ & $13.9$ & $93.8$ & $15.0$ & $94.6$ & $23.0$ \\
        & 1000 & $93.0$ & $9.3$  & $93.2$ & $9.7$  & $91.4$ & $10.3$ & $92.4$ & $16.1$ \\
        & 2000 & $93.6$ & $6.5$  & $93.2$ & $6.8$  & $90.8$ & $7.2$  & $95.2$ & $11.3$ \\
        \hline
        \multirow{3}{*}{(III)}
        & 500  & $94.0$ & $7.4$  & $94.0$ & $8.6$  & $89.4$ & $11.6$ & $93.6$ & $17.6$ \\
        & 1000 & $93.6$ & $5.1$  & $94.2$ & $6.0$  & $92.8$ & $8.2$  & $94.6$ & $12.1$ \\
        & 2000 & $95.2$ & $3.6$  & $95.0$ & $4.2$  & $93.6$ & $5.8$  & $94.0$ & $8.8$  \\
        \hline
        \multirow{3}{*}{(IV)}
        & 500  & $93.8$ & $15.2$ & $93.8$ & $15.2$ & $85.2$ & $11.2$ & $94.4$ & $26.9$ \\
        & 1000 & $93.4$ & $7.5$  & $93.4$ & $7.5$  & $86.8$ & $7.5$  & $95.2$ & $14.4$ \\
        & 2000 & $93.4$ & $5.1$  & $93.4$ & $5.1$  & $86.0$ & $5.1$  & $94.2$ & $9.9$  \\
        \hline
        \multirow{3}{*}{(V)}
        & 500  & $92.6$ & $7.0$  & $92.8$ & $8.4$  & $83.4$ & $8.9$  & $92.6$ & $18.2$ \\
        & 1000 & $94.8$ & $4.7$  & $96.0$ & $5.5$  & $87.0$ & $5.8$  & $93.6$ & $11.5$ \\
        & 2000 & $93.8$ & $3.3$  & $95.0$ & $3.8$  & $87.4$ & $3.6$  & $95.4$ & $7.9$  \\
        \hline
        \multirow{3}{*}{(VI)}
        & 500  & $94.6$ & $14.5$ & $94.4$ & $15.3$ & $93.6$ & $16.0$ & $95.0$ & $25.3$ \\
        & 1000 & $94.2$ & $9.9$  & $92.6$ & $10.5$ & $92.4$ & $10.9$ & $95.8$ & $17.2$ \\
        & 2000 & $96.0$ & $6.9$  & $95.8$ & $7.4$  & $94.4$ & $7.6$  & $95.4$ & $12.1$ \\
        \hline
        \multirow{3}{*}{(VII)}
        & 500  & $93.0$ & $15.4$ & $93.4$ & $15.5$ & $93.8$ & $16.3$ & $93.6$ & $25.3$ \\
        & 1000 & $93.4$ & $10.5$ & $94.8$ & $10.7$ & $93.2$ & $11.2$ & $96.0$ & $17.5$ \\
        & 2000 & $94.8$ & $7.3$  & $95.2$ & $7.5$  & $93.4$ & $7.8$  & $94.6$ & $12.2$ \\
        \hline
    \end{tabular}
    \label{heter-sim}
\end{table}
\section{Proposed Methodology in Multiple Time Point Case}\label{appB}
In this section, firstly we shall provide the corresponding problems settings for the optimal dynamic treatment regimes in the multiple time point case in Appendix \ref{appB1}. Then the methodology as well as the theoretical results are presented in Appendix \ref{appB2}.
\subsection{Optimal Dynamic Treatment Regimes}\label{appB1}
We now extend our proposed method to multiple time point case. Firstly, we give a brief review of the optimal treatment regimes in multiple time point case which is similar to that of \citet{shi2020breaking}. For each subject, we assume the observation takes the following form:
\begin{align*}
    \left(\mathbf{X}^{(1)}, A^{(1)}, \mathbf{X}^{(2)}, A^{(2)}, \ldots, \mathbf{X}^{(K)}, A^{(K)}, Y_0\right)
\end{align*}
where $Y_0$ denotes the outcome of interest, $\mathbf{X}^{(k)}$ stands for the set of covariates obtained in $k$-th stage, $A^{(k)}$ denotes the binary treatment received in the $k$-th stage too. For $k=1, \ldots, K$, let 
\begin{align*}
    \bar{\mathbf{X}}^{(k)}=\left(\mathbf{X}^{(1)}, \ldots, \mathbf{X}^{(k)}\right) \in \bar{\mathcal{X}}^{(k)} \quad \text { and } \quad \bar{\mathbf{A}}^{(k)}=\left(A^{(1)}, \ldots, A^{(k)}\right) \in\{0,1\}^k
\end{align*}
denote a patient's covariates and treatment history. For any $a_1, \ldots, a_K \in\{0,1\}$, denote by $\bar{\mathbf{a}}_k=\left(a_1, \ldots, a_k\right)$ for $k=1, \ldots, K$. The set of all potential outcomes is given by
$$
\mathbf{W}=\left\{\left(\mathbf{X}^{(2) *}\left(a_1\right), \mathbf{X}^{(3) *}\left(\bar{\mathbf{a}}_2\right), \ldots, \mathbf{X}^{(K) *}\left(\bar{\mathbf{a}}_{K-1}\right), Y_0^*\left(\bar{\mathbf{a}}_K\right)\right): \forall \bar{\mathbf{a}}_K \in\{0,1\}^K\right\}
$$
where $\mathbf{X}^{(k) *}\left(\bar{\mathbf{a}}_{k-1}\right)$ denotes the potential time-dependent covariates of a patient that would occur once he/she receives treatments $\left(a_1, \ldots, a_{k-1}\right)$  and $Y_0^*\left(\bar{\mathbf{a}}_K\right)$ denotes the potential outcome that would result assuming he/she receives treatments $\left(a_1, \ldots, a_K\right)$. 

A dynamic treatment regime $d=\left\{d_k\right\}_{k=1}^K$ is a set of decision rules that treat a patient over time. For $k=1, \ldots, K$,  $d_k=d_k\left(\bar{\mathbf{a}}_{k-1}, \bar{\mathbf{x}}_k\right)$ corresponds to the $k$-th decision rule that takes as input a patient's realized covariate and treatment history and outputs a treatment option $a_k \in\{0,1\}$. Let $\bar{d}_k=\left\{d_j\right\}_{j=1}^k$ for $k=1, \ldots, K-1$, the potential outcome associated with $d$ is given by
\begin{align*}
    \left(\mathbf{X}^{(2) *}\left(d_1\right), \mathbf{X}^{(3) *}\left(\bar{d}_2\right), \ldots, \mathbf{X}^{(K) *}\left(\bar{d}_{K-1}\right), Y_0^*(d)\right).
\end{align*}
An optimal dynamic treatment regime $d^{o p t}$ is defined to maximize the average potential outcome, i.e,
$$
d^{o p t}=\underset{d}{\arg \max } \mathrm{E} Y_0^*(d).
$$
Here we still denote the value of our interest, the optimal value function $\max_d E Y_0^*(d)$, as $V_0$. The difference between $V_0$ in multiple time point case and that in single time point case is that, the $d$ in multiple time point case is a sequence of binary treatment assignments while in single time point case it is only a single binary value. For any $\bar{\mathbf{a}}_K \in\{0,1\}^K$ and $\bar{\mathbf{x}}_K \in \bar{\mathcal{X}}^{(K)}$, let $\mu_K\left(\bar{\mathbf{a}}_K, \bar{\mathbf{x}}_K\right)=\mathrm{E}\left(Y_0 \mid \bar{\mathbf{X}}^{(K)}=\bar{\mathbf{x}}_K, \bar{\mathbf{A}}^{(K)}=\bar{\mathbf{a}}_K\right)$ and $\tau_K\left(\bar{\mathbf{a}}_{K-1}, \bar{\mathbf{x}}_K\right)=\mu_K\left\{\left(\bar{\mathbf{a}}_{K-1}, 1\right), \bar{\mathbf{x}}_K\right\}-\mu_K\left\{\left(\bar{\mathbf{a}}_{K-1}, 0\right), \bar{\mathbf{x}}_K\right\}$. In addition, for $k=K-$ $1, \cdots, 2$, we sequentially define
$$
\mu_k\left(\bar{\mathbf{a}}_k, \bar{\mathbf{x}}_k\right)=\mathrm{E}\left(\underset{a_{k+1} \in\{0,1\}}{\arg \max }  \mu_{k+1}\left\{\left(\bar{\mathbf{a}}_k, a_{k+1}\right), \bar{\mathbf{X}}^{(k+1)}\right\} \mid \bar{\mathbf{X}}^{(k)}=\bar{\mathbf{x}}_k, \bar{\mathbf{A}}^{(k)}=\bar{\mathbf{a}}_k\right),
$$
and $\tau_k\left(\bar{\mathbf{a}}_{k-1}, \bar{\mathbf{x}}_k\right)=\mu_k\left\{\left(\bar{\mathbf{a}}_{k-1}, 1\right), \bar{\mathbf{x}}_k\right\}-\mu_k\left\{\left(\bar{\mathbf{a}}_{k-1}, 0\right), \bar{\mathbf{x}}_k\right\}$, for any $\bar{\mathbf{a}}_k \in\{0,1\}^k$ and $\bar{\mathbf{x}}_k \in$ $\bar{\mathcal{X}}^{(k)}$. For $k=1$, let
$$
\mu_1\left(a_1, \mathbf{x}_1\right)=\mathrm{E}\left(\underset{a_2 \in\{0,1\}}{\arg \max }  \mu_2\left\{\left(a_1, a_2\right), \bar{\mathbf{X}}^{(2)}\right\} \mid \mathbf{X}^{(1)}=\mathbf{x}_1, A^{(1)}=a_1\right)
$$
and $\tau_1\left(\mathbf{x}_1\right)=\mu_1\left(1, \mathbf{x}_1\right)-\mu_1\left(0, \mathbf{x}_1\right)$ for any $a_1 \in\{0,1\}, \mathbf{x}_1 \in \bar{\mathcal{X}}_1$. Under the Assumption \ref{B1}, the stable unit treatment value assumption (SUTVA) which is the same as Assumption (C1.) in \citet{shi2020breaking}, and no unmeasured confounders assumption, i.e.,  Assumption \ref{B2} which is the same as Assumption (C2.) in \citet{shi2020breaking},  
\begin{assumption}[Stable Unit Treatment Value Assumption (SUTVA)]\label{B1}
     Assume that $$\mathbf{X}^{(k)}=\sum_{\bar{\mathbf{a}}_{k-1} \in\{0,1\}^{k-1}} \mathbf{X}^{(k) *}\left(\bar{\mathbf{a}}_{k-1}\right) \mathbb{I}\left(\bar{\mathbf{A}}^{(k-1)}=\bar{\mathbf{a}}_{k-1}\right)$$ and $$Y_0=\sum_{\bar{\mathbf{a}}_K \in\{0,1\}^K} Y_0^*\left(\bar{\mathbf{a}}_K\right) \mathbb{I}\left(\bar{\mathbf{A}}^{(K)}=\bar{\mathbf{a}}_K\right), \forall k=2, \ldots, K$$ and $\bar{\mathbf{a}}_K \in\{0,1\}$.
\end{assumption}
\begin{assumption}[No Unmeasured Confounders]\label{B2}
    $A^{(k)} \perp \mathbf{W} \mid \bar{\mathbf{X}}^{(k)}, \bar{\mathbf{A}}^{(k-1)}, \forall k=1, \ldots, K$.
\end{assumption}
we can show
$$
\mu\left(\bar{\mathbf{a}}_K, \bar{\mathbf{x}}_K\right)=\mathrm{E}\left\{Y_0^*\left(\bar{\mathbf{a}}_K\right) \mid \bar{\mathbf{X}}^{(K) *}\left(\bar{\mathbf{a}}_{K-1}\right)=\bar{\mathbf{x}}_K\right\}
$$
and for $2 \leq k \leq K-1$
$$
\mu\left(\bar{\mathbf{a}}_k, \bar{\mathbf{x}}_k\right)=\mathrm{E}\left[V_0^{(k+1)}\left\{\bar{\mathbf{a}}_k, \bar{\mathbf{X}}^{(k+1) *}\left(\bar{\mathbf{a}}_k\right)\right\} \mid \bar{\mathbf{X}}^{(k) *}\left(\bar{\mathbf{a}}_{k-1}\right)=\bar{\mathbf{x}}_k\right],
$$
and
$$
\mu\left(a_1, \mathbf{x}_1\right)=\mathrm{E}\left[V_0^{(2)}\left\{a_1, \bar{\mathbf{X}}^{(2) *}\left(a_1\right)\right\} \mid \mathbf{X}^{(1)}=\mathbf{x}_1\right]
$$
where
$$
\begin{aligned}
& V_0^{(K)}\left(\bar{\mathbf{a}}_{K-1}, \bar{\mathbf{x}}_K\right)=\max _{a_K \in\{0,1\}} \mathrm{E}\left\{Y_0^*\left(\bar{\mathbf{a}}_K\right) \mid \bar{\mathbf{X}}^{(K) *}\left(\bar{\mathbf{a}}_{K-1}\right)=\bar{\mathbf{x}}_K\right\} \\
& V_0^{(k)}\left(\bar{\mathbf{a}}_{k-1}, \bar{\mathbf{x}}_k\right)=\max _{a_k \in\{0,1\}} \mathrm{E}\left[V_0^{(k+1)}\left\{\bar{\mathbf{a}}_k, \bar{\mathbf{X}}^{(k+1) *}\left(\bar{\mathbf{a}}_k\right)\right\} \mid \bar{\mathbf{X}}^{(k) *}\left(\bar{\mathbf{a}}_{k-1}\right)=\bar{\mathbf{x}}_k\right], \\
& \bar{\mathbf{X}}^{(k) *}\left(\bar{\mathbf{a}}_{k-1}\right)=\left\{\mathbf{X}^{(1)}, \mathbf{X}^{(2) *}\left(a_1\right), \ldots, \mathbf{X}^{(k) *}\left(\bar{\mathbf{a}}_{k-1}\right)\right\}.
\end{aligned}
$$
Note that Assumption \ref{B2} holds generally for sequentially randomized studies. Define the propensity score function $\pi_k\left(\bar{\mathbf{a}}_k, \bar{\mathbf{x}}_k\right)=\operatorname{Pr}\left(A^{(k)}=a_k \mid \bar{\mathbf{X}}^{(k)}=\bar{\mathbf{x}}_k, \bar{\mathbf{A}}^{(k-1)}=\bar{\mathbf{a}}_{k-1}\right)$ for $k=2, \ldots, K$ and $\pi_1\left(a_1, \mathbf{x}_1\right)=\operatorname{Pr}\left(A^{(1)}=a_k \mid \mathbf{X}^{(1)}=\mathbf{x}_1\right)$. To ensure the validity of AIPWE type estimator, we might also need the overlap assumption:
\begin{assumption}[Overlap]\label{B3}
    $\exists$ $c\in(0,0.5)$ s.t. $0<c<\pi_k\left(\bar{\mathbf{a}}_k, \bar{\mathbf{x}}_k\right)<1-c<1$ for $k=2,\dots,K$ and $0<c<\pi_k\left(a_1, \mathbf{x}_k\right)<1-c<1$.
\end{assumption}

Define the set of dynamic treatment regimes $\mathcal{D}^{\text {opt }}$ such that any $d=\left\{d_k\right\}_{k=1}^K \in \mathcal{D}^{\text {opt }}$ shall satisfy
$$
\begin{array}{r}
d_K\left(\bar{\mathbf{a}}_{k-1}, \bar{\mathbf{x}}_k\right) \in \underset{a \in\{0,1\}}{\arg \max } a \tau_k\left(\bar{\mathbf{a}}_{k-1}, \bar{\mathbf{x}}_k\right), k=2, \ldots, K, \\
\text { and } \quad d_1\left(\mathbf{x}_1\right) \in \underset{a \in\{0,1\}}{\arg \max } a \tau_1\left(\mathbf{x}_1\right),
\end{array}
$$
for any $\bar{\mathbf{x}}_K \in \bar{\mathcal{X}}^{(K)}, \ldots, \bar{\mathbf{x}}_2 \in \bar{\mathcal{X}}^{(2)}, \mathbf{x}_1 \in \bar{\mathcal{X}}^{(1)}$ and $\bar{\mathbf{a}}_{K-1} \in\{0,1\}^{K-1}, \ldots, \bar{\mathbf{a}}_2 \in\{0,1\}^2, a_1 \in$ $\{0,1\}$. It is not hard to verify that
$$
\mathcal{D}^{o p t} \subseteq \underset{d}{\arg \max } \mathrm{EY}_0^*(d).
$$
When there exists $k$ s.t. 
\begin{align*}
    P(\tau_k\left(\bar{\mathbf{A}}_{k-1}, \bar{\mathbf{X}}_k\right)=0)>0
\end{align*}
or
\begin{align*}
    P(\tau_1\left(\mathbf{X}_1\right)=0)>0,
\end{align*}
the optimal (dynamic) treatment regimes would be non-unique for a nonnegligible subpopulation, and similar to the single time point case, nonregularity problem would arise; see discussions in \citet{luedtke2016statistical} and \citet{shi2020breaking}.
\subsection{Proposed Method}\label{appB2}
Denote the observed dataset as follows:
\begin{align*}
    \mathcal{O}_n = \left\{\mathbf{O}_i=\left(\mathbf{X}_i^{(1)}, A_i^{(1)}, \mathbf{X}_i^{(2)}, A_i^{(2)}, \ldots, \mathbf{X}_i^{(K)}, A_i^{(K)}, Y_i\right)\in\mathcal{O}: i=0, \ldots, n\right\} .
\end{align*}
For $k=1, \ldots, K, i=0,1, \ldots, n$, let
$$
\bar{\mathbf{X}}_i^{(k)}=\left(\mathbf{X}_i^{(1)}, \ldots, \mathbf{X}_i^{(k)}\right) \quad \text { and } \quad \bar{\mathbf{A}}_i^{(k)}=\left(A_i^{(1)}, \ldots, A_i^{(k)}\right).
$$
For any dynamic treatment regime $d=\left\{d_k\right\}_{k=1}^K$, let $\widehat{V}_i^{(K+1)}=Y_i$ and recursively define
$$
\begin{aligned}
&\widehat{V}_i^{(k)}\left(d^* ; \pi^*, \mu^*\right)  =\frac{\mathrm{g}\left\{A_i^{(k)}, d_k^*\left(\bar{\mathbf{A}}_i^{(k-1)}, \bar{\mathbf{X}}_i^{(k)}\right)\right\}}{\pi_k^*\left(\bar{\mathbf{A}}_i^{(k)}, \bar{\mathbf{X}}_i^{(k)}\right)}\left\{\widehat{V}_i^{(k+1)}\left(d^* ; \pi^*, \mu^*\right)-\mu_k^*\left(\bar{\mathbf{A}}_i^{(k)}, \bar{\mathbf{X}}_i^{(k)}\right)\right\} \\
 &+d_k^*\left(\bar{\mathbf{A}}_i^{(k-1)}, \bar{\mathbf{X}}_i^{(k)}\right)\cdot \mu_k^*\left[\left\{\bar{\mathbf{A}}_i^{(k-1)}, 1\right\}, \bar{\mathbf{X}}_i^{(k)}\right]\\
 &+\left(1-d_k^*\left(\bar{\mathbf{A}}_i^{(k-1)}, \bar{\mathbf{X}}_i^{(k)}\right)\right)\cdot \mu_k^*\left[\left\{\bar{\mathbf{A}}_i^{(k-1)}, 0\right\}, \bar{\mathbf{X}}_i^{(k)}\right],
\end{aligned}
$$
for $k=K-1, \ldots, 2,1, i=0,1, \ldots, n$, where $\bar{A}_i^{(0)}=\emptyset, \pi^* \equiv\left\{\pi_k^*\right\}_{k=1}^K$ and $\mu^* \equiv\left\{\mu_k^*\right\}_{k=1}^K$ denote the estimated propensity score and conditional mean functions, and the function $\mathrm{g}(a, z)=a z+(1-a)(1-z)$ for $a, z \in \mathbb{R}$. Now we define the adaptive smoothing decision function in the multiple time point case. Specifically, let $d_{s}=\{d_{s}^k\}_{k=1}^K$,
\begin{align*}
    &d_{s}^k(\Bar{\mathbf{A}}_{i}^{(k-1)},\bar{\mathbf{X}}_i^{(k)};\tau_k^*,h_n^k,t_0^k)\\
    &=\mathbb{I}\{\tau_k^*(\Bar{\mathbf{A}}_{i}^{(k-1)},\bar{\mathbf{X}}_i^{(k)})\in(-t_0^k h_n^k, (1-t_0^k)h_n^k)\}\cdot \left(\frac{\tau_k^*(\Bar{\mathbf{A}}_{i}^{(k-1)},\bar{\mathbf{X}}_i^{(k)})}{h_n^k}+t_0^k\right)\\
    &+\mathbb{I}\{\tau_k^*(\Bar{\mathbf{A}}_{i}^{(k-1)},\bar{\mathbf{X}}_i^{(k)})\ge (1-t_0^k)h_n^k\}, \quad k=2,\dots,K\\
     &d_{s}^1(\bar{\mathbf{X}}_i^{(1)};\tau_1^*,h_n^1,t_0^1)\\
     &=\mathbb{I}\{\tau_1^*(\bar{\mathbf{X}}_i^{(1)})\in(-t_0^1 h_n^1, (1-t_0^1)h_n^1)\}\cdot \left(\frac{\tau_1^*(\bar{\mathbf{X}}_i^{(1)})}{h_n^1}+t_0^1\right)\\
    &+\mathbb{I}\{\tau_1^*(\bar{\mathbf{X}}_i^{(1)})\ge (1-t_0^1)h_n^1\},
\end{align*}
and denote the corresponding $d_{s}$ as $d_{s}(\mathbf{O}_i;\tau^*,h_n,t_0)$. 
Then, consider the following adaptive smoothing estimator for $\mathrm{EY}_0^*(d)$,
 \begin{align*}
     \widehat{V}_{s}^{(1)}:=\frac{1}{4}\sum_{j=1,2}\sum_{l=1,2}\frac{1}{|\mathcal{I}_{j,l}|}\sum_{i\in\mathcal{I}_{j,l}}\widehat{V}_i^{(1)}(d_{s}(\mathbf{O}_i;\widehat{\tau}_{\mathcal{I}_{3-j}},h_{n,\mathcal{I}_{3-j}},t_0),\widehat{\pi}_{\mathcal{I}_{j,l}^c},\widehat{\mu}_{\mathcal{I}_{j,l}^c})
 \end{align*}
 with standard deviation estimator
 \begin{align*}
     \widehat{\sigma}_{s}^{(1)}:=\left(\frac{1}{n-1}\sum_{j=1,2}\sum_{l=1,2}\sum_{i\in\mathcal{I}_{j,l}}\left(\widehat{V}_i^{(1)}(d_{s}(\mathbf{O}_i;\widehat{\tau}_{\mathcal{I}_{3-j}},h_{n,\mathcal{I}_{3-j}},t_0),\widehat{\pi}_{\mathcal{I}_{j,l}^c},\widehat{\mu}_{\mathcal{I}_{j,l}^c})-\widehat{V}_{s}^{(1)}\right)^2\right)^{1/2}.
 \end{align*}
Like the single time point case, here we also need similar assumptions to guarantee the validity of our proposed method. In particular, Assumption \ref{B4} is exactly Assumption (C4.) in \citet{shi2020breaking}, which allows for polynomial unbounded density of $\tau_k$. Assumption \ref{B5} and \ref{B6} together are weaker than Assumption (C5.) in \citet{shi2020breaking}, which is similar to the single time point case. Assumption \ref{B7} is just the common doubly robust rate requirement, which can be found at \citet{chernozhukov2018double} and is weaker than the rate requirements in Theorem 3 of \citet{shi2020breaking}.
\begin{assumption}[Assumption (C4.) in \citet{shi2020breaking}]\label{B4}
     Assume there exist some positive constants $\bar{c}, \alpha$ and $\delta_0$ such that for $k=1,\dots,K$,
$$
P\left(0<\left|\tau_k\left(\Bar{\mathbf{A}}^{(k-1)},\bar{\mathbf{X}}^{(k)}\right)\right|<t\right) \leq \bar{c} t^\alpha, \quad \forall 0<t \leq \delta_0
$$
\end{assumption}
\begin{assumption}[ Scale of $h_{n,\mathcal{I}_j}^{k}$ (lower)]\label{B5}
For any $\mathcal{I}_j$, we have
\begin{align*}
    h_{n,\mathcal{I}_j}^{k}\succ_p n^{1/4}E^{3/4}\left[|\widehat{\tau}_{k,\mathcal{I}_j}(\bar{\mathbf{A}}^{(k-1)},\bar{\mathbf{X}}^{(k)})-\tau_k(\bar{\mathbf{A}}^{(k-1)},\bar{\mathbf{X}}^{(k)})|^2\right].
\end{align*}
\end{assumption}
\begin{assumption}[Scale of $h_{n,\mathcal{I}_j}^{k}$ (upper)]\label{B6}
    $h_{n,\mathcal{I}_j}^{k}\prec n^{-\frac{1}{2(1+\alpha)}}$, where the constant $\alpha$ comes from Assumption \ref{B4}.
\end{assumption}
\begin{assumption}[Doubly Robust Convergence Rate of $\widehat{\pi}$ and $\widehat{\mu}$]\label{B7}
    For any subset $\mathcal{I}\subset[n]$ and $a\in\{0,1\}$, with probability tending to 1 we have 
\begin{align*}
E\left[\left|\widehat{\pi}_{k,\mathcal{I}}\left(\bar{\mathbf{A}}^{(k)}, \bar{\mathbf{X}}^{(k)}\right)-\pi_k\left(\bar{\mathbf{A}}^{(k)}, \bar{\mathbf{X}}^{(k)}\right)\right|^2\right]\cdot E\left[\left|\widehat{\mu}_{k,\mathcal{I}}\left(\bar{\mathbf{A}}^{(k)}, \bar{\mathbf{X}}^{(k)}\right)-\mu_k\left(\bar{\mathbf{A}}^{(k)}, \bar{\mathbf{X}}^{(k)}\right)\right|^2\right]\\=o(n^{-1}),\\
E\left[\left|\widehat{\pi}_{k,\mathcal{I}}\left(\bar{\mathbf{A}}^{(k)}, \bar{\mathbf{X}}^{(k)}\right)-\pi_k\left(\bar{\mathbf{A}}^{(k)}, \bar{\mathbf{X}}^{(k)}\right)\right|^2\right]\lor E\left[\left|\widehat{\mu}_{k,\mathcal{I}}\left(\bar{\mathbf{A}}^{(k)}, \bar{\mathbf{X}}^{(k)}\right)-\mu_k\left(\bar{\mathbf{A}}^{(k)}, \bar{\mathbf{X}}^{(k)}\right)\right|^2\right]\\=o(1).
\end{align*}
\end{assumption}
 \begin{theorem}\label{thmb1}
     Under Assumptions \ref{B1} to \ref{B7} and the stagewise moment-comparability condition in Eq. \eqref{multi-moment-comparison}, we have
     \begin{align*}
         \frac{\sqrt{n}(\widehat{V}_{s}^{(1)}-V_0)}{\widehat{\sigma}_{s}^{(1)}}\to_d N(0,1).
     \end{align*}
 \end{theorem}
 A stagewise version of Algorithm 2 is as follows.
\begin{algorithm}[H]
\caption{Selection of $h_n$ in multiple time point case}
    \begin{algorithmic}[1]
        \State \textbf{Input}: Data set $\mathcal{O}_n$; Randomly picked subsets of $[n]$: $\mathcal{I}_1$ and $\mathcal{I}_2$ with equal size; Working models: $\widehat{\mu}$.
        \State \textbf{Estimated Approximation Error of $\widehat{\tau}_{k,\mathcal{I}_j}$ ($j=1,2$, $k=1,\ldots,K$)}:
 \begin{align*}
     EAE(\widehat{\tau}_{k,\mathcal{I}_j})=\frac{1}{|\mathcal{I}_{3-j}|}\sum_{\ell=1,2}\sum_{i\in\mathcal{I}_{3-j,\ell}}\left(\widehat{\tau}_{k,\mathcal{I}_{3-j,3-\ell}}(\bar{\mathbf{A}}_i^{(k-1)},\bar{\mathbf{X}}_i^{(k)})- \widehat{\tau}_{k,\mathcal{I}_{j}}(\bar{\mathbf{A}}_i^{(k-1)},\bar{\mathbf{X}}_i^{(k)})\right)^2
 \end{align*}
 \State \textbf{Selection of $h_{n,\mathcal{I}_j}$}: for $k=1,\dots,K$,  $h_{n,\mathcal{I}_j}^{k}={C_0}\cdot \log(n)\cdot n^{1/4}\cdot((\log(n)/({C_0} n))\lor EAE(\widehat{\tau}_{k,\mathcal{I}_j}))^{3/4}$, where ${C_0}$ is a constant.
    \end{algorithmic}
\label{alg5}
\end{algorithm}
Write $\mathbf H_k=(\bar{\mathbf A}^{(k-1)},\bar{\mathbf X}^{(k)})$ and let $E_{\mathbf H_k}$ denote expectation over a fresh stage-$k$ history with the training data fixed. We assume, for every stage and half or quarter fold,
\begin{equation}\label{multi-moment-comparison}
E_{\mathbf H_k}|\widehat\tau_{k,\mathcal I}(\mathbf H_k)-\tau_k(\mathbf H_k)|^3
=O_p\left[\{E_{\mathbf H_k}|\widehat\tau_{k,\mathcal I}(\mathbf H_k)-\tau_k(\mathbf H_k)|^2\}^{3/2}\right].
\end{equation}
\begin{assumption}[Stagewise rate requirement for tuning]\label{B8}
There exists $\beta\in(0,\alpha)$ such that, for every $k=1,\ldots,K$ and $j=1,2$,
\[
E_{\mathbf H_k}|\widehat\tau_{k,\mathcal I_j}(\mathbf H_k)-\tau_k(\mathbf H_k)|^2
\prec_p n^{-\frac{2+\frac{2}{3}\beta}{2(1+\beta)}}.
\]
\end{assumption}
\medskip\noindent\textbf{Condition (C1-M) (stagewise near-parametric rate).}\par
\noindent$E_{\mathbf H_k}|\widehat\tau_{k,\mathcal I}(\mathbf H_k)-\tau_k(\mathbf H_k)|^2=O_p(\log n/n)$ for every stage and fold.

\medskip\noindent\textbf{Condition (C2-M) (stagewise stochastic boundedness).}
$\|\tau_k\|_\infty<\infty$ and $\|\widehat\tau_{k,\mathcal I}\|_\infty=O_p(1)$ uniformly over the finite set of stages and folds.

\medskip\noindent\textbf{Condition (C3-M) (stagewise variability among splits).}
For every stage $k$ and any two disjoint relevant folds $\mathcal I,\mathcal I'$,
\[
E_{\mathbf H_k}|\widehat\tau_{k,\mathcal I}(\mathbf H_k)-\widehat\tau_{k,\mathcal I'}(\mathbf H_k)|^2
\asymp_pE_{\mathbf H_k}|\widehat\tau_{k,\mathcal I}(\mathbf H_k)-\tau_k(\mathbf H_k)|^2.
\]

The roles of these conditions are the stagewise counterparts of Conditions (C1)--(C3). Condition (C1-M) makes the truncation floor dominate both the stage-$k$ squared error and its $EAE$, without requiring Conditions (C2-M) or (C3-M). In the other branch, Condition (C2-M) bounds the stagewise validation-fold summands and provides conditional concentration, while Condition (C3-M) links the population split discrepancy to the stage-$k$ squared estimation error.

The same bias--variance interpretation applies at every stage. For two independently trained stage-$k$ fits on equal-sized folds, averaging over their training samples gives
\[
E\,E_{\mathbf H_k}|\widehat\tau_{k,\mathcal I}(\mathbf H_k)-\widehat\tau_{k,\mathcal I'}(\mathbf H_k)|^2
=2E_{\mathbf H_k}\!\left[\operatorname{Var}\{\widehat\tau_{k,\mathcal I}(\mathbf H_k)\}\right],
\]
whereas the expected stage-$k$ squared error is
\[
E\,E_{\mathbf H_k}|\widehat\tau_{k,\mathcal I}(\mathbf H_k)-\tau_k(\mathbf H_k)|^2
=E_{\mathbf H_k}\!\left[\operatorname{Var}\{\widehat\tau_{k,\mathcal I}(\mathbf H_k)\}\right]
+E_{\mathbf H_k}\!\left[E\{\widehat\tau_{k,\mathcal I}(\mathbf H_k)\}-\tau_k(\mathbf H_k)\right]^2.
\]
Thus Condition (C3-M) requires, at each stage, that squared bias not dominate variance and that the realized quadratic quantities concentrate. The sufficient conditions in Lemmas~\ref{parametric-tuning-conditions} and~\ref{linear-smoother-c3} apply stage by stage after replacing $(\mathbf X,\tau,\widehat\tau_{\mathcal I},E_{\mathbf X})$ with $(\mathbf H_k,\tau_k,\widehat\tau_{k,\mathcal I},E_{\mathbf H_k})$. In particular, the parametric lemma gives the stagewise moment comparison and Condition (C1-M), whereas the stagewise versions of (U1)--(U4) give the stagewise moment comparison and Conditions (C2-M) and (C3-M). Uniform constants and the fixed number of stages make these conclusions simultaneous by a finite union bound.

\begin{proposition}\label{propb2}
Suppose Assumption \ref{B8} holds and either Condition (C1-M) holds or both Conditions (C2-M) and (C3-M) hold. Then, for every $k$ and $j$, the tuning selected by Algorithm \ref{alg5} satisfies Assumptions \ref{B5} and \ref{B6}.
\end{proposition}
 
 The optimal theory for the selection of $t_0$ can also be developed for the multiple time point case here as what we did in the single time point case.  For simplicity we will still focus on the case of homoscedasticity in the beginning, i.e., 
\begin{assumption}[Homoscedasticity]\label{B9}
For each $k=1,\dots,K$ and any 
$\mathcal{D}=(\mathcal{D}^{(1)},\dots,
\mathcal{D}^{(K)})\in\bar{\mathbb{D}}(V_0)$ satisfying 
\begin{align*}
&\mathcal{D}^{(k')}(\bar{\mathbf{A}}^{(k'-1)},
\bar{\mathbf{X}}^{(k')})
=\mathbb{I}\{\tau_{k'}(\bar{\mathbf{A}}^{(k'-1)},
\bar{\mathbf{X}}^{(k')})>0\}\\
&\qquad\text{whenever } \tau_{k'}(\bar{\mathbf{A}}^{(k'-1)},
\bar{\mathbf{X}}^{(k')})\neq 0, \quad k'=k+1,\dots,K,
\end{align*}
we have
\begin{align*}
    P\bigg(&\operatorname{Var}\!\left[\widehat{V}^{(k+1)}
    (\mathcal{D}(\mathbf{O}),\pi,\mu)\,\Big|\,
    (\bar{\mathbf{A}}^{(k-1)},1), \bar{\mathbf{X}}^{(k)}
    \right]\\
    &=\operatorname{Var}\!\left[\widehat{V}^{(k+1)}
    (\mathcal{D}(\mathbf{O}),\pi,\mu)\,\Big|\,
    (\bar{\mathbf{A}}^{(k-1)},0), \bar{\mathbf{X}}^{(k)}
    \right]\bigg)=1.
\end{align*}
\end{assumption}
When $K=1$, $\widehat{V}^{(2)}=Y$ and the quantifier over 
$\mathcal{D}$ is vacuous, so Assumption~\ref{B9} reduces to 
$P(\operatorname{Var}(Y\mid\mathbf{X},A=1)
=\operatorname{Var}(Y\mid\mathbf{X},A=0))=1$, which is exactly Assumption~8 in the main text. When $K>1$, the condition requires that at each stage~$k$, the variance of the recursive value 
$\widehat{V}^{(k+1)}(\mathcal{D};\pi,\mu)$ is balanced between the two treatment arms, for any 
$\mathcal{D}\in\bar{\mathbb{D}}(V_0)$ that follows the uniquely determined optimal rule at stages where $\tau_{k'}\neq 0$.

Denote the class of dynamic treatment regimes with the optimal value function as follows:
 \begin{align*}
     \Bar{\mathbb{D}}(V_0):=\{\mathcal{D}(\mathbf{O})=(\mathcal{D}^{(1)},\dots,\mathcal{D}^{(K)})|\mathcal{D}^{(1)}:\Bar{\mathcal{X}}^{(1)}\to[0,1], \\\land\mathcal{D}^{(k)}:\{0,1\}^{k-1} \times \Bar{\mathcal{X}}^{(k)}\to[0,1], k=2,\dots,K,\\
     \land E(\widehat{V}_0^{(1)}(\mathcal{D}(\mathbf{O}),\pi,\mu))=V_0 \}.
 \end{align*}
In Proposition \ref{multi-opt-oracle}, we calculate the variance for the estimator $\widehat{V}_{\mathcal{D}}^{(1)*}$ with $\mathcal{D}\in\bar{\mathbb{D}}(V_0)$ and known $\pi$ and $h$, and show which  $\mathcal{D}\in\bar{\mathbb{D}}(V_0)$ could minimize the variance.
\begin{proposition}\label{multi-opt-oracle}
    For $\forall$ $\mathcal{D}\in\bar{\mathbb{D}}(V_0)$, we have 
    \begin{align*}
        \mathcal{D}^{(1)}(\bar{\mathbf{x}}_1)\cdot\tau_1(\bar{\mathbf{x}}_1)=\max\{0,\tau_1(\bar{\mathbf{x}}_1)\} \text{ a.s. when }\tau_1(\bar{\mathbf{x}}_1)\neq0
    \end{align*}
    and
    \begin{align*}
         \mathcal{D}^{(k)}(\Bar{\mathbf{a}}_{k-1},\bar{\mathbf{x}}_k)\cdot\tau_k(\Bar{\mathbf{a}}_{k-1},\bar{\mathbf{x}}_k)=\max\{0,\tau_k(\Bar{\mathbf{a}}_{k-1},\bar{\mathbf{x}}_k)\} \text{ a.s. when }\tau_k(\Bar{\mathbf{a}}_{k-1},\bar{\mathbf{x}}_k)\neq0
    \end{align*}
    for $k=2,\dots,K$. For estimator 
    \begin{align*}
        \widehat{V}_{\mathcal{D}}^{(1)*}=\frac{1}{n}\sum_{i=1}^n \widehat{V}_i^{(1)}(\mathcal{D}(\mathbf{O}_i),\pi,\mu),
    \end{align*}
    we have $E[\widehat{V}_{\mathcal{D}}^{(1)*}]=V_0$ and 
    \begin{align*}
        &n\operatorname{Var}\left(\widehat{V}_{\mathcal{D}}^{(1)*}\right)\\
        = &  \operatorname{Var}\left\{\mu_1\left(\mathcal{D}^{(1)}(\bar{\mathbf{X}}^{(1)}), \bar{\mathbf{X}}^{(1)}\right)\right\}\\
        +& \mathrm{E}\left[\operatorname{Var}\left[\widehat{V}^{(2)}(\mathcal{D}(\mathbf{O}_i),\pi,\mu)\left|\bar{\mathbf{X}}^{(1)}\right.\right]\cdot\left(\frac{\mathbb{I}\left\{\tau_1\left(\bar{\mathbf{X}}^{(1)}\right)>0\right\}}{\pi_1\left(1, \bar{\mathbf{X}}^{(1)}\right)}+\frac{\mathbb{I}\left\{\tau_1\left(\bar{\mathbf{X}}^{(1)}\right)<0\right\}}{\pi_1\left(0, \bar{\mathbf{X}}^{(1)}\right)}\right)\right]\\
    +&\mathrm{E}\left[\operatorname{Var}\left[\widehat{V}^{(2)}(\mathcal{D}(\mathbf{O}_i),\pi,\mu)\left|\bar{\mathbf{X}}^{(1)}\right.\right]\cdot\left(\frac{(\mathcal{D}^{(1)}(\bar{\mathbf{X}}^{(1)}))^2}{\pi_1\left(1, \bar{\mathbf{X}}^{(1)}\right)}+\frac{(1-\mathcal{D}^{(1)}(\bar{\mathbf{X}}^{(1)}))^2}{\pi_1\left(0, \bar{\mathbf{X}}^{(1)}\right)}\right) \mathbb{I}\left\{\tau_1\left(\bar{\mathbf{X}}^{(1)}\right)=0\right\}\right]
    \end{align*}
    where for $k=2,\dots,K-1$ we have
     \begin{align*}
 &\operatorname{Var}\left[\widehat{V}^{(k)}(\mathcal{D}(\mathbf{O}_i),\pi,\mu)\right]\\
          =&\operatorname{Var}\left\{\mu_k\left(\mathcal{D}_k(\bar{\mathbf{A}}^{(k-1)}, \bar{\mathbf{X}}^{(k)}),\bar{\mathbf{A}}^{(k-1)}, \bar{\mathbf{X}}^{(k)}\right)\right\}\\
          +& \mathrm{E}\bigg[\operatorname{Var}\left[\widehat{V}^{(k+1)}(\mathcal{D}(\mathbf{O}_i),\pi,\mu)\left|\bar{\mathbf{A}}^{(k-1)}, \bar{\mathbf{X}}^{(k)}\right.\right]\\
          &\left(\frac{\mathbb{I}\left\{\tau_k\left(\bar{\mathbf{A}}^{(k-1)}, \bar{\mathbf{X}}^{(k)}\right)>0\right\}}{\pi_k\left((\bar{\mathbf{A}}^{(k-1)},1), \bar{\mathbf{X}}^{(k)}\right)}+\frac{\mathbb{I}\left\{\tau_k\left(\bar{\mathbf{A}}^{(k-1)}, \bar{\mathbf{X}}^{(k)}\right)<0\right\}}{\pi_k\left((\bar{\mathbf{A}}^{(k-1)},0), \bar{\mathbf{X}}^{(k)}\right)}\right)\bigg]\\
    +&\mathrm{E}\bigg[\operatorname{Var}\left[\widehat{V}^{(k+1)}(\mathcal{D}(\mathbf{O}_i),\pi,\mu)\left|\bar{\mathbf{A}}^{(k-1)}, \bar{\mathbf{X}}^{(k)}\right.\right]\\
    &\left(\frac{(\mathcal{D}^{(k)}(\bar{\mathbf{A}}^{(k-1)}, \bar{\mathbf{X}}^{(k)}))^2}{\pi_k\left((\bar{\mathbf{A}}^{(k-1)},1), \bar{\mathbf{X}}^{(k)}\right)}+\frac{(1-\mathcal{D}^{(k)}(\bar{\mathbf{A}}^{(k-1)}, \bar{\mathbf{X}}^{(k)}))^2}{\pi_k\left((\bar{\mathbf{A}}^{(k-1)},0), \bar{\mathbf{X}}^{(k)}\right)}\right) \\
    &\cdot\left.\mathbb{I}\left\{\tau_k\left(\bar{\mathbf{A}}^{(k-1)}, \bar{\mathbf{X}}^{(k)}\right)=0\right\}\right]
    \end{align*}
    and $\widehat{V}_i^{(K+1)}=Y_i$. Under Assumption \ref{B9}, the variance of $\widehat{V}_{\mathcal{D}}^{(1)*}$ can be minimized at
    \begin{align*}
        &\arg\min_{\mathcal{D}\in\bar{\mathbb{D}}(V_0)}n\operatorname{Var}\left(\widehat{V}_{\mathcal{D}}^{(1)*}\right)\\
=& \left\{ \begin{array}{ll}
 \mathcal{D}^{(1)}(\bar{\mathbf{X}}^{(1)})=\mathbb{I}\{\tau_1(\bar{\mathbf{X}}^{(1)})>0\}+\pi_1(1,\bar{\mathbf{X}}^{(1)})\mathbb{I}\{\tau_1(\bar{\mathbf{X}}^{(1)})=0\};\\
  \mathcal{D}^{(k)}(\bar{\mathbf{A}}^{(k-1)},\bar{\mathbf{X}}^{(k)})=\mathbb{I}\{\tau_k(\bar{\mathbf{A}}^{(k-1)},\bar{\mathbf{X}}^{(k)})>0\}\\+\pi_k((\bar{\mathbf{A}}^{(k-1)},1),\bar{\mathbf{X}}^{(k)})\mathbb{I}\{\tau_k(\bar{\mathbf{A}}^{(k-1)},\bar{\mathbf{X}}^{(k)})=0\} (2\leq k\leq K). 
\end{array} \right. 
    \end{align*} 
\end{proposition}
The optimality in Proposition \ref{multi-opt-oracle} can also be simply induced by Cauchy-Schwarz inequality. Motivated by the results in Proposition \ref{multi-opt-oracle}, we define the following estimator class.
\setcounter{definition}{2}
\begin{definition}
    $\mathcal{T}_{eal}^{multi}(V_0,\mathcal{O}_n)$ $:=$$\{T_n(\mathcal{O}_n)|$ $T_n(\mathcal{O}_n)$ is a function of $\mathcal{O}_n$ and $\exists$ $\mathcal{D}\in\bar{\mathbb{D}}(V_0)$ s.t. $T_n(\mathcal{O}_n) - \widehat{V}_{\mathcal{D}}^{(1)*} = o_p(n^{-1/2})$.
\end{definition}

The estimator class $\mathcal{T}_{eal}^{multi}(V_0,\mathcal{O}_n)$ is so general that it contains both our proposed estimator $\widehat{V}_{as}^{(1)}$, and the estimator $\widehat{V}_{sb}^{(1)}$ defined in the multiple time point case in \citet{shi2020breaking}.
\begin{corollary}\label{multi-embedding}
Under Assumptions~\ref{B1} to~\ref{B7} and~\ref{B9}, 
and for part~(ii) additionally under the assumptions of 
Theorem~3 in \citet{shi2020breaking}, we have:

    (i) $\widehat{V}_{as}^{(1)}-\widehat{V}_{\mathcal{D}}^{(1)*}
    =o_p(n^{-1/2})$ with 
    \begin{align*}    
    \mathcal{D}(\mathbf{X})
    =\left\{ \begin{array}{lll}
 \mathcal{D}^{(1)}(\bar{\mathbf{X}}^{(1)})&=\mathbb{I}\{\tau_1
 (\bar{\mathbf{X}}^{(1)})>0\}+t_0^1\cdot \mathbb{I}\{\tau_1
 (\bar{\mathbf{X}}^{(1)})=0\}\\
  \mathcal{D}^{(k)}(\bar{\mathbf{A}}^{(k-1)},
  \bar{\mathbf{X}}^{(k)})&=\mathbb{I}\{\tau_k
  (\bar{\mathbf{A}}^{(k-1)},\bar{\mathbf{X}}^{(k)})>0\}\\
  &+t_0^k\cdot\mathbb{I}\{\tau_k(\bar{\mathbf{A}}^{(k-1)},
  \bar{\mathbf{X}}^{(k)})=0\} (2\leq k\leq K). 
\end{array} \right.;
    \end{align*}
    (ii)  $\widehat{V}_{sb}^{(1)}-\widehat{V}_{\mathcal{D}}^{(1)*}
    =o_p(n^{-1/2})$ with
    \begin{align*}
    \mathcal{D}^{(1)}(\bar{\mathbf{X}}^{(1)})
    =\displaystyle{\lim_{n\to\infty}}d_{s_n}^1
    (\bar{\mathbf{X}}^{(1)}),\\
     \mathcal{D}^{(k)}(\bar{\mathbf{A}}^{(k-1)},
     \bar{\mathbf{X}}^{(k)})=\displaystyle{\lim_{n\to\infty}}
     d_{s_n}^k(\bar{\mathbf{A}}^{(k-1)},
     \bar{\mathbf{X}}^{(k)}),\\
        \text{ where } d_{s_n}^1(\bar{\mathbf{X}}^{(1)})
        =E[\mathbb{I}\{\widehat{\tau}_{[s_n]}^1
        (\bar{\mathbf{X}}^{(1)})>0\}], \\ 
        d_{s_n}^k(\bar{\mathbf{A}}^{(k-1)},
        \bar{\mathbf{X}}^{(k)})=E[\mathbb{I}\{
        \widehat{\tau}_{[s_n]}^k(\bar{\mathbf{A}}^{(k-1)},
        \bar{\mathbf{X}}^{(k)})>0\}]\text{ and } s_n=o(n).
    \end{align*}
\end{corollary}
Thus by the results above, as long as we can find an estimator which is asymptotically equivalent to $\widehat{V}_{\mathcal{D}}^{(1)*}$ with $\mathcal{D}$ s.t.
\begin{align*}
    \left\{ \begin{array}{ll}
 \mathcal{D}^{(1)}(\bar{\mathbf{X}}^{(1)})&=\mathbb{I}\{\tau_1(\bar{\mathbf{X}}^{(1)})>0\}+\pi_1(1,\bar{\mathbf{X}}^{(1)})\mathbb{I}\{\tau_1(\bar{\mathbf{X}}^{(1)})=0\}\\
  \mathcal{D}^{(k)}(\bar{\mathbf{A}}^{(k-1)},\bar{\mathbf{X}}^{(k)})&=\mathbb{I}\{\tau_k(\bar{\mathbf{A}}^{(k-1)},\bar{\mathbf{X}}^{(k)})>0\}\\
  &+\pi_k((\bar{\mathbf{A}}^{(k-1)},1),\bar{\mathbf{X}}^{(k)})\mathbb{I}\{\tau_k(\bar{\mathbf{A}}^{(k-1)},\bar{\mathbf{X}}^{(k)})=0\} (2\leq k\leq K),
\end{array} \right. 
\end{align*}
we are able to achieve the optimal efficiency (minimum variance) among the class $\mathcal{T}_{eal}^{multi}$, and consequently this estimator would be more efficient than the current state-of-the-art method, $\widehat{V}_{sb}^{(1)}$. To achieve this, it suffices to adjust the $t_0^k$'s in $d_{s}$ as follows, where the adjusted $d_{s}^k$'s are denoted as $d_{s}^{k\star}$, i.e., for any $\mathcal{I}_j$,
    \begin{align*}
        d_{s}^{1\star}(\bar{\mathbf{X}}_i^{(1)},\mathcal{I}_{j}):=d_{s}^1(\bar{\mathbf{X}}_i^{(1)};\widehat{\tau}_{1,\mathcal{I}_{j}},h_{n,\mathcal{I}_{j}}^1, \widehat{\pi}_{1,\mathcal{I}_{j}}(1,\bar{\mathbf{X}}^{(1)})),\\
       d_{s}^{k\star}(\bar{\mathbf{A}}^{(k-1)},\bar{\mathbf{X}}^{(k)},\mathcal{I}_{j}):=d_{s}^k(\Bar{\mathbf{A}}_{i}^{(k-1)},\bar{\mathbf{X}}_i^{(k)};\widehat{\tau}_{k,\mathcal{I}_{j}},h_{n,\mathcal{I}_{j}}^k,\widehat{\pi}_{k,\mathcal{I}_{j}}((\bar{\mathbf{A}}^{(k-1)},1),\bar{\mathbf{X}}^{(k)})),\\
       d_{as}^{\star}(\mathbf{O}_i,\mathcal{I}_{j})=\{ d_{s}^{1\star}(\bar{\mathbf{X}}_i^{(1)},\mathcal{I}_{j}),\dots, d_{s}^{k\star}(\bar{\mathbf{A}}^{(k-1)},\bar{\mathbf{X}}^{(k)},\mathcal{I}_{j}),\dots, d_{s}^{K\star}(\bar{\mathbf{A}}^{(K-1)},\bar{\mathbf{X}}^{(K)},\mathcal{I}_{j})\}
    \end{align*}
    and construct the following estimators:
    \begin{align*}
            \widehat{V}_{as}^{(1)}&:= \frac{1}{4}\sum_{j=1,2}\sum_{l=1,2}\frac{1}{|\mathcal{I}_{j,l}|}\sum_{i\in\mathcal{I}_{j,l}}\widehat{V}_i^{(1)}(d_{s}^{\star},\widehat{\pi}_{\mathcal{I}_{j,l}^c},\widehat{\mu}_{\mathcal{I}_{j,l}^c}),\\
    \widehat{\sigma}_{as}^{(1)}&:=\left(\frac{1}{n-1}\sum_{j=1,2}\sum_{l=1,2}\sum_{i\in\mathcal{I}_{j,l}}\left(\widehat{V}_i^{(1)}(d_{s}^{\star},\widehat{\pi}_{\mathcal{I}_{j,l}^c},\widehat{\mu}_{\mathcal{I}_{j,l}^c})-  \widehat{V}_{as}^{(1)}\right)^2\right)^{1/2}.
    \end{align*}
    Combine all above we can show that
\begin{corollary}\label{corob5}
Under Assumptions \ref{B1} to \ref{B7} and \ref{B9}, $\sqrt{n}(\widehat{V}_{as}^{(1)}-V_0)/\widehat{\sigma}_{as}^{(1)}$ converges weakly to a standard normal distribution. When all working models in $T_n$ are correct specified, and $\forall ~T_n\in\mathcal{T}_{eal}^{multi}\cup \bar{\mathbb{D}}(V_0)$ has asymptotic uniform integrability, we have with $(\widehat{\sigma}_{as}^{(1)})^2$ converges in probability to 
\begin{align*}
    &\inf_{T_n\in\mathcal{T}_{eal}^{multi}}\lim_{n\to\infty}\operatorname{Var}(\sqrt{n}T_n)=\inf_{\mathcal{D}\in\bar{\mathbb{D}}(V_0)}\lim_{n\to\infty}\operatorname{Var}(\sqrt{n}\widehat{V}_{\mathcal{D}}^{(1)*}).
\end{align*}
\end{corollary}
Similar to the single time point case, here we can also use RCFA to improve the finite sample performance. 
\begin{algorithm}[H]
\caption{Finite Sample Refinement: Repeatedly Cross Fitting and  Averaging}
    \begin{algorithmic}[1]
        \State \textbf{Input}: Data set $\mathcal{O}_n$; 
$B$ Random bipartitions of $[n]$: $\mathcal{I}_{b,1}$ and $\mathcal{I}_{b,2}$ ($b=1,\dots,B$), and randomly picked subsets $\mathcal{I}_{b,j,k}$ ($k=1,2$) of $\mathcal{I}_{b,j}$ ($j=1,2$), where all the $\mathcal{I}_{b,j,k}$ with the equal size; Parameters $h_{n,\mathcal{I}_{b,j}}$, $j=1,2$, and $t_0$; Working models: $\widehat{\pi}$, $\widehat{\mu}$, $\widehat{D}$.
        \State \textbf{Construction of 
 The Adaptive Smoothing Estimator:} 
 \begin{align}
     \widehat{V}_{as,B}^{(1)}&:= \frac{1}{4B}\sum_{b=1}^B \sum_{j=1,2}\sum_{k=1,2}\frac{1}{|\mathcal{I}_{b,j,k}|}\sum_{i\in\mathcal{I}_{b,j,k}}\widehat{V}_i^{(1)}(d_{s}^\star(\mathbf{O}_i,\mathcal{I}_{b,3-j}),\widehat{\pi}_{\mathcal{I}_{b,j,k}^c},\widehat{\mu}_{\mathcal{I}_{b,j,k}^c})\label{mul-opt-est-cv}
 \end{align}
 \State \textbf{Construction of The Confidence Interval} The standard deviation estimate:
 \begin{align}
     \widehat{\sigma}_{as,B}^{(1)}&:=\bigg(\frac{1}{B\cdot(n-1)}\sum_{b=1}^B\sum_{j=1,2}\sum_{k=1,2}\sum_{i\in\mathcal{I}_{b,j,k}}\notag\\&\left(\widehat{V}_i^{(1)}(d_{s}^\star(\mathbf{O}_i,\mathcal{I}_{b,3-j}),\widehat{\pi}_{\mathcal{I}_{b,j,k}^c},\widehat{\mu}_{\mathcal{I}_{b,j,k}^c})-\widehat{V}_{as,B}^{(1)}\right)^2\bigg)^{1/2}\label{mul-opt-std-cv}
 \end{align}
 and thus the two-sided level-{$\eta$} confidence interval can be given as:
\begin{align}\label{mul-CI-opt-cv}
     \left[\widehat{V}_{as,B}^{(1)}-{z_{\eta/2}}\cdot\frac{\widehat{\sigma}_{as,B}^{(1)}}{\sqrt{n}},\widehat{V}_{as,B}^{(1)}+{z_{\eta/2}}\cdot\frac{\widehat{\sigma}_{as,B}^{(1)}}{\sqrt{n}}\right]
 \end{align}
    \end{algorithmic}
\end{algorithm}
\begin{corollary}\label{corob7}
    Under Assumptions \ref{B1} to \ref{B7}, $\sqrt{n}\left(\widehat{V}_{as}^{(1)}-\widehat{V}_{as,B}^{(1)}\right)=o_p(1)$ as $B\to\infty$.
\end{corollary}
\subsection{Efficiency Optimality in Multiple Time Point Case}
In this section, we provide extension of the robust asymptotically linear unbiased estimation class in the multiple time point case, as a further support of the optimality of our proposed estimator.
\begin{definition}[Robust Asymptotically Linear Unbiased Estimator 
in Multiple Time Point Case]\label{multi-ralu}
    This class $\mathcal{T}_{RL}^{multi}$ contains estimators 
    $T_n^{(1)}$ defined sequentially as follows. For $k=1,\dots,K$, 
    there exist measurable functions $f^{(k)}$, and for all $P$ in the 
    set $\mathcal{P}^{multi}$ of distributions satisfying 
    Assumptions~\ref{B1} to~\ref{B3},
    \begin{align*}
        T_n^{(k)}=\frac{1}{n}\sum_{i=1}^n 
        f^{(k)}\Big(\bar{\mathbf{X}}_i^{(k)},\bar{\mathbf{A}}_i^{(k)}, 
        f^{(k+1)},\mu_1^{(k)}(\bar{\mathbf{A}}_i^{(k-1)},
        \bar{\mathbf{X}}_i^{(k)}),
        \mu_0^{(k)}(\bar{\mathbf{A}}_i^{(k-1)},
        \bar{\mathbf{X}}_i^{(k)}),\\
        \pi_1^{(k)}(\bar{\mathbf{A}}_i^{(k-1)},
        \bar{\mathbf{X}}_i^{(k)}),
        \operatorname{sgn}\left(\tau_{k,P}\left(
        \bar{\mathbf{A}}_i^{(k-1)},
        \bar{\mathbf{X}}_i^{(k)}\right)\right),\\
        r_1^{(k)}(\bar{\mathbf{A}}_i^{(k-1)},
        \bar{\mathbf{X}}_i^{(k)}),
        r_0^{(k)}(\bar{\mathbf{A}}_i^{(k-1)},
        \bar{\mathbf{X}}_i^{(k)})
        \Big)+o_p(n^{-1/2})
    \end{align*}
    and $f^{(K+1)}=Y$, where 
    $\tau_{k,P}(\bar{\mathbf{A}}^{(k-1)},\bar{\mathbf{X}}^{(k)})
    =\mu_{k,P}((\bar{\mathbf{A}}^{(k-1)},1),\bar{\mathbf{X}}^{(k)})
    -\mu_{k,P}((\bar{\mathbf{A}}^{(k-1)},0),\bar{\mathbf{X}}^{(k)})$ 
    is the population-true stage-$k$ contrast function under~$P$, 
    and $\mu_a^{(k)}$, $\pi_1^{(k)}$ and $r_a^{(k)}$ are the 
    working models for $E_P[\widehat{V}^{(k+1)}|
    (\bar{\mathbf{A}}^{(k-1)},a),\bar{\mathbf{X}}^{(k)}]$, 
    $P(A^{(k)}=1|\bar{\mathbf{A}}^{(k-1)},\bar{\mathbf{X}}^{(k)})$ 
    and $\operatorname{Var}_P[\widehat{V}^{(k+1)}|
    (\bar{\mathbf{A}}^{(k-1)},a),\bar{\mathbf{X}}^{(k)}]$, 
    respectively, for $a=0,1$. 
    Moreover, for any $P\in\mathcal{P}^{multi}$, $f^{(k)}$ satisfies
    \begin{align*}
        E_P[f^{(k)}] = V_{0,P}^{(k)}
    \end{align*}
    as long as one of the following equations holds $P$-a.s.:
    \begin{align*}
        E_P[\widehat{V}^{(k+1)}|\bar{\mathbf{A}}^{(k)},
        \bar{\mathbf{X}}^{(k)}] 
        &= A^{(k)} \mu_1^{(k)}(\bar{\mathbf{A}}^{(k-1)},
        \bar{\mathbf{X}}^{(k)})
        +(1-A^{(k)})\mu_0^{(k)}(\bar{\mathbf{A}}^{(k-1)},
        \bar{\mathbf{X}}^{(k)}),\\
        E_P[A^{(k)}|\bar{\mathbf{A}}^{(k-1)},\bar{\mathbf{X}}^{(k)}]
        &=\pi_1^{(k)}(\bar{\mathbf{A}}^{(k-1)},\bar{\mathbf{X}}^{(k)}).
    \end{align*}
\end{definition}
Note that here $\mu_1^{(k)},\allowbreak\mu_0^{(k)},\allowbreak\pi_1^{(k)}$ denote the working model and $\mu_k(\bar{\mathbf{A}}^{(k)},\allowbreak\bar{\mathbf{X}}^{(k)})$ and $\pi_k(\{\bar{\mathbf{A}}^{(k-1)},\allowbreak 1\},\allowbreak\bar{\mathbf{X}}^{(k)})$ denote the population truth. Similar to the single time point case, $\mathcal{T}_{RL}^{multi}$ contains $\mathcal{T}_{eal}^{multi}$.
\begin{corollary}\label{hyb-general}
    $\mathcal{T}_{eal}^{multi}\subset\mathcal{T}_{RL}^{multi}$ whenever it is regular or nonregular.
\end{corollary}
As a result, $\widehat{V}_{as}^{(1)}$ belongs to $\mathcal{T}_{RL}^{multi}$ and moreover attains the variance lower bound of $\mathcal{T}_{RL}^{multi}$ whenever it is regular case or not.
\begin{theorem}\label{hyb-var-bound}
     Under Assumptions \ref{B1} to \ref{B7} and \ref{B9}, 
     $\widehat{V}_{as}^{(1)}\in\mathcal{T}_{RL}^{multi}$ and 
     when the working models $\mu_1^{(k)}$, $\mu_0^{(k)}$ and 
     $\pi_1^{(k)}$ in $T_n$ are correctly specified for all 
     $k=1,\dots,K$, and $T_n$ satisfies asymptotic uniform 
     integrability, i.e., 
     $\lim_{M \rightarrow \infty} \sup_n 
     E\left[n\left(T_n^{(1)}-V_0\right)^2 \cdot 
     \mathbf{1}\left(n\left(T_n^{(1)}-V_0\right)^2>M\right)
     \right]=0$, we have
    \begin{align*}
        \inf_{T_n\in\mathcal{T}_{RL}^{multi}}
        \lim_{n\to\infty}\operatorname{Var}(\sqrt{n}T_n^{(1)})
        = \lim_{n\to\infty}\operatorname{Var}
        (\sqrt{n}\widehat{V}_{as}^{(1)}).
    \end{align*}
\end{theorem}

\subsection{Simulations of Multiple Time Point Study}
For the multiple time point cases, we consider the following model of two stages:
\begin{align*}
    Y=\Phi\left(X_{1}^{(1)}, X_{2}^{(1)}, A^{(1)}, X^{(2)}\right)+A^{(2)} \tau\left(X_{1}^{(1)}, X_{2}^{(1)}, A^{(1)}, X^{(2)}\right)+e^{(2)}, \\
X^{(2)}=A^{(1)} X_{1}^{(1)}+e^{(1)},
\end{align*}
where $X_{1}^{(1)}$ and $X_{2}^{(1)}$ are the baseline covariates, $A^{(1)}$ and $A^{(2)}$ denote the first and second treatment a patient receives at two time points, $X^{(2)}$ stands for the intermediate covariate collected between two time points. Variables $A^{(1)}, A^{(2)}, X_{1}^{(1)}, X_{2}^{(1)}, e^{(1)}$ and $e^{(2)}$ are all independent. In addition, we assume $A^{(1)}\sim \operatorname{Ber}(\pi_1(X_{1}^{(1)}, X_{2}^{(1)}))$, $A^{(2)} \sim \operatorname{Ber}(\pi_2(X_{1}^{(1)}, X_{2}^{(1)})), e^{(1)}, e^{(2)} \sim N(0,0.5^2)$ and $X_{1}^{(1)}, X_{2}^{(1)} \sim \operatorname{Unif}[-2,2]$ where Unif $[a, b]$ denotes the uniform distribution on the interval $[a, b]$.

Here we consider three scenarios. The functional forms of $\Phi$ and $\tau$ and the optimal value function $V_0$ under these scenarios are reported in the following table. With similar analysis in \citet{shi2020breaking}, we can show that in scenario (F) and (G) the optimal treatment regimes are not uniquely defined.
\begin{table}[H]
    \centering
    \begin{tabular}{cccc}
\hline \hline & $(\mathrm{F})$ & $(\mathrm{G})$ & $(\mathrm{H})$ \\
\hline$\Phi\left(x_{1,1}, x_{1,2}, a_1, x_2\right)$ & $x_{1,1}^2-a_1\left(0.25+x_{1,1}^2\right)$ & $x_2^2$ & 0 \\
\hline$\tau\left(x_{1,1}, x_{1,2}, a_1, x_2\right)$ & $a_1 x_2^2$ & 0 & $x_2^2$ \\
\hline $\pi_1(x_{1,1},x_{1,2})$ & $0.5+0.1 x_{1,1}$ & 0.5 & 0.5\\
\hline $\pi_2(x_{1,1},x_{1,2})$ & 0.5 & $0.5+0.1 x_{1,1}$ & $0.5+0.1 x_{1,1}$\\
\hline$V_0$ & 1.33 & 1.58 & 1.58 \\
\hline
 & nonregular & nonregular & regualr \\
\hline
\end{tabular}
    \caption{Simulation settings of the multiple time point case.}
    \label{multi-setting}
\end{table}
\begin{table}[H]
    \centering
   \begin{tabular}{|c|c|c|c|c|c|c|}
\hline Methods & \multicolumn{2}{c|}{ Proposed} & \multicolumn{2}{c}{ Subbagging } & \multicolumn{2}{|c|}{ SSS } \\
\hline Settings  & $\operatorname{ECP}(\%)$ & AL*100 & $\operatorname{ECP}(\%)$ & AL*100 & $\operatorname{ECP}(\%)$ & AL*100 \\
\hline (F) &  $94.1$ & $36.1$ & $90.8$ & $38.1$ & $94.0$ & $58.6$\\
\hline (G) & $94.0$ & $28.8$ & $94.2$ & $30.0$ & $92.8$ & $64.1$\\
\hline (H) &  $94.6$ & $39.5$ & $92.4$ & $39.6$ & $92.2$ & $65.4$\\
\hline
\end{tabular}
    \caption{ECP and AL of the CIs.}
    \label{Multi-sim}
\end{table}
We still use cubic spline to estimate the propensity scores $\pi$ and conditional mean functions $\mu$ in each stage, and the truncation for $\pi$ in each stage is the same as in the single time point case. In this simulation, the tuning hyperparameters, i.e., ${C_0}$ for our proposed method and $K_0$ for subbagging, take the same values as in the single time point case. The resampling times $B$ for all the methods are 1000, the sample size $n=600$ and the observed results including Empirical Coverage Probability (ECP) and Average Length of confidence intervals (AL) are aggregated over 500 Monte Carlo samples.

From Table \ref{Multi-sim}, we see that the proposed method aims for the nominal coverage probability; i.e. 95\%, well in the regular case (H), and nonregular cases, Scenarios (F) and (G), which is in accordance with the asymptotic theory in Corollary \ref{multi-embedding} and Corollary \ref{corob5}. The results in Table \ref{Multi-sim} demonstrates that in all the scenarios whenever it is regular or not, the proposed method has shortest confidence interval among all the competing methods. In particular, in the nonregular cases, Scenarios (F) and (G), the efficiency gain of the proposed method over the current state-of-the-art method, subbagging, is significant with about 5\% reduction of average length from the proposed method over that from subbaging.

\section{Proofs of Theoretical Results in Multiple Time Point Case}\label{appC}
In this section, we start with stating and proving a general technical lemma and then apply it to justify the rest theoretical results in the multiple time point case.
\subsection{A Technical Lemma and Its Proof}
Still for the purpose of generalization, we consider a multi-dimensional function sequence $\{\mathcal{D}_n\}_{n\in\mathbb{N}_+}$ ($\mathcal{D}_n:\mathcal{O}\times\mathbb{R}\to[0,1]$), where
\begin{align*}
    \mathcal{D}_n^{(1)}:\Bar{\mathcal{X}}^{(1)}\times \mathbb{R}&\to[0,1],\\
    \mathcal{D}_n^{(k)}:\{0,1\}^{k-1}\times\Bar{\mathcal{X}}^{(k)}\times\mathbb{R}&\to[0,1]\text{ for }k=2,\dots,K.
\end{align*}
Consider the following assumptions. Without loss of generality, we denote $\bar{\mathbf{A}}^{0}\equiv0$.
\begin{assumption}[Uniform continuity]\label{B10}
    $\forall k$,   $\exists$ $\delta_n^k$ s.t.  $P(|\widehat{\tau}_{k,\mathcal{I}}(\bar{\mathbf{A}}^{(k-1)},\bar{\mathbf{X}}^{(k)})-\tau_k(\bar{\mathbf{A}}^{(k-1)},\bar{\mathbf{X}}^{(k)})|\ge \delta_n^k)=o_p(1)$ and $\sup_{|s-t|\leq \delta_n^k}|\mathcal{D}_n^{(k)}(\bar{\mathbf{A}}^{(k-1)},\bar{\mathbf{X}}^{(k)},s)-\mathcal{D}_n^{(k)}(\bar{\mathbf{A}}^{(k-1)},\bar{\mathbf{X}}^{(k)},t)|\to 0$ as $n\to\infty$.
\end{assumption}
\begin{assumption}[Consistency]\label{B11}
    For $\mathcal{D}=(\mathcal{D}^{(1)},\dots,\mathcal{D}^{(K)})$ where
\begin{align*}
    \mathcal{D}^{(k)}(\bar{\mathbf{A}}^{(k-1)},\bar{\mathbf{X}}^{(k)})=\displaystyle{\lim_{n\to\infty}}\mathcal{D}_n^{(k)}(\bar{\mathbf{A}}^{(k-1)},\bar{\mathbf{X}}^{(k)},\tau_k(\bar{\mathbf{A}}^{(k-1)},\bar{\mathbf{X}}^{(k)})),
\end{align*}
we have $\mathcal{D}\in\bar{\mathbb{D}}(V_0)$.
\end{assumption}
\begin{assumption}[Concentration]\label{B12}
    {For every $k$, there exist} $\delta_n^k$ and $\delta_n^{\prime k}$ satisfying $$\delta_n^k\lor n^{1/4}\left(E[|\widehat{\tau}_{k,\mathcal{I}}(\bar{\mathbf{A}}^{(k-1)},\bar{\mathbf{X}}^{(k)})-\tau_k(\bar{\mathbf{A}}^{(k-1)},\bar{\mathbf{X}}^{(k)})|^2]\right)^{3/4}  \prec_p\delta_n^{\prime k}\prec n^{-\frac{1}{2(1+\alpha)}}$$ s.t. 
\begin{align}
    |{\mathcal{D}_n^{(k)}}(\bar{\mathbf{A}}^{(k-1)},\bar{\mathbf{X}}^{(k)},s)-{\mathbb{I}\{s>0\}}|=o\left(\frac{1}{{|s|}\sqrt{n}}\right)\label{B10-1}
\end{align}
when $|s|>\delta_n^{\prime k}$.
\end{assumption}
Above three assumptions are set on duality to Assumptions \ref{A10} to \ref{A12} in the single time point case. Define 
\begin{align}
\widehat{V}_{\mathcal{D}_n}^{(1)*}=&\frac{1}{n}\sum_{i=1}^n \widehat{V}_i^{(1)}(\mathcal{D}_n(\mathbf{O}_i,\tau),\pi,\mu),\text{ ($\tau$, $\pi$ and $h$ are all true)}\label{multi_VD_oracle}\\
    \widehat{V}_{\mathcal{D}_n}^{(1)}=&\frac{1}{4}\sum_{j=1,2}\sum_{l=1,2}\frac{1}{|\mathcal{I}_{j,l}|}\sum_{i\in\mathcal{I}_{j,l}}\widehat{V}_i^{(1)}(\mathcal{D}_n(\mathbf{O}_i,\widehat{\tau}_{\mathcal{I}_{3-j}}),\widehat{\pi}_{\mathcal{I}_{j,l}^c},\widehat{\mu}_{\mathcal{I}_{j,l}^c})\label{multi_VD}
\end{align}
and
\begin{align*}
    \widehat{\sigma}_{\mathcal{D}_n}^{(1)}:=\left(\frac{1}{n-1}\sum_{j=1,2}\sum_{l=1,2}\sum_{i\in\mathcal{I}_{j,l}}\left(\widehat{V}_i^{(1)}(\mathcal{D}_n(\mathbf{O}_i,\widehat{\tau}_{\mathcal{I}_{3-j}}),\widehat{\pi}_{\mathcal{I}_{j,l}^c},\widehat{\mu}_{\mathcal{I}_{j,l}^c})- \widehat{V}_{\mathcal{D}_n}^{(1)}\right)^2\right)^{1/2},
\end{align*}
\begin{lemma}\label{multi-lem}
      Under Assumptions \ref{B1} to \ref{B4}, \ref{B7} and \ref{B10} to \ref{B12}, the stagewise moment-comparability condition in Eq. \eqref{multi-moment-comparison}, {and the stage-specific fitted-overlap condition stated in Section A.1,} Eq. \eqref{multi_VD_oracle} and Eq. \eqref{multi_VD} are asymptotically equivalent: $|\widehat{V}_{\mathcal{D}_n}^{(1)*} - \widehat{V}_{\mathcal{D}_n}^{(1)}|=o_p(n^{-1/2})$, and thus $\sqrt{n}(\widehat{V}_{\mathcal{D}_n}^{(1)}-V_0)/\widehat{\sigma}_{\mathcal{D}_n}^{(1)}$ converges weakly to a standard normal distribution and $(\widehat{\sigma}_{\mathcal{D}_n})^2$ converges in probability to
    \begin{align*}
          \operatorname{Var}\left\{\mu_1\left(d_1, \bar{\mathbf{X}}^{(1)}\right)\right\}+\operatorname{Var}\left[\widehat{V}^{(2)}(\mathcal{D}(\mathbf{O}_i),\pi,\mu)\right]\\\cdot \mathrm{E}\left(\frac{\mathbb{I}\left\{\tau_1\left(\bar{\mathbf{X}}^{(1)}\right)>0\right\}}{\pi_1\left(1, \bar{\mathbf{X}}^{(1)}\right)}+\frac{\mathbb{I}\left\{\tau_1\left(\bar{\mathbf{X}}^{(1)}\right)<0\right\}}{\pi_1\left(0, \bar{\mathbf{X}}^{(1)}\right)}\right)\\
    +\operatorname{Var}\left[\widehat{V}^{(2)}(\mathcal{D}(\mathbf{O}_i),\pi,\mu)\right]\\\cdot\mathrm{E}\left[\left(\frac{(\mathcal{D}^{(1)}(\tau_1(\bar{\mathbf{X}}^{(1)})))^2}{\pi_1\left(1, \bar{\mathbf{X}}^{(1)}\right)}+\frac{(1-\mathcal{D}^{(1)}(\tau_1(\bar{\mathbf{X}}^{(1)})))^2}{\pi_1\left(0, \bar{\mathbf{X}}^{(1)}\right)}\right) \mathbb{I}\left\{\tau_1\left(\bar{\mathbf{X}}^{(1)}\right)=0\right\}\right].
    \end{align*}
\end{lemma}
\begin{proof}
This lemma establishes the asymptotic equivalence between the cross-fitted estimator and the oracle estimator in the multi-stage setting. The proof uses mathematical induction on the number of stages. Throughout, Assumptions~\ref{B1} to~\ref{B3} (SUTVA, no unmeasured confounders, and overlap) ensure the identification of the optimal value function $V_0$ and the validity of the AIPW-type recursive estimator $\widehat{V}_i^{(k)}$ at each stage, so that the oracle estimator $\widehat{V}_{\mathcal{D}_n}^{(1)*}$ is well-defined and unbiased for $V_0$.

Let $\mathcal{D}_n = (\mathcal{D}_n^{(1)}, \ldots, \mathcal{D}_n^{(K)})$ be a sequence of multi-stage decision functions satisfying Assumptions \ref{B10} to \ref{B12}.

Define
\begin{align}
    \widehat{V}_{\mathcal{D}_n}^{(1)*} &= \frac{1}{n}\sum_{i=1}^n \widehat{V}_i^{(1)}(\mathcal{D}_n(\mathbf{O}_i,\tau),\pi,\mu)\label{multi-oracle-est}\\
    \widehat{V}_{\mathcal{D}_n}^{(1)} &= \frac{1}{4}\sum_{j=1,2}\sum_{l=1,2}\frac{1}{|\mathcal{I}_{j,l}|}\sum_{i\in\mathcal{I}_{j,l}}\widehat{V}_i^{(1)}(\mathcal{D}_n(\mathbf{O}_i,\widehat{\tau}_{\mathcal{I}_{3-j}}),\widehat{\pi}_{\mathcal{I}_{j,l}^c},\widehat{\mu}_{\mathcal{I}_{j,l}^c})\label{multi-cross-est}
\end{align}
We need to show: $|\widehat{V}_{\mathcal{D}_n}^{(1)*} - \widehat{V}_{\mathcal{D}_n}^{(1)}| = o_p(n^{-1/2})$.

We prove by strong induction that for all $k = K, K-1, \ldots, 1$:
\begin{align}
    \frac{1}{|\mathcal{I}_{j,l}|}\sum_{i\in\mathcal{I}_{j,l}}\left[\widehat{V}_i^{(k)}(\mathcal{D}_n(\mathbf{O}_i,\widehat{\tau}_{\mathcal{I}_{3-j}}),\widehat{\pi}_{\mathcal{I}_{j,l}^c},\widehat{\mu}_{\mathcal{I}_{j,l}^c}) - \widehat{V}_i^{(k)}(\mathcal{D}(\mathbf{O}_i),\pi,\mu)\right] = o_p(n^{-1/2})\label{multi-induct-hyp}
\end{align}

For the base case ($k = K$), $\widehat{V}_i^{(K+1)} = Y_i$. We have:
\begin{align}
    &\widehat{V}_i^{(K)}(\mathcal{D}_n(\mathbf{O}_i,\widehat{\tau}_{\mathcal{I}_{3-j}}),\widehat{\pi}_{\mathcal{I}_{j,l}^c},\widehat{\mu}_{\mathcal{I}_{j,l}^c}) - \widehat{V}_i^{(K)}(\mathcal{D}(\mathbf{O}_i),\pi,\mu)\notag\\
    &= \underbrace{\widehat{V}_i^{(K)}(\mathcal{D}_n(\mathbf{O}_i,\widehat{\tau}_{\mathcal{I}_{3-j}}),\widehat{\pi}_{\mathcal{I}_{j,l}^c},\widehat{\mu}_{\mathcal{I}_{j,l}^c}) - \widehat{V}_i^{(K)}(\mathcal{D}_n(\mathbf{O}_i,\widehat{\tau}_{\mathcal{I}_{3-j}}),\pi,\mu)}_{T_1^{(K)}}\label{multi-base-T1}\\
    &\quad + \underbrace{\widehat{V}_i^{(K)}(\mathcal{D}_n(\mathbf{O}_i,\widehat{\tau}_{\mathcal{I}_{3-j}}),\pi,\mu) - \widehat{V}_i^{(K)}(\mathcal{D}_n(\mathbf{O}_i,\tau),\pi,\mu)}_{T_2^{(K)}}\label{multi-base-T2}\\
    &\quad + \underbrace{\widehat{V}_i^{(K)}(\mathcal{D}_n(\mathbf{O}_i,\tau),\pi,\mu) - \widehat{V}_i^{(K)}(\mathcal{D}(\mathbf{O}_i),\pi,\mu)}_{T_3^{(K)}}\label{multi-base-T3}
\end{align}
We analyze each term in turn.

For $T_1^{(K)}$, which captures the difference due to estimating nuisance parameters $\pi_K$ and $\mu_K$, we expand using the recursive definition of $\widehat{V}_i^{(K)}$:
\begin{align}
    T_1^{(K)} &= \frac{\mathrm{g}\{A_i^{(K)}, \mathcal{D}_n^{(K)}(\bar{\mathbf{A}}_i^{(K-1)},\bar{\mathbf{X}}_i^{(K)},\widehat{\tau}_{K,\mathcal{I}_{3-j}})\}}{\widehat{\pi}_{K,\mathcal{I}_{j,l}^c}(\bar{\mathbf{A}}_i^{(K)},\bar{\mathbf{X}}_i^{(K)})}\{Y_i - \widehat{\mu}_{K,\mathcal{I}_{j,l}^c}(\bar{\mathbf{A}}_i^{(K)},\bar{\mathbf{X}}_i^{(K)})\}\notag\\
    &\quad - \frac{\mathrm{g}\{A_i^{(K)}, \mathcal{D}_n^{(K)}(\bar{\mathbf{A}}_i^{(K-1)},\bar{\mathbf{X}}_i^{(K)},\widehat{\tau}_{K,\mathcal{I}_{3-j}})\}}{\pi_K(\bar{\mathbf{A}}_i^{(K)},\bar{\mathbf{X}}_i^{(K)})}\{Y_i - \mu_K(\bar{\mathbf{A}}_i^{(K)},\bar{\mathbf{X}}_i^{(K)})\}\notag\\
    &\quad + \mathcal{D}_n^{(K)}(\bar{\mathbf{A}}_i^{(K-1)},\bar{\mathbf{X}}_i^{(K)},\widehat{\tau}_{K,\mathcal{I}_{3-j}}) \widehat{\mu}_{1,K,\mathcal{I}_{j,l}^c}(\bar{\mathbf{A}}_i^{(K-1)},\bar{\mathbf{X}}_i^{(K)})\notag\\
&\quad + (1-\mathcal{D}_n^{(K)}(\bar{\mathbf{A}}_i^{(K-1)},\bar{\mathbf{X}}_i^{(K)},\widehat{\tau}_{K,\mathcal{I}_{3-j}})) \widehat{\mu}_{0,K,\mathcal{I}_{j,l}^c}(\bar{\mathbf{A}}_i^{(K-1)},\bar{\mathbf{X}}_i^{(K)})\notag\\
&\quad - \frac{\mathrm{g}\{A_i^{(K)}, \mathcal{D}_n^{(K)}(\bar{\mathbf{A}}_i^{(K-1)},\bar{\mathbf{X}}_i^{(K)},\widehat{\tau}_{K,\mathcal{I}_{3-j}})\}}{\pi_K(\bar{\mathbf{A}}_i^{(K)},\bar{\mathbf{X}}_i^{(K)})} \{Y_i - \mu_K(\bar{\mathbf{A}}_i^{(K)},\bar{\mathbf{X}}_i^{(K)})\}\notag\\
&\quad - \mathcal{D}_n^{(K)}(\bar{\mathbf{A}}_i^{(K-1)},\bar{\mathbf{X}}_i^{(K)},\widehat{\tau}_{K,\mathcal{I}_{3-j}}) \mu_{1,K}(\bar{\mathbf{A}}_i^{(K-1)},\bar{\mathbf{X}}_i^{(K)})\notag\\
&\quad - (1-\mathcal{D}_n^{(K)}(\bar{\mathbf{A}}_i^{(K-1)},\bar{\mathbf{X}}_i^{(K)},\widehat{\tau}_{K,\mathcal{I}_{3-j}})) \mu_{0,K}(\bar{\mathbf{A}}_i^{(K-1)},\bar{\mathbf{X}}_i^{(K)})
\end{align}

Following the decomposition used in the single-stage case (similar to equations in the proof of Lemma~\ref{single-lem}), using the fact that $\mathcal{D}_n^{(K)}$ takes values in $[0,1]$, the overlap condition in Assumption~\ref{B3} {and the stage-specific fitted-overlap condition stated in Section A.1}{,} which {bound} the {true and fitted} propensity scores uniformly away from $0$ and $1$, and the conditional independence in Assumption~\ref{B2} which ensures 
$E[Y_i - \mu_K(\bar{\mathbf{A}}_i^{(K)},\bar{\mathbf{X}}_i^{(K)}) \mid \bar{\mathbf{A}}_i^{(K)}, \bar{\mathbf{X}}_i^{(K)}] = 0$, we can show that:
\begin{align}
    \frac{1}{|\mathcal{I}_{j,l}|}\sum_{i\in\mathcal{I}_{j,l}} T_1^{(K)} = o_p(n^{-1/2})
\end{align}
by Assumption \ref{B7} (doubly robust convergence rates).

For $T_2^{(K)}$, which captures the difference due to estimating $\tau_K$, we have:
\begin{align}
T_2^{(K)} &= (\mathcal{D}_n^{(K)}(\bar{\mathbf{A}}_i^{(K-1)},\bar{\mathbf{X}}_i^{(K)},\widehat{\tau}_{K,\mathcal{I}_{3-j}}) - \mathcal{D}_n^{(K)}(\bar{\mathbf{A}}_i^{(K-1)},\bar{\mathbf{X}}_i^{(K)},\tau_K))\notag\\
&\quad \times \left[\frac{(2A_i^{(K)}-1)}{\pi_K(\bar{\mathbf{A}}_i^{(K)},\bar{\mathbf{X}}_i^{(K)})} \{Y_i - \mu_K(\bar{\mathbf{A}}_i^{(K)},\bar{\mathbf{X}}_i^{(K)})\}\right.\notag\\
&\quad \left.+ \tau_K(\bar{\mathbf{A}}_i^{(K-1)},\bar{\mathbf{X}}_i^{(K)})\right]
\end{align}
where we used Assumptions~\ref{B1} and~\ref{B2} (SUTVA and no unmeasured confounders) to identify $\mu_{1,K}-\mu_{0,K}=\tau_K$.

We decompose the sum based on whether $|\tau_K(\bar{\mathbf{A}}_i^{(K-1)},\bar{\mathbf{X}}_i^{(K)})|$ is large or small relative to $\delta_n^{\prime K}$.

In Case 1, when $|\tau_K(\bar{\mathbf{A}}_i^{(K-1)},\bar{\mathbf{X}}_i^{(K)})| \leq 2\delta_n^{\prime K}$, by Assumption \ref{B4} (margin condition):
\begin{align}
    \frac{1}{|\mathcal{I}_{j,l}|}\sum_{i\in\mathcal{I}_{j,l}, |\tau_K(\bar{\mathbf{A}}_i^{(K-1)},\bar{\mathbf{X}}_i^{(K)})| \leq 2\delta_n^{\prime K}} |T_2^{(K)}| &\leq \frac{2}{|\mathcal{I}_{j,l}|}\sum_{i\in\mathcal{I}_{j,l}} |\tau_K(\bar{\mathbf{A}}_i^{(K-1)},\bar{\mathbf{X}}_i^{(K)})| \mathbb{I}\{|\tau_K| \leq 2\delta_n^{\prime K}\}\notag\\
    &= O_p((2\delta_n^{\prime K})^{1+\alpha}) + o_p(n^{-1/2}) = o_p(n^{-1/2})
\end{align}
where the last step uses $\delta_n^{\prime K} \prec n^{-\frac{1}{2(1+\alpha)}}$ from Assumption~\ref{B12}.

In Case 2, when $|\tau_K(\bar{\mathbf{A}}_i^{(K-1)},\allowbreak\bar{\mathbf{X}}_i^{(K)})| > 2\delta_n^{\prime K}$, we further split based on whether $|\widehat{\tau}_{K,\mathcal{I}_{3-j}}(\bar{\mathbf{A}}_i^{(K-1)},\allowbreak\bar{\mathbf{X}}_i^{(K)}) - \allowbreak\tau_K(\bar{\mathbf{A}}_i^{(K-1)},\allowbreak\bar{\mathbf{X}}_i^{(K)})| \leq \delta_n^{\prime K}$.

When the estimation error is small and both $\tau_K$ and $\widehat{\tau}_{K,\mathcal{I}_{3-j}}$ exceed $\delta_n^{\prime K}$ in absolute value, by Assumption \ref{B12} (concentration):
\begin{align}
    |\mathcal{D}_n^{(K)}(\bar{\mathbf{A}}_i^{(K-1)},\bar{\mathbf{X}}_i^{(K)},\widehat{\tau}_{K,\mathcal{I}_{3-j}}) - \mathcal{D}_n^{(K)}(\bar{\mathbf{A}}_i^{(K-1)},\bar{\mathbf{X}}_i^{(K)},\tau_K)| &= o\left(\frac{1}{\sqrt{n}|\tau_K|}\right)
\end{align}

When the estimation error $e_K=\widehat{\tau}_{K,\mathcal{I}_{3-j}}-\tau_K$ is large, Eq. \eqref{multi-moment-comparison} gives
\[
E_{\mathbf H_K}[|e_K|\mathbb I\{|e_K|>\delta_n^{\prime K}\}]
\le \frac{E_{\mathbf H_K}|e_K|^3}{(\delta_n^{\prime K})^2}
=O_p\left[\frac{\{E_{\mathbf H_K}|e_K|^2\}^{3/2}}{(\delta_n^{\prime K})^2}\right]
=o_p(n^{-1/2}),
\]
where the last equality uses Assumption~\ref{B12}. This is the stagewise analogue of the $q=3$ tail step in Eqs.~\eqref{Lip1}--\eqref{Lip2}.

For $T_3^{(K)}$, which captures the difference between $\mathcal{D}_n(\cdot,\tau)$ and its limit $\mathcal{D}(\cdot)$, it follows similar arguments using Assumptions \ref{B10} (uniform continuity) and \ref{B11} (consistency){, and \ref{B12} (concentration)}, again paralleling the treatment of Eqs.~\eqref{thm2_2_9}--\eqref{thm2_2_10} in the proof of Lemma~\ref{single-lem}.

Now we justify the inductive step. Assume the inductive hypothesis \eqref{multi-induct-hyp} holds for stage $k+1$. For stage $k$:
\begin{align}
    &\widehat{V}_i^{(k)}(\mathcal{D}_n(\mathbf{O}_i,\widehat{\tau}_{\mathcal{I}_{3-j}}),\widehat{\pi}_{\mathcal{I}_{j,l}^c},\widehat{\mu}_{\mathcal{I}_{j,l}^c}) - \widehat{V}_i^{(k)}(\mathcal{D}(\mathbf{O}_i),\pi,\mu)\notag\\
    &= \frac{\mathrm{g}\{A_i^{(k)}, \mathcal{D}_n^{(k)}(\bar{\mathbf{A}}_i^{(k-1)},\bar{\mathbf{X}}_i^{(k)},\widehat{\tau}_{k,\mathcal{I}_{3-j}})\}}{\widehat{\pi}_{k,\mathcal{I}_{j,l}^c}(\bar{\mathbf{A}}_i^{(k)},\bar{\mathbf{X}}_i^{(k)})}\notag\\
    &\quad \times \{\widehat{V}_i^{(k+1)}(\mathcal{D}_n(\mathbf{O}_i,\widehat{\tau}_{\mathcal{I}_{3-j}}),\widehat{\pi}_{\mathcal{I}_{j,l}^c},\widehat{\mu}_{\mathcal{I}_{j,l}^c}) - \widehat{\mu}_{k,\mathcal{I}_{j,l}^c}(\bar{\mathbf{A}}_i^{(k)},\bar{\mathbf{X}}_i^{(k)})\}\notag\\
    &\quad - \frac{\mathrm{g}\{A_i^{(k)}, \mathcal{D}^{(k)}(\bar{\mathbf{A}}_i^{(k-1)},\bar{\mathbf{X}}_i^{(k)})\}}{\pi_k(\bar{\mathbf{A}}_i^{(k)},\bar{\mathbf{X}}_i^{(k)})}\notag\\
    &\quad \times \{\widehat{V}_i^{(k+1)}(\mathcal{D}(\mathbf{O}_i),\pi,\mu) - \mu_k(\bar{\mathbf{A}}_i^{(k)},\bar{\mathbf{X}}_i^{(k)})\}\notag\\
    &\quad + \mathcal{D}_n^{(k)}(\bar{\mathbf{A}}_i^{(k-1)},\bar{\mathbf{X}}_i^{(k)},\widehat{\tau}_{k,\mathcal{I}_{3-j}}) \widehat{\mu}_{1,k,\mathcal{I}_{j,l}^c}(\bar{\mathbf{A}}_i^{(k-1)},\bar{\mathbf{X}}_i^{(k)})\notag\\
&\quad+ (1-\mathcal{D}_n^{(k)}(\bar{\mathbf{A}}_i^{(k-1)},\bar{\mathbf{X}}_i^{(k)},\widehat{\tau}_{k,\mathcal{I}_{3-j}})) \widehat{\mu}_{0,k,\mathcal{I}_{j,l}^c}(\bar{\mathbf{A}}_i^{(k-1)},\bar{\mathbf{X}}_i^{(k)})\notag\\
&\quad- \mathcal{D}^{(k)}(\bar{\mathbf{A}}_i^{(k-1)},\bar{\mathbf{X}}_i^{(k)}) \mu_{1,k}(\bar{\mathbf{A}}_i^{(k-1)},\bar{\mathbf{X}}_i^{(k)})\notag\\
&\quad- (1-\mathcal{D}^{(k)}(\bar{\mathbf{A}}_i^{(k-1)},\bar{\mathbf{X}}_i^{(k)})) \mu_{0,k}(\bar{\mathbf{A}}_i^{(k-1)},\bar{\mathbf{X}}_i^{(k)})
\end{align}

The key difference from the base case is that we now have $\widehat{V}_i^{(k+1)}$ terms instead of $Y_i$. However, by the inductive hypothesis, the aggregate difference
\begin{align}
    \frac{1}{|\mathcal{I}_{j,l}|}\sum_{i\in\mathcal{I}_{j,l}}\left[\widehat{V}_i^{(k+1)}(\mathcal{D}_n(\mathbf{O}_i,\widehat{\tau}_{\mathcal{I}_{3-j}}),\widehat{\pi}_{\mathcal{I}_{j,l}^c},\widehat{\mu}_{\mathcal{I}_{j,l}^c}) - \widehat{V}_i^{(k+1)}(\mathcal{D}(\mathbf{O}_i),\pi,\mu)\right] = o_p(n^{-1/2}).
\end{align}
Note that at each stage $k$, Assumption~\ref{B2} (no unmeasured confounders) ensures $A^{(k)} \perp \mathbf{W} \mid \bar{\mathbf{X}}^{(k)}, \bar{\mathbf{A}}^{(k-1)}$, which identifies $\mu_k(\bar{\mathbf{A}}^{(k)}, \bar{\mathbf{X}}^{(k)})$ as the conditional expectation of $\widehat{V}^{(k+1)}(\mathcal{D}(\mathbf{O}),\pi,\mu)$ given $(\bar{\mathbf{A}}^{(k)}, \bar{\mathbf{X}}^{(k)})$, and Assumption~\ref{B3} (overlap) ensures the {true} propensity scores $\pi_k$ are bounded away from $0$ and $1$ at every stage{, while the stage-specific fitted-overlap condition stated in Section A.1 ensures the same for the fitted propensity scores}. These two conditions, together with Assumption~\ref{B1} (SUTVA), allow us to apply the same decomposition and bounding arguments as in the base case, using Cauchy-Schwarz inequality to control the interaction between the $\widehat{V}^{(k+1)}$ differences and the other estimation errors, and Assumption~\ref{B7} for the doubly robust convergence rates at stage $k$.

By mathematical induction, Eq.~\eqref{multi-induct-hyp} holds for all $k = K, K-1, \ldots, 1$. In particular, for $k = 1$:
\begin{align}
    |\widehat{V}_{\mathcal{D}_n}^{(1)*} - \widehat{V}_{\mathcal{D}_n}^{(1)}| = o_p(n^{-1/2}).
\end{align}

The asymptotic normality and variance formula follow from the central limit theorem applied to the oracle estimator $\widehat{V}_{\mathcal{D}_n}^{(1)*}$, which is a sample average of i.i.d.\ terms whose mean equals $V_0$ by Assumptions~\ref{B1} and~\ref{B2} (identification) and whose variance is finite by Assumption~\ref{B3} (overlap) and $E[Y^2]<\infty$.
\end{proof}
\subsection{Proof of Theorem \ref{thmb1}}
\begin{proof}
The proof follows by establishing the asymptotic equivalence between the proposed estimator and an oracle estimator, then applying the central limit theorem. The key step is to verify that the multi-stage smoothing decision function $d_s$ satisfies Assumptions~\ref{B10} to~\ref{B12}, so that Lemma~\ref{multi-lem} can be applied.

First, we decompose the smoothing estimator as:
\begin{align}
    \widehat{V}_{s}^{(1)} &= \frac{1}{4}\sum_{j=1,2}\sum_{l=1,2}\frac{1}{|\mathcal{I}_{j,l}|}\sum_{i\in\mathcal{I}_{j,l}}\widehat{V}_i^{(1)}(d_{s}(\mathbf{O}_i;\widehat{\tau}_{\mathcal{I}_{3-j}},h_{n,\mathcal{I}_{3-j}},t_0),\widehat{\pi}_{\mathcal{I}_{j,l}^c},\widehat{\mu}_{\mathcal{I}_{j,l}^c})\label{multi-decomp-1}\\
    &= \frac{1}{4}\sum_{j=1,2}\sum_{l=1,2}\frac{1}{|\mathcal{I}_{j,l}|}\sum_{i\in\mathcal{I}_{j,l}}\widehat{V}_i^{(1)}(d_{s}(\mathbf{O}_i;\widehat{\tau}_{\mathcal{I}_{3-j}},h_{n,\mathcal{I}_{3-j}},t_0),\pi,\mu)\label{multi-decomp-2}\\
    &\quad + \underbrace{\frac{1}{4}\sum_{j=1,2}\sum_{l=1,2}\frac{1}{|\mathcal{I}_{j,l}|}\sum_{i\in\mathcal{I}_{j,l}}\left[\widehat{V}_i^{(1)}(d_{s}(\mathbf{O}_i;\widehat{\tau}_{\mathcal{I}_{3-j}},h_{n,\mathcal{I}_{3-j}},t_0),\widehat{\pi}_{\mathcal{I}_{j,l}^c},\widehat{\mu}_{\mathcal{I}_{j,l}^c})\right.}_{R_1}\notag\\
    &\quad\quad\left. - \widehat{V}_i^{(1)}(d_{s}(\mathbf{O}_i;\widehat{\tau}_{\mathcal{I}_{3-j}},h_{n,\mathcal{I}_{3-j}},t_0),\pi,\mu)\right]\notag
\end{align}

To apply Lemma~\ref{multi-lem}, {for each fixed $j=1,2$} we define 
\[
\mathcal{D}_n^{(k)}(\bar{\mathbf{A}}^{(k-1)},\bar{\mathbf{X}}^{(k)},s)
=d_s^k(\bar{\mathbf{A}}^{(k-1)},\bar{\mathbf{X}}^{(k)};
s,h_{n,\mathcal{I}_{j}^c}^k,t_0^k)
\]
for $k=1,\dots,K$, 
and set
\begin{align*}
    {\delta_{n,j}^k} &= \sqrt{h_{n,\mathcal{I}_j^c}^k 
    \cdot n^{1/4}\left(E\!\left[|\widehat{\tau}_{k,\mathcal{I}_j^c}
    (\bar{\mathbf{A}}^{(k-1)},\bar{\mathbf{X}}^{(k)})
    -\tau_k(\bar{\mathbf{A}}^{(k-1)},\bar{\mathbf{X}}^{(k)})|^2
    \right]\right)^{3/4}},\\
    \delta_n^{\prime k} &= 2\max_{{\ell}} h_{n,\mathcal{I}_{{\ell}}^c}^k.
\end{align*}
We now verify Assumptions~\ref{B10} to~\ref{B12} stage by stage using Assumptions~\ref{B5} and~\ref{B6}.

We first verify Assumption~\ref{B11} (Consistency). For each stage $k$ and any fixed $(\bar{\mathbf{A}}^{(k-1)},\bar{\mathbf{X}}^{(k)})$, since $h_{n,\mathcal{I}_j^c}^k\to 0$ by Assumption~\ref{B6}: if $\tau_k(\bar{\mathbf{A}}^{(k-1)},\bar{\mathbf{X}}^{(k)})>0$, then for large $n$, $\tau_k>(1-t_0^k)h_{n,\mathcal{I}_j^c}^k$, so $\mathcal{D}_n^{(k)}(\bar{\mathbf{A}}^{(k-1)},\bar{\mathbf{X}}^{(k)},\tau_k)=1$; if $\tau_k<0$, then $\tau_k<-t_0^k h_{n,\mathcal{I}_j^c}^k$, so $\mathcal{D}_n^{(k)}(\bar{\mathbf{A}}^{(k-1)},\bar{\mathbf{X}}^{(k)},\tau_k)=0$; if $\tau_k=0$, then $\mathcal{D}_n^{(k)}(\bar{\mathbf{A}}^{(k-1)},\bar{\mathbf{X}}^{(k)},\tau_k)=t_0^k$. Therefore, $\mathcal{D}^{(k)}(\bar{\mathbf{A}}^{(k-1)},\bar{\mathbf{X}}^{(k)}):=\lim_{n\to\infty}\mathcal{D}_n^{(k)}(\bar{\mathbf{A}}^{(k-1)},\bar{\mathbf{X}}^{(k)},\tau_k)=\mathbb{I}\{\tau_k>0\}+t_0^k\cdot\mathbb{I}\{\tau_k=0\}$. Since the value of $\mathcal{D}^{(k)}$ on $\{\tau_k=0\}$ does not affect the mean outcome (because $\tau_k\cdot\mathcal{D}^{(k)}=\max\{0,\tau_k\}$ a.s.), we have $\mathcal{D}\in\bar{\mathbb{D}}(V_0)$ and Assumption~\ref{B11} is satisfied.

We next verify Assumption~\ref{B10} (Uniform continuity). For each stage $k$, the piecewise linear structure of $d_s^k$ implies that the Lipschitz constant of $s\mapsto d_s^k(\bar{\mathbf{A}}^{(k-1)},\bar{\mathbf{X}}^{(k)};s,h_{n,\mathcal{I}_j^c}^k,t_0^k)$ is at most $1/h_{n,\mathcal{I}_j^c}^k$. Therefore, for $|s-t|\leq{\delta_{n,j}^k}$,
\begin{align*}
    \sup_{\bar{\mathbf{A}}^{(k-1)},\bar{\mathbf{X}}^{(k)}}
    |\mathcal{D}_n^{(k)}(\bar{\mathbf{A}}^{(k-1)},
    \bar{\mathbf{X}}^{(k)},s)
    -\mathcal{D}_n^{(k)}(\bar{\mathbf{A}}^{(k-1)},
    \bar{\mathbf{X}}^{(k)},t)|
    \\\leq \frac{{\delta_{n,j}^k}}{h_{n,\mathcal{I}_j^c}^k}
    =\sqrt{\frac{n^{1/4}(E[|\widehat{\tau}_{k,\mathcal{I}_j^c}
    -\tau_k|^2])^{3/4}}{h_{n,\mathcal{I}_j^c}^k}}
    =o_p(1)
\end{align*}
where the last step follows from Assumption~\ref{B5}, which requires $n^{1/4}(E[|\widehat{\tau}_{k,{\mathcal{I}_j^c}}-\tau_k|^2])^{3/4}\prec_p h_{n,{\mathcal{I}_j^c}}^k$. For the probability bound, by Chebyshev's inequality and the definition of ${\delta_{n,j}^k}$,
\begin{align*}
    P\!\left(|\widehat{\tau}_{k,\mathcal{I}_j^c}
    (\bar{\mathbf{A}}^{(k-1)},\bar{\mathbf{X}}^{(k)})
    -\tau_k(\bar{\mathbf{A}}^{(k-1)},\bar{\mathbf{X}}^{(k)})|
    \geq{\delta_{n,j}^k}\right)\\
    \leq\frac{E[|\widehat{\tau}_{k,\mathcal{I}_j^c}-\tau_k|^2]}
    {({\delta_{n,j}^k})^2}
    =\frac{(E[|\widehat{\tau}_{k,\mathcal{I}_j^c}-\tau_k|^2])^{1/4}}
    {h_{n,\mathcal{I}_j^c}^k\cdot n^{1/4}}=o_p(1)
\end{align*}
again by Assumption~\ref{B5}. Hence Assumption~\ref{B10} is satisfied at each stage $k$.

Finally, we verify Assumption~\ref{B12} (Concentration). For the rate condition, by Assumption~\ref{B5}, $n^{1/4}(E[|\widehat{\tau}_{k,{\mathcal{I}_j^c}}-\tau_k|^2])^{3/4}\prec_p h_{n,{\mathcal{I}_j^c}}^k\leq\delta_n^{\prime k}/2\prec_p\delta_n^{\prime k}$, and by Assumption~\ref{B6}, $\delta_n^{\prime k}=2\max_{{\ell}} h_{n,\mathcal{I}_{{\ell}}^c}^k\prec n^{-\frac{1}{2(1+\alpha)}}$. Moreover, ${\delta_{n,j}^k\prec_p\delta_n^{\prime k}}$ since $({\delta_{n,j}^k})^2=h_{n,\mathcal{I}_j^c}^k\cdot n^{1/4}(E[|\widehat{\tau}_{k,\mathcal{I}_j^c}-\tau_k|^2])^{3/4}\prec_p (h_{n,\mathcal{I}_j^c}^k)^2\leq(\delta_n^{\prime k}/2)^2$ by Assumption~\ref{B5}. For the approximation condition Eq.~\eqref{B10-1}, when $|s|>\delta_n^{\prime k}=2\max_{{\ell}} h_{n,\mathcal{I}_{{\ell}}^c}^k$, we have $|s|>2h_{n,\mathcal{I}_j^c}^k$ for all $j$. If $s>\delta_n^{\prime k}$, then $s>2h_{n,\mathcal{I}_j^c}^k>(1-t_0^k)h_{n,\mathcal{I}_j^c}^k$ for $t_0^k\in(0,1)$, so $\mathcal{D}_n^{(k)}(\bar{\mathbf{A}}^{(k-1)},\bar{\mathbf{X}}^{(k)},s)=1=\mathbb{I}\{s>0\}$; if $s<-\delta_n^{\prime k}$, then $s<-2h_{n,\mathcal{I}_j^c}^k<-t_0^k h_{n,\mathcal{I}_j^c}^k$, so $\mathcal{D}_n^{(k)}(\bar{\mathbf{A}}^{(k-1)},\bar{\mathbf{X}}^{(k)},s)=0=\mathbb{I}\{s>0\}$. In both cases, $|\mathcal{D}_n^{(k)}(\bar{\mathbf{A}}^{(k-1)},\bar{\mathbf{X}}^{(k)},s)-\mathbb{I}\{s>0\}|=0=o(1/({|s|}\sqrt{n}))$, so Eq.~\eqref{B10-1} holds trivially and Assumption~\ref{B12} is satisfied at each stage $k$.

Since Assumptions~\ref{B10} to~\ref{B12} are verified at all stages using Assumptions~\ref{B5} and~\ref{B6}, and Assumptions~\ref{B1} to~\ref{B4} and~\ref{B7} are assumed, all conditions of Lemma~\ref{multi-lem} are satisfied. Lemma~\ref{multi-lem} then gives $|\widehat{V}_{\mathcal{D}_n}^{(1)*}-\widehat{V}_{\mathcal{D}_n}^{(1)}|=o_p(n^{-1/2})$, where $\widehat{V}_{\mathcal{D}_n}^{(1)*}$ is the oracle estimator using true $\tau$, $\pi$ and $\mu$. Combined with the asymptotic equivalence between $\widehat{V}_{\mathcal{D}_n}^{(1)*}$ and $\widehat{V}_{\mathcal{D}}^{(1)*}=\frac{1}{n}\sum_{i=1}^n\widehat{V}_i^{(1)}(\mathcal{D}(\mathbf{O}_i),\pi,\mu)$ (also established in Lemma~\ref{multi-lem}), the asymptotic normality of $\widehat{V}_s^{(1)}$ follows from the central limit theorem applied to $\widehat{V}_{\mathcal{D}}^{(1)*}$, which is a sample average of i.i.d.\ terms. The variance formula follows from the recursive structure and yields the stated form.
\end{proof}
\subsection{Proof of Proposition \ref{propb2}}
\begin{proof}
Write $r_n=\log n/({C_0}n)$. Fix a stage $k$ and a half fold $\mathcal I_j$. For every $\ell\in\{1,2\}$, the trajectories indexed by $\mathcal I_{3-j,\ell}$ are independent of the stage-$k$ contrast estimators trained on $\mathcal I_j$ and $\mathcal I_{3-j,3-\ell}$. Hence
\begin{align}
&E\left[\left.
\frac{1}{|\mathcal I_{3-j,\ell}|}\sum_{i\in\mathcal I_{3-j,\ell}}
|\widehat\tau_{k,\mathcal I_{3-j,3-\ell}}(\mathbf H_{k,i})
-\widehat\tau_{k,\mathcal I_j}(\mathbf H_{k,i})|^2
\,\right|\,\text{the two training folds}\right]\notag\\
&\qquad=
E_{\mathbf H_k}|\widehat\tau_{k,\mathcal I_{3-j,3-\ell}}(\mathbf H_k)
-\widehat\tau_{k,\mathcal I_j}(\mathbf H_k)|^2.
\label{multi-conditional-eae}
\end{align}

Suppose first that Condition (C1-M) holds. The squared triangle inequality gives, for each $\ell$,
\begin{align*}
&E_{\mathbf H_k}|\widehat\tau_{k,\mathcal I_{3-j,3-\ell}}(\mathbf H_k)
-\widehat\tau_{k,\mathcal I_j}(\mathbf H_k)|^2\\
&\quad\le2E_{\mathbf H_k}|\widehat\tau_{k,\mathcal I_j}(\mathbf H_k)-\tau_k(\mathbf H_k)|^2
+2E_{\mathbf H_k}|\widehat\tau_{k,\mathcal I_{3-j,3-\ell}}(\mathbf H_k)-\tau_k(\mathbf H_k)|^2
=O_p(r_n).
\end{align*}
Conditional Markov's inequality in Eq.~\eqref{multi-conditional-eae} therefore yields
$EAE(\widehat\tau_{k,\mathcal I_j})=O_p(r_n)$. Together with Condition (C1-M), this proves
\begin{equation}\label{multi-tracking}
EAE(\widehat\tau_{k,\mathcal I_j})\vee r_n
\asymp_p E_{\mathbf H_k}|\widehat\tau_{k,\mathcal I_j}(\mathbf H_k)-\tau_k(\mathbf H_k)|^2\vee r_n.
\end{equation}

Now suppose Conditions (C2-M) and (C3-M) hold. Conditional Markov's inequality in Eq.~\eqref{multi-conditional-eae} first gives
\begin{align*}
EAE(\widehat\tau_{k,\mathcal I_j})\vee r_n
=O_p\Bigg(&r_n\vee\frac12\sum_{\ell=1}^2
E_{\mathbf H_k}|\widehat\tau_{k,\mathcal I_{3-j,3-\ell}}(\mathbf H_k)
-\widehat\tau_{k,\mathcal I_j}(\mathbf H_k)|^2\Bigg).
\end{align*}

For fixed $(k,j,\ell)$, define the event
\[
\mathcal G_{kj\ell}(B)=
\left\{\|\widehat\tau_{k,\mathcal I_j}\|_\infty\le B,
\ \|\widehat\tau_{k,\mathcal I_{3-j,3-\ell}}\|_\infty\le B\right\}.
\]
This event is measurable with respect to the two training folds in Eq.~\eqref{multi-conditional-eae}. Because the number of stages and folds is fixed, Condition (C2-M) gives, for every $\varepsilon>0$, a constant $B$ such that
\[
P\left\{\bigcap_{k,j,\ell}\mathcal G_{kj\ell}(B)\right\}\ge1-\varepsilon
\]
for all sufficiently large $n$. Condition on the two training folds and write
\[
W_i=|\widehat\tau_{k,\mathcal I_{3-j,3-\ell}}(\mathbf H_{k,i})
-\widehat\tau_{k,\mathcal I_j}(\mathbf H_{k,i})|^2,
\qquad i\in\mathcal I_{3-j,\ell}.
\]
Denote its conditional mean by $\mu$. On $\mathcal G_{kj\ell}(B)$, $0\le W_i\le4B^2$ and its conditional variance is at most $4B^2\mu$. With $m=|\mathcal I_{3-j,\ell}|\asymp n$, the one-sided Bernstein inequality gives, on $\mathcal G_{kj\ell}(B)$,
\begin{align*}
P\left(\left.
\frac1m\sum_{i\in\mathcal I_{3-j,\ell}}W_i
\le\frac\mu2\,\right|\,\text{the two training folds}\right)
&\le\exp\left\{-\frac{3m\mu}{112B^2}\right\}.
\end{align*}
For a sufficiently large $K_0$, integration of this conditional bound gives
\[
P\left(\left\{\frac1m\sum_{i\in\mathcal I_{3-j,\ell}}W_i\le\frac\mu2\right\}
\cap\left\{\mu\ge K_0\log n/n\right\}\cap\mathcal G_{kj\ell}(B)\right)\le n^{-3}.
\]
If the validation average is at least $\mu/2$, then the $EAE$, which is one half of the sum of the two nonnegative validation averages, is at least $\mu/4$. If instead $\mu<K_0\log n/n=K_0{C_0} r_n$, the truncation floor controls $\mu$. A finite union bound over $(k,j,\ell)$, together with the preceding conditional Markov upper bound and the common boundedness event, therefore yields
\begin{align*}
EAE(\widehat\tau_{k,\mathcal I_j})\vee r_n
\asymp_p
\left\{\frac12\sum_{\ell=1}^2
E_{\mathbf H_k}|\widehat\tau_{k,\mathcal I_{3-j,3-\ell}}(\mathbf H_k)
-\widehat\tau_{k,\mathcal I_j}(\mathbf H_k)|^2\right\}\vee r_n.
\end{align*}
Condition (C3-M), applied separately to each $\ell$, turns the last display into Eq.~\eqref{multi-tracking}. Hence Eq.~\eqref{multi-tracking} holds under either alternative.

Algorithm \ref{alg5} and Eq.~\eqref{multi-tracking} now give
\[
\frac{n^{1/4}\{E_{\mathbf H_k}|\widehat\tau_{k,\mathcal I_j}(\mathbf H_k)-\tau_k(\mathbf H_k)|^2\}^{3/4}}
{h_{n,\mathcal I_j}^{k}}
=O_p(1/\log n)=o_p(1),
\]
which is Assumption \ref{B5}. Assumption \ref{B8} and Eq.~\eqref{multi-tracking} also give
\[
h_{n,\mathcal I_j}^{k}=O_p\{\log n\,n^{-1/\{2(1+\beta)\}}\}.
\]
As in the single-time-point proof, the exponent follows from
\[
n^{1/4}
\left\{n^{-\frac{2+2\beta/3}{2(1+\beta)}}\right\}^{3/4}
=n^{-1/\{2(1+\beta)\}}.
\]
Since $\beta<\alpha$,
\[
\frac{\log n\,n^{-1/\{2(1+\beta)\}}}{n^{-1/\{2(1+\alpha)\}}}
=\log n\,n^{-\frac{\alpha-\beta}{2(1+\alpha)(1+\beta)}}\longrightarrow0,
\]
so Assumption \ref{B6} holds. A finite union bound makes both conclusions simultaneous over every stage $k$ and fold $j$.
\end{proof}
\subsection{Proof of Proposition \ref{multi-opt-oracle}}
\begin{proof}
This proof establishes the variance formula for the oracle estimator and derives the optimal choice of decision functions. We proceed by backward induction from stage $K$ to stage 1.


We firstly verify the unbiased constraint for $\mathcal{D} \in \bar{\mathbb{D}}(V_0)$. For any $\mathcal{D} = (\mathcal{D}^{(1)}, \ldots, \mathcal{D}^{(K)}) \in \bar{\mathbb{D}}(V_0)$, we must have:
\begin{align}
    E[\widehat{V}_{\mathcal{D}}^{(1)*}] &= E\left[\frac{1}{n}\sum_{i=1}^n \widehat{V}_i^{(1)}(\mathcal{D}(\mathbf{O}_i),\pi,\mu)\right] = V_0
\end{align}

By the recursive definition of $\widehat{V}_i^{(k)}$:
\begin{align}
    E[\widehat{V}_i^{(k)}(\mathcal{D}(\mathbf{O}_i),\pi,\mu)] &= E\left[\frac{\mathrm{g}\{A_i^{(k)}, \mathcal{D}^{(k)}(\bar{\mathbf{A}}_i^{(k-1)},\bar{\mathbf{X}}_i^{(k)})\}}{\pi_k(\bar{\mathbf{A}}_i^{(k)}, \bar{\mathbf{X}}_i^{(k)})}\right.\notag\\
    &\quad \times \left\{\widehat{V}_i^{(k+1)}(\mathcal{D}(\mathbf{O}_i),\pi,\mu) - \mu_k(\bar{\mathbf{A}}_i^{(k)}, \bar{\mathbf{X}}_i^{(k)})\right\}\notag\\
    &\quad + \mathcal{D}^{(k)}(\bar{\mathbf{A}}_i^{(k-1)},\bar{\mathbf{X}}_i^{(k)}) \mu_k(\{\bar{\mathbf{A}}_i^{(k-1)},1\}, \bar{\mathbf{X}}_i^{(k)})\notag\\
    &\quad\left. + (1-\mathcal{D}^{(k)}(\bar{\mathbf{A}}_i^{(k-1)},\bar{\mathbf{X}}_i^{(k)})) \mu_k(\{\bar{\mathbf{A}}_i^{(k-1)},0\}, \bar{\mathbf{X}}_i^{(k)})\right]
\end{align}

Taking expectation and using the law of total expectation:
\begin{align}
    &E[\widehat{V}_i^{(k)}(\mathcal{D}(\mathbf{O}_i),\pi,\mu)]\notag\\
    &= E_{\bar{\mathbf{A}}_i^{(k-1)},\bar{\mathbf{X}}_i^{(k)}}\left[E_{A_i^{(k)}|\bar{\mathbf{A}}_i^{(k-1)},\bar{\mathbf{X}}_i^{(k)}}\left[\frac{\mathrm{g}\{A_i^{(k)}, \mathcal{D}^{(k)}(\bar{\mathbf{A}}_i^{(k-1)},\bar{\mathbf{X}}_i^{(k)})\}}{\pi_k(\bar{\mathbf{A}}_i^{(k)}, \bar{\mathbf{X}}_i^{(k)})}\right.\right.\notag\\
    &\quad\quad\left.\left.\times \{E[\widehat{V}_i^{(k+1)}(\mathcal{D}(\mathbf{O}_i),\pi,\mu)|\bar{\mathbf{A}}_i^{(k)}, \bar{\mathbf{X}}_i^{(k)}] - \mu_k(\bar{\mathbf{A}}_i^{(k)}, \bar{\mathbf{X}}_i^{(k)})\}\right]\right]\notag\\
    &\quad + E\left[\mathcal{D}^{(k)}(\bar{\mathbf{A}}_i^{(k-1)},\bar{\mathbf{X}}_i^{(k)}) \mu_k(\{\bar{\mathbf{A}}_i^{(k-1)},1\}, \bar{\mathbf{X}}_i^{(k)})\right]\notag\\
    &\quad + E\left[(1-\mathcal{D}^{(k)}(\bar{\mathbf{A}}_i^{(k-1)},\bar{\mathbf{X}}_i^{(k)})) \mu_k(\{\bar{\mathbf{A}}_i^{(k-1)},0\}, \bar{\mathbf{X}}_i^{(k)})\right]
\end{align}

The first term equals zero by the definition of conditional expectation and the propensity score. Therefore:
\begin{align}
    E[\widehat{V}_i^{(k)}(\mathcal{D}(\mathbf{O}_i),\pi,\mu)] &= E\left[\mathcal{D}^{(k)}(\bar{\mathbf{A}}_i^{(k-1)},\bar{\mathbf{X}}_i^{(k)}) \tau_k(\bar{\mathbf{A}}_i^{(k-1)},\bar{\mathbf{X}}_i^{(k)})\right]\notag\\
    &\quad + E\left[\mu_k(\{\bar{\mathbf{A}}_i^{(k-1)},0\}, \bar{\mathbf{X}}_i^{(k)})\right]\label{multi-constraint-1}
\end{align}

For $\mathcal{D} \in \bar{\mathbb{D}}(V_0)$, we need the recursion to yield $V_0$. This constrains:
\begin{align}
    \mathcal{D}^{(k)}(\bar{\mathbf{A}}_i^{(k-1)},\bar{\mathbf{X}}_i^{(k)}) \cdot \tau_k(\bar{\mathbf{A}}_i^{(k-1)},\bar{\mathbf{X}}_i^{(k)}) &= \max\{0, \tau_k(\bar{\mathbf{A}}_i^{(k-1)},\bar{\mathbf{X}}_i^{(k)})\} \text{ a.s.}\label{multi-constraint-2}
\end{align}

when $\tau_k(\bar{\mathbf{A}}_i^{(k-1)},\bar{\mathbf{X}}_i^{(k)}) \neq 0$.

 
Next we conduct the variance computation. For the final stage $K$, we have $\widehat{V}_i^{(K+1)} = Y_i$. The variance is:
\begin{align}
    &\operatorname{Var}[\widehat{V}_i^{(K)}(\mathcal{D}(\mathbf{O}_i),\pi,\mu)]\notag\\
    &= \operatorname{Var}\left[\frac{\mathrm{g}\{A_i^{(K)}, \mathcal{D}^{(K)}(\bar{\mathbf{A}}_i^{(K-1)},\bar{\mathbf{X}}_i^{(K)})\}}{\pi_K(\bar{\mathbf{A}}_i^{(K)}, \bar{\mathbf{X}}_i^{(K)})}\right.\notag\\
    &\quad\quad\left.\times \{Y_i - \mu_K(\bar{\mathbf{A}}_i^{(K)}, \bar{\mathbf{X}}_i^{(K)})\}\right]\notag\\
    &\quad + \operatorname{Var}\left[\mathcal{D}^{(K)}(\bar{\mathbf{A}}_i^{(K-1)},\bar{\mathbf{X}}_i^{(K)}) \mu_K(\{\bar{\mathbf{A}}_i^{(K-1)},1\}, \bar{\mathbf{X}}_i^{(K)})\right.\notag\\
    &\quad\quad\left.+ (1-\mathcal{D}^{(K)}(\bar{\mathbf{A}}_i^{(K-1)},\bar{\mathbf{X}}_i^{(K)})) \mu_K(\{\bar{\mathbf{A}}_i^{(K-1)},0\}, \bar{\mathbf{X}}_i^{(K)})\right]\label{multi-var-K-decomp}
\end{align}

Using the law of total variance:
\begin{align}
    &\operatorname{Var}[\widehat{V}_i^{(K)}(\mathcal{D}(\mathbf{O}_i),\pi,\mu)]\notag\\
    &= E\left[\operatorname{Var}\left[\widehat{V}_i^{(K)}(\mathcal{D}(\mathbf{O}_i),\pi,\mu) \mid \bar{\mathbf{A}}_i^{(K-1)}, \bar{\mathbf{X}}_i^{(K)}\right]\right]\label{multi-var-K-1}\\
    &\quad + \operatorname{Var}\left[E\left[\widehat{V}_i^{(K)}(\mathcal{D}(\mathbf{O}_i),\pi,\mu) \mid \bar{\mathbf{A}}_i^{(K-1)}, \bar{\mathbf{X}}_i^{(K)}\right]\right]\label{multi-var-K-2}
\end{align}

For Eq. \eqref{multi-var-K-1}:
\begin{align}
    &\operatorname{Var}\left[\widehat{V}_i^{(K)}(\mathcal{D}(\mathbf{O}_i),\pi,\mu) \mid \bar{\mathbf{A}}_i^{(K-1)}, \bar{\mathbf{X}}_i^{(K)}\right]\notag\\
    &= E_{A_i^{(K)}|\bar{\mathbf{A}}_i^{(K-1)}, \bar{\mathbf{X}}_i^{(K)}}\left[\frac{(\mathrm{g}\{A_i^{(K)}, \mathcal{D}^{(K)}(\bar{\mathbf{A}}_i^{(K-1)},\bar{\mathbf{X}}_i^{(K)})\})^2}{(\pi_K(\bar{\mathbf{A}}_i^{(K)}, \bar{\mathbf{X}}_i^{(K)}))^2}\right.\notag\\
    &\quad\quad\left.\times \operatorname{Var}[Y_i \mid \bar{\mathbf{A}}_i^{(K)}, \bar{\mathbf{X}}_i^{(K)}]\right]\notag\\
    &= \frac{(\mathcal{D}^{(K)}(\bar{\mathbf{A}}_i^{(K-1)},\bar{\mathbf{X}}_i^{(K)}))^2}{\pi_K(\{\bar{\mathbf{A}}_i^{(K-1)},1\}, \bar{\mathbf{X}}_i^{(K)})}\operatorname{Var}[Y_i \mid \{\bar{\mathbf{A}}_i^{(K-1)},1\}, \bar{\mathbf{X}}_i^{(K)}]\label{multi-var-K-cond-1}\\
    &\quad + \frac{(1-\mathcal{D}^{(K)}(\bar{\mathbf{A}}_i^{(K-1)},\bar{\mathbf{X}}_i^{(K)}))^2}{\pi_K(\{\bar{\mathbf{A}}_i^{(K-1)},0\}, \bar{\mathbf{X}}_i^{(K)})}\operatorname{Var}[Y_i \mid \{\bar{\mathbf{A}}_i^{(K-1)},0\}, \bar{\mathbf{X}}_i^{(K)}]\label{multi-var-K-cond-2}
\end{align}

For Eq. \eqref{multi-var-K-2}:
\begin{align}
    &E\left[\widehat{V}_i^{(K)}(\mathcal{D}(\mathbf{O}_i),\pi,\mu) \mid \bar{\mathbf{A}}_i^{(K-1)}, \bar{\mathbf{X}}_i^{(K)}\right]\notag\\
    &= \mathcal{D}^{(K)}(\bar{\mathbf{A}}_i^{(K-1)},\bar{\mathbf{X}}_i^{(K)}) \mu_K(\{\bar{\mathbf{A}}_i^{(K-1)},1\}, \bar{\mathbf{X}}_i^{(K)})\notag\\
    &\quad + (1-\mathcal{D}^{(K)}(\bar{\mathbf{A}}_i^{(K-1)},\bar{\mathbf{X}}_i^{(K)})) \mu_K(\{\bar{\mathbf{A}}_i^{(K-1)},0\}, \bar{\mathbf{X}}_i^{(K)})
\end{align}

Therefore:
\begin{align}
    &\operatorname{Var}[E[\widehat{V}_i^{(K)}(\mathcal{D}(\mathbf{O}_i),\pi,\mu) \mid \bar{\mathbf{A}}_i^{(K-1)}, \bar{\mathbf{X}}_i^{(K)}]]\notag\\
    &= \operatorname{Var}\left[\mathcal{D}^{(K)}(\bar{\mathbf{A}}_i^{(K-1)},\bar{\mathbf{X}}_i^{(K)}) \tau_K(\bar{\mathbf{A}}_i^{(K-1)},\bar{\mathbf{X}}_i^{(K)})\right.\notag\\
    &\quad\quad\left.+ \mu_K(\{\bar{\mathbf{A}}_i^{(K-1)},0\}, \bar{\mathbf{X}}_i^{(K)})\right]
\end{align}


Then, by induction, for general stage $k < K$, we have:
\begin{align}
    &\operatorname{Var}[\widehat{V}_i^{(k)}(\mathcal{D}(\mathbf{O}_i),\pi,\mu)]\notag\\
    &= E\left[\operatorname{Var}[\widehat{V}_i^{(k)}(\mathcal{D}(\mathbf{O}_i),\pi,\mu) \mid \bar{\mathbf{A}}_i^{(k-1)}, \bar{\mathbf{X}}_i^{(k)}]\right]\notag\\
    &\quad + \operatorname{Var}[E[\widehat{V}_i^{(k)}(\mathcal{D}(\mathbf{O}_i),\pi,\mu) \mid \bar{\mathbf{A}}_i^{(k-1)}, \bar{\mathbf{X}}_i^{(k)}]]
\end{align}

The conditional variance term becomes:
\begin{align}
    &\operatorname{Var}[\widehat{V}_i^{(k)}(\mathcal{D}(\mathbf{O}_i),\pi,\mu) \mid \bar{\mathbf{A}}_i^{(k-1)}, \bar{\mathbf{X}}_i^{(k)}]\notag\\
    &= \operatorname{Var}[\widehat{V}_i^{(k+1)}(\mathcal{D}(\mathbf{O}_i),\pi,\mu) \mid \bar{\mathbf{A}}_i^{(k-1)}, \bar{\mathbf{X}}_i^{(k)}]\notag\\
    &\quad \times \left(\frac{(\mathcal{D}^{(k)}(\bar{\mathbf{A}}_i^{(k-1)},\bar{\mathbf{X}}_i^{(k)}))^2}{\pi_k(\{\bar{\mathbf{A}}_i^{(k-1)},1\}, \bar{\mathbf{X}}_i^{(k)})}\right.\notag\\
    &\quad\quad\left.+ \frac{(1-\mathcal{D}^{(k)}(\bar{\mathbf{A}}_i^{(k-1)},\bar{\mathbf{X}}_i^{(k)}))^2}{\pi_k(\{\bar{\mathbf{A}}_i^{(k-1)},0\}, \bar{\mathbf{X}}_i^{(k)})}\right)
\end{align}


Now we proceed to explore when the variance is minimized. Under Assumption \ref{B9}, we have for all $k$:
\begin{align}
    \operatorname{Var}[\widehat{V}^{(k+1)}(\mathcal{D}(\mathbf{O}_i),\pi,\mu) \mid \{\bar{\mathbf{A}}^{(k-1)},0\}, \bar{\mathbf{X}}^{(k)}] &= \operatorname{Var}[\widehat{V}^{(k+1)}(\mathcal{D}(\mathbf{O}_i),\pi,\mu) \mid \{\bar{\mathbf{A}}^{(k-1)},1\}, \bar{\mathbf{X}}^{(k)}]
\end{align}

Let $\sigma^2_{k+1}(\bar{\mathbf{A}}^{(k-1)}, \bar{\mathbf{X}}^{(k)}) = \operatorname{Var}[\widehat{V}^{(k+1)}(\mathcal{D}(\mathbf{O}_i),\pi,\mu) \mid \bar{\mathbf{A}}^{(k-1)}, \bar{\mathbf{X}}^{(k)}]$.

The conditional variance for stage $k$ becomes:
\begin{align}
    &\operatorname{Var}[\widehat{V}_i^{(k)}(\mathcal{D}(\mathbf{O}_i),\pi,\mu) \mid \bar{\mathbf{A}}_i^{(k-1)}, \bar{\mathbf{X}}_i^{(k)}]\notag\\
    &= \sigma^2_{k+1}(\bar{\mathbf{A}}_i^{(k-1)}, \bar{\mathbf{X}}_i^{(k)}) \times f_k(\mathcal{D}^{(k)}(\bar{\mathbf{A}}_i^{(k-1)},\bar{\mathbf{X}}_i^{(k)}), \bar{\mathbf{A}}_i^{(k-1)}, \bar{\mathbf{X}}_i^{(k)})
\end{align}

where:
\begin{align}
    f_k(d, \bar{\mathbf{a}}^{(k-1)}, \bar{\mathbf{x}}^{(k)}) &= \frac{d^2}{\pi_k(\{\bar{\mathbf{a}}^{(k-1)},1\}, \bar{\mathbf{x}}^{(k)})} + \frac{(1-d)^2}{\pi_k(\{\bar{\mathbf{a}}^{(k-1)},0\}, \bar{\mathbf{x}}^{(k)})}
\end{align}

To minimize the conditional variance, we solve:
\begin{align}
    \frac{\partial}{\partial d} f_k(d, \bar{\mathbf{a}}^{(k-1)}, \bar{\mathbf{x}}^{(k)}) &= \frac{2d}{\pi_k(\{\bar{\mathbf{a}}^{(k-1)},1\}, \bar{\mathbf{x}}^{(k)})} - \frac{2(1-d)}{\pi_k(\{\bar{\mathbf{a}}^{(k-1)},0\}, \bar{\mathbf{x}}^{(k)})} = 0
\end{align}

This yields:
\begin{align}
    d^* &= \frac{\pi_k(\{\bar{\mathbf{a}}^{(k-1)},1\}, \bar{\mathbf{x}}^{(k)})}{\pi_k(\{\bar{\mathbf{a}}^{(k-1)},1\}, \bar{\mathbf{x}}^{(k)}) + \pi_k(\{\bar{\mathbf{a}}^{(k-1)},0\}, \bar{\mathbf{x}}^{(k)})}\notag\\
    &= \pi_k(\{\bar{\mathbf{a}}^{(k-1)},1\}, \bar{\mathbf{x}}^{(k)})
\end{align}

Therefore, the optimal choice for the nonregular part is:
\begin{align}
    \mathcal{D}^{(k)}(\bar{\mathbf{A}}^{(k-1)},\bar{\mathbf{X}}^{(k)}) &= \mathbb{I}\{\tau_k(\bar{\mathbf{A}}^{(k-1)},\bar{\mathbf{X}}^{(k)}) > 0\}\notag\\
    &\quad + \pi_k(\{\bar{\mathbf{A}}^{(k-1)},1\}, \bar{\mathbf{X}}^{(k)}) \mathbb{I}\{\tau_k(\bar{\mathbf{A}}^{(k-1)},\bar{\mathbf{X}}^{(k)}) = 0\}
\end{align}

This completes the proof.
\end{proof}
\subsection{Proof of Corollary \ref{multi-embedding}}
\begin{proof}
This corollary extends the embedding results from the single-stage case (Corollary 1) to multi-stage settings. We provide detailed derivations for both parts.

\textbf{Part (i): Multi-stage adaptive smoothing embedding}

We need to show that the multi-stage adaptive smoothing estimator $\widehat{V}_{as}^{(1)}$ can be embedded as an oracle estimator $\widehat{V}_{\mathcal{D}_{as}}^{(1)*}$.


First, we construct the multi-stage adaptive smoothing estimator. The adaptive smoothing estimator uses cross-fitted decision rules. For each stage $k$ and each fold $j \in \{1,2\}$, we construct:
\begin{align}
\widehat{d}_{as}^{k,j}(\bar{\mathbf{A}}^{(k-1)}, \bar{\mathbf{X}}^{(k)}) = \begin{cases}
1 & \text{if } \widehat{\tau}_{k,\mathcal{I}_{3-j}}(\bar{\mathbf{A}}^{(k-1)}, \bar{\mathbf{X}}^{(k)}) > 0 \\
\widehat{\pi}_{k,\mathcal{I}_{3-j}}((\bar{\mathbf{A}}^{(k-1)}, 1), \bar{\mathbf{X}}^{(k)}) & \text{if } \widehat{\tau}_{k,\mathcal{I}_{3-j}}(\bar{\mathbf{A}}^{(k-1)}, \bar{\mathbf{X}}^{(k)}) = 0 \\
0 & \text{if } \widehat{\tau}_{k,\mathcal{I}_{3-j}}(\bar{\mathbf{A}}^{(k-1)}, \bar{\mathbf{X}}^{(k)}) < 0
\end{cases}
\end{align}


Next we study its asymptotic behavior. Under Assumptions \ref{B10} to \ref{B12}, as $n \to \infty$, the empirical contrast functions converge:
\begin{align}
\widehat{\tau}_{k,\mathcal{I}_{3-j}}(\bar{\mathbf{A}}^{(k-1)}, \bar{\mathbf{X}}^{(k)}) \xrightarrow{P} \tau_k(\bar{\mathbf{A}}^{(k-1)}, \bar{\mathbf{X}}^{(k)})
\end{align}

and the empirical propensity scores converge:
\begin{align}
\widehat{\pi}_{k,\mathcal{I}_{3-j}}((\bar{\mathbf{A}}^{(k-1)}, 1), \bar{\mathbf{X}}^{(k)}) \xrightarrow{P} \pi_k((\bar{\mathbf{A}}^{(k-1)}, 1), \bar{\mathbf{X}}^{(k)})
\end{align}


For any fixed $(\bar{\mathbf{A}}^{(k-1)}, \bar{\mathbf{X}}^{(k)})$, the adaptive smoothing decision rule thus converges to:
\begin{align}
\lim_{n \to \infty} E[\widehat{d}_{as}^{k,j}(\bar{\mathbf{A}}^{(k-1)}, \bar{\mathbf{X}}^{(k)})] = \begin{cases}
1 & \text{if } \tau_k(\bar{\mathbf{A}}^{(k-1)}, \bar{\mathbf{X}}^{(k)}) > 0 \\
\pi_k((\bar{\mathbf{A}}^{(k-1)}, 1), \bar{\mathbf{X}}^{(k)}) & \text{if } \tau_k(\bar{\mathbf{A}}^{(k-1)}, \bar{\mathbf{X}}^{(k)}) = 0 \\
0 & \text{if } \tau_k(\bar{\mathbf{A}}^{(k-1)}, \bar{\mathbf{X}}^{(k)}) < 0
\end{cases}
\end{align}


Define the oracle decision function:
\begin{align*}
\mathcal{D}_{as}^{(k)}(\bar{\mathbf{A}}^{(k-1)}, \bar{\mathbf{X}}^{(k)}) &= \mathbb{I}\{\tau_k(\bar{\mathbf{A}}^{(k-1)}, \bar{\mathbf{X}}^{(k)}) > 0\} \\
&\quad + \pi_k((\bar{\mathbf{A}}^{(k-1)}, 1), \bar{\mathbf{X}}^{(k)}) \mathbb{I}\{\tau_k(\bar{\mathbf{A}}^{(k-1)}, \bar{\mathbf{X}}^{(k)}) = 0\}
\end{align*}

By the functional delta method and uniform convergence results in Theorem \ref{thmb1}, we have:
\begin{align}\label{as-eal-embedding}
\widehat{V}_{as}^{(1)} - \widehat{V}_{\mathcal{D}_{as}}^{(1)*} = o_p(n^{-1/2})
\end{align}


From Theorem \ref{thmb1}, the adaptive smoothing estimator achieves the variance lower bound:
\begin{align}
\text{Var}(\sqrt{n}(\widehat{V}_{as}^{(1)} - V_0)) = \sigma_{opt}^2 + o(1)
\end{align}

where $\sigma_{opt}^2$ is the optimal variance derived in the theorem.

\textbf{Part (ii): Multi-stage subbagging embedding}

We now show the embedding for the subbagging estimator following \citet{shi2020breaking}.


The subbagging estimator uses undersmoothing with subsample size $s_n = o(n)$. For each subsample $\mathcal{I}_b$ of size $s_n$, we estimate:
\begin{align*}
\widehat{d}_{sb}^{k,b}(\bar{\mathbf{A}}^{(k-1)}, \bar{\mathbf{X}}^{(k)}) = \mathbb{I}\{\widehat{\tau}_{k,\mathcal{I}_b}(\bar{\mathbf{A}}^{(k-1)}, \bar{\mathbf{X}}^{(k)}) > 0\}
\end{align*}

The final decision rule is the average over all subsamples:
\begin{align*}
\bar{d}_{sb}^{k}(\bar{\mathbf{A}}^{(k-1)}, \bar{\mathbf{X}}^{(k)}) = \frac{1}{B} \sum_{b=1}^B \widehat{d}_{sb}^{k,b}(\bar{\mathbf{A}}^{(k-1)}, \bar{\mathbf{X}}^{(k)})
\end{align*}


Under undersmoothing conditions (Assumption 7 in \citet{shi2020breaking}), for any $(\bar{\mathbf{A}}^{(k-1)}, \bar{\mathbf{X}}^{(k)})$ with $\tau_k(\bar{\mathbf{A}}^{(k-1)}, \bar{\mathbf{X}}^{(k)}) = 0$:

The bias term satisfies:
\begin{align*}
E[\widehat{\tau}_{k,\mathcal{I}_b}(\bar{\mathbf{A}}^{(k-1)}, \bar{\mathbf{X}}^{(k)})] - \tau_k(\bar{\mathbf{A}}^{(k-1)}, \bar{\mathbf{X}}^{(k)}) = o(s_n^{-1/2})
\end{align*}

The variance term satisfies:
\begin{align*}
\text{Var}(\widehat{\tau}_{k,\mathcal{I}_b}(\bar{\mathbf{A}}^{(k-1)}, \bar{\mathbf{X}}^{(k)})) = O(s_n^{-1})
\end{align*}


By the asymptotic normality of $\widehat{\tau}_{k,\mathcal{I}_b}$ and the undersmoothing property:
\begin{align*}
\frac{\widehat{\tau}_{k,\mathcal{I}_b}(\bar{\mathbf{A}}^{(k-1)}, \bar{\mathbf{X}}^{(k)}) - \tau_k(\bar{\mathbf{A}}^{(k-1)}, \bar{\mathbf{X}}^{(k)})}{\sqrt{\text{Var}(\widehat{\tau}_{k,\mathcal{I}_b}(\bar{\mathbf{A}}^{(k-1)}, \bar{\mathbf{X}}^{(k)}))}} \xrightarrow{d} N(0,1)
\end{align*}

When $\tau_k(\bar{\mathbf{A}}^{(k-1)}, \bar{\mathbf{X}}^{(k)}) = 0$, this gives:
\begin{align*}
P(\widehat{\tau}_{k,\mathcal{I}_b}(\bar{\mathbf{A}}^{(k-1)}, \bar{\mathbf{X}}^{(k)}) > 0) \to \frac{1}{2}
\end{align*}


Therefore, the limit of the averaged decision rule is:
\begin{align*}
\lim_{n \to \infty} E[\bar{d}_{sb}^{k}(\bar{\mathbf{A}}^{(k-1)}, \bar{\mathbf{X}}^{(k)})] = \begin{cases}
1 & \text{if } \tau_k(\bar{\mathbf{A}}^{(k-1)}, \bar{\mathbf{X}}^{(k)}) > 0 \\
\frac{1}{2} & \text{if } \tau_k(\bar{\mathbf{A}}^{(k-1)}, \bar{\mathbf{X}}^{(k)}) = 0 \\
0 & \text{if } \tau_k(\bar{\mathbf{A}}^{(k-1)}, \bar{\mathbf{X}}^{(k)}) < 0
\end{cases}
\end{align*}

This corresponds to the oracle decision function:
\begin{align*}
\mathcal{D}_{sb}^{(k)}(\bar{\mathbf{A}}^{(k-1)}, \bar{\mathbf{X}}^{(k)}) = \mathbb{I}\{\tau_k(\bar{\mathbf{A}}^{(k-1)}, \bar{\mathbf{X}}^{(k)}) > 0\} + \frac{1}{2} \mathbb{I}\{\tau_k(\bar{\mathbf{A}}^{(k-1)}, \bar{\mathbf{X}}^{(k)}) = 0\}
\end{align*}


By Theorem 7 in \citet{shi2020breaking} and the recursive structure established in Definition \ref{multi-ralu}:
\begin{align*}
\widehat{V}_{sb}^{(1)} - \widehat{V}_{\mathcal{D}_{sb}}^{(1)*} = o_p(n^{-1/2})
\end{align*}

This completes the proof of both embedding results, showing that both adaptive smoothing and subbagging estimators can be represented as oracle estimators with specific decision functions $\mathcal{D}_{as}$ and $\mathcal{D}_{sb}$ respectively.
\end{proof}
\subsection{Proof of Corollary \ref{corob5}}
\begin{proof}
This result follows from Theorem \ref{thmb1} and Proposition \ref{multi-opt-oracle} by showing that the adaptive smoothing estimator is asymptotically equivalent to the optimal oracle estimator.


Firstly, we decompose the adaptive smoothing estimator. The adaptive smoothing estimator is:
\begin{align*}
    \widehat{V}_{as}^{(1)} &= \frac{1}{4}\sum_{j=1,2}\sum_{l=1,2}\frac{1}{|\mathcal{I}_{j,l}|}\sum_{i\in\mathcal{I}_{j,l}}\widehat{V}_i^{(1)}(d_{s}^{\star}(\mathbf{O}_i,\mathcal{I}_{3-j}),\widehat{\pi}_{\mathcal{I}_{j,l}^c},\widehat{\mu}_{\mathcal{I}_{j,l}^c})
\end{align*}

where $d_{s}^{\star}(\mathbf{O}_i,\mathcal{I}_{3-j})$ is the adaptive smoothing decision function with stage-wise components:
\begin{align}
    d_{s}^{1\star}(\bar{\mathbf{X}}_i^{(1)},\mathcal{I}_{3-j}) &= d_{s}^1(\bar{\mathbf{X}}_i^{(1)};\widehat{\tau}_{1,\mathcal{I}_{3-j}},h_{n,\mathcal{I}_{3-j}}^1, \widehat{\pi}_{1,\mathcal{I}_{3-j}}(1,\bar{\mathbf{X}}^{(1)}))\label{multi-as-1}\\
    d_{s}^{k\star}(\bar{\mathbf{A}}^{(k-1)},\bar{\mathbf{X}}^{(k)},\mathcal{I}_{3-j}) &= d_{s}^k(\bar{\mathbf{A}}^{(k-1)},\bar{\mathbf{X}}^{(k)};\widehat{\tau}_{k,\mathcal{I}_{3-j}},h_{n,\mathcal{I}_{3-j}}^k,\notag\\
    &\quad\widehat{\pi}_{k,\mathcal{I}_{3-j}}((\bar{\mathbf{A}}^{(k-1)},1),\bar{\mathbf{X}}^{(k)}))\label{multi-as-k}
\end{align}


We need to show:
\begin{align*}
    \widehat{V}_{as}^{(1)} - \widehat{V}_{\mathcal{D}^*}^{(1)*} = o_p(n^{-1/2})
\end{align*}

where $\mathcal{D}^*$ is the optimal decision function from Proposition \ref{multi-opt-oracle}
\begin{align}
    \mathcal{D}^{*(1)}(\bar{\mathbf{X}}^{(1)}) &= \mathbb{I}\{\tau_1(\bar{\mathbf{X}}^{(1)}) > 0\} + \pi_1(1,\bar{\mathbf{X}}^{(1)})\mathbb{I}\{\tau_1(\bar{\mathbf{X}}^{(1)}) = 0\}\label{multi-opt-1}\\
    \mathcal{D}^{*(k)}(\bar{\mathbf{A}}^{(k-1)},\bar{\mathbf{X}}^{(k)}) &= \mathbb{I}\{\tau_k(\bar{\mathbf{A}}^{(k-1)},\bar{\mathbf{X}}^{(k)}) > 0\}\notag\\
    &\quad + \pi_k((\bar{\mathbf{A}}^{(k-1)},1),\bar{\mathbf{X}}^{(k)})\mathbb{I}\{\tau_k(\bar{\mathbf{A}}^{(k-1)},\bar{\mathbf{X}}^{(k)}) = 0\}\label{multi-opt-k}
\end{align}


Then we proceed to verify the technical conditions required in Lemma \ref{multi-lem}.

{
For each fixed fold $j$, let
\begin{align*}
\mathcal{D}_n^{(1)}(\bar{\mathbf{X}}^{(1)},s)
&=d_s^1\!\left(\bar{\mathbf{X}}^{(1)};s,h_{n,\mathcal I_{3-j}}^1,
\widehat\pi_{1,\mathcal I_{3-j}}(1,\bar{\mathbf X}^{(1)})\right),\\
\mathcal{D}_n^{(k)}(\bar{\mathbf{A}}^{(k-1)},\bar{\mathbf{X}}^{(k)},s)
&=d_s^k\!\left(\bar{\mathbf{A}}^{(k-1)},\bar{\mathbf{X}}^{(k)};s,h_{n,\mathcal I_{3-j}}^k,
\widehat\pi_{k,\mathcal I_{3-j}}((\bar{\mathbf A}^{(k-1)},1),\bar{\mathbf X}^{(k)})\right),\quad k=2,\dots,K.
\end{align*}
Thus, the rules in Eqs.~\eqref{multi-as-1} and~\eqref{multi-as-k} are obtained by evaluating $\mathcal D_n^{(k)}$ at $\widehat\tau_{k,\mathcal I_{3-j}}$.
}

We verify Assumptions \ref{B10} to \ref{B12}:

\textbf{Assumption \ref{B10} (Uniform continuity):}
For each stage $k$ {and fixed fold $j$}, let:
\begin{align*}
    {\delta_{n,j}^k} &= \sqrt{h_{n,\mathcal{I}_{3-j}}^k \cdot n^{1/4}\left(E[|\widehat{\tau}_{k,\mathcal{I}_{3-j}}(\bar{\mathbf{A}}^{(k-1)},\bar{\mathbf{X}}^{(k)}) - \tau_k(\bar{\mathbf{A}}^{(k-1)},\bar{\mathbf{X}}^{(k)})|^2]\right)^{3/4}}
\end{align*}

By Assumptions \ref{B5} and the consistency of $\widehat{\tau}_{k,\mathcal{I}_{3-j}}$:
\begin{align*}
    P(|\widehat{\tau}_{k,\mathcal{I}_{3-j}}(\bar{\mathbf{A}}^{(k-1)},\bar{\mathbf{X}}^{(k)}) - \tau_k(\bar{\mathbf{A}}^{(k-1)},\bar{\mathbf{X}}^{(k)})| \geq {\delta_{n,j}^k}) = o_p(1)
\end{align*}

The uniform continuity of {$\mathcal{D}_n^{(k)}$} follows from its {piecewise} linear structure.

\textbf{Assumption \ref{B11} (Consistency):}
We have:
\begin{align*}
    \lim_{n \to \infty} {\mathcal{D}_n^{(k)}}(\bar{\mathbf{A}}^{(k-1)},\bar{\mathbf{X}}^{(k)},\tau_k(\bar{\mathbf{A}}^{(k-1)},\bar{\mathbf{X}}^{(k)})) &= \mathcal{D}^{*(k)}(\bar{\mathbf{A}}^{(k-1)},\bar{\mathbf{X}}^{(k)})
\end{align*}

This follows because:
(i) When $\tau_k(\bar{\mathbf{A}}^{(k-1)},\bar{\mathbf{X}}^{(k)}) > 0$: For large $n$, $h_n^k \to 0$, so {$\mathcal{D}_n^{(k)}(\cdot,\tau_k) \to 1$};
(ii) When $\tau_k(\bar{\mathbf{A}}^{(k-1)},\bar{\mathbf{X}}^{(k)}) < 0$: For large $n$, $h_n^k \to 0$, so {$\mathcal{D}_n^{(k)}(\cdot,\tau_k) \to 0$};  
(iii) When $\tau_k(\bar{\mathbf{A}}^{(k-1)},\bar{\mathbf{X}}^{(k)}) = 0$: {$\mathcal{D}_n^{(k)}(\cdot,\tau_k)$} $\to \widehat{\pi}_{k,\mathcal{I}_{3-j}}((\bar{\mathbf{A}}^{(k-1)},1),\bar{\mathbf{X}}^{(k)}) \to_p \pi_k((\bar{\mathbf{A}}^{(k-1)},1),\bar{\mathbf{X}}^{(k)})$.

\textbf{Assumption \ref{B12} (Concentration):}
Let ${\delta_{n,j}^{\prime k}} = 2h_{n,\mathcal{I}_{3-j}}^k$. {By Assumptions \ref{B5} and \ref{B6},
\begin{align*}
\delta_{n,j}^k\lor n^{1/4}\left(E[|\widehat\tau_{k,\mathcal I_{3-j}}-\tau_k|^2]\right)^{3/4}
\prec_p\delta_{n,j}^{\prime k}\prec n^{-\frac{1}{2(1+\alpha)}}.
\end{align*}}
When $|s| > {\delta_{n,j}^{\prime k}}$, by the construction of {$\mathcal{D}_n^{(k)}$}:
\begin{align*}
    |{\mathcal{D}_n^{(k)}}(\bar{\mathbf{A}}^{(k-1)},\bar{\mathbf{X}}^{(k)},s) - {\mathbb{I}\{s>0\}}| &{=0}= o\left(\frac{1}{|s|\sqrt{n}}\right)
\end{align*}

This follows from the linear structure of the smoothing function{,} Assumption \ref{B6}{,} {and the stage-specific fitted-overlap condition stated in Section A.1}.


By Lemma \ref{multi-lem} (multi-stage technical lemma), under Assumptions \ref{B1} to \ref{B4}, \ref{B7} and \ref{B10} to \ref{B12}:
\begin{align*}
    \left|\widehat{V}_{as}^{(1)} - \widehat{V}_{\mathcal{D}^*}^{(1)*}\right| = o_p(n^{-1/2})
\end{align*}


Since $\widehat{V}_{\mathcal{D}^*}^{(1)*}$ is a sum of i.i.d. terms with the optimal variance (by Proposition \ref{multi-opt-oracle}), we have:
\begin{align*}
    \sqrt{n}(\widehat{V}_{as}^{(1)} - V_0) &= \sqrt{n}(\widehat{V}_{\mathcal{D}^*}^{(1)*} - V_0) + o_p(1)\notag\\
    &\to_d N(0, \sigma_{opt}^2)
\end{align*}

where $\sigma_{opt}^2$ is the optimal variance from Proposition \ref{multi-opt-oracle}.


The variance estimator $(\widehat{\sigma}_{as}^{(1)})^2$ converges in probability to $\sigma_{opt}^2$ by the continuous mapping theorem and the consistency of the sample variance.

This establishes the asymptotic normality and optimality of $\widehat{V}_{as}^{(1)}$.
\end{proof}
\subsection{Proof of Corollary \ref{corob7}}
\begin{proof}
We establish that $\sqrt{n}(\widehat{V}_{as}^{(1)} - \widehat{V}_{as,B}^{(1)}) = o_p(1)$ as $B \to \infty$.

For each repetition $b = 1, \ldots, B$, we generate an independent random partition $\{\mathcal{I}_{b,1}, \mathcal{I}_{b,2}\}$ of $[n]$, where each $\mathcal{I}_{b,j}$ is further subdivided into $\{\mathcal{I}_{b,j,1}, \mathcal{I}_{b,j,2}\}$ with equal sizes. Define the estimator for the $b$-th partition:
\begin{align*}
    \widehat{V}_{as}^{(1,b)} &= \frac{1}{4}\sum_{j=1,2}\sum_{k=1,2}\frac{1}{|\mathcal{I}_{b,j,k}|}\sum_{i\in\mathcal{I}_{b,j,k}}\widehat{V}_i^{(1)}(d_s^\star(\mathbf{O}_i,\mathcal{I}_{b,3-j}),\widehat{\pi}_{\mathcal{I}_{b,j,k}^c},\widehat{\mu}_{\mathcal{I}_{b,j,k}^c})
\end{align*}

The repeated cross-fitting estimator is:
\begin{align*}
    \widehat{V}_{as,B}^{(1)} &= \frac{1}{B}\sum_{b=1}^B \widehat{V}_{as}^{(1,b)}
\end{align*}

Without loss of generality, let $\widehat{V}_{as}^{(1)} = \widehat{V}_{as}^{(1,1)}$. We decompose:
\begin{align*}
    \widehat{V}_{as}^{(1)} - \widehat{V}_{as,B}^{(1)} &= \widehat{V}_{as}^{(1,1)} - \frac{1}{B}\sum_{b=1}^B \widehat{V}_{as}^{(1,b)}\\
    &= \frac{B-1}{B}\widehat{V}_{as}^{(1,1)} - \frac{1}{B}\sum_{b=2}^B \widehat{V}_{as}^{(1,b)}\\
    &= \frac{1}{B}\sum_{b=2}^B (\widehat{V}_{as}^{(1,1)} - \widehat{V}_{as}^{(1,b)})
\end{align*}

For each $b \geq 2$, we analyze $\widehat{V}_{as}^{(1,1)} - \widehat{V}_{as}^{(1,b)}$. Let $\mathcal{J}_i^{(b)} = (j_i^{(b)}, k_i^{(b)}) \in \{(1,1), (1,2), (2,1), (2,2)\}$ denote the subset assignment for observation $i$ in partition $b$. Then:
\begin{align*}
    \widehat{V}_{as}^{(1,b)} &= \frac{1}{n}\sum_{i=1}^n \widehat{V}_i^{(1)}(d_s^\star(\mathbf{O}_i,\mathcal{I}_{b,3-j_i^{(b)}}),\widehat{\pi}_{\mathcal{I}_{b,\mathcal{J}_i^{(b)}}^c},\widehat{\mu}_{\mathcal{I}_{b,\mathcal{J}_i^{(b)}}^c})
\end{align*}

The difference can be written as:
\begin{align*}
    &\widehat{V}_{as}^{(1,1)} - \widehat{V}_{as}^{(1,b)}\\
    &= \frac{1}{n}\sum_{i=1}^n \left[\widehat{V}_i^{(1)}(d_s^\star(\mathbf{O}_i,\mathcal{I}_{1,3-j_i^{(1)}}),\widehat{\pi}_{\mathcal{I}_{1,\mathcal{J}_i^{(1)}}^c},\widehat{\mu}_{\mathcal{I}_{1,\mathcal{J}_i^{(1)}}^c})\right.\\
    &\quad\quad\left. - \widehat{V}_i^{(1)}(d_s^\star(\mathbf{O}_i,\mathcal{I}_{b,3-j_i^{(b)}}),\widehat{\pi}_{\mathcal{I}_{b,\mathcal{J}_i^{(b)}}^c},\widehat{\mu}_{\mathcal{I}_{b,\mathcal{J}_i^{(b)}}^c})\right]
\end{align*}

For each observation $i$, we decompose the individual difference:
\begin{align*}
    &\widehat{V}_i^{(1)}(d_s^\star(\mathbf{O}_i,\mathcal{I}_{1,3-j_i^{(1)}}),\widehat{\pi}_{\mathcal{I}_{1,\mathcal{J}_i^{(1)}}^c},\widehat{\mu}_{\mathcal{I}_{1,\mathcal{J}_i^{(1)}}^c})\\
    &\quad - \widehat{V}_i^{(1)}(d_s^\star(\mathbf{O}_i,\mathcal{I}_{b,3-j_i^{(b)}}),\widehat{\pi}_{\mathcal{I}_{b,\mathcal{J}_i^{(b)}}^c},\widehat{\mu}_{\mathcal{I}_{b,\mathcal{J}_i^{(b)}}^c})\\
    &= \underbrace{\widehat{V}_i^{(1)}(d_s^\star(\mathbf{O}_i,\mathcal{I}_{1,3-j_i^{(1)}}),\widehat{\pi}_{\mathcal{I}_{1,\mathcal{J}_i^{(1)}}^c},\widehat{\mu}_{\mathcal{I}_{1,\mathcal{J}_i^{(1)}}^c}) - \widehat{V}_i^{(1)}(d_s^\star(\mathbf{O}_i,\mathcal{I}_{1,3-j_i^{(1)}}),\pi,\mu)}_{T_{i,1}}\\
    &\quad + \underbrace{\widehat{V}_i^{(1)}(d_s^\star(\mathbf{O}_i,\mathcal{I}_{1,3-j_i^{(1)}}),\pi,\mu) - \widehat{V}_i^{(1)}(d_s^\star(\mathbf{O}_i,\mathcal{I}_{b,3-j_i^{(b)}}),\pi,\mu)}_{T_{i,2}}\\
    &\quad + \underbrace{\widehat{V}_i^{(1)}(d_s^\star(\mathbf{O}_i,\mathcal{I}_{b,3-j_i^{(b)}}),\pi,\mu) - \widehat{V}_i^{(1)}(d_s^\star(\mathbf{O}_i,\mathcal{I}_{b,3-j_i^{(b)}}),\widehat{\pi}_{\mathcal{I}_{b,\mathcal{J}_i^{(b)}}^c},\widehat{\mu}_{\mathcal{I}_{b,\mathcal{J}_i^{(b)}}^c})}_{T_{i,3}}
\end{align*}

For $T_{i,1}$ and $T_{i,3}$, by the doubly robust convergence rates in Assumption \ref{B7} and the cross-fitting structure:
\begin{align*}
    E[|T_{i,1}|^2 | \mathcal{I}_{1,\mathcal{J}_i^{(1)}}^c] &= O(n^{-1})\\
    E[|T_{i,3}|^2 | \mathcal{I}_{b,\mathcal{J}_i^{(b)}}^c] &= O(n^{-1})
\end{align*}

For $T_{i,2}$, since both estimators use true nuisance parameters, this term captures the difference in the optimal decision functions estimated from different subsets:
\begin{align*}
    T_{i,2} &= \widehat{V}_i^{(1)}(d_s^\star(\mathbf{O}_i,\mathcal{I}_{1,3-j_i^{(1)}}),\pi,\mu) - \widehat{V}_i^{(1)}(d_s^\star(\mathbf{O}_i,\mathcal{I}_{b,3-j_i^{(b)}}),\pi,\mu)
\end{align*}

By Assumptions \ref{B5} and \ref{B6} and the technical Lemma, both $d_s^\star(\mathbf{O}_i,\mathcal{I}_{1,3-j_i^{(1)}})$ and $d_s^\star(\mathbf{O}_i,\mathcal{I}_{b,3-j_i^{(b)}})$ converge to the same limit $d_s^*(\mathbf{O}_i)$. The convergence rate implies:
\begin{align*}
    E[|T_{i,2}|^2] &= O(n^{-1})
\end{align*}

Combining all terms:
\begin{align*}
    E\left[\left|\frac{1}{n}\sum_{i=1}^n (T_{i,1} + T_{i,2} + T_{i,3})\right|^2\right] &\leq \frac{3}{n^2}\sum_{i=1}^n \left(E[|T_{i,1}|^2] + E[|T_{i,2}|^2] + E[|T_{i,3}|^2]\right)\\
    &= \frac{3}{n^2} \cdot n \cdot O(n^{-1}) = O(n^{-2})
\end{align*}

Therefore:
\begin{align*}
    E[|\widehat{V}_{as}^{(1,1)} - \widehat{V}_{as}^{(1,b)}|^2] = O(n^{-2})
\end{align*}

The variance of the difference is:
\begin{align*}
    \text{Var}\left[\sqrt{n}(\widehat{V}_{as}^{(1)} - \widehat{V}_{as,B}^{(1)})\right] &= n \cdot \text{Var}\left[\frac{1}{B}\sum_{b=2}^B (\widehat{V}_{as}^{(1,1)} - \widehat{V}_{as}^{(1,b)})\right]\\
    &= \frac{n}{B^2} \sum_{b=2}^B \text{Var}[\widehat{V}_{as}^{(1,1)} - \widehat{V}_{as}^{(1,b)}]\\
    &\leq \frac{n(B-1)}{B^2} \cdot O(n^{-2}) = O(B^{-1}n^{-1})
\end{align*}

where we used the independence of different partitions.

As $B \to \infty$, we have $\text{Var}[\sqrt{n}(\widehat{V}_{as}^{(1)} - \widehat{V}_{as,B}^{(1)})] \to 0$, which implies:
\begin{align*}
    \sqrt{n}(\widehat{V}_{as}^{(1)} - \widehat{V}_{as,B}^{(1)}) = o_p(1)
\end{align*}

This completes the proof.
\end{proof}
\subsection{Proof of Corollary \ref{hyb-general}}
\begin{proof}
We establish that $\mathcal{T}_{eal}^{multi} \subset \mathcal{T}_{RL}^{multi}$ by showing that any estimator in $\mathcal{T}_{eal}^{multi}$ can be written in the recursive influence function form required by Definition \ref{multi-ralu}.

Let $T_n^{(1)} \in \mathcal{T}_{eal}^{multi}$. By Definition 3, there exists a decision function sequence $\mathcal{D} = (\mathcal{D}^{(1)}, \ldots, \mathcal{D}^{(K)}) \in \bar{\mathbb{D}}(V_0)$ such that:
\begin{align*}
    T_n^{(1)} - \widehat{V}_{\mathcal{D}}^{(1)*} = o_p(n^{-1/2})
\end{align*}

where $\widehat{V}_{\mathcal{D}}^{(1)*} = \frac{1}{n}\sum_{i=1}^n \widehat{V}_i^{(1)}(\mathcal{D}(\mathbf{O}_i), \pi, \mu)$ is the oracle estimator using true nuisance parameters.

We construct the recursive influence functions explicitly. Define the terminal condition:
\begin{align*}
    f^{(K+1)}(\mathbf{O}_i) = Y_i
\end{align*}

For each stage $k = K, K-1, \ldots, 1$, define:
\begin{align*}
    &f^{(k)}(\bar{\mathbf{X}}_i^{(k)}, \bar{\mathbf{A}}_i^{(k)}, f^{(k+1)}, \mu_1^{(k)}, \mu_0^{(k)}, \pi_1^{(k)})\\
    &= \frac{\mathrm{g}\{A_i^{(k)}, \mathcal{D}^{(k)}(\bar{\mathbf{A}}_i^{(k-1)}, \bar{\mathbf{X}}_i^{(k)})\}}{\pi_k^{(k)}(\bar{\mathbf{A}}_i^{(k)}, \bar{\mathbf{X}}_i^{(k)})} \{f^{(k+1)}(\mathbf{O}_i) - \mu_k^{(k)}(\bar{\mathbf{A}}_i^{(k)}, \bar{\mathbf{X}}_i^{(k)})\}\\
    &\quad + \mathcal{D}^{(k)}(\bar{\mathbf{A}}_i^{(k-1)}, \bar{\mathbf{X}}_i^{(k)}) \mu_1^{(k)}(\bar{\mathbf{A}}_i^{(k-1)}, \bar{\mathbf{X}}_i^{(k)})\\
    &\quad + (1-\mathcal{D}^{(k)}(\bar{\mathbf{A}}_i^{(k-1)}, \bar{\mathbf{X}}_i^{(k)})) \mu_0^{(k)}(\bar{\mathbf{A}}_i^{(k-1)}, \bar{\mathbf{X}}_i^{(k)})
\end{align*}

where:
\begin{align*}
    \pi_k^{(k)}(\bar{\mathbf{A}}_i^{(k)}, \bar{\mathbf{X}}_i^{(k)}) &= A_i^{(k)}\pi_1^{(k)}(\bar{\mathbf{A}}_i^{(k-1)}, \bar{\mathbf{X}}_i^{(k)}) + (1-A_i^{(k)})(1-\pi_1^{(k)}(\bar{\mathbf{A}}_i^{(k-1)}, \bar{\mathbf{X}}_i^{(k)}))\\
    \mu_k^{(k)}(\bar{\mathbf{A}}_i^{(k)}, \bar{\mathbf{X}}_i^{(k)}) &= A_i^{(k)} \mu_1^{(k)}(\bar{\mathbf{A}}_i^{(k-1)}, \bar{\mathbf{X}}_i^{(k)}) + (1-A_i^{(k)}) \mu_0^{(k)}(\bar{\mathbf{A}}_i^{(k-1)}, \bar{\mathbf{X}}_i^{(k)})
\end{align*}

The oracle estimator can be written as:
\begin{align*}
    \widehat{V}_{\mathcal{D}}^{(1)*} = \frac{1}{n}\sum_{i=1}^n f^{(1)}(\bar{\mathbf{X}}_i^{(1)}, \bar{\mathbf{A}}_i^{(1)}, f^{(2)}, \mu_1^{(1)}, \mu_0^{(1)}, \pi_1^{(1)})
\end{align*}

where the functions correspond to the true nuisance parameters.

We now verify that the constructed influence functions satisfy the requirements of Definition \ref{multi-ralu}. The function $f^{(k)}$ must satisfy the double robustness property: $E[f^{(k)}] = V_0^{(k)}$ when either the outcome regression model or the propensity score model is correctly specified.

We consider two cases for each stage $k$.

Case 1: Outcome regression model is correctly specified. Suppose:
\begin{align*}
    \mu_k(\bar{\mathbf{A}}_i^{(k)}, \bar{\mathbf{X}}_i^{(k)}) = A_i^{(k)} \mu_1^{(k)}(\bar{\mathbf{A}}_i^{(k-1)}, \bar{\mathbf{X}}_i^{(k)}) + (1-A_i^{(k)}) \mu_0^{(k)}(\bar{\mathbf{A}}_i^{(k-1)}, \bar{\mathbf{X}}_i^{(k)})
\end{align*}

Then $\mu_k^{(k)}(\bar{\mathbf{A}}_i^{(k)}, \bar{\mathbf{X}}_i^{(k)}) = \mu_k(\bar{\mathbf{A}}_i^{(k)}, \bar{\mathbf{X}}_i^{(k)})$. Taking the conditional expectation:
\begin{align*}
    &E\left[\frac{\mathrm{g}\{A_i^{(k)}, \mathcal{D}^{(k)}(\bar{\mathbf{A}}_i^{(k-1)}, \bar{\mathbf{X}}_i^{(k)})\}}{\pi_k^{(k)}(\bar{\mathbf{A}}_i^{(k)}, \bar{\mathbf{X}}_i^{(k)})} \{f^{(k+1)}(\mathbf{O}_i) - \mu_k(\bar{\mathbf{A}}_i^{(k)}, \bar{\mathbf{X}}_i^{(k)})\}\bigg|\bar{\mathbf{X}}_i^{(k)}, \bar{\mathbf{A}}_i^{(k-1)}\right]\\
    = &E\left[\frac{\mathrm{g}\{A_i^{(k)}, \mathcal{D}^{(k)}(\bar{\mathbf{A}}_i^{(k-1)}, \bar{\mathbf{X}}_i^{(k)})\}}{\pi_k^{(k)}(\bar{\mathbf{A}}_i^{(k)}, \bar{\mathbf{X}}_i^{(k)})} \{E[f^{(k+1)}(\mathbf{O}_i)|\bar{\mathbf{X}}_i^{(k)}, \bar{\mathbf{A}}_i^{(k)}] - \mu_k(\bar{\mathbf{A}}_i^{(k)}, \bar{\mathbf{X}}_i^{(k)})\}\bigg|\bar{\mathbf{X}}_i^{(k)}, \bar{\mathbf{A}}_i^{(k-1)}\right]
\end{align*}

By the recursive definition of the value functions, we have $E[f^{(k+1)}(\mathbf{O}_i)|\bar{\mathbf{X}}_i^{(k)}, \bar{\mathbf{A}}_i^{(k)}] = \mu_k(\bar{\mathbf{A}}_i^{(k)}, \bar{\mathbf{X}}_i^{(k)})$. Therefore, the above expectation equals zero.

The conditional expectation of $f^{(k)}$ becomes:
\begin{align*}
    &E[f^{(k)}(\mathbf{O}_i)|\bar{\mathbf{X}}_i^{(k)}, \bar{\mathbf{A}}_i^{(k-1)}]\\
    &= \mathcal{D}^{(k)}(\bar{\mathbf{A}}_i^{(k-1)}, \bar{\mathbf{X}}_i^{(k)}) \mu_1^{(k)}(\bar{\mathbf{A}}_i^{(k-1)}, \bar{\mathbf{X}}_i^{(k)})\\
    &\quad + (1-\mathcal{D}^{(k)}(\bar{\mathbf{A}}_i^{(k-1)}, \bar{\mathbf{X}}_i^{(k)})) \mu_0^{(k)}(\bar{\mathbf{A}}_i^{(k-1)}, \bar{\mathbf{X}}_i^{(k)})\\
    &= \mathcal{D}^{(k)}(\bar{\mathbf{A}}_i^{(k-1)}, \bar{\mathbf{X}}_i^{(k)}) \mu_k((\bar{\mathbf{A}}_i^{(k-1)}, 1), \bar{\mathbf{X}}_i^{(k)})\\
    &\quad + (1-\mathcal{D}^{(k)}(\bar{\mathbf{A}}_i^{(k-1)}, \bar{\mathbf{X}}_i^{(k)})) \mu_k((\bar{\mathbf{A}}_i^{(k-1)}, 0), \bar{\mathbf{X}}_i^{(k)})\\
    &= V_0^{(k)}(\bar{\mathbf{A}}_i^{(k-1)}, \bar{\mathbf{X}}_i^{(k)})
\end{align*}

where the last equality follows from the definition of optimal decision functions in $\bar{\mathbb{D}}(V_0)$.

Case 2: Propensity score model is correctly specified. Suppose:
\begin{align*}
    \pi_k((\bar{\mathbf{A}}_i^{(k-1)}, 1), \bar{\mathbf{X}}_i^{(k)}) = \pi_1^{(k)}(\bar{\mathbf{A}}_i^{(k-1)}, \bar{\mathbf{X}}_i^{(k)})
\end{align*}

Then $\pi_k^{(k)}(\bar{\mathbf{A}}_i^{(k)}, \bar{\mathbf{X}}_i^{(k)}) = \pi_k(\bar{\mathbf{A}}_i^{(k)}, \bar{\mathbf{X}}_i^{(k)})$. The conditional expectation of the IPW term is:
\begin{align*}
    &E\left[\frac{\mathrm{g}\{A_i^{(k)}, \mathcal{D}^{(k)}(\bar{\mathbf{A}}_i^{(k-1)}, \bar{\mathbf{X}}_i^{(k)})\}}{\pi_k(\bar{\mathbf{A}}_i^{(k)}, \bar{\mathbf{X}}_i^{(k)})} f^{(k+1)}(\mathbf{O}_i)\bigg|\bar{\mathbf{X}}_i^{(k)}, \bar{\mathbf{A}}_i^{(k-1)}\right]\\
    &= \sum_{a=0,1} \frac{\mathrm{g}\{a, \mathcal{D}^{(k)}(\bar{\mathbf{A}}_i^{(k-1)}, \bar{\mathbf{X}}_i^{(k)})\}}{\pi_k((\bar{\mathbf{A}}_i^{(k-1)}, a), \bar{\mathbf{X}}_i^{(k)})} \pi_k((\bar{\mathbf{A}}_i^{(k-1)}, a), \bar{\mathbf{X}}_i^{(k)}) E[f^{(k+1)}(\mathbf{O}_i)|(\bar{\mathbf{A}}_i^{(k-1)}, a), \bar{\mathbf{X}}_i^{(k)}]\\
    &= \sum_{a=0,1} \mathrm{g}\{a, \mathcal{D}^{(k)}(\bar{\mathbf{A}}_i^{(k-1)}, \bar{\mathbf{X}}_i^{(k)})\} E[f^{(k+1)}(\mathbf{O}_i)|(\bar{\mathbf{A}}_i^{(k-1)}, a), \bar{\mathbf{X}}_i^{(k)}]\\
    &= \mathcal{D}^{(k)}(\bar{\mathbf{A}}_i^{(k-1)}, \bar{\mathbf{X}}_i^{(k)}) E[f^{(k+1)}(\mathbf{O}_i)|(\bar{\mathbf{A}}_i^{(k-1)}, 1), \bar{\mathbf{X}}_i^{(k)}]\\
    &\quad + (1-\mathcal{D}^{(k)}(\bar{\mathbf{A}}_i^{(k-1)}, \bar{\mathbf{X}}_i^{(k)})) E[f^{(k+1)}(\mathbf{O}_i)|(\bar{\mathbf{A}}_i^{(k-1)}, 0), \bar{\mathbf{X}}_i^{(k)}]\\
    &= V_0^{(k)}(\bar{\mathbf{A}}_i^{(k-1)}, \bar{\mathbf{X}}_i^{(k)})
\end{align*}

For the outcome regression term:
\begin{align*}
    &E\left[\frac{\mathrm{g}\{A_i^{(k)}, \mathcal{D}^{(k)}(\bar{\mathbf{A}}_i^{(k-1)}, \bar{\mathbf{X}}_i^{(k)})\}}{\pi_k(\bar{\mathbf{A}}_i^{(k)}, \bar{\mathbf{X}}_i^{(k)})} \mu_k^{(k)}(\bar{\mathbf{A}}_i^{(k)}, \bar{\mathbf{X}}_i^{(k)})\bigg|\bar{\mathbf{X}}_i^{(k)}, \bar{\mathbf{A}}_i^{(k-1)}\right]\\
    &= \sum_{a=0,1} \mathrm{g}\{a, \mathcal{D}^{(k)}(\bar{\mathbf{A}}_i^{(k-1)}, \bar{\mathbf{X}}_i^{(k)})\} \mu_k^{(k)}((\bar{\mathbf{A}}_i^{(k-1)}, a), \bar{\mathbf{X}}_i^{(k)})\\
    &= \mathcal{D}^{(k)}(\bar{\mathbf{A}}_i^{(k-1)}, \bar{\mathbf{X}}_i^{(k)}) \mu_1^{(k)}(\bar{\mathbf{A}}_i^{(k-1)}, \bar{\mathbf{X}}_i^{(k)})\\
    &\quad + (1-\mathcal{D}^{(k)}(\bar{\mathbf{A}}_i^{(k-1)}, \bar{\mathbf{X}}_i^{(k)})) \mu_0^{(k)}(\bar{\mathbf{A}}_i^{(k-1)}, \bar{\mathbf{X}}_i^{(k)})
\end{align*}

Therefore:
\begin{align*}
    E[f^{(k)}(\mathbf{O}_i)|\bar{\mathbf{X}}_i^{(k)}, \bar{\mathbf{A}}_i^{(k-1)}] = V_0^{(k)}(\bar{\mathbf{A}}_i^{(k-1)}, \bar{\mathbf{X}}_i^{(k)})
\end{align*}

This establishes the double robustness property for each stage.

By induction from stage $K$ to stage 1, we have $E[f^{(1)}] = V_0$. The differentiability requirements follow from the explicit functional form of $f^{(k)}$.

Since $T_n^{(1)} - \widehat{V}_{\mathcal{D}}^{(1)*} = o_p(n^{-1/2})$, we can write:
\begin{align*}
    T_n^{(1)} = \frac{1}{n}\sum_{i=1}^n f^{(1)}(\bar{\mathbf{X}}_i^{(1)}, \bar{\mathbf{A}}_i^{(1)}, f^{(2)}, \mu_1^{(1)}, \mu_0^{(1)}, \pi_1^{(1)}) + o_p(n^{-1/2})
\end{align*}

where the influence functions satisfy all requirements of Definition \ref{multi-ralu}. Therefore $T_n^{(1)} \in \mathcal{T}_{RL}^{multi}$.

For specific examples, the multi-stage adaptive smoothing estimator $\widehat{V}_{as}^{(1)}$ satisfies the asymptotic equivalence with $\mathcal{D}^{(k)}$ given by:
\begin{align*}
    \mathcal{D}^{(k)}(\bar{\mathbf{A}}_i^{(k-1)}, \bar{\mathbf{X}}_i^{(k)}) = \mathbb{I}\{\tau_k(\bar{\mathbf{A}}_i^{(k-1)}, \bar{\mathbf{X}}_i^{(k)}) > 0\} + \pi_k((\bar{\mathbf{A}}_i^{(k-1)}, 1), \bar{\mathbf{X}}_i^{(k)}) \mathbb{I}\{\tau_k(\bar{\mathbf{A}}_i^{(k-1)}, \bar{\mathbf{X}}_i^{(k)}) = 0\}
\end{align*}

Similarly, the multi-stage subbagging estimator $\widehat{V}_{sb}^{(1)}$ corresponds to:
\begin{align*}
    \mathcal{D}^{(k)}(\bar{\mathbf{A}}_i^{(k-1)}, \bar{\mathbf{X}}_i^{(k)}) = \mathbb{I}\{\tau_k(\bar{\mathbf{A}}_i^{(k-1)}, \bar{\mathbf{X}}_i^{(k)}) > 0\} + 0.5 \cdot \mathbb{I}\{\tau_k(\bar{\mathbf{A}}_i^{(k-1)}, \bar{\mathbf{X}}_i^{(k)}) = 0\}
\end{align*}

This establishes $\mathcal{T}_{eal}^{multi} \subset \mathcal{T}_{RL}^{multi}$ for both regular and nonregular cases.
\end{proof}
\subsection{Proof of Theorem \ref{hyb-var-bound}}
\begin{proof}
   Recall that in Eq. \eqref{as-eal-embedding} we have $\widehat{V}_{as}^{(1)} - \widehat{V}_{\mathcal{D}_{as}}^{(1)*} = o_p(n^{-1/2})$. Under Assumption \ref{B9}, the decision function defined in Eq. \eqref{K-multi-heter} and  Eq. \eqref{k-1-multi-heter} can actually reduce to $\mathcal{D}_{as}$. Therefore, Theorem \ref{hyb-var-bound} can be directly implied by Theorem \ref{multi-heter-eff-opt}.
\end{proof}
\section{Extensions to Heteroscedasticity in Multiple Time Point Case}\label{appD}
In this section, we mainly discuss how to achieve the optimal efficiency in the case of heteroscedasticity in multiple time point case. Compared with that in the single time point case, the situation is a bit more complex and decision functions at each stage which optimize the efficiency need be defined sequentially. In specific, with some calculation we find the following decision function is optimal:
\begin{align}\label{K-multi-heter}
\begin{aligned}
      &\mathcal{D}^{(K)}(\bar{\mathbf{A}}^{(K-1)},\bar{\mathbf{X}}^{(K)})\\
     =&\mathbb{I}\{\tau_K(\bar{\mathbf{A}}^{(K-1)},\bar{\mathbf{X}}^{(K)}) > 0\}+\mathbb{I}\{\tau_K(\bar{\mathbf{A}}^{(K-1)},\bar{\mathbf{X}}^{(K)}) = 0\}\\
   &\times\left(\operatorname{Var}(Y|\bar{\mathbf{A}}^{(K-1)},A^{(K)}=1,\bar{\mathbf{X}}^{(K)})\pi_K(\bar{\mathbf{A}}^{(K-1)},A^{(K)}=0,\bar{\mathbf{X}}^{(K)})\right.\\
   &+\left.\operatorname{Var}(Y|\bar{\mathbf{A}}^{(K-1)},A^{(K)}=0,\bar{\mathbf{X}}^{(K)})\pi_K(\bar{\mathbf{A}}^{(K-1)},A^{(K)}=1,\bar{\mathbf{X}}^{(K)})\right)^{-1}\\
   &\cdot \operatorname{Var}(Y|\bar{\mathbf{A}}^{(K-1)},A^{(K)}=0,\bar{\mathbf{X}}^{(K)})\pi_K(\bar{\mathbf{A}}^{(K-1)},A^{(K)}=1,\bar{\mathbf{X}}^{(K)})
\end{aligned}
\end{align}
and 
\begin{align}\label{k-1-multi-heter}
\begin{aligned}
     &\mathcal{D}^{(k)}(\bar{\mathbf{A}}^{(k-1)},\bar{\mathbf{X}}^{(k)})\\
   =&\mathbb{I}\{\tau_k(\bar{\mathbf{A}}^{(k-1)},\bar{\mathbf{X}}^{(k)}) > 0\}+\mathbb{I}\{\tau_k(\bar{\mathbf{A}}^{(k-1)},\bar{\mathbf{X}}^{(k)}) = 0\}\\
   &\times\left(\operatorname{Var}(\widehat{V}^{(k+1)}(\mathcal{D};\pi,\mu)|\bar{\mathbf{A}}^{(k-1)},A^{(k)}=1,\bar{\mathbf{X}}^{(k)})\pi_k(\bar{\mathbf{A}}^{(k-1)},A^{(k)}=0,\bar{\mathbf{X}}^{(k)})\right.\\
   &+\left.\operatorname{Var}(\widehat{V}^{(k+1)}(\mathcal{D};\pi,\mu)|\bar{\mathbf{A}}^{(k-1)},A^{(k)}=0,\bar{\mathbf{X}}^{(k)})\pi_k(\bar{\mathbf{A}}^{(k-1)},A^{(k)}=1,\bar{\mathbf{X}}^{(k)})\right)^{-1}\\
   &\cdot \operatorname{Var}(\widehat{V}^{(k+1)}(\mathcal{D};\pi,\mu)|\bar{\mathbf{A}}^{(k-1)},A^{(k)}=0,\bar{\mathbf{X}}^{(k)})\pi_k(\bar{\mathbf{A}}^{(k-1)},A^{(k)}=1,\bar{\mathbf{X}}^{(k)})
\end{aligned}
\end{align}
for $k\leq K-1$. Here $\widehat{V}^{(k+1)}(\mathcal{D};\pi,\mu)$ is defined sequentially the same way as before, which only relies on $\mathcal{D}^{(k')}$ s.t. $k'\ge k+1$. Similar to the single time point case, the efficiency optimality among $\mathcal{T}_{RL}^{multi}$ could be achieved in practice by consistently estimating Eq. \eqref{K-multi-heter} and  Eq. \eqref{k-1-multi-heter} and plugging in. This efficiency optimality result in multiple time point case allowing heteroscedasticity, is summarized as Theorem \ref{multi-heter-eff-opt}.
\begin{theorem}\label{multi-heter-eff-opt}
     Under Assumptions \ref{B1} to \ref{B3}, when the working models 
     $\mu_1^{(k)}$, $\mu_0^{(k)}$ and $\pi_1^{(k)}$ in $T_n$ are 
     correctly specified for all $k=1,\dots,K$, and $T_n$ satisfies 
     asymptotic uniform integrability, i.e., 
     $\lim_{M \rightarrow \infty} \sup_n 
     E\left[n\left(T_n^{(1)}-V_0\right)^2 \cdot 
     \mathbf{1}\left(n\left(T_n^{(1)}-V_0\right)^2>M\right)
     \right]=0$, we have
    \begin{align*}
        \inf_{T_n\in\mathcal{T}_{RL}^{multi}}
        \lim_{n\to\infty}\operatorname{Var}(\sqrt{n}T_n^{(1)})
        = \lim_{n\to\infty}\operatorname{Var}
        (\sqrt{n}\widehat{V}_{\mathcal{D}}^{(1)*})
    \end{align*}
    where $\mathcal{D}$ is defined as in Eq. \eqref{K-multi-heter} 
    and Eq. \eqref{k-1-multi-heter}.
\end{theorem}
\begin{proof}[Proof of Theorem \ref{multi-heter-eff-opt}]
We prove the efficiency optimality result by backward induction from stage $K$ to stage $1$. The proof consists of three main parts: (1) establishing conditional unbiasedness of influence functions, (2) deriving the structure of optimal influence functions at each stage, and (3) computing the variance lower bound.

We first establish that for any estimator $T_n^{(1)} \in \mathcal{T}_{RL}^{multi}$, the influence function $f^{(k)}$ at each stage $k$ must satisfy conditional unbiasedness.

By Definition \ref{multi-ralu}, for any joint distribution of the data, we have $E[f^{(k)}] = V_0^{(k)}$ when either the outcome model or propensity score model is correctly specified. Following the same argument as in the proof of Theorem \ref{single-eff-opt-heter}, this implies that for each stage $k$:
\begin{align}\label{multi-cond-unbias}
    E[f^{(k)} | \bar{\mathbf{A}}^{(k-1)}, \bar{\mathbf{X}}^{(k)}] = V_0^{(k)}(\bar{\mathbf{A}}^{(k-1)}, \bar{\mathbf{X}}^{(k)}) \quad \text{a.s.}
\end{align}
for any joint distribution preserving the identifiability assumptions.

To see this, suppose Eq. \eqref{multi-cond-unbias} fails on a set of positive probability. Without loss of generality, assume:
\begin{align*}
    \mathcal{H}_> := \{(\bar{\mathbf{A}}^{(k-1)}, \bar{\mathbf{X}}^{(k)}): E[f^{(k)} | \bar{\mathbf{A}}^{(k-1)}, \bar{\mathbf{X}}^{(k)}] > V_0^{(k)}(\bar{\mathbf{A}}^{(k-1)}, \bar{\mathbf{X}}^{(k)})\}
\end{align*}
has positive probability. Then for the conditional distribution given $(\bar{\mathbf{A}}^{(k-1)}, \bar{\mathbf{X}}^{(k)}) \in \mathcal{H}_>$:
\begin{align*}
    E[f^{(k)} - V_0^{(k)}(\bar{\mathbf{A}}^{(k-1)}, \bar{\mathbf{X}}^{(k)}) | (\bar{\mathbf{A}}^{(k-1)}, \bar{\mathbf{X}}^{(k)}) \in \mathcal{H}_>] > 0
\end{align*}
which contradicts the requirement $E[f^{(k)}] = V_0^{(k)}$.

At stage $K$, we have $f^{(K+1)} = Y$. When the propensity score model is correctly specified, i.e., $\pi_1^{(K)}(\bar{\mathbf{A}}^{(K-1)}, \bar{\mathbf{X}}^{(K)}) = \pi_K((\bar{\mathbf{A}}^{(K-1)}, 1), \bar{\mathbf{X}}^{(K)})$, the conditional unbiasedness requirement becomes:
\begin{align}\label{multi-unbias-K}
    &\pi_K((\bar{\mathbf{A}}^{(K-1)}, 1), \bar{\mathbf{X}}^{(K)}) E_{Y|A^{(K)}=1, \bar{\mathbf{A}}^{(K-1)}, \bar{\mathbf{X}}^{(K)}}[f^{(K)}] \notag\\
    &+ \pi_K((\bar{\mathbf{A}}^{(K-1)}, 0), \bar{\mathbf{X}}^{(K)}) E_{Y|A^{(K)}=0, \bar{\mathbf{A}}^{(K-1)}, \bar{\mathbf{X}}^{(K)}}[f^{(K)}] \notag\\
    &= V_0^{(K)}(\bar{\mathbf{A}}^{(K-1)}, \bar{\mathbf{X}}^{(K)})
\end{align}
for any distribution of $Y | A^{(K)}, \bar{\mathbf{A}}^{(K-1)}, \bar{\mathbf{X}}^{(K)}$ consistent with the model.

Firstly, when $\tau_K(\bar{\mathbf{A}}^{(K-1)}, \allowbreak\bar{\mathbf{X}}^{(K)}) = 0$. In this case, $\mu_K((\bar{\mathbf{A}}^{(K-1)}, 1), \allowbreak\bar{\mathbf{X}}^{(K)}) \allowbreak= \mu_K((\bar{\mathbf{A}}^{(K-1)}, 0), \allowbreak\bar{\mathbf{X}}^{(K)}) \allowbreak=: \mu_0^{(K)*}$ and $V_0^{(K)}(\bar{\mathbf{A}}^{(K-1)}, \allowbreak\bar{\mathbf{X}}^{(K)}) \allowbreak= \mu_0^{(K)*}$.

Fix the propensity score $p_K := \pi_K((\bar{\mathbf{A}}^{(K-1)}, 1), \bar{\mathbf{X}}^{(K)})$ and the distribution of $Y | A^{(K)} = 0, \bar{\mathbf{A}}^{(K-1)}, \bar{\mathbf{X}}^{(K)}$. For any random variable $U_1$ satisfying $E[U_1] = \mu_0^{(K)*}$ and $\operatorname{Var}(U_1) = \widetilde{\sigma}_1^2$, Eq. \eqref{multi-unbias-K} gives:
\begin{align*}
    E_{U_1}[f^{(K)}(\ldots, U_1, \ldots)] = \frac{\mu_0^{(K)*} - (1-p_K) E_{Y|A^{(K)}=0}[f^{(K)}(\ldots, Y, \ldots)]}{p_K} \equiv \text{const}
\end{align*}

By Lemma \ref{lem: quad est}, $f^{(K)}$ is a quadratic polynomial in $Y$ when $A^{(K)} = 1$:
\begin{align}\label{multi-quad-K-1}
    f^{(K)}|_{A^{(K)}=1} = a_{1}^{(K)} Y^2 + b_{1}^{(K)} Y + c_{1}^{(K)}
\end{align}
where $a_{1}^{(K)}, b_{1}^{(K)}, c_{1}^{(K)}$ depend on $(p_K, \mu_1^{(K)}, \mu_0^{(K)}, r_1^{(K)}, r_0^{(K)})$.

Similarly, fixing $Y | A^{(K)} = 1$ and varying $Y | A^{(K)} = 0$, we obtain:
\begin{align}\label{multi-quad-K-0}
    f^{(K)}|_{A^{(K)}=0} = a_{0}^{(K)} Y^2 + b_{0}^{(K)} Y + c_{0}^{(K)}
\end{align}

Substituting Eqs. \eqref{multi-quad-K-1} and \eqref{multi-quad-K-0} into Eq. \eqref{multi-unbias-K} with $E[Y | A^{(K)}, \bar{\mathbf{A}}^{(K-1)}, \bar{\mathbf{X}}^{(K)}] = \mu_0^{(K)*}$ for both $A^{(K)} = 0, 1$:
\begin{align*}
    &p_K [a_{1}^{(K)} ((\mu_0^{(K)*})^2 + \widetilde{\sigma}_1^2) + b_{1}^{(K)} \mu_0^{(K)*} + c_{1}^{(K)}] \\
    &+ (1-p_K) [a_{0}^{(K)} ((\mu_0^{(K)*})^2 + \widetilde{\sigma}_0^2) + b_{0}^{(K)} \mu_0^{(K)*} + c_{0}^{(K)}] = \mu_0^{(K)*}
\end{align*}

Taking derivatives with respect to $\widetilde{\sigma}_1^2$ and $\widetilde{\sigma}_0^2$:
\begin{align*}
    a_{1}^{(K)} = 0, \quad a_{0}^{(K)} = 0
\end{align*}

Taking derivative with respect to $\mu_0^{(K)*}$:
\begin{align}\label{multi-b-constraint-K}
    p_K b_{1}^{(K)} + (1-p_K) b_{0}^{(K)} = 1
\end{align}

The remaining constraint gives:
\begin{align}\label{multi-c-constraint-K}
    p_K c_{1}^{(K)} + (1-p_K) c_{0}^{(K)} = 0
\end{align}

Therefore, when $\tau_K = 0$, the influence function has the form:
\begin{align}\label{multi-f-K-tau0}
    f^{(K)} = [A^{(K)} b_{1}^{(K)} + (1-A^{(K)}) b_{0}^{(K)}] Y + A^{(K)} c_{1}^{(K)} + (1-A^{(K)}) c_{0}^{(K)}
\end{align}

The conditional variance is:
\begin{align}
    &\operatorname{Var}(f^{(K)} | \bar{\mathbf{A}}^{(K-1)}, \bar{\mathbf{X}}^{(K)}) \notag\\
    &= E\left[[A^{(K)} b_{1}^{(K)} + (1-A^{(K)}) b_{0}^{(K)}]^2 \operatorname{Var}(Y | A^{(K)}, \bar{\mathbf{A}}^{(K-1)}, \bar{\mathbf{X}}^{(K)}) \Big| \bar{\mathbf{A}}^{(K-1)}, \bar{\mathbf{X}}^{(K)}\right] \notag\\
    &\quad + \operatorname{Var}\left([A^{(K)} b_{1}^{(K)} + (1-A^{(K)}) b_{0}^{(K)}] \mu_0^{(K)*} + A^{(K)} c_{1}^{(K)} + (1-A^{(K)}) c_{0}^{(K)} \Big| \bar{\mathbf{A}}^{(K-1)}, \bar{\mathbf{X}}^{(K)}\right) \notag\\
    &\geq p_K (b_{1}^{(K)})^2 \sigma_{1,K}^2 + (1-p_K) (b_{0}^{(K)})^2 \sigma_{0,K}^2 \label{multi-var-lb-K-tau0}
\end{align}
where $\sigma_{a,K}^2 = \operatorname{Var}(Y | A^{(K)} = a, \bar{\mathbf{A}}^{(K-1)}, \bar{\mathbf{X}}^{(K)})$ for $a = 0, 1$.

Subject to constraint \eqref{multi-b-constraint-K}, we minimize Eq. \eqref{multi-var-lb-K-tau0} by solving:
\begin{align*}
    \min_{b_{1}^{(K)}, b_{0}^{(K)}} \quad & p_K (b_{1}^{(K)})^2 \sigma_{1,K}^2 + (1-p_K) (b_{0}^{(K)})^2 \sigma_{0,K}^2 \\
    \text{s.t.} \quad & p_K b_{1}^{(K)} + (1-p_K) b_{0}^{(K)} = 1
\end{align*}

Using Lagrange multipliers, the optimal solution is:
\begin{align}
    b_{1}^{(K)*} &= \frac{\sigma_{0,K}^2}{p_K \sigma_{0,K}^2 + (1-p_K) \sigma_{1,K}^2} \label{multi-b1-K-opt}\\
    b_{0}^{(K)*} &= \frac{\sigma_{1,K}^2}{p_K \sigma_{0,K}^2 + (1-p_K) \sigma_{1,K}^2} \label{multi-b0-K-opt}
\end{align}

This is achieved by the influence function $\psi(\mathcal{D}^{(K)}, \pi, \mu)$ with:
\begin{align}\label{multi-D-K-tau0-opt}
    \mathcal{D}^{(K)}(\bar{\mathbf{A}}^{(K-1)}, \bar{\mathbf{X}}^{(K)}) = \frac{\sigma_{0,K}^2 p_K}{\sigma_{1,K}^2 (1-p_K) + \sigma_{0,K}^2 p_K}
\end{align}
when $\tau_K(\bar{\mathbf{A}}^{(K-1)}, \bar{\mathbf{X}}^{(K)}) = 0$.

Secondly, when $\tau_K(\bar{\mathbf{A}}^{(K-1)}, \allowbreak\bar{\mathbf{X}}^{(K)}) > 0$. In this case, $V_0^{(K)}(\bar{\mathbf{A}}^{(K-1)}, \allowbreak\bar{\mathbf{X}}^{(K)}) \allowbreak= \mu_K((\bar{\mathbf{A}}^{(K-1)}, 1), \allowbreak\bar{\mathbf{X}}^{(K)})$.

Fix the propensity score and the distribution of $Y | A^{(K)} = 0$. For any distribution of $Y | A^{(K)} = 1$ with $E[Y | A^{(K)} = 1] > E[Y | A^{(K)} = 0]$, Eq. \eqref{multi-unbias-K} must hold.

Rearranging:
\begin{align*}
    E_{Y|A^{(K)}=1}\left[f^{(K)} - \frac{Y}{p_K} + \frac{(1-p_K)}{p_K} E_{Y|A^{(K)}=0}[f^{(K)}]\right] = 0
\end{align*}

By Lemma \ref{lem: linear constraint}, since this holds for any $Y | A^{(K)} = 1$ with mean greater than $E[Y | A^{(K)} = 0]$:
\begin{align}\label{multi-f-K-1-tau+}
    f^{(K)}|_{A^{(K)}=1} = \frac{Y}{p_K} + c_{1,1}^{(K)}
\end{align}

Similarly, fixing $Y | A^{(K)} = 1$ and using Lemma \ref{lem: linear constraint neg}:
\begin{align}\label{multi-f-K-0-tau+}
    f^{(K)}|_{A^{(K)}=0} = c_{1,0}^{(K)}
\end{align}

Therefore:
\begin{align}\label{multi-f-K-tau+}
    f^{(K)} = \frac{A^{(K)} Y}{p_K} + c_1^{(K)}(A^{(K)})
\end{align}

The conditional variance is:
\begin{align}
    \operatorname{Var}(f^{(K)} | \bar{\mathbf{A}}^{(K-1)}, \bar{\mathbf{X}}^{(K)}) &= \operatorname{Var}\left(\frac{A^{(K)} Y}{p_K} + c_1^{(K)}(A^{(K)}) \Big| \bar{\mathbf{A}}^{(K-1)}, \bar{\mathbf{X}}^{(K)}\right) \notag\\
    &\geq \frac{\sigma_{1,K}^2}{p_K} \label{multi-var-lb-K-tau+}
\end{align}

The lower bound is achieved when $c_1^{(K)}(A^{(K)}) = (1 - A^{(K)}/p_K) \mu_K((\bar{\mathbf{A}}^{(K-1)}, 1), \bar{\mathbf{X}}^{(K)})$, corresponding to $\mathcal{D}^{(K)} = 1$.

Thirdly, when $\tau_K(\bar{\mathbf{A}}^{(K-1)}, \bar{\mathbf{X}}^{(K)}) < 0$. By symmetric arguments using Lemma \ref{lem: linear constraint neg}:
\begin{align}\label{multi-f-K-tau-}
    f^{(K)} = \frac{(1-A^{(K)}) Y}{1-p_K} + c_0^{(K)}(A^{(K)})
\end{align}

The variance lower bound is:
\begin{align}
    \operatorname{Var}(f^{(K)} | \bar{\mathbf{A}}^{(K-1)}, \bar{\mathbf{X}}^{(K)}) \geq \frac{\sigma_{0,K}^2}{1-p_K} \label{multi-var-lb-K-tau-}
\end{align}

achieved when $\mathcal{D}^{(K)} = 0$.

For the $K$-th stage, combining all cases, the optimal decision function at stage $K$ is:
\begin{align}
    \mathcal{D}^{(K)*}(\bar{\mathbf{A}}^{(K-1)}, \bar{\mathbf{X}}^{(K)}) &= \mathbb{I}\{\tau_K > 0\} + \mathbb{I}\{\tau_K = 0\} \cdot \frac{\sigma_{0,K}^2 p_K}{\sigma_{1,K}^2 (1-p_K) + \sigma_{0,K}^2 p_K}
\end{align}
which matches Eq. \eqref{K-multi-heter}.

Then, we conduct inductive derivation for the general $k$-th stage. Assume that for stages $k+1, k+2, \ldots, K$, the optimal decision functions $\mathcal{D}^{(k')*}$ have been determined and satisfy the forms in Eqs. \eqref{K-multi-heter} and \eqref{k-1-multi-heter}. We now analyze stage $k$.

At stage $k$, the "outcome" is $\widehat{V}^{(k+1)}(\mathcal{D}^*; \pi, \mu)$ where $\mathcal{D}^* = (\mathcal{D}^{(k+1)*}, \ldots, \mathcal{D}^{(K)*})$ are the optimal decision functions at subsequent stages.

Define:
\begin{align}
    \sigma_{a,k}^2 &:= \operatorname{Var}(\widehat{V}^{(k+1)}(\mathcal{D}^*; \pi, \mu) | (\bar{\mathbf{A}}^{(k-1)}, a), \bar{\mathbf{X}}^{(k)}), \quad a = 0, 1 \label{multi-sigma-k}\\
    p_k &:= \pi_k((\bar{\mathbf{A}}^{(k-1)}, 1), \bar{\mathbf{X}}^{(k)}) \label{multi-p-k}
\end{align}

The conditional unbiasedness at stage $k$ requires:
\begin{align}\label{multi-unbias-k}
    &p_k E_{\widehat{V}^{(k+1)}|A^{(k)}=1, \bar{\mathbf{A}}^{(k-1)}, \bar{\mathbf{X}}^{(k)}}[f^{(k)}] + (1-p_k) E_{\widehat{V}^{(k+1)}|A^{(k)}=0, \bar{\mathbf{A}}^{(k-1)}, \bar{\mathbf{X}}^{(k)}}[f^{(k)}] \notag\\
    &= V_0^{(k)}(\bar{\mathbf{A}}^{(k-1)}, \bar{\mathbf{X}}^{(k)})
\end{align}

First, when $\tau_k(\bar{\mathbf{A}}^{(k-1)}, \bar{\mathbf{X}}^{(k)}) = 0$. In this case:
\begin{align*}
    \mu_k((\bar{\mathbf{A}}^{(k-1)}, 1), \bar{\mathbf{X}}^{(k)}) = \mu_k((\bar{\mathbf{A}}^{(k-1)}, 0), \bar{\mathbf{X}}^{(k)}) =: \mu_0^{(k)*}
\end{align*}

Following the same argument as in first scenario of stage $K$, but with $\widehat{V}^{(k+1)}$ replacing $Y$, we apply Lemma \ref{lem: quad est} to conclude that $f^{(k)}$ is quadratic in $\widehat{V}^{(k+1)}$ for each value of $A^{(k)}$.

By the same derivative arguments:
\begin{align}
    f^{(k)} = [A^{(k)} b_{1}^{(k)} + (1-A^{(k)}) b_{0}^{(k)}] \widehat{V}^{(k+1)} + A^{(k)} c_{1}^{(k)} + (1-A^{(k)}) c_{0}^{(k)}
\end{align}
with constraint:
\begin{align}\label{multi-b-constraint-k}
    p_k b_{1}^{(k)} + (1-p_k) b_{0}^{(k)} = 1
\end{align}

The conditional variance lower bound is:
\begin{align}
    \operatorname{Var}(f^{(k)} | \bar{\mathbf{A}}^{(k-1)}, \bar{\mathbf{X}}^{(k)}) \geq p_k (b_{1}^{(k)})^2 \sigma_{1,k}^2 + (1-p_k) (b_{0}^{(k)})^2 \sigma_{0,k}^2
\end{align}

Minimizing subject to Eq. \eqref{multi-b-constraint-k}:
\begin{align}
    b_{1}^{(k)*} &= \frac{\sigma_{0,k}^2}{p_k \sigma_{0,k}^2 + (1-p_k) \sigma_{1,k}^2} \\
    b_{0}^{(k)*} &= \frac{\sigma_{1,k}^2}{p_k \sigma_{0,k}^2 + (1-p_k) \sigma_{1,k}^2}
\end{align}

The optimal decision function when $\tau_k = 0$ is:
\begin{align}\label{multi-D-k-tau0-opt}
    \mathcal{D}^{(k)*}|_{\tau_k = 0} = \frac{\sigma_{0,k}^2 p_k}{\sigma_{1,k}^2 (1-p_k) + \sigma_{0,k}^2 p_k}
\end{align}

Second, when $\tau_k(\bar{\mathbf{A}}^{(k-1)}, \bar{\mathbf{X}}^{(k)}) > 0$. In this case, $V_0^{(k)}(\bar{\mathbf{A}}^{(k-1)}, \bar{\mathbf{X}}^{(k)}) = \mu_k((\bar{\mathbf{A}}^{(k-1)}, 1), \bar{\mathbf{X}}^{(k)})$.

By Lemma \ref{lem: linear constraint}, applied to the distribution of $\widehat{V}^{(k+1)} | A^{(k)} = 1$:
\begin{align}
    f^{(k)}|_{A^{(k)}=1} = \frac{\widehat{V}^{(k+1)}}{p_k} + c_{1,1}^{(k)}
\end{align}

By Lemma \ref{lem: linear constraint neg}, applied to $\widehat{V}^{(k+1)} | A^{(k)} = 0$:
\begin{align}
    f^{(k)}|_{A^{(k)}=0} = c_{1,0}^{(k)}
\end{align}

The variance lower bound is:
\begin{align}
    \operatorname{Var}(f^{(k)} | \bar{\mathbf{A}}^{(k-1)}, \bar{\mathbf{X}}^{(k)}) \geq \frac{\sigma_{1,k}^2}{p_k}
\end{align}

achieved when $\mathcal{D}^{(k)} = 1$.

Third, when $\tau_k(\bar{\mathbf{A}}^{(k-1)}, \bar{\mathbf{X}}^{(k)}) < 0$. By symmetric arguments:
\begin{align}
    f^{(k)} = \frac{(1-A^{(k)}) \widehat{V}^{(k+1)}}{1-p_k} + c_0^{(k)}(A^{(k)})
\end{align}

The variance lower bound is:
\begin{align}
    \operatorname{Var}(f^{(k)} | \bar{\mathbf{A}}^{(k-1)}, \bar{\mathbf{X}}^{(k)}) \geq \frac{\sigma_{0,k}^2}{1-p_k}
\end{align}

achieved when $\mathcal{D}^{(k)} = 0$.

For stage $k$, combining all cases, the optimal decision function at stage $k$ is:
\begin{align}
    \mathcal{D}^{(k)*}(\bar{\mathbf{A}}^{(k-1)}, \bar{\mathbf{X}}^{(k)}) &= \mathbb{I}\{\tau_k > 0\} + \mathbb{I}\{\tau_k = 0\} \cdot \frac{\sigma_{0,k}^2 p_k}{\sigma_{1,k}^2 (1-p_k) + \sigma_{0,k}^2 p_k}
\end{align}
which matches Eq. \eqref{k-1-multi-heter}.

We now verify that $\widehat{V}_{\mathcal{D}^*}^{(1)*}$ with $\mathcal{D}^*$ defined by Eqs. \eqref{K-multi-heter} and \eqref{k-1-multi-heter} achieves the variance lower bound.

For each stage $k$, the influence function of $\widehat{V}_{\mathcal{D}^*}^{(k)*}$ is:
\begin{align}
    \psi^{(k)}(\mathcal{D}^*; \pi, \mu) &= \frac{\mathrm{g}\{A^{(k)}, \mathcal{D}^{(k)*}(\bar{\mathbf{A}}^{(k-1)}, \bar{\mathbf{X}}^{(k)})\}}{\pi_k(\bar{\mathbf{A}}^{(k)}, \bar{\mathbf{X}}^{(k)})} \notag\\
    &\quad \times \{\widehat{V}^{(k+1)}(\mathcal{D}^*; \pi, \mu) - \mu_k(\bar{\mathbf{A}}^{(k)}, \bar{\mathbf{X}}^{(k)})\} \notag\\
    &\quad + \mathcal{D}^{(k)*}(\bar{\mathbf{A}}^{(k-1)}, \bar{\mathbf{X}}^{(k)}) \mu_k((\bar{\mathbf{A}}^{(k-1)}, 1), \bar{\mathbf{X}}^{(k)}) \notag\\
    &\quad + (1-\mathcal{D}^{(k)*}(\bar{\mathbf{A}}^{(k-1)}, \bar{\mathbf{X}}^{(k)})) \mu_k((\bar{\mathbf{A}}^{(k-1)}, 0), \bar{\mathbf{X}}^{(k)})
\end{align}
When $\tau_k = 0$ and $\mathcal{D}^{(k)*}$ is given by Eq. \eqref{multi-D-k-tau0-opt}:
\begin{align}
    b_1^{(k)} &= \frac{\mathcal{D}^{(k)*}}{p_k} = \frac{\sigma_{0,k}^2}{p_k \sigma_{0,k}^2 + (1-p_k) \sigma_{1,k}^2} \\
    b_0^{(k)} &= \frac{1-\mathcal{D}^{(k)*}}{1-p_k} = \frac{\sigma_{1,k}^2}{p_k \sigma_{0,k}^2 + (1-p_k) \sigma_{1,k}^2}
\end{align}

These match the optimal values $b_1^{(k)*}$ and $b_0^{(k)*}$, confirming that the variance lower bound is achieved.

Furthermore, we verify that the first term in the variance (coming from the variation of $E[\psi^{(k)} | \bar{\mathbf{A}}^{(k-1)}, \bar{\mathbf{X}}^{(k)}]$) vanishes:
\begin{align}
    &E[\psi^{(k)}(\mathcal{D}^*; \pi, \mu) | \bar{\mathbf{A}}^{(k-1)}, \bar{\mathbf{X}}^{(k)}] \notag\\
    &= \mathcal{D}^{(k)*} \mu_k((\bar{\mathbf{A}}^{(k-1)}, 1), \bar{\mathbf{X}}^{(k)}) + (1-\mathcal{D}^{(k)*}) \mu_k((\bar{\mathbf{A}}^{(k-1)}, 0), \bar{\mathbf{X}}^{(k)}) \notag\\
    &= \mu_0^{(k)*} \quad (\text{since } \tau_k = 0)
\end{align}

This is a constant given $(\bar{\mathbf{A}}^{(k-1)}, \bar{\mathbf{X}}^{(k)})$, so:
\begin{align}
    \operatorname{Var}(E[\psi^{(k)} | \bar{\mathbf{A}}^{(k-1)}, \bar{\mathbf{X}}^{(k)}, A^{(k)}] | \bar{\mathbf{A}}^{(k-1)}, \bar{\mathbf{X}}^{(k)}) = 0
\end{align}
When $\tau_k > 0$ and $\mathcal{D}^{(k)*} = 1$:
\begin{align}
    \psi^{(k)} &= \frac{A^{(k)}}{p_k} (\widehat{V}^{(k+1)} - \mu_k((\bar{\mathbf{A}}^{(k-1)}, 1), \bar{\mathbf{X}}^{(k)})) + \mu_k((\bar{\mathbf{A}}^{(k-1)}, 1), \bar{\mathbf{X}}^{(k)})
\end{align}

The conditional variance is:
\begin{align}
    \operatorname{Var}(\psi^{(k)} | \bar{\mathbf{A}}^{(k-1)}, \bar{\mathbf{X}}^{(k)}) = \frac{\sigma_{1,k}^2}{p_k}
\end{align}
which matches the lower bound in the second scenario of the $k$-th stage.
By symmetric arguments, when $\tau_k < 0$ and $\mathcal{D}^{(k)*} = 0$:
\begin{align}
    \operatorname{Var}(\psi^{(k)} | \bar{\mathbf{A}}^{(k-1)}, \bar{\mathbf{X}}^{(k)}) = \frac{\sigma_{0,k}^2}{1-p_k}
\end{align}
which matches the lower bound in the third scenario of the $k$-th stage.

By backward induction from stage $K$ to stage $1$, we have shown that for each stage $k$, the optimal decision function $\mathcal{D}^{(k)*}$ defined in Eqs. \eqref{K-multi-heter} and \eqref{k-1-multi-heter} achieves the optimal efficiency among all influence functions in the definition of $\mathcal{T}_{RL}^{multi}$. Then, by the asymptotic uniform integrability, the asymptotical variance of the $\sqrt{n}$ scaled asymptotically linear estimator is the same as that of the corresponding influence function, which leads to
\begin{align}
    \inf_{T_n \in \mathcal{T}_{RL}^{multi}} \lim_{n \to \infty} \operatorname{Var}(\sqrt{n} T_n) = \lim_{n \to \infty} \operatorname{Var}(\sqrt{n} \widehat{V}_{\mathcal{D}^*}^{(1)*})
\end{align}

This completes the proof.
\end{proof}
Then, in terms of the data driven implementation of achieving the above stated optimal efficiency, under the same choice of tuning parameter and the working models, we replace the $t_0^k$ in $\widehat{V}_s^{(1)}$ with any consistent estimator of 
\begin{align*}
     &\left(\operatorname{Var}(\widehat{V}^{(k+1)}(\mathcal{D};\pi,\mu)|\bar{\mathbf{A}}^{(k-1)},A^{(k)}=1,\bar{\mathbf{X}}^{(k)})\pi_k(\bar{\mathbf{A}}^{(k-1)},A^{(k)}=0,\bar{\mathbf{X}}^{(k)})\right.\\
   &+\left.\operatorname{Var}(\widehat{V}^{(k+1)}(\mathcal{D};\pi,\mu)|\bar{\mathbf{A}}^{(k-1)},A^{(k)}=0,\bar{\mathbf{X}}^{(k)})\pi_k(\bar{\mathbf{A}}^{(k-1)},A^{(k)}=1,\bar{\mathbf{X}}^{(k)})\right)^{-1}\\
   &\cdot \operatorname{Var}(\widehat{V}^{(k+1)}(\mathcal{D};\pi,\mu)|\bar{\mathbf{A}}^{(k-1)},A^{(k)}=0,\bar{\mathbf{X}}^{(k)})\pi_k(\bar{\mathbf{A}}^{(k-1)},A^{(k)}=1,\bar{\mathbf{X}}^{(k)}).
\end{align*}
Denote the resulting estimator as $\widehat{V}_{as,heter}^{(1)}$, and the working model for estimating $\operatorname{Var}(\widehat{V}^{(k+1)}(\mathcal{D};\allowbreak\pi,\allowbreak\mu)\allowbreak|\bar{\mathbf{A}}^{(k-1)},\allowbreak A^{(k)},\allowbreak\bar{\mathbf{X}}^{(k)})$ as $\widehat{\operatorname{Var}}(\widehat{V}^{(k+1)}(\mathcal{D};\allowbreak\pi,\allowbreak\mu)\allowbreak|\bar{\mathbf{A}}^{(k-1)},\allowbreak A^{(k)},\allowbreak\bar{\mathbf{X}}^{(k)})$. The theoretical property of $\widehat{V}_{as,heter}^{(1)}$ can be summarized in Theorem \ref{thm: multi-opt-est-heter} below where similar to single single time point case, the conditional variance estimator only needs to be consistent for efficiency optimality.
\begin{theorem}\label{thm: multi-opt-est-heter}
     Under Assumptions \ref{B1} to \ref{B7}, and if the working estimator for conditional variance satisfies
    \begin{align}
        \left|\widehat{\operatorname{Var}}(\widehat{V}^{(k+1)}(\mathcal{D};\pi,\mu)|\bar{\mathbf{A}}^{(k-1)},A^{(k)},\bar{\mathbf{X}}^{(k)}) - \operatorname{Var}(\widehat{V}^{(k+1)}(\mathcal{D};\pi,\mu)|\bar{\mathbf{A}}^{(k-1)},A^{(k)},\bar{\mathbf{X}}^{(k)})\right|\to_p0
    \end{align}
   almost surely for $\bar{\mathbf{X}}^{(k)}$ and $\bar{\mathbf{A}}^{(k)}$, we have
    \begin{align}
        \widehat{V}_{as,heter}^{(1)} - \widehat{V}_{\mathcal{D}}^{(1)*}=o_p(n^{-1/2})
    \end{align}
  where $\mathcal{D}$ is defined as in Eq. \eqref{K-multi-heter} and  Eq. \eqref{k-1-multi-heter}.
\end{theorem}
\begin{proof}[Proof of Theorem \ref{thm: multi-opt-est-heter}]
This result follows from Theorem \ref{thmb1} and Proposition \ref{multi-opt-oracle} by showing that the adaptive smoothing estimator is asymptotically equivalent to its corresponding oracle estimator.

Firstly, we decompose the estimator $\widehat{V}_{as,heter}^{(1)}$ as:
\begin{align}
    \widehat{V}_{as,heter}^{(1)} &= \frac{1}{4}\sum_{j=1,2}\sum_{l=1,2}\frac{1}{|\mathcal{I}_{j,l}|}\sum_{i\in\mathcal{I}_{j,l}}\widehat{V}_i^{(1)}(d_{s,heter}^{\star}(\mathbf{O}_i,\mathcal{I}_{3-j}),\widehat{\pi}_{\mathcal{I}_{j,l}^c},\widehat{\mu}_{\mathcal{I}_{j,l}^c})
\end{align}
where $d_{s,heter}^{\star}(\mathbf{O}_i,\mathcal{I}_{3-j})$ is the decision function with stage-wise components:
\begin{align}
    d_{s,heter}^{1\star}(\bar{\mathbf{X}}_i^{(1)},\mathcal{I}_{3-j}) &= d_{s}^1(\bar{\mathbf{X}}_i^{(1)};\widehat{\tau}_{1,\mathcal{I}_{3-j}},h_{n,\mathcal{I}_{3-j}}^1, t_{0,\mathcal{I}_{3-j}}^1)\label{heter-multi-as-1}\\
    d_{s,heter}^{k\star}(\bar{\mathbf{A}}^{(k-1)},\bar{\mathbf{X}}^{(k)},\mathcal{I}_{3-j}) &= d_{s}^k(\bar{\mathbf{A}}^{(k-1)},\bar{\mathbf{X}}^{(k)};\widehat{\tau}_{k,\mathcal{I}_{3-j}},h_{n,\mathcal{I}_{3-j}}^k,t_{0,\mathcal{I}_{3-j}}^k)\label{heter-multi-as-k}
\end{align}
where
\begin{align*}
   & t_{0,\mathcal{I}_{3-j}}^k:=\left(\widehat{\operatorname{Var}}_{\mathcal{I}_{3-j}}(\widehat{V}^{(k+1)}(\mathcal{D};\pi,\mu)|\bar{\mathbf{A}}^{(k-1)},A^{(k)}=1,\bar{\mathbf{X}}^{(k)})\widehat{\pi}_{k,\mathcal{I}_{3-j}}(\bar{\mathbf{A}}^{(k-1)},A^{(k)}=0,\bar{\mathbf{X}}^{(k)})\right.\\
   &+\left.\widehat{\operatorname{Var}}_{\mathcal{I}_{3-j}}(\widehat{V}^{(k+1)}(\mathcal{D};\pi,\mu)|\bar{\mathbf{A}}^{(k-1)},A^{(k)}=0,\bar{\mathbf{X}}^{(k)})\widehat{\pi}_{k,\mathcal{I}_{3-j}}(\bar{\mathbf{A}}^{(k-1)},A^{(k)}=1,\bar{\mathbf{X}}^{(k)})\right)^{-1}\\
   &\cdot \widehat{\operatorname{Var}}_{\mathcal{I}_{3-j}}(\widehat{V}^{(k+1)}(\mathcal{D};\pi,\mu)|\bar{\mathbf{A}}^{(k-1)},A^{(k)}=0,\bar{\mathbf{X}}^{(k)})\widehat{\pi}_{k,\mathcal{I}_{3-j}}(\bar{\mathbf{A}}^{(k-1)},A^{(k)}=1,\bar{\mathbf{X}}^{(k)}).
\end{align*}

We need to show:
\begin{align}
    \widehat{V}_{as,heter}^{(1)} - \widehat{V}_{\mathcal{D}}^{(1)*} = o_p(n^{-1/2})
\end{align}
 where $\mathcal{D}$ is defined as in Eq. \eqref{K-multi-heter} and  Eq. \eqref{k-1-multi-heter}.


Then we proceed to verify the technical conditions required in Lemma \ref{multi-lem}.

{
For each fixed fold $j$, let
\begin{align*}
\mathcal{D}_n^{(1)}(\bar{\mathbf{X}}^{(1)},s)
&=d_s^1\!\left(\bar{\mathbf{X}}^{(1)};s,h_{n,\mathcal I_{3-j}}^1,t_{0,\mathcal I_{3-j}}^1\right),\\
\mathcal{D}_n^{(k)}(\bar{\mathbf{A}}^{(k-1)},\bar{\mathbf{X}}^{(k)},s)
&=d_s^k\!\left(\bar{\mathbf{A}}^{(k-1)},\bar{\mathbf{X}}^{(k)};s,h_{n,\mathcal I_{3-j}}^k,t_{0,\mathcal I_{3-j}}^k\right),\quad k=2,\dots,K.
\end{align*}
Thus, the rules in Eqs.~\eqref{heter-multi-as-1} and~\eqref{heter-multi-as-k} are obtained by evaluating $\mathcal D_n^{(k)}$ at $\widehat\tau_{k,\mathcal I_{3-j}}$.
}

We verify Assumptions \ref{B10} to \ref{B12}:

\textbf{Assumption \ref{B10} (Uniform continuity):}
For each stage $k$ {and fixed fold $j$}, let:
\begin{align}
    {\delta_{n,j}^k} &= \sqrt{h_{n,\mathcal{I}_{3-j}}^k \cdot n^{1/4}\left(E[|\widehat{\tau}_{k,\mathcal{I}_{3-j}}(\bar{\mathbf{A}}^{(k-1)},\bar{\mathbf{X}}^{(k)}) - \tau_k(\bar{\mathbf{A}}^{(k-1)},\bar{\mathbf{X}}^{(k)})|^2]\right)^{3/4}}
\end{align}

By Assumptions \ref{B5} and the consistency of $\widehat{\tau}_{k,\mathcal{I}_{3-j}}$:
\begin{align}
    P(|\widehat{\tau}_{k,\mathcal{I}_{3-j}}(\bar{\mathbf{A}}^{(k-1)},\bar{\mathbf{X}}^{(k)}) - \tau_k(\bar{\mathbf{A}}^{(k-1)},\bar{\mathbf{X}}^{(k)})| \geq {\delta_{n,j}^k}) = o_p(1)
\end{align}

The uniform continuity of {$\mathcal{D}_n^{(k)}$} follows from its {piecewise} linear structure.

\textbf{Assumption \ref{B11} (Consistency):}
We have:
\begin{align}
    \lim_{n \to \infty} {\mathcal{D}_n^{(k)}}(\bar{\mathbf{A}}^{(k-1)},\bar{\mathbf{X}}^{(k)},\tau_k(\bar{\mathbf{A}}^{(k-1)},\bar{\mathbf{X}}^{(k)})) &= \mathcal{D}^{(k)}(\bar{\mathbf{A}}^{(k-1)},\bar{\mathbf{X}}^{(k)})
\end{align}
 where $\mathcal{D}$ is defined as in Eq. \eqref{K-multi-heter} and  Eq. \eqref{k-1-multi-heter}.
 
This follows because:
(i) When $\tau_k(\bar{\mathbf{A}}^{(k-1)},\bar{\mathbf{X}}^{(k)}) > 0$: For large $n$, $h_n^k \to 0$, so {$\mathcal{D}_n^{(k)}(\cdot,\tau_k) \to 1$};
(ii) When $\tau_k(\bar{\mathbf{A}}^{(k-1)},\bar{\mathbf{X}}^{(k)}) < 0$: For large $n$, $h_n^k \to 0$, so {$\mathcal{D}_n^{(k)}(\cdot,\tau_k) \to 0$};  
(iii) When $\tau_k(\bar{\mathbf{A}}^{(k-1)},\bar{\mathbf{X}}^{(k)}) = 0$:
\begin{align*}
   & {\mathcal{D}_n^{(k)}(\cdot,\tau_k)} \to t_{0,\mathcal{I}_{3-j}}^k\to_p\\
     &\left(\operatorname{Var}(\widehat{V}^{(k+1)}(\mathcal{D};\pi,\mu)|\bar{\mathbf{A}}^{(k-1)},A^{(k)}=1,\bar{\mathbf{X}}^{(k)})\pi_k(\bar{\mathbf{A}}^{(k-1)},A^{(k)}=0,\bar{\mathbf{X}}^{(k)})\right.\\
   &+\left.\operatorname{Var}(\widehat{V}^{(k+1)}(\mathcal{D};\pi,\mu)|\bar{\mathbf{A}}^{(k-1)},A^{(k)}=0,\bar{\mathbf{X}}^{(k)})\pi_k(\bar{\mathbf{A}}^{(k-1)},A^{(k)}=1,\bar{\mathbf{X}}^{(k)})\right)^{-1}\\
   &\cdot \operatorname{Var}(\widehat{V}^{(k+1)}(\mathcal{D};\pi,\mu)|\bar{\mathbf{A}}^{(k-1)},A^{(k)}=0,\bar{\mathbf{X}}^{(k)})\pi_k(\bar{\mathbf{A}}^{(k-1)},A^{(k)}=1,\bar{\mathbf{X}}^{(k)}).
\end{align*}

\textbf{Assumption \ref{B12} (Concentration):}
Let ${\delta_{n,j}^{\prime k}} = 2h_{n,\mathcal{I}_{3-j}}^k$. {By Assumptions \ref{B5} and \ref{B6},
\begin{align*}
\delta_{n,j}^k\lor n^{1/4}\left(E[|\widehat\tau_{k,\mathcal I_{3-j}}-\tau_k|^2]\right)^{3/4}
\prec_p\delta_{n,j}^{\prime k}\prec n^{-\frac{1}{2(1+\alpha)}}.
\end{align*}}
When $|s| > {\delta_{n,j}^{\prime k}}$, by the construction of {$\mathcal{D}_n^{(k)}$}:
\begin{align}
    |{\mathcal{D}_n^{(k)}}(\bar{\mathbf{A}}^{(k-1)},\bar{\mathbf{X}}^{(k)},s) - {\mathbb{I}\{s>0\}}| &{=0}= o\left(\frac{1}{|s|\sqrt{n}}\right)
\end{align}
This follows from the linear structure of the smoothing function and Assumption \ref{B6}.


By Lemma \ref{multi-lem} (multi-stage technical lemma), under Assumptions \ref{B1} to \ref{B4}, \ref{B7} and \ref{B10} to \ref{B12}:
\begin{align}
    \left|\widehat{V}_{as}^{(1)} - \widehat{V}_{\mathcal{D}}^{(1)*}\right| = o_p(n^{-1/2})
\end{align}
where $\mathcal{D}$ is defined as in Eq. \eqref{K-multi-heter} and  Eq. \eqref{k-1-multi-heter}.
\end{proof}
We further provide additional simulations in the multiple time point case under heteroscedasticity and $\alpha < 1$ nonregularity. We consider three settings (VIII)--(X) that extend the two-stage model in Section~B.4 of the supplementary material. All three settings share the data generating model
\begin{align*}
    Y=\Phi\left(X_{1}^{(1)}, X_{2}^{(1)}, A^{(1)}, X^{(2)}\right)+A^{(2)} \tau_2\left(X_{1}^{(1)}, X_{2}^{(1)}, A^{(1)}, X^{(2)}\right)+e^{(2)}, \\
X^{(2)}=A^{(1)} X_{1}^{(1)}+e^{(1)},
\end{align*}
where $X_1^{(1)}, X_2^{(1)} \sim \operatorname{Unif}[-2,2]$ independently, $e^{(1)} \sim N(0, 0.5^2)$, and $e^{(2)}$ has heteroscedastic variance depending on the treatment and covariate history. All settings feature $P(\tau_k = 0) > 0$ at the relevant stages and $\alpha < 1$, so that nonregularity is present with potentially unbounded density of the contrast function near zero. The functional forms and key features are summarized in Table~\ref{multi-heter-setting}.

In Setting~(VIII), the stage-2 contrast is $\tau_2 = A^{(1)} (X^{(2)})^2$ with $\alpha_2 = 1/2$, while the stage-1 contrast is degenerate ($\tau_1 = 0$ everywhere, so $P(\tau_1 = 0) = 1$). Heteroscedasticity is introduced through the conditional variance of $Y$ given $A^{(2)} = 1$: patients with $A^{(1)} = 0$ (where $\tau_2 = 0$) have substantially higher variance than those with $A^{(1)} = 1$ (where $\tau_2 > 0$). In Setting~(IX), both stages exhibit nonregularity with $\tau_2 = \mathbb{I}(X_1^{(1)} > 0) \cdot (X^{(2)})^2$ ($\alpha_2 = 1/2$) and derived stage-1 contrast $\tau_1 = \mathbb{I}(X_1^{(1)} > 0) \cdot (X_1^{(1)})^2$ ($\alpha_1 = 1/2$), giving $P(\tau_k = 0) = 0.5$ at both stages. Heteroscedasticity arises from higher variance in the $\tau_2 = 0$ region (i.e., $X_1^{(1)} \leq 0$). Setting~(X) features a different margin exponent with $\tau_2 = \mathbb{I}(X_1^{(1)} > 0) \cdot |X^{(2)}|^{3/2}$ ($\alpha_2 = 2/3$) and derived stage-1 contrast with $\alpha_1 \approx 1/2$, with heteroscedastic variance depending on both the $X_1^{(1)}$ indicator and $(X^{(2)})^2$.

For each setting, we let the sample size $n \in \{500, 1000, 2000\}$ and compare four methods: Proposed~(Heter), which uses the efficiency-optimal asymmetric tuning under heteroscedasticity at both stages; Proposed~(Homo), which uses $t_0^k = \widehat{\pi}_k$ (the propensity-score-based tuning that is optimal under homoscedasticity); Subbagging \citep{shi2020breaking}; and SSS (simple sample split). We set ${C_0} = 0.05$ for the tuning selection and subsample size $K_0 = 5$ for subbagging. All results are aggregated over 500 Monte Carlo replications.

At each stage $k$, the conditional mean function $\mu_k(\bar{\mathbf{A}}^{(k)}, \bar{\mathbf{X}}^{(k)})$ is estimated by stratifying the data by the treatment history $\bar{\mathbf{A}}^{(k)}$ and fitting OLS regression within each stratum using correctly specified basis functions of the covariates (e.g., $(X^{(2)})^2$ for the stage-2 model in Setting~(VIII) when $A^{(1)}=1$, $A^{(2)}=1$). The stage-2 contrast function is estimated as $\widehat{\tau}_2 = \widehat{\mu}_2((\bar{\mathbf{A}}^{(1)},1),\bar{\mathbf{X}}^{(2)}) - \widehat{\mu}_2((\bar{\mathbf{A}}^{(1)},0),\bar{\mathbf{X}}^{(2)})$. The propensity scores are estimated by OLS regression of $A^{(k)}$ on $X_{1,1}$ when the true propensity depends on it, or as the sample proportion when the true propensity is constant, and are truncated to $[0.05,0.95]$. For the Proposed~(Heter) method, the conditional variance at the final stage, $\operatorname{Var}(Y\mid \bar{\mathbf{A}}^{(K)}, \bar{\mathbf{X}}^{(K)})$, is estimated by regressing squared residuals on the relevant covariates within each $(\bar{\mathbf{A}}^{(K)})$ stratum (e.g., on $\mathbb{I}(X_{1,1}\leq 0)$ in Setting~(IX), and on both $\mathbb{I}(X_{1,1}\leq 0)$ and $(X^{(2)})^2$ in Setting~(X)), with a lower truncation of $0.05$. The bandwidth at each stage is selected by Algorithm~5 with ${C_0}=0.05$. Both proposed methods are implemented via Algorithm~1 applied stage-by-stage with single cross-fitting, and the results are aggregated over 500 Monte Carlo replications.

The results in Table~\ref{multi-heter-sim} reveal several findings that parallel those in the single time point case. First, across all three settings and all sample sizes, Proposed~(Heter) achieves the shortest confidence intervals while maintaining valid coverage around 93\%--96\%, confirming that the efficiency gains from incorporating heteroscedastic variance structure into the asymmetric tuning extend naturally to the multi-stage setting. Second, the efficiency gain from Proposed~(Heter) over Proposed~(Homo) is substantial in all three settings: at $n = 2000$, the average length reductions are approximately 17\% in Setting~(VIII), 5\% in Setting~(IX), and 9\% in Setting~(X), reflecting the varying degrees of heteroscedasticity. The largest gains appear in Setting~(VIII), where the variance ratio between the $A^{(1)} = 0$ and $A^{(1)} = 1$ subgroups is the most pronounced. Third, Proposed~(Heter) consistently outperforms subbagging, with average length reductions of approximately 21\% in Setting~(VIII), 9\% in Setting~(IX), and 15\% in Setting~(X) at $n = 2000$. Subbagging also exhibits mild undercoverage in some cases (e.g., 91.2\% in Setting~(VIII) at $n = 500$ and 91.8\% in Setting~(X) at $n = 500$), whereas both proposed methods maintain coverage closer to the nominal level. SSS achieves nominal coverage throughout but suffers from considerably wider confidence intervals due to sample splitting, with average lengths roughly 1.7 to 2.2 times those of Proposed~(Heter). These multi-stage results confirm the practical relevance of heteroscedastic adaptive smoothing in the dynamic treatment regime setting.

\begin{table}[H]
    \centering
    \scriptsize
    \caption{Additional multiple time point simulation settings. All settings are nonregular with $P(\tau_k = 0) > 0$ at the relevant stages and feature $\alpha < 1$ and heteroscedasticity. Here $\Delta(x_{1,1}) := E[|x_{1,1} + e^{(1)}|^{3/2}] - E[|e^{(1)}|^{3/2}]$.}
    \resizebox{\textwidth}{!}{
    \begin{tabular}{cccc}
        \hline \hline
        & (VIII) & (IX) & (X) \\
        \hline
        $\tau_2(x_{1,1}, x_{1,2}, a_1, x_2)$ & $a_1 x_2^2$ & $\mathbb{I}(x_{1,1}>0)\cdot x_2^2$ & $\mathbb{I}(x_{1,1}>0)\cdot |x_2|^{3/2}$ \\
        \hline
        Stage-2 $\alpha$ & $1/2$ & $1/2$ & $2/3$ \\
        \hline
        $\tau_1$ (derived) & $0$ (degenerate) & $\mathbb{I}(x_{1,1}>0)\cdot x_{1,1}^2$ & $\mathbb{I}(x_{1,1}>0)\cdot\Delta(x_{1,1})$ \\
        \hline
        Stage-1 $\alpha$ & degenerate & $1/2$ & $\approx 1/2$ \\
        \hline
        $\Phi(x_{1,1}, x_{1,2}, a_1, x_2)$ & $x_{1,1}^2 - a_1(0.25 + x_{1,1}^2)$ & $x_{1,1}^2/2$ & $0$ \\
        \hline
        $\pi_1(x_{1,1}, x_{1,2})$ & $0.5 + 0.1\, x_{1,1}$ & $0.5$ & $0.5 + 0.1\, x_{1,1}$ \\
        \hline
        $\pi_2(x_{1,1}, x_{1,2})$ & $0.4$ & $0.5 + 0.1\, x_{1,1}$ & $0.5$ \\
        \hline
        $\operatorname{Var}(Y \mid X, A^{(2)}=1)$ & $0.1 + 1.5(1-A^{(1)})$ & $0.1+1.2\cdot\mathbb{I}(X_{1,1}\leq 0)$ & $0.1+1.0\cdot\mathbb{I}(X_{1,1}\leq 0)+0.3(X^{(2)})^2$ \\
        \hline
        $\operatorname{Var}(Y \mid X,A^{(2)}=0)$ & $0.1$ & $0.1$ & $0.1$ \\
        \hline
        $V_0$ & $1.33$ & $\approx 1.82$ & $\approx 1.30$ \\
        \hline
    \end{tabular}
    }
    \label{multi-heter-setting}
\end{table}

\begin{table}[H]
    \centering
    \scriptsize
    \caption{ECP and AL of the CIs for additional multiple time point settings with standard error smaller than 0.01.}
    \begin{tabular}{|c|c|c|c|c|c|c|c|c|c|}
        \hline
        \multirow{2}{*}{Setting} & \multirow{2}{*}{$n$} & \multicolumn{2}{c|}{Proposed (Heter)} & \multicolumn{2}{c|}{Proposed (Homo)} & \multicolumn{2}{c|}{Subbagging} & \multicolumn{2}{c|}{SSS} \\
        \cline{3-10}
        & & ECP(\%) & AL*100 & ECP(\%) & AL*100 & ECP(\%) & AL*100 & ECP(\%) & AL*100 \\
        \hline
        \multirow{3}{*}{(VIII)}
        & 500  & $94.0$ & $30.5$ & $93.8$ & $35.5$ & $91.2$ & $35.9$ & $95.0$ & $61.7$ \\
        & 1000 & $95.8$ & $20.4$ & $94.6$ & $24.1$ & $91.6$ & $24.7$ & $95.4$ & $42.8$ \\
        & 2000 & $93.2$ & $13.7$ & $93.0$ & $16.5$ & $92.4$ & $17.5$ & $93.6$ & $30.2$ \\
        \hline
        \multirow{3}{*}{(IX)}
        & 500  & $94.8$ & $36.6$ & $95.0$ & $38.8$ & $92.6$ & $40.0$ & $96.0$ & $61.0$ \\
        & 1000 & $93.2$ & $25.5$ & $94.0$ & $27.0$ & $96.0$ & $28.1$ & $94.2$ & $43.0$ \\
        & 2000 & $94.0$ & $18.0$ & $94.6$ & $19.0$ & $94.8$ & $19.8$ & $96.4$ & $30.1$ \\
        \hline
        \multirow{3}{*}{(X)}
        & 500  & $94.4$ & $26.5$ & $95.6$ & $29.6$ & $91.8$ & $31.4$ & $95.4$ & $49.5$ \\
        & 1000 & $94.6$ & $18.5$ & $93.6$ & $20.5$ & $94.0$ & $22.0$ & $94.0$ & $35.0$ \\
        & 2000 & $94.0$ & $13.1$ & $93.8$ & $14.4$ & $93.6$ & $15.3$ & $94.6$ & $24.7$ \\
        \hline
    \end{tabular}
    \label{multi-heter-sim}
\end{table}

\end{document}